\documentclass[traditabstract]{aa}  

\usepackage{graphicx}
\usepackage{natbib}
\usepackage{lscape}
\usepackage{supertabular}
\usepackage{xcolor}

\usepackage{txfonts}
\begin{document}

\title{The CARMENES search for exoplanets around M dwarfs}
\subtitle{Period search in H$\alpha$, \ion{Na}{i} D, and \ion{Ca}{ii} IRT lines}

\author{B. Fuhrmeister\inst{1}, S. Czesla\inst{1}, J. H. M. M. Schmitt\inst{1}
  \and  E.~N.~Johnson\inst{2}
  \and  P.~Sch\"ofer\inst{2}
  \and  S.~V.~Jeffers\inst{2}
  \and  J.~A.~Caballero\inst{3}
  \and M.~Zechmeister\inst{2}
  \and   A.~Reiners\inst{2}
  \and   I.~Ribas\inst{4,5}
  \and P.~J.~Amado\inst{6}
  \and  A.~Quirrenbach\inst{7}
  \and  F.~Bauer\inst{2}
  \and  V.~J.~S.~B\'ejar\inst{8,9}
  \and  M.~Cort\'es-Contreras\inst{3}
  \and  E.~D\'iez Alonso\inst{10,11}
  \and  S.~Dreizler\inst{2}
  \and  D.~Galad\'{\i}-Enr\'{\i}quez\inst{12}
  \and  E.~W.~Guenther\inst{13,8}
  \and  A. Kaminski\inst{7}
  \and  M.~K\"urster\inst{14}
  \and  M.~Lafarga\inst{4,5}
  \and  D.~Montes\inst{10}}

\institute{Hamburger Sternwarte, Universit\"at Hamburg, Gojenbergsweg 112, 21029 Hamburg, Germany\\
  \email{bfuhrmeister@hs.uni-hamburg.de}%01
        \and
        Institut f\"ur Astrophysik, Friedrich-Hund-Platz 1, D-37077 G\"ottingen, Germany %10
        \and
        Centro de Astrobiolog\'{\i}a (CSIC-INTA), ESAC, Camino Bajo del Castillo s/n, E-28692 Villanueva de la Ca\~nada, Madrid, Spain %04
        \and
        Institut de Ci\`encies de l'Espai (ICE, CSIC), Campus UAB, c/ de Can Magrans s/n, E-08193 Bellaterra, Barcelona, Spain
        \and
        Institut d'Estudis Espacials de Catalunya (IEEC), E-08034 Barcelona, Spain
           \and 
        Instituto de Astrof\'isica de Andaluc\'ia (CSIC), Glorieta de la Astronom\'ia s/n, E-18008 Granada, Spain %02
        \and 
        Landessternwarte, Zentrum f\"ur Astronomie der Universit\"at Heidelberg, K\"onigstuhl 12, D-69117 Heidelberg, Germany %09 
        \and
        Instituto de Astrof\'{\i}sica de Canarias, c/ V\'{\i}a L\'actea s/n, E-38205 La Laguna, Tenerife, Spain
        \and
        Departamento de Astrof\'{\i}sica, Universidad de La Laguna, E-38206 Tenerife, Spain %03}
        \and
        Departamento de F\'{i}sica de la Tierra y Astrof\'{i}sica 
        and UPARCOS-UCM (Unidad de F\'{i}sica de Part\'{i}culas y del Cosmos de la
        UCM), Facultad de Ciencias F\'{\i}sicas, Universidad Complutense de Madrid, E-28040 Madrid, Spain %08
        \and
        Departamento de Explotaci\'on y Prospecc\'on de Minas, Escuela de Minas, Energ\'{\i}a y Materiales, Universidad de Oviedo, E-33003 Oviedo, Asturias, Spain
        \and
        Centro Astron\'omico Hispano-Alem\'an (MPG-CSIC), Observatorio Astron\'omico de Calar Alto, Sierra de los Filabres, E-04550 G\'ergal, Almer\'{\i}a, Spain %05
        \and
        Th\"uringer Landessternwarte Tautenburg, Sternwarte 5, D-07778 Tautenburg, Germany %06
        \and
        Max-Planck-Institut f\"ur Astronomie, K\"onigstuhl 17, D-69117 Heidelberg, Germany %07
       }
        
\date{Received dd/10/2018; accepted dd/mm/2018}

\abstract
{We use spectra from CARMENES, the Calar Alto
  high-Resolution search for M dwarfs with Exo-earths with Near-infrared and optical 
  Echelle Spectrographs, to search for periods in chromospheric indices in 16 M0 to M2 dwarfs.
  We measure spectral indices in the H$\alpha$, the \ion{Ca}{ii} infrared triplet (IRT), and
  the \ion{Na}{i} D lines to study which of these indices are best-suited to find rotation periods
  in these stars. Moreover, we test a number of different period-search algorithms, namely the string length method,
  the phase dispersion minimisation, the generalized Lomb-Scargle  periodogram, 
  and the Gaussian process regression with quasi-periodic kernel. We find periods in four stars using H$\alpha$ and in five stars
  using the \ion{Ca}{ii} IRT, two of which have not been found before. Our results show
  that both H$\alpha$ and  the \ion{Ca}{ii} IRT lines are well suited for
  period searches, with the \ion{Ca}{ii} IRT index performing slightly better than H$\alpha$.
  Unfortunately, the \ion{Na}{i} D lines are strongly affected by telluric airglow, and we could not find
  any rotation period using this index. Further, different definitions of the
  line indices have no major impact on the results. Comparing the different search methods, the string length
  method and the phase dispersion minimisation perform worst, while Gaussian process models produce the smallest
numbers of false positives and non-detections.}

\titlerunning{Period search in M dwarfs' CARMENES spectra}
\authorrunning{B. Fuhrmeister et~al.}
\maketitle

%-----------------------------------

\section{Introduction}
Magnetic activity in late-type dwarfs manifests itself in a plethora of observed phenomena 
originating in different stellar atmospheric layers. Some of these different
activity indicators are coupled to the rotational variability of the
star and can therefore be used to measure stellar rotation periods.
Knowledge of the rotation period of a given star is crucial for an understanding of 
its magnetic activity, but also in the context of searches for exoplanets, since
inhomogeneities such as spots in the photosphere or plages in the chromosphere
in combination with stellar rotation can easily mimic a feigned planetary signal
\citep[see e.g.][]{Hatzes2013}. 

The rotation period of a star can be measured using photometric data, using spots on
its surface which cause cyclic variability of the light curve of the star while rotating
with the star; for field M dwarf stars this method has been extensively used, for example 
by \citet{DA18, Newton, SM16, West2015,Irwin11}, and many others. 
Alternatively, spectral data can be used,
such as index or equivalent width (EW) time series of 
chromospheric lines. Such index time series carry a rotational imprint if the emitting plage regions are not
homogeneously distributed on the stellar surface \citep[see e.g.][]{Mittag2017, SM15}. 
Traditionally, for such spectroscopic studies in solar-like dwarfs the \ion{Ca}{ii} H and K doublet lines are
used \citep{Noyes}. However, for M dwarfs these lines are hard to access since the pseudo-continuum
is very low at these wavelengths, and therefore other spectral indices have been tested for correlation 
with \ion{Ca}{ii} H and K or for use as rotation diagnostics. For example, 
\citet{Diaz2007} find that the \ion{Na}{i} D index correlates with H$\alpha$ for the most active stars
(Balmer lines in emission).
\citet{Walkowicz2009} also find correlations between the \ion{Ca}{ii} K and H$\alpha$ emissions
for active stars, while for
inactive or medium active stars (H$\alpha$ in absorption) the relation between the two indicators remains ambiguous.
On the other hand, in the framework of cycle studies for M dwarf stars, \citet{GomesdaSilva}
find a correlation between \ion{Ca}{ii}~H and K and \ion{Na}{i} D also for less-active stars.
Moreover, \citet{Martin} find a good correlation between \ion{Ca}{ii}~H and K and the infrared triplet (IRT) of
this ion, located at (vacuum) wavelengths of 8500.35, 8544.44, and 8664.52~\AA.
A recent study by \citet{Patrick}, concentrating on M dwarfs using CARMENES \citep{CARMENES1}  data,
shows a correlation between H$\alpha$ and \ion{He}{i} D$_{3}$
as well as the \ion{Ca}{ii} IRT.

Besides the use of different spectral lines for the search of rotation periods, a variety of search methods
are available; a comparison of 11 different period-search algorithms for photometric data and seven different object
classes (e.g. eruptive, pulsating, rotating) was performed by \citet{Graham}, who find
that the recovery of the signal strongly depends on the quality of the light curve and that
especially for lower-quality light curves the signal recovery depends sensitively on
the amplitude off the signal. The best-performing algorithm identified by \citet{Graham} was a conditional entropy-based algorithm.

The purpose of this study is to explore the effects of using different spectral lines and 
search algorithms on finding rotation periods in M dwarfs. 
We use a large set of CARMENES high-resolution spectra of 16 relatively
inactive early M dwarfs and use spectral indices
of H$\alpha$, the \ion{Ca}{ii} IRT, and \ion{Na}{i} D. Moreover, we compare the following period-search
algorithm: Generalized Lomb-Scargle (GLS) periodogram,  the string length (SL) method,
phase dispersion (PD) minimisation, and Gaussian process (GP) regression with quasi-periodic kernel.
Our paper is structured as follows: in Sect.~\ref{obs} we describe the data we have used, the stellar
sample is described in Sect.~\ref{sec:sample}, and in Sect.~\ref{sec:char} we define the different activity
indices. In Sect.~\ref{sec:algorithms} we give some details on the search algorithms that we use and present our results. Finally, a discussion and our conclusions are presented in Sects.~\ref{sec:discussion} and \ref{sec:conclusion}, respectively.

%--------------------------------

\section{Observations and data reduction}\label{obs}

All spectra discussed in this paper were taken with the CARMENES spectrograph
\citep{CARMENES1} at the 3.5-m Calar Alto 
telescope during guaranteed time observations for the
radial-velocity exo-planet survey. The main scientific objective of 
CARMENES is the search for low-mass planets  
orbiting M  dwarfs in their habitable zone.
For that, the CARMENES consortium is conducting a 750-night 
survey, targeting $\sim$330 M~dwarfs \citep{AF15a, Reiners2017}.
To date, CARMENES has obtained more than 10\,000 high-resolution visible and near-infrared (NIR) spectra.
The CARMENES spectrograph is a
two-channel, fibre-fed spectrograph covering the wavelength range from 0.52 to 0.96\,$\mu$m 
in the visual channel (VIS) and from 0.96 to 1.71\,$\mu$m in the NIR
channel  with spectral resolution of R $\sim$ 94\,600 in VIS and R $\sim$ 80\,400
in NIR. 
%All data are given in vacuum wavelength.

The CARMENES spectra were reduced using the CARMENES reduction pipeline
and retrieved from the CARMENES data archive \citep{Caballero2, Zechmeister2017}.  
To simplify measurements we shifted all spectral lines to laboratory wavelength by correcting
  for barycentric and stellar
radial-velocity shifts. Since we are not interested in high-precision radial-velocity
measurements in our context, we use the mean radial velocity for each star as given in the Carmencita database \citep{carmencita}
and used the \textit{python} package \textit{helcorr} of PyAstronomy\footnote{\tt https://github.com/sczesla/PyAstronomy} to compute the barycentric shift.

We do not correct for telluric lines, which are
normally weak in the vicinity of the H$\alpha$ and \ion{Ca}{ii} IRT lines.
We do however carefully select our reference bands, avoiding regions affected by
telluric lines. Since the
airglow of \ion{Na}{i} D lines can be several times stronger than the stellar
contribution and thus poses a serious challenge to analysis, we excluded
the contaminated spectra (for more details see Sect. \ref{sec:char}).
The exposure times for our program stars vary from about 100 s for the brightest stars
to a few hundred seconds, but are typically below 500 s.

\section{The stellar sample}\label{sec:sample}

Our sample selection starts with all 95 dwarfs of spectral types from M0.0 to M2.0 observed by the CARMENES survey, 
for which we retrieved all available spectra as of 1~Feb 2018. In this study we investigate periodicity and
  period search algorithms and focus on early M dwarfs.
  Later M dwarfs show increasingly frequent flaring, which strongly affects the chromospheric
  emission lines studied here. 
Moreover, we limit the sample to stars with
more than 50 CARMENES observations without flaring according to our criterion (Sect. \ref{flaring}).
This choice ensures a mean sampling rate of less than or approximately one every 14~days
(more than 50 observations in about 700~days).
Nevertheless, the irregular sampling allows for additional studies of periods below the formal Nyquist limit.
In our initial sample,  only six stars
have between 40 and 50 observations distributed over $700$~days and the majority of stars have
less than 30 observations, which we consider insufficient for the
period analysis. Thus,
our selection leaves us with a final target sample of 16 stars of which nine have 100 or more observations.

In Table~\ref{stars} we list the CARMENES identifier and a more common name, together with the number of 
spectra used for our period analysis. In addition, we provide
relevant information retrieved from Carmencita \citep{carmencita}, such as
spectral type, rotational velocity, membership to kinematic populations, and, for nine of them,
previously derived rotation periods from photometry or spectroscopy.

Notably, neither M2 nor halo stars fulfil our target-selection criteria.
All target stars can be attributed to the thick-disc, disc, or young-disc populations.
From CARMENES spectra, \citet{Reiners2017} deduced an upper limit of 2.0~km\,s$^{-1}$
to the rotational velocity of 11 stars, and measured $v \sin{i}$ of five M0.0 and M0.5 dwarfs
(see also \citet{Jeffers2018}). For these five stars,
we compute an estimated (maximum) rotation period $P_{\rm est}$ assuming an inclination of $90^{\circ}$
and using individual radii from \citet{Schweitzer}, which are all between 0.57 and 0.64~$R_{\odot}$. These correspond well
to a radius of 0.62~$R_{\odot}$ by
\citet{Kaltenegger}, which is applicable to M0.0\,V stars.

While most spectral observations
are separated by at least one night, some spectra were obtained at shorter cadence on the same night for engineering purposes.
The respective number of these additional densely sampled observations is also given in Table~\ref{stars}. Notable cases are
GX~And, HD~119850, BD+21~652, and Lalande~21185, for which $53$\,\%, $18$\,\%, $7.5$\,\%, and $7.5$\,\% of the available
spectra were obtained in such single-night runs; these cases are discussed 
in some detail in Sect.~\ref{multiobs}.

Our sample consists of stars with low to modest activity as reflected in the H$\alpha$ line profiles.
None of our sample stars show H$\alpha$ lines in true emission. The most active star is  
V2689~Ori, whose H$\alpha$ line is just below the level where it would go into emission. On the other hand, BD+21~652 exhibits
the deepest H$\alpha$ absorption line and is therefore among the stars with the lowest activity of our sample.
More details on the activity of our sample stars are provided in the following section.

\begin{table*}
\caption{\label{stars} Properties of the sample stars. }
\footnotesize
\begin{tabular}[h!]{llrl crcclclcl}
\hline
\hline
\noalign{\smallskip}
Karmn & Name  &No.    &No.   & Sp.  & Ref. & Pop$^{c}$ &  $v \sin{i}$\,$^{d}$ & $P_{\rm est}$ & $P_{\rm lit}$ & Ref$^{e}$\\ 
      &       &spec.$^a$   &short$^b$&type &(SpT)         & & [km\,s$^{-1}$] &[d] & [d] & ($P_{\rm lit}$)\\
\noalign{\smallskip}
\hline
\noalign{\smallskip}
J00183+440 & GX And    & 190 & 112& M1.0 & AF15a&  D &      <2.0& ... & ... & ... \\
J01026+623 & BD+61 195 & 68 & 2   & M1.5 & AF15a &   YD&   <2.0   &... &19.9$\pm$ 0.4, 18.4$\pm$ 0.7  & DA18, SM17 \\ 
J04290+219 & BD+21 652 & 161 & 23 & M0.5 & Gra06 &  D&  3.9 $\pm$1.4& 8.3 & 25.4$\pm$ 0.3& DA18\\ 
J04376+528 & BD+52 857 & 78 & 11   &M0.0 & Gra03 &  D &3.4 $\pm$0.6 & 8.5 & ... & ... \\
J05314-036 & HD 36395  & 87 & 7   &M1.5 & AF15a &  D &    <2.0   &... &33.8$\pm$ 0.6, 33.61& DA18, KS07\\ 
J05365+113 & V2689 Ori & 50 & 12   &M0.0 & Lep13 &  YD&    3.8 $\pm$0.4& 7.5 & 12.3$\pm$ 0.1, 12.02 & DA18, Kir12 \\ 
J09144+526 & HD 79211  & 115 & 9  &M0.0 & AF15a &  YD&    2.3 $\pm$0.4& 12.7 &... & ...\\
J09561+627 & BD+63 869 & 57 & 7   &M0.0 & PMSU  &  YD&    <2.0   & ... & ... & ...\\
J10122-037 & AN Sex    & 61 & 6   &M1.5 & PMSU  &  YD&    <2.0   & ...&21.6$\pm$  0.2,21.56& DA18,Kir12\\ 
J11033+359 & Lalande 21185   & 147 &21 & M1.5 & AF15a &  TD&    <2.0 & ... & 48.0     &  No84 \\ 
J11054+435 & BD+44 2051A& 100& 7  &M1.0 & AF15a & TD-D&   <2.0   & ... & ... & ...\\
J13299+102 & BD+11 2576&  64& 7   &M0.5 & PMSU  &  D &   <2.0    &... &28.0$\pm$ 2.9  & SM15\\ 
J13457+148 & HD 119850 &  223& 58 &M1.5 & PMSU  &  D &    <2.0   &... &52.3$\pm$ 1.7  & SM15\\ 
J16167+672S & HD 147379& 124 & 4  &M0.0 & AF15a & YD&   2.7 $\pm$0.2& 11.6 & ... & ...\\
J20533+621 & HD 199305 & 133 & 2  &M1.0 & Lep13 &  D &    <2.0   & ... & ... & ...\\
J22565+165 & HD 216899 & 203 & 14  &M1.5 & PMSU  & TD&    <2.0   & ... & 39.5$\pm$ 0.2, 37.5$\pm$ 0.1& DA18, SM15\\ 

\noalign{\smallskip}
\hline

\end{tabular}
\tablebib{
  AF15a:~\citet{AF15a}; DA18:~\citet{DA18}; Gra03:~\citet{Gra03}; Gra06:~\citet{Gra06};
  Kir12:~\citet{Kir12}; KS07:~\citet{KS07}; Lep13:~\citet{Lepine}; No84:~\citet{Noyes};
PMSU:~\citet{PMSU};  SM15:~\citet{SM15}; SM17:~\citet{SuarezMascareno2017} \\
$^{a}$ Total number of spectra used in the analysis.\\
$^{b}$ Number of spectra obtained at short sub-day cadence. \\
$^{c}$ All kinematic population designations are measured by  \citet{CC16}: abbreviations: YD: young disc, D: disc, TD: thick disc, TD-D: thin-to-thick transition disc; there are no halo stars in our sample.\\
$^{d}$ All rotational velocities are measured by  \citet{Reiners2017} and are included in the
catalogue of \citet{Jeffers2018}.\\
$^{e}$ No84 and SM15 determined spectroscopically, all other periods determined photometrically.
}
\normalsize
\end{table*}

\section{Indices and time series}\label{sec:char}

To characterise the stellar activity state for each spectrum, we define activity indices following
the method of \citet{Robertson}. In particular, we use the mean flux density in a
spectral band centred on the spectral feature (the so-called line band) and
divide by the sum of the average flux density in two reference bands
\begin{equation}
I_{\rm line}=w \, \left(1-\frac{2\cdot\overline{F_{\rm line}}}{\overline{F_{\rm ref1}}+\overline{F_{\rm
ref2}}}\right) \; .
\end{equation}
Here, $w$ is the width of the line band, and
$\overline{F_{\rm line}}$, $\overline{F_{\rm ref1}}$, and $\overline{F_{\rm ref2}}$ are the mean flux densities in the
line and the two reference bands, respectively. The thus-defined index resembles a pseudo-EW (pEW) with the difference
that we use the mean flux densities and not integrals.

To compute statistical errors for $I_{\rm line}$, we use the pipeline-generated errors on the flux density
and a Monte Carlo simulation. Specifically, we compute the index 100\,000 times for each line and each spectrum
using flux density values randomly
drawn from the error distribution and, thus, estimate the standard deviation of the respective index.
We consider indices for  H$\alpha$, \ion{Na}{i}, and the \ion{Ca}{ii} IRT. The line and
reference bands used are provided in Table \ref{Table:indices}.

\begin{table}
\caption{\label{Table:indices} Properties of the indices' line and reference bands. }
\footnotesize
\begin{tabular}[h!]{lcccc}
\hline
\hline
\noalign{\smallskip}
Line      & \multicolumn{2}{c}{Line band} & \multicolumn{2}{c}{Reference bands}\\ 
          &central & full  & blue & red \\
&wavel.  & width &            \\
& [\AA] & [\AA] & [\AA] & [\AA]  \\
\noalign{\smallskip}
\hline
\noalign{\smallskip}
H$\alpha$ & 6564.62 & 1.6 & 6547.4 -- 6557.9 & 6577.9 -- 6586.4\\
\ion{Ca}{ii} 1 & 8500.35 & 0.5 & 8476.3 -- 8486.3 & 8552.4 -- 8554.4\\
\ion{Ca}{ii} 2 & 8544.44 & 0.5 & 8476.3 -- 8486.3 & 8552.4 -- 8554.4\\
\ion{Na}{i}  1 & 5891.58 & 0.5 & 5870.0 -- 5874.6 & 5910.0 -- 5914.0\\
\ion{Na}{i}  2 & 5897.56 & 0.5 & 5870.0 -- 5874.6 & 5910.0 -- 5914.0\\
\noalign{\smallskip}
\hline

\end{tabular}
\normalsize
\end{table}

We follow \citet{Robertson} and \citet{GomesdaSilva} in adopting a width $w$ of 
1.6~\AA\ for the (full) width of the H$\alpha$ line band.
While this is insufficient during strong flares, when the line width can exceed 5~\AA\ \citep{asym},
this situation does not occur among our sample of rather inactive stars.
 In Table \ref{activityproperties} we give the medians of our H$\alpha$ index measurements, which are all positive (i.e. in absorption). 
Visual inspection shows that the H$\alpha$
line is in fact in absorption for all sample stars but V2689~Ori, where it shows weak emission in the wings
alongside an absorption signature in the centre (see below).

For the \ion{Ca}{ii} IRT lines we adopt 0.5~\AA\, for the width of the line band $w$
because the lines are narrower than the H$\alpha$ line. We expect the three components of the \ion{Ca}{ii}~IRT to behave similarly with respect to activity \citep[e.g.][]{Martin}.
Here we consider only the two bluest ones, which are located in the same \'{e}chelle order of CARMENES.  
The resulting indices are referred to as $I_{\rm Ca,1}$ and $I_{\rm Ca,2}$.

We also compute activity indices for the  \ion{Na}{i} D doublet for which we fixed the
width of the line band to be 0.5~\AA. These indices are referred to as $I_{\rm Na,1}$
and $I_{\rm Na,2}$.

In the analysis of the sodium lines, airglow is a particular problem.  To avoid airglow contamination
we only use spectra in our period
search for which the position of the \ion{Na}{i}~D airglow  is at least 0.3~\AA\,
away from the central wavelength of the stellar \ion{Na}{i}~D line. Since the airglow lines
are narrow, we consider this sufficient to ensure an acceptably low level of airglow contamination in the
line band. This procedure reduces the number of spectra considered for these lines
(see Table~\ref{activityproperties}).

\subsection{Spectra taken during flares}\label{flaring}

A further problem in the period search are spectra whose indices are influenced by flares.
To prevent  contamination by flares in our period search,
we identify those spectra possibly affected by flaring based on the $I_{\rm H\alpha}$ index.
Specifically, we flag spectra with $I_{\rm H\alpha}$ lower than $0.7 {\rm median}(I_{\rm{H}\alpha})$ as being
affected. This remains a coarse classification criterion that is insensitive to smaller flares or 
spectra taken toward the end of a flare-decay phase. However, 
the lack of continuous time series renders flare detection rather difficult.
In Table~\ref{stars} we give the number of stellar spectra remaining after the exclusion of spectra affected by flaring.
Additionally, we provide the number of identified flares in Table~\ref{activityproperties}. Since the
stars are all rather inactive, the flaring rate remains low and many stars show no detectable flare events
during the observations according to our criterion. The only exception is V2689~Ori, the most active star in the
sample, for which 14 out of 64 spectra were excluded due to possible flaring. We investigate the effect of
this high exclusion rate in the period analysis (Sect.~\ref{Sec:flaring}).

\subsection{Scatter and correlation}

We use the time series of the above-defined  indices to
further characterise the activity levels of our sample stars.
In Table~\ref{activityproperties} we list the medians
of the measured indices $I_{\rm{H}\alpha}$, $I_{\rm Ca,1}$, and $I_{\rm Na,1}$ for all stars.
Our median H$\alpha$ indices are all positive, indicating that the stars
are rather inactive. 

\begin{figure*}
\begin{center}
\includegraphics[width=0.5\textwidth, clip]{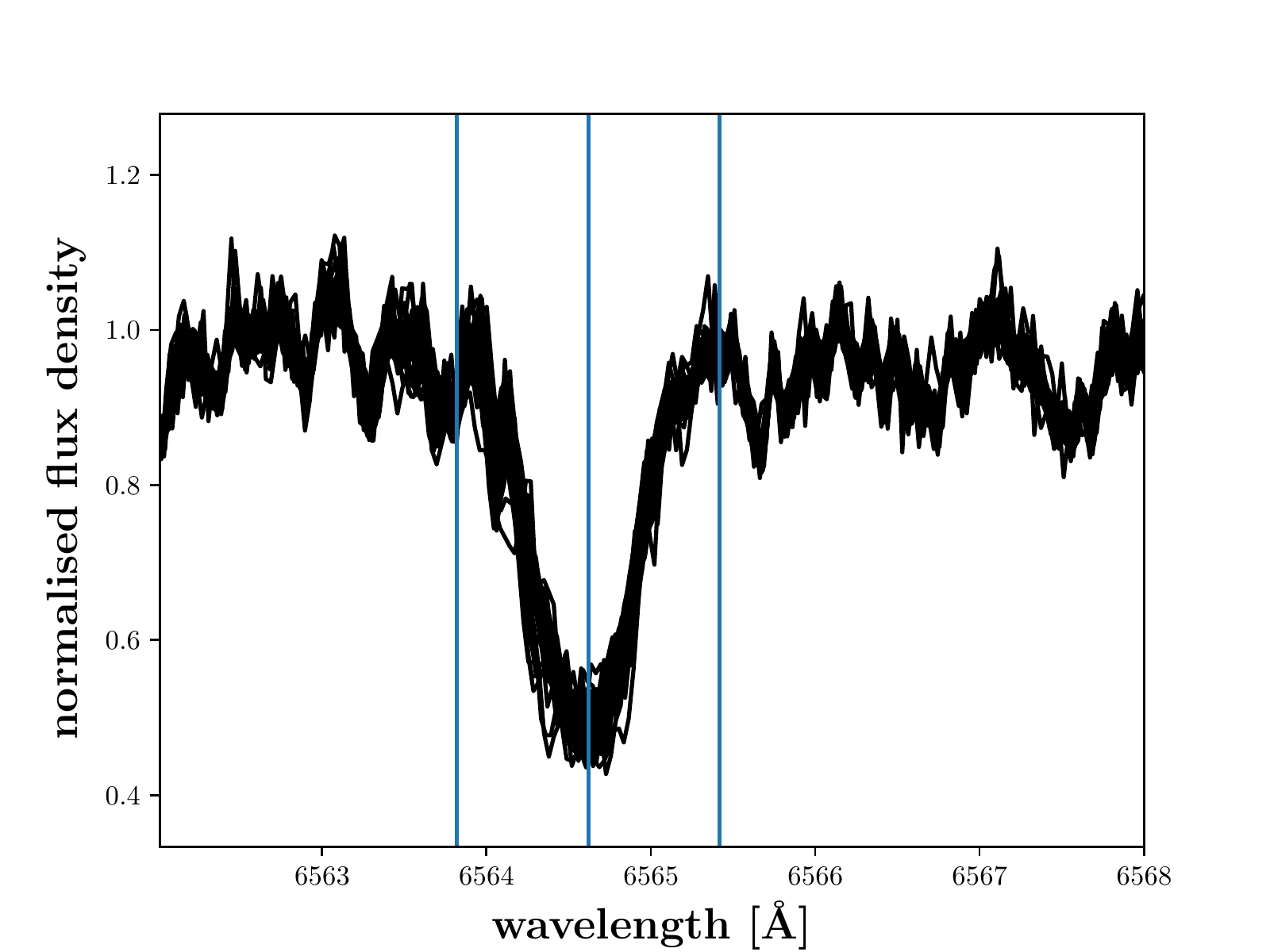}%\vspace{-0.4mm}\\
\includegraphics[width=0.5\textwidth, clip]{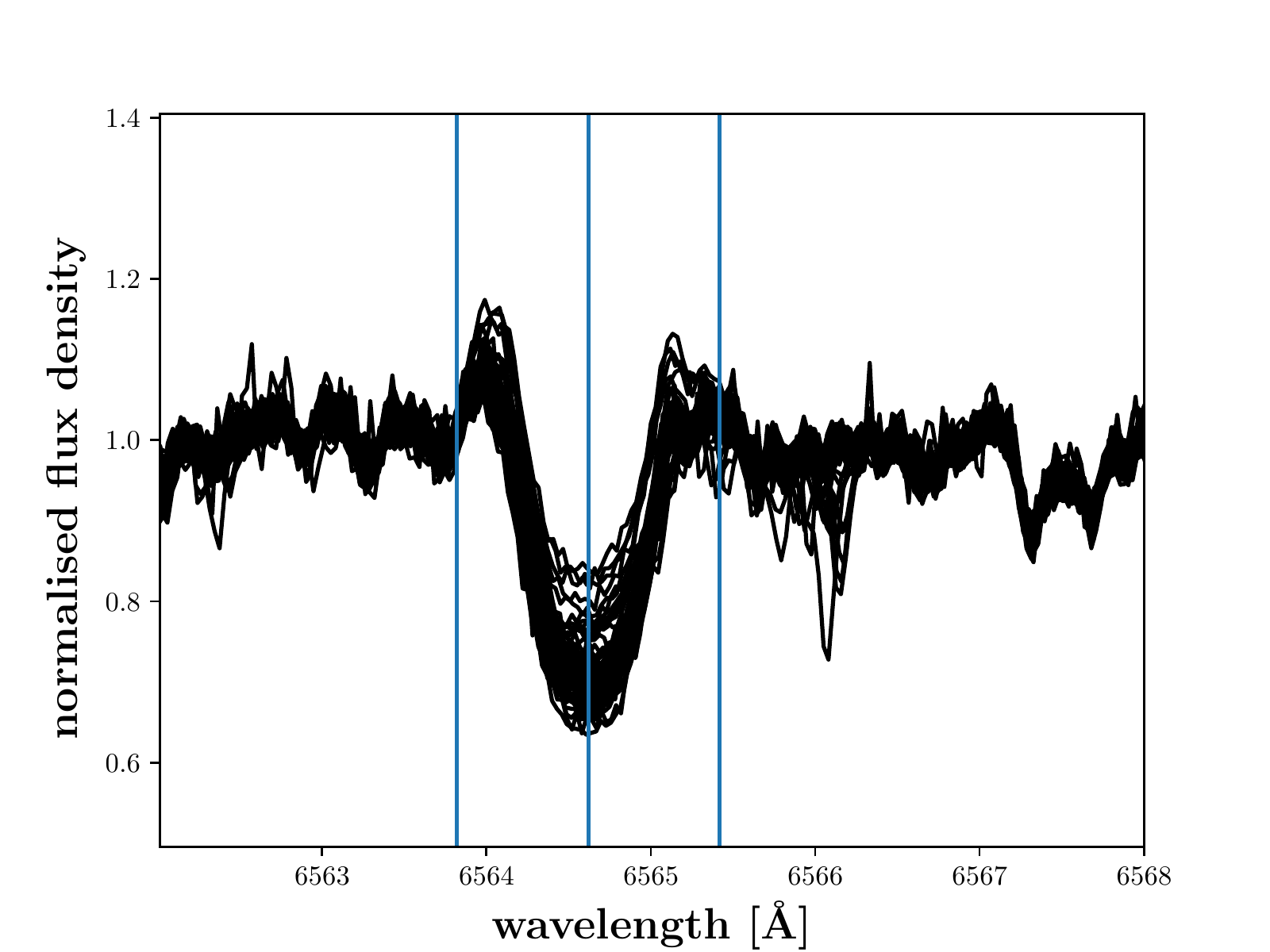}
\caption{\label{halphaline} Two typical H$\alpha$ line profiles for our sample stars: 
BD+44~2051A (left panel), one of the least active stars and V2689 Ori (right panel), 
the most active star in our sample.
All usable spectra of the two stars are shown to demonstrate that the chromospheric line shows relatively
  small variations. The vertical lines correspond to the borders of our central wavelength interval.}
\end{center}
\end{figure*}

In Fig.~\ref{halphaline} we show the H$\alpha$ line of V2689~Ori and 
BD+44~2051A.
% which represent the most active and one of the most inactive stars in our sample according
% to $I_{\rm Ca,1}$.
While V2689~Ori is clearly the most active star in our sample, there are several candidates
for the most inactive member. In terms of $I_{\rm Ca,1}$, Lalande~21285, BD+44~2051A, and GX~And
all show indistinguishable low activity states, i.e., the numerically largest value. The star with the deepest
H$\alpha$ line is BD+21~652, which is also quite inactive according to $I_{\rm Ca,1}$.
In both stars (cf., in Fig.~\ref{halphaline}) the H$\alpha$ lines remain relatively stable compared to more active
M~stars that show an H$\alpha$ emission line with considerable variations in amplitude and line shape
on various timescales \citep{asym}.

For reference, the median value and median average deviations (MAD, see following paragraph) of
$I_{\rm{H}\alpha}$ for the highly active star YZ~CMi are $-7.2$~\AA\, and
$0.6$~\AA. The latter exceeds the highest value observed in our current sample by a factor of 20
(see Table~\ref{activityproperties}). 

To quantify the scatter, we computed the median average deviations (MADs), which
is a robust estimate somewhat similar to the standard deviation \citep{Rousseeuw1993, Czesla2018},
for H$\alpha$, the bluest line of the \ion{Ca}{ii} IRT, and the blue \ion{Na}{i} D line.
The level of variability in the H$\alpha$ and the \ion{Ca}{ii} IRT lines is well correlated 
% \tbd{Siehe Abb. Hamad\_camad.png. Sollen wir
% das bringen?} 
and is consistent with low activity levels. In contrast, the correlation between the MADs of the \ion{Ca}{ii} IRT and
\ion{Na}{i} D is non-significant. While the MAD(Na) is normally equal or lower than the MAD(Ca), for
Lalande~21185 this is not the case. In general, H$\alpha$ shows the highest degree of variability,
followed by the \ion{Ca}{ii} IRT lines, while \ion{Na}{i} D is the least variable.

To study the mutual relation between the behaviour of the individual activity indicators, 
we compute Pearson's correlation coefficient obtained from the time series (with the 
flare spectra being excluded)
and give the values
in Table~\ref{activityproperties}. Values not significant at the 1 \,\% level are shown in square brackets.
In particular, we relate $I_{\rm H\alpha}$ to $I_{\rm Ca,1}$ and $I_{\rm Na}$
and also investigate the correlation
coefficient corresponding to the first and second \ion{Ca}{ii} IRT lines and between the
two lines of the \ion{Na}{i}~D doublet. 

As expected,
the indices of the individual \ion{Na}{i}~D doublet lines are correlated with each other; similarly,
the two bluer \ion{Ca}{ii} triplet lines are significantly correlated in all stars.
The correlation between 
$I_{\rm{H}\alpha}$ and $I_{\rm Ca,1}$ is also strong for the majority of the sample stars and remains 
non-significant in only four of
them, which are all among the least active according to their median $I_{\rm Ca,1}$ index.
We attribute the lack of a significant correlation to the low level of variability in these stars
and our resulting lack of sensitivity  rather than a physical mechanism.
All significant correlation coefficients are positive and rather large with seven
exceeding $0.8$, which indicates a similar response of these two indices to changes in the activity level. 

% we consider a lack of sensitivity likely responsible for this finding
% a too low level of variability to be detected.

The rate of significant correlation coefficients between
$I_{\rm{H}\alpha}$ and $I_{\rm Na,1}$ is lower compared to the previous comparison with $I_{\rm Ca,1}$.
Only five stars show a significant correlation between these indices, and whenever it is significant, the
coefficient is always lower than for the correlation between $I_{\rm{H}\alpha}$ and $I_{\rm Ca,1}$.
While this result may also be partially explained by a lack of sensitivity, which would have to affect
primarily the $I_{\rm Na,1}$ index, 
% Again we attribute
% this weak correlation between $I_{\rm{H}\alpha}$  and $I_{\rm Na,1}$ at least partially to the
% low level of variability in these stars. But
this argument is certainly inapplicable in the case of V2689~Ori, which shows detectable variability in all
studied indices. We attribute the  correlation coefficients with lower significance and the high rate of
non-significant values to
differences in chromospheric activity in the different line-formation regions, 
that is, the higher chromosphere in the case of H$\alpha$ and the
lower chromosphere for the \ion{Na}{i}~D doublet \citep{AndrettaDoyle1997}.

For the two stars AN~Sex and V2689~Ori, for which we excluded the highest
number of flaring spectra (4 and 14, respectively), we also computed the correlation coefficient
between $I_{\rm{H}\alpha}$ and $I_{\rm Na,1}$ including the flaring
spectra. In this case, we found significant correlation coefficients of $0.53$ for AN~Sex and $0.79$ 
for V2689~Ori. This result is  driven by the flare spectra and shows that
strong changes in the activity level are similarly traced by both lines.

\subsection{Alternative indices and robustness}
\label{sec:alternative}

To study the impact of the index definition on the results, we introduce alternative indices for the
H$\alpha$ and the bluest \ion{Ca}{ii} IRT line. 
This allows us to
test, for example, whether or not the width of the central wavelength interval influences the
period analysis.
To this end, we define four alternative indices, which we dub A1--4. Specifically, we
use (i) the pseudo equivalent width (pEW) of the H$\alpha$ line based on a central wavelength band of
6563.9 -- 6565.5 \AA, referred to in the following as A1 pEW; (ii) a pEW covering only the core from 6564.4 -- 6564.85 \AA\, (A2 pEW);
(iii) a pEW with broader central wavelength range of 6560 -- 6570 \AA\, (the latter would account for flares -- A3 pEW); and
(iv) a pEW with variable central bandwidth which is defined by the foot points of the line (A4 pEW).
For these pEW time series we used three-sigma clipping to exclude flares and other outliers.

The variable pEW measurements (A4  pEW) rely on locating the maxima and minima  in the ranges
6563.6 to 6564.4 and 6564.85 to 6565.7 \AA. The minimum values form the range for A4 pEW and the maximum
values are used to calculate the pEW of the H$\alpha$ central reversal in stars that have H$\alpha$ in
emission. All pEW(A1-A4) use the reference regions of 6550 to 6555 and 6576 to 6581\AA, which
are different from the reference regions used for $I_{\rm{H}\alpha}$. However, also for A1 to A4, none of the
spectra were corrected for tellurics as the used regions have only very weak telluric lines. These  weak lines have
an influence on the resultant pEW, which is significantly less than the margin of error.

Of these alternative indices, A1 is most similar to $I_{\rm{H}\alpha}$; only the reference bands differ
and the method of calculation (integral instead of mean).

\begin{table*}

\caption{\label{activityproperties} Measured activity properties of our sample stars. }
\footnotesize
\begin{tabular}[h!]{llll lllllllll}
\hline
\hline
\noalign{\smallskip}
Name  &No.    & No. &   Median &  Median& Median & Corr  & Corr  & Corr$^{a}$  & Corr$^{b}$ &MAD  & MAD  & MAD\\
      & spec & spec&($I_{\rm{H}\alpha}$) & ($I_{\rm Ca,1}$) & ($I_{\rm Na,1}$)& H$\alpha$ - & H$\alpha$ -& Ca - &Na -&H$\alpha$ &Ca &Na\\   
      & Na D     & flare &     [\AA]   &  [\AA]   & [\AA] & Ca & Na & Ca& Na& [$10^{-2}$]& [$10^{-2}$] & [$10^{-2}$]\\
\noalign{\smallskip}
\hline
\noalign{\smallskip}
GX And    & 145 & 1 & 0.37 & 0.27 & 0.45 & [-0.026] & [-0.144] &  0.55 &0.96& 0.6 & 0.1&0.2  \\
BD+61 195 & 37  & 3 & 0.31 & 0.18 & 0.40 & 0.96 & 0.60 & 0.99 & 0.87 & 3.0 & 0.7 &0.4\\
BD+21 652 & 114 & 0 & 0.62 & 0.25 & 0.45 & 0.75 & 0.38 & 0.91 & 0.93 &0.8 & 0.3 &0.2\\
BD+52 857 & 58  & 0 & 0.52 & 0.20 & 0.43 & 0.83 & [0.29] & 0.96 & 0.85 &1.4 & 0.5 &0.3 \\
HD 36395  & 65  & 1 & 0.46 & 0.21 & 0.42 & 0.96 & 0.37 & 0.98 & 0.87 & 2.9 & 0.8 &0.4\\
V2689 Ori & 35  & 14 & 0.16 & 0.08 & 0.41 & 0.91 & [0.46] & 0.96 &0.88& 3.0 & 0.8&0.2 \\
HD 79211  & 53  & 0 & 0.51 & 0.20 & 0.43 & 0.73 & [0.01] & 0.97 & 0.78 & 1.1 & 0.4 &0.2\\
BD+63 869 & 26  & 0 & 0.48 & 0.20 & 0.43 & 0.93 & 0.65 & 0.97 & 0.86 & 1.8 & 0.6&0.3 \\
AN Sex    & 25 &  4 & 0.34 & 0.20 & 0.40 & 0.90 & [0.09] & 0.95 & 0.86 & 1.5 & 0.4& 0.3\\
Lalande 21185   & 145 & 2 &0.27 & 0.27 & 0.46 & [-0.16] &[-0.12] & 0.31 & 0.98  & 1.1 & 0.2 &0.5 \\
BD+44 2051A&  99 & 1 & 0.33 & 0.27 & 0.46 & [-0.09]& [-0.24]& 0.83 &0.86& 1.1 & 0.3 &0.3\\
BD+11 2576&  19 & 2 & 0.48 & 0.25 & 0.43 & 0.42 & [0.09] & 0.87& 0.75 & 1.1 & 0.2 &0.1\\
HD 119850 &  107 &0 & 0.45 & 0.26 & 0.43 & [0.14] & [0.19] & 0.86 & 0.81 & 1.3 & 0.2& 0.1 \\
HD 147379&  71 & 0 & 0.60 & 0.23 & 0.44 & 0.89 & 0.35 & 0.98& 0.45 & 0.8 & 0.3 &0.1 \\
HD 199305 & 90 & 2 & 0.46 & 0.23 & 0.43 & 0.66 &[-0.05] & 0.93 & 0.83 & 1.0 & 0.3&0.1 \\
HD 216899 & 115 & 3 &0.42 & 0.22 & 0.42 &0.76 & [-0.05]& 0.91 & 0.94 & 1.5 & 0.3 &0.3\\

\noalign{\smallskip}
\hline

\end{tabular}\\
Note: The Pearson correlation coefficients given in square brackets have a corresponding p-value higher than 0.01.\\
$^{a}$: The correlation between $I_{\rm Ca,1}$ (at 8500.35 \AA) and $I_{\rm Ca,2}$ (at 8544.44 \AA).\\
$^{b}$:The correlation between the indices of the two \ion{Na}{i} D lines.\\
\end{table*}

\section{Period search}\label{sec:algorithms}
\subsection{Methods}

We use several methods to search for periodic variations in the
the previously defined time series of activity indices. Since the CARMENES data are not
evenly spaced, some widely used period-search
algorithms, such as many variants of Fourier transforms (FT), are not applicable.
Here we focus on 
the following methods: Generalized Lomb-Scargle (GLS) periodogram, 
phase dispersion (PD) minimisation, string length (SL) method, and
Gaussian process (GP) modelling with quasi-periodic kernel -- hereafter
referred to as GP.

Before the index time series are analysed, we detrend them (flare spectra excluded)
using a second-order polynomial fit. This procedure
yields a reasonable correction of long-term trends for most time series. 
A notable exception is the H$\alpha$ time series of HD~119850,
which shows a complex evolution as  shown in Fig.~\ref{hd119850}.
This procedure  also possibly suppresses periodic variations due to activity cycles on scales
of years and
beyond the observed timescale, to which we are therefore not sensitive.
Detrending is not used for the GP analysis, for which  it is not required. 

Generally, we search for periods between 1 and 150 days, with the lower limit given by
the observing scheme, which normally does not lead to more than one observation per night, and
the upper limit suggested by the length of one observing block introduced by the seasonal visibility
of the objects.

%Our results are summarized in Tables~\ref{periodsha} and \ref{periodsca} and
%Sect.~\ref{sec:results}.

\subsubsection{Generalized Lomb-Scargle periodogram}

\begin{figure}
\begin{center}
\includegraphics[width=0.5\textwidth, clip]{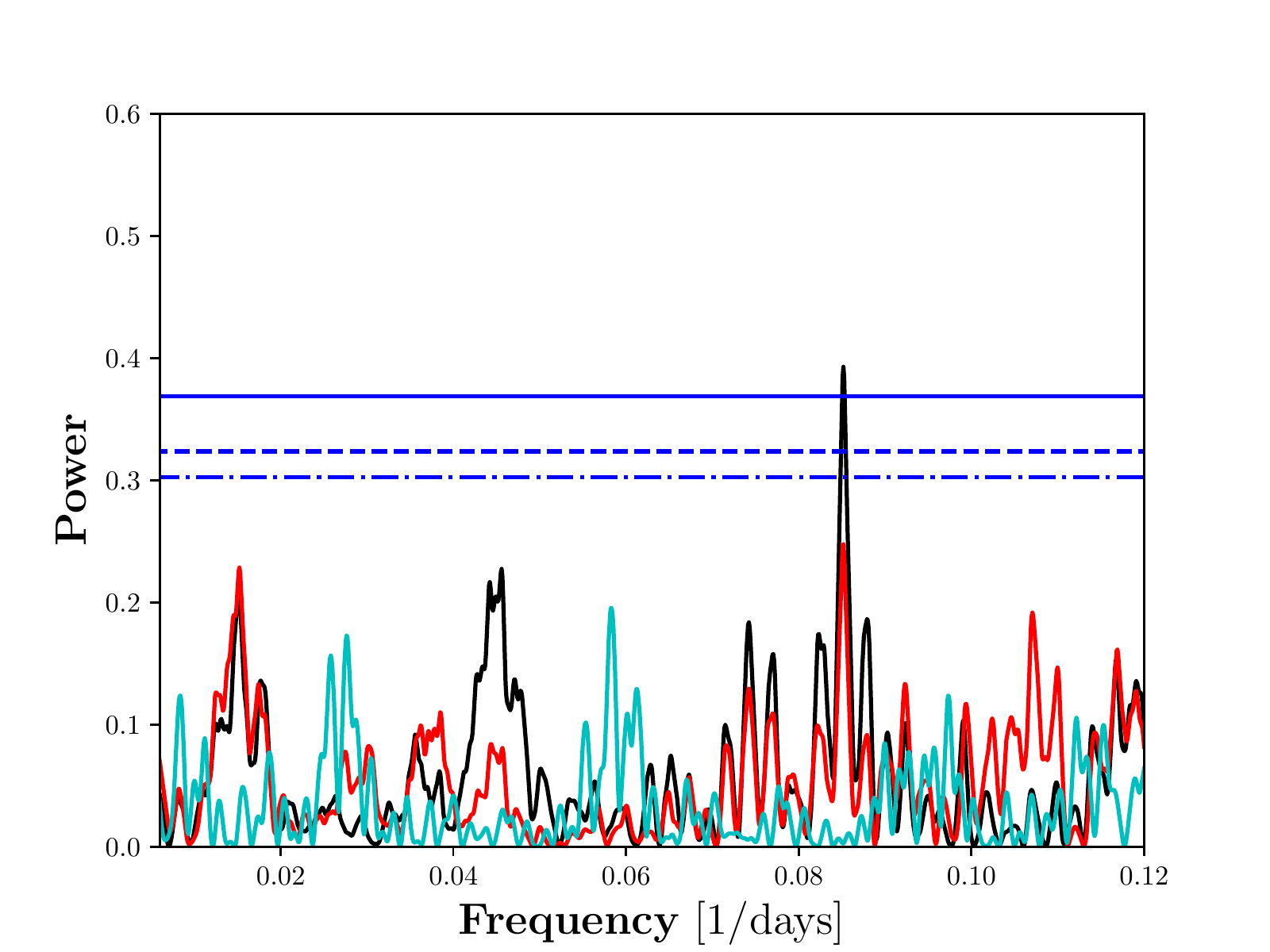}%\vspace{-0.4mm}\\
\caption{\label{fig:gls} A representative GLS periodogram, here for the star V2689~Ori. While the horizontal dashed lines
  mark the 10 (blue/dot-dashed), 5 (blue/dashed), and 1\,\% FAP levels (blue/solid), the power is given in black for $I_{\rm H\alpha}$ and
  in red for $I_{\rm Ca,1}$. The cyan line denotes the window function.}
\end{center}
\end{figure}

We use the generalized Lomb-Scargle (GLS)
periodogram as implemented in PyAstronomy\footnote{\tt https://github.com/sczesla/PyAstronomy}
\citep{Zechmeister2009, Scargle1982, Lomb1976} %  on all individual index time series 
%  Before applying the GLS, we detrended the time series using a second-order polynomial, which
%  suppresses long-term evolution, possibly including periodic variation on scales beyond the
%  observed timescale.
to compute periodograms for the index time series of H$\alpha$, the bluest \ion{Ca}{ii} IRT line,
and the bluer \ion{Na}{i}~D line, adopting the normalisation to unity introduced by \citet{Zechmeister2009} in their Eq. 4.
In the case of H$\alpha$, we consider the four different
pEWs introduced in Sect.~\ref{sec:alternative} to study the impact of the definition.
Along with the periodograms, we derive false alarm probability (FAP) levels of 10,
5, and 1\,\% and also the window function. 
In our analysis, we consider only peaks with power values exceeding the 1\,\% FAP level.
A typical example of a GLS periodogram is shown in Fig.~\ref{fig:gls} for the star
V2689~Ori. While the main power peak of the H$\alpha$ time series is significant, the
one for \ion{Ca}{ii} IRT is not.  Both agree well with the literature
value for the period of 12.3 days \citep{DA18}. In H$\alpha$ a (non-significant) peak is seen also at
half the frequency (or twice the period) indicating the importance of harmonics for this star.

\subsubsection{Phase dispersion minimisation and string length method}

The phase dispersion minimisation and string length
methods are both based on phase folding the time series with a series of 
trial periods and analysing the folded time series.
The basic idea is that the phased time series shows a well defined `curve'
only if the data are folded with the correct period that is 
also present in the data.
This rather intuitive ansatz can be quantified by measuring
(i) the length of the line connecting all data points, the so-called 
string length, $L$, \citep{Dworetsky} or 
(ii) the variance of the folded time series estimated in several phase 
bins compared to the sample variance, $\Theta$ \citep{Stellingwerf}.
Both quantities are minimised when the trial period matches the period of a 
signal.  The advantage of both of these methods is that they do not
rely on specific functional forms of the signal (such as sine waves in the 
case of GLS) or equidistant sampling. 
 
To compute the phase dispersion minimisation (PDM) statistic, $\Theta$, as a 
function of a trial period,
we use the implementation provided by PyAstronomy, which is
based on the method introduced by \citet{Stellingwerf}. 
The investigated trial periods range from 1 to 150~days 
with an increment of 0.05~days. Each phase diagram is then divided into ten bins and three
phase-shifted bin-sets are used to estimate $\Theta$.

For computing the string length, $L$, we fold the index time series with 20\,000 trial 
periods between 1 and 150~days.
Prior to folding, we re-scaled the data following the procedure by \citet{Dworetsky}.
For each trial period, we compute the resulting string length and,
finally, both $\Theta$ and $L$ are normalised by their
median values to facilitate an easier comparison between the different indices used. 
We dub the resulting statistics $\Theta_n$ and $L_n$.

We applied both methods to investigate the index time series of H$\alpha$, the 
bluest \ion{Ca}{ii} IRT line, and the blue \ion{Na}{i}~D line. 
To identify appropriate FAP levels,
we applied a bootstrap procedure. In particular, 
we randomly shuffle the index time series 10\,000 times, which should destroy any temporal
coherence, and compute new values of
$\Theta_n$ and $L_n$, respectively.
Based on their resulting distributions, we defined 1\,\% FAP levels, which are tabulated for each star
in Table~\ref{faplevel} and are used as cut-off in our analysis.  

To identify the significant periods, we visually inspected the diagrams and screened them for dips
exceeding the previously defined cut-off. It appears that
both methods are rather prone to aliasing, that is, they yield peaks at frequency multiples of the (unknown)
true period, which complicates the interpretation of the resulting
string length and phase dispersion diagrams.
If more than one dip exceeding the 1\,\% FAP level was found, we picked the most pronounced and also report the second in strength.
Only when a series of likely aliases was detected, did we opt for the smallest  period in this series, even if
this did not correspond to the peak with the highest formal significance. If this peak did not reach the
1\,\% FAP level we give it in square brackets in our results in Tables~\ref{periodsha} and \ref{periodsca}.

\begin{table}
  \caption{\label{faplevel} Monte Carlo simulated values of string length or phase dispersion
    corresponding to 1\,\% FAP for each star. }
  \footnotesize

\begin{tabular}[h!]{lcc}
\hline
\hline
\noalign{\smallskip}
 Star  & $L_n$ cut off & $\Theta_n$ cut-off\\

\noalign{\smallskip}
\hline
\noalign{\smallskip}
GX And    & 0.85 & 0.88\\
BD+61 195 & 0.69 & 0.70\\
BD+21 652 & 0.78 & 0.86\\
BD+52 857 & 0.70 & 0.73\\
HD 36395  & 0.74 & 0.77\\
V2689 Ori & 0.65 & 0.60\\
HD 79211  & 0.78 & 0.79\\
BD+63 869 & 0.67 & 0.66\\
AN Sex    & 0.72 & 0.69 \\
Lalande 21185   & 0.82 & 0.88\\
BD+44 2051A&  0.73 & 0.77 \\
BD+11 2576& 0.72 & 0.69 \\ 
HD 119850 & 0.82 & 0.86 \\
HD 147379&  0.76 & 0.83 \\
HD 199305 & 0.77 & 0.84 \\
HD 216899 & 0.81 & 0.88 \\

\noalign{\smallskip}
\hline

\end{tabular}\\
\end{table}

\begin{figure}
\begin{center}
\includegraphics[width=0.5\textwidth, clip]{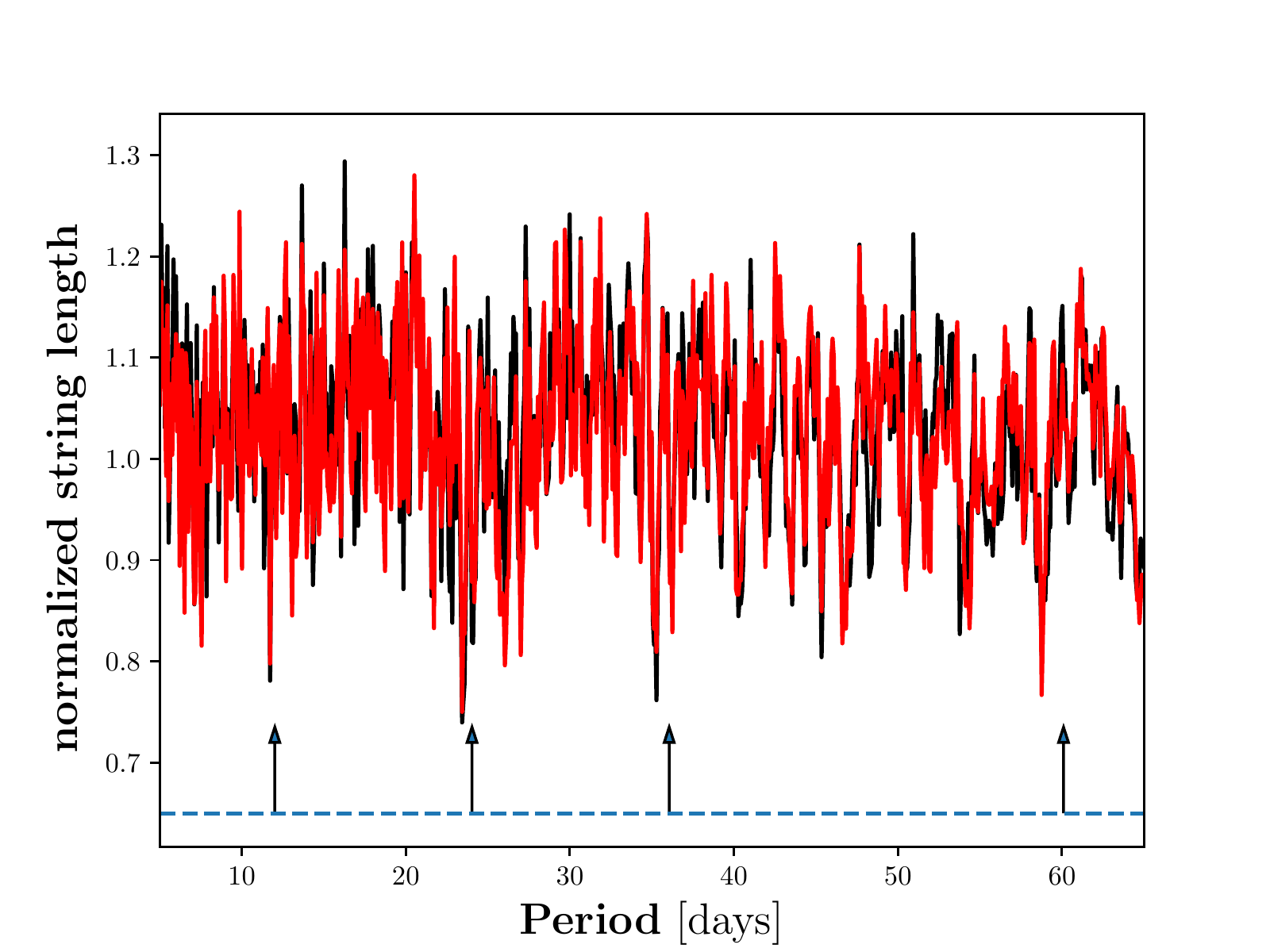}\\%\vspace{-0.4mm}\\
\includegraphics[width=0.5\textwidth, clip]{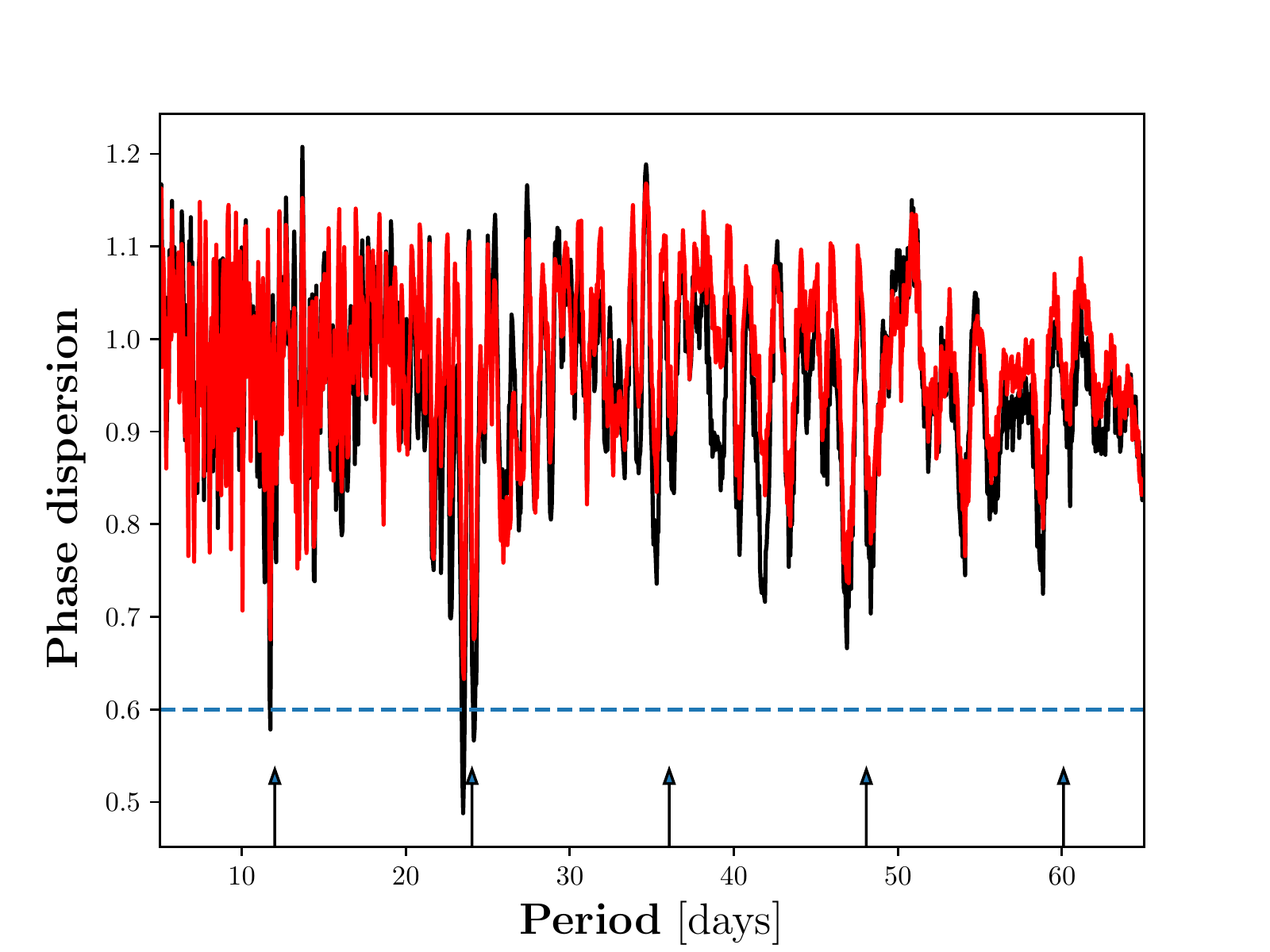}%\vspace{-0.4mm}\\
\caption{\label{fig:stringlength} "Representative string length (top) and phase
dispersion (bottom) diagrams for the star V2689~Ori.
  The black line denotes the computed $L_n$ and $\Theta_n$  for $I_{\rm H\alpha}$, respectively.
  The red line denotes $L_n$ and $\Theta_n$ for $I_{\rm Ca,1}$, respectively.
  The blue dashed horizontal line denotes the 1\,\%
  FAP level below which a period would be significant. The arrows mark the photometrically measured period of 12.02 days
\citep{Kir12} and multiples.}
\end{center}
\end{figure}

As an example, we show the computed string length  and the phase dispersion
for V2689~Ori in Fig.~\ref{fig:stringlength}.
For this specific light curve the double, triple, and even quintuple of the known
period of 12 days actually produce a lower string length than the true period.
For \ion{Na}{i}~D the correct period is not found, while the alias corresponding to twice the actual
period is detected.
The string length and phase dispersion curves for all stars are found in the Appendix.

\subsubsection{Gaussian process regression with quasi-periodic kernel}

A Gaussian process (GP) consists of sets of random variables of which any finite set has a joint normal distribution
\citep{Rasmussen2005}. A GP is completely specified by a mean and a co-variance function, and the latter
can be used to search for periodic variation.
The actual distribution of the errors in the data is rarely known. 
Nonetheless, the assumption of normality is often very helpful and appropriate 
in data modelling \citep[e.g.]
[]{Jaynes}, which makes GPs a highly valuable tool for data analysis.
At the heart of GP modelling lies the chosen kernel function, which describes the
co-variance of the data.  The choice of kernel is essentially {\it ad hoc} and depends
on the data to be modelled.

\citet{Rasmussen2005} use the product of a squared exponential
and a periodic kernel to model the temporal variation of the CO$_{2}$ concentration
measured in Hawaii. Here the squared exponential term accounts for secular
variation and the periodic term describes the seasonal modulation. The
construction
of the kernel is such that a decay away from strict periodicity is allowed.

\citet{Angus2017} use this kernel to model rotational stellar variability
in photometric time series. Again, the kernel accounts for a periodic
term, in this case caused by spot-induced variation, superimposed on a long-term
secular evolution. The periodic term here accounts for spot-induced
rotational variation while the squared exponential term models
long-term changes in the activity state.

The main difficulty in GP modelling is due to the fact that the
co-variance matrix of the data needs to be inverted many times; for large data sets
containing thousands of data points this is a true challenge.
For our analysis we use the
\textit{python} interface of the \texttt{celerite} algorithm, which is capable of highly efficiently
carrying out a GP regression in one dimension \citep{celerite}.   We use
a co-variance function of the form
\begin{equation}
  k(t_n, t_m)=\frac{a}{2+b}e^{-c\tau}\left[\cos\left(\frac{2\pi \tau}{P}\right) + (1+b)\right] + \delta_{nm} \sigma^2 \;,
  %k(\tau)=\frac{a}{2+b}e^{-c\tau}[\cos(\frac{2\pi \tau}{P}+ (1+b)]
  \label{eq:kern1}
\end{equation}  
where $\tau = |t_n - t_m|$ and $t_i$ denotes the individual points on the time axis.
This choice corresponds to the proposal by \citet{celerite} in their Eq.~56, extended by
a ``jitter'' term, that is, additional, normally distributed white noise.
The co-variance function requires that all of the parameters $a, b, c$, and $P$ are positive.
The \citet{celerite}  kernel is characterised by an exponential decay on
timescale $\tau$ and includes a periodic term with a period, $P$, and a secular
term. The amplitude of the co-variance and the relative impact of the
periodic and secular terms are determined by combinations of the values of the
parameters b and c.  This \citet{celerite} kernel is similar (but not identical) to the one
used by \citet{Rasmussen2005} and \citet{Angus2017}.

In our modelling, the four parameters $a, b, c$, and $P$ are varied to find the maximum of the
likelihood function, which
is accomplished using the ``limited memory Broyden-Fletcher-Goldfarb-Shanno with box
constraints'' (L-BFGS-B) algorithm, an iterative optimisation algorithm belonging to the group of the 
quasi-Newton methods, as implemented in \texttt{SciPy} \citep{scipylib}.

Some value for the period, $P$, will always maximise the likelihood; however, this does not
automatically imply that it is reasonable to consider this value in a physical interpretation. This brings
us into the realm of hypothesis testing. As an alternative to the periodic co-variance function as specified
by Eq.~\ref{eq:kern1}, we can set up a hypothesis without any periodic behaviour. The periodic term in
Eq.~\ref{eq:kern1} is suppressed if the period $P$ is large or, for that matter, infinitely long. 
Taking the limit, Eq.~\ref{eq:kern1} then becomes
\begin{equation}
  k(t_n, t_m)= a e^{-c\tau} + \delta_{nm} \sigma^2 \;.
  %k(\tau)=\frac{a}{2+b}e^{-c\tau}[\cos(\frac{2\pi \tau}{P}+ (1+b)]
  \label{eq:kern2}
\end{equation} 

The relative plausibility of these hypotheses is expressed by the posterior odds, which is
equivalent to the Bayes factor if impartial prior odds are assumed. Evaluating the Bayes
factor requires  a numerical value for the marginal likelihoods to be obtained, which is usually
a troublesome enterprise. However, it can be shown that the so-called
% which
% To estimate how reliable the hypothethis of a periodic behaviour in the data is,
% we computed besides our model with a periodic term a second model with a kernel consisting only of a real
% term and a jitter term.
Schwarz Criterion, 
\begin{equation}
S = \max(\mathcal{\log\,L}_1) - \max(\mathcal{\log\,L}_2) - \frac{1}{2} d \log\,n
,\end{equation}
which asymptotically approximates the Bayes factor \citep[e.g.][]{Kass}. Here, $\max(\log(\mathcal{L}))$ denotes the
maximum of the (logarithm of the) likelihood function under the respective hypothesis, and $d$ represents the difference
in dimensionality, which we take to be two in our case, and $n$ is the number of data points for the specific star.   
A value of about zero for $S$ translates into posterior odds of around one. In that case, the data provide similar
support for both hypotheses.
Positive $S$ indicates support for the periodic model in our case. At which point to accept the periodic
hypothesis depends on the assumed loss associated with a false decision. We here adopt a conservative
limit of $\log(S) = 4.6$, which corresponds to posterior odds of $100 : 1$ in favour of the periodic term.
% the periodic model is hundred times as probable as the non-periodic model. We choose that as our threshold.
In Tables~\ref{periodsha} and \ref{periodsca} we report the resulting periods. Values in agreement with the
results of other methods but unacceptable according to our criterion are given in 
square brackets. The $S$ values are provided in Table \ref{significanceGP}.

\begin{figure}
\begin{center}
\includegraphics[width=0.5\textwidth, clip]{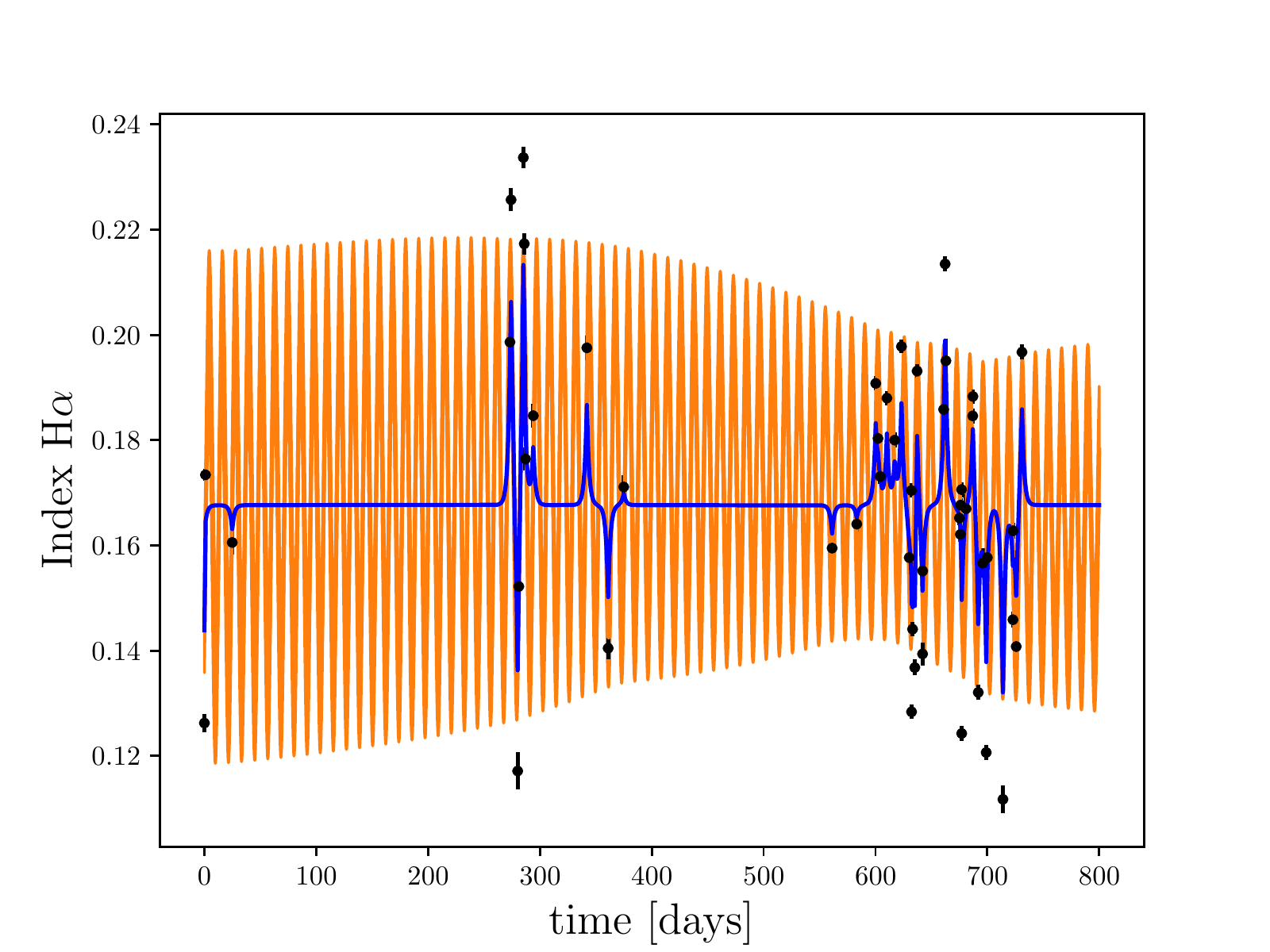}%\vspace{-0.4mm}\\
\caption{\label{fig:gp} H$\alpha$ time index series of V2689~Ori (black data points) along with maximum a
posteriori GP model prediction and standard deviation indicated by
orange/light grey shade. In blue/dark grey the best GP model without periodic term.}
\end{center}
\end{figure}

As an example, we show our GP modelling of 
the H$\alpha$ index time series of V2689~Ori along
with the maximum-likelihood predictions for the mean and standard deviation in Fig.~\ref{fig:gp}.
Moreover we show in blue the best model prediction with the non-periodic kernel.
%\tbd{Sollen wir auch das Ergebnis des nicht-periodischen Ansatzes zeigen?}
This example demonstrates
the ability of the model to account for more complicated long-term trends, which do not have to be removed before
applying the method.
In the Appendix we show similar graphs for all stars.% along with the power spectral distribution (PSD)
%of the GP kernel also comparing it to the PSD for Gaussian noise and a real term %\tbd{Ist hier der andere Kernel gemeint?}.
%TBD:  Sollen wir da nicht Fig. A6 zeigen ?  Wie signifikant ist denn jetzt die Periode f??r V2689 ?

\subsection{Results}\label{sec:results}

Naturally, the period search outcomes provided by the various techniques differ somewhat.  
We claim a solid detection of a periodic signal in a specific star, if we find the same period 
with at least four of the methods (SL, PD, GLS, GP, GLS and GP applied to the alternative index A1) allowing 
for a margin of 10\,\% in the numerical value of the period.
A tentative period is reported in square brackets if the same period is found using three 
methods (allowing also for tentative detections given in brackets).
Otherwise, we consider the signal to be non-significant.

According to this criterion, we report a solid detection of a periodic signal in four out of the 
16 sample stars based on
the H$\alpha$ index and in five stars using the \ion{Ca}{ii}~IRT index.
The GLS periodogram and GP were applied to the line index and the alternative indices defined in Sect.~\ref{sec:alternative}
Our results are summarised in
Table~\ref{periodsha} for the H$\alpha$ indices and in Table~\ref{periodsca} for the \ion{Ca}{ii}~IRT indices.
We also give an
error estimate for the mean periods computed as the standard deviation of the mean of the different methods.
In addition to the found periods we give the difference $\delta P_{\rm lit}$ to the
literature period values given in Table \ref{stars}. In all cases, we find agreement between our periods and those
previously reported to within 3$\sigma$.
Corresponding to the GP modelling we give the Schwarz criterion $S$ in Table \ref{significanceGP}.
For all of our sample stars using the H$\alpha$ line index we show the GLS periodogram, the string length and
phase dispersion distributions, and the modelling of the GP in the Appendix (Figs. \ref{gxand}
to \ref{hd216899}).

All measured periods using the H$\alpha$ and \ion{Ca}{ii}~IRT
indices agree to within 2$\sigma$. This is also the case for the three stars HD~79211, AN~Sex, and BD+11~2576
where we found tentative periods with both H$\alpha$ and the \ion{Ca}{ii}~IRT indices.  Although many methods
do not find the period with the demanded significance, the finding of the same period in two different
indices strongly suggests that this is the correct period. The  case is different for the stars BD+52~857
and HD~216899 where a tentative period is found only in the \ion{Ca}{ii}~IRT indices but not in the H$\alpha$
line indices. Nevertheless, the tentative period of HD~216899 is in agreement with the literature value.

For the stars HD~79211, BD+63 869, and HD~147379
we could derive previously unknown rotation periods, with HD~79211 being a tentative detection and HD~147379
being detected only in \ion{Ca}{ii} IRT.

For the three stars BD+61~195, HD~36395, and BD+63~869 consideration of the H$\alpha$ and \ion{Ca}{ii} indices and
nearly all methods led to the individual period. For V2689~Ori and AN~Sex the same applies, except that we found the same period or multiples thereof
with all methods, and not every method reached the significance threshold (though gave the right period).
These five stars are among the most active in our sample as measured by their median $I_{\rm Ca,1}$ index. Not surprisingly, they
are also the most variable as, for example, measured by
the MAD of $I_{\rm{H}\alpha}$. These stars are members of the young disc population with the only exception being HD~36395, which
is classified as a thin disc member.

For GX And, Lalande~21185, and BD+44~2051A the algorithms in our analysis had the lowest success rate in finding periods.
These three stars are the least active in the sample according to their median $I_{\rm Ca,1}$ index. Expectedly, they are also among
the least variable ones as measured by the MADs of $I_{\rm Ca,1}$ and $I_{\rm{H}\alpha}$ (Table~\ref{activityproperties}).
All three stars are members of the disc or the thick-disc populations (Table~\ref{stars}).
In contrast to the cases of Lalande~21185 and BD+44~2051A, our period search in 
GX~And failed because it shows many formally significant periods instead of none.
This behaviour is caused by the special temporal sampling with many observations taken on a single night
and is discussed in more detail in Sect.~\ref{multiobs}.

For four stars with detected periods, we have also computed expected (maximum) periods from the rotational velocity
(see Table~\ref{stars}). For V2689 Ori and HD~79211 the expected period is slightly smaller than the measured
period. For HD~147379 we find twice the expected period,
which could indicate that by using chromospheric indicators we actually detected an alias of the true period.
In the case of BD+21~652, we recover the photometrically measured period, which is, however, about four times larger
than the one based on the projected rotation velocity, which we attribute to the uncertainty of the
$v\sin i$ value. Lowering $v\sin i$ by two standard errors reconciles the values, which
appears statistically plausible.

The stars Lalande~21185 and HD~119850 have known spectroscopically determined periods, which we could not recover.

We attribute this
to their low activity levels and the resulting
low variability amplitudes. A spectroscopic rotation period has also been determined using the Ca~H and K lines
and the H$\alpha$ line for HD~216899 \citep{SM15}. For this star we
find only a tentative period and this only in $I_{\rm Ca,1}$. Since the star is not as inactive as the preceding two,
we are probably not limited by sensitivity in this case. Instead, a change in the surface spot configuration
provides a possible explanation for the lack of a detection in our data set.

\begin{table*}

  \caption{\label{periodsha} Results of the different period search methods for each star using H$\alpha$. }
\footnotesize
\begin{tabular}[h!]{llll llllll}
\hline
\hline
\noalign{\smallskip}
Star       &  SL      & PD    & GLS             &GP     & GLS A1   & GP A1    & Found period& $\Delta P_{\rm lit}$\\
&   [d]               &   [d]         & [d]            & [d]    & [d]      & [d]      & [d]  & [d]\\
\hline
GX And     &   94.9,86.4         & many          & many           & ...    & many      & ...     & no period & ...\\   
BD+61 195  &   18.9              & 18.8          & 18.8,18.3      & 18.8   & 18.9      & 18.8    & 18.8 $\pm$ 0.2 & 0.4, 1.1\\
BD+21 652  &   ...                & 50.5          & 24.3,25.5,27.7 & [24.5] & 56.6,25.1 & [25.0]  & [25.3$\pm$ 1.1] & 0.1\\
BD+52 857  &   ...               & ...           & 28.3,23.8,22.2 & ...    & 16.5,19.9 & ...     & no period & ...\\
HD 36395   &   ...               & 33.4          & 34.0           & 34.1   & 34.1,33.9 & 33.9    & 33.9 $\pm$ 0.4 & 0.1, 0.3\\
V2689 Ori  &   [11.8]            & 11.7, 23.4     & 11.8           & 11.8   & 12.2,11.7 & 11.7    & 11.8 $\pm$ 0.2 & 0.2, 0.5\\
HD 79211   &   ...               &52.3, [16.7]   & 23.6,33.3      & [17.4] & 16.5      & [16.4]    & [16.6 $\pm$ 0.5]& ...\\
BD+63 869  &   17.8              &17.8           & 17.8           & 17.8   & 17.7      & 17.8    & 17.9 $\pm$ 0.3& ...\\
AN Sex     &   ...               &[21.4]         & 21.3           & [21.7] & 21.3,17.9 & [19.4]    & [20.1 $\pm$ 1.5]& 1.5, 1.5\\
Lalande 21185&      5.8,7.9      & 97.5          & 150            & ...    & ...       & ...   & no period & ...\\
BD+44 2051A&   ...               & ...           & ...            & ...     & ...       & ...        & no period& ...\\
BD+11 2576 &   30.7              &30.7           & [31.3]         & [31.0] & ...       & ...     & [30.8 $\pm$ 0.3]& 2.8\\
HD 119850  &   many              &88.8           & ...            & ...     & 46.2,51.4 & ...     & no period & ...\\
HD 147379  &   ...               & 73.2          & ...            & [21.9] & 20.8      & [21.5]    & no period & ...\\
HD 199305  &   ...                & ...            & ...            & ...     & ...       & ...   & no period & ...\\
HD 216899  &   ...               &89.2,39.1      & ...            & ...     & 18.3,19.2 & ...   & no period & ...\\

\noalign{\smallskip}
\hline

\end{tabular}
\normalsize

\end{table*}

\begin{table*}

  \caption{\label{periodsca} Results of the different period search methods for each star using \ion{Ca}{ii} IRT. }
\footnotesize
\begin{tabular}[h!]{llll llllll}
\hline
\hline
\noalign{\smallskip}
Star       &  SL     & PD    & GLS             & GP   & GLS A1    & GP A1  &Found period& $\Delta P_{\rm lit}$\\
           &   [d]              & [d]           & [d]            & [d]   & [d]       & [d]    & [d]  & [d] \\
\hline
GX And     &   ...              & ...           & many           & ...  & many       & ...   & no period & ...\\   
BD+61 195  &   [18.9]           & 18.8          & 18.8,18.3,20.4 & 18.7 & 18.9       & 18.9    & 19.1$\pm$0.7 & 0.8,0.7\\
BD+21 652  &   137.2            & 27.6, 27.4    & 27.7 +sidemax  & 24.7 & 25.6       & [25.4]    & 25.9$\pm$1.0 & 0.5\\
BD+52 857  &   ...              & 69.9,22.0,15.8& 22.2,23.8,28.3 & 16.7 & 30.3       & [16.7]    & [16.5$\pm$ 0.3]& ...\\
HD 36395   &   33.4,34.2        & 34.1          & 34.3           & 34.3 & 34.5       & 34.3    & 34.1$\pm$ 0.5 & 0.3, 0.5\\
V2689 Ori  &   [11.8]           & [11.7]        & [11.8]         & 11.8 & 11.8       & 12.2    & [11.9 $\pm$ 0.2] & 0.4, 0.1\\
HD 79211   &   ...              &49.6,29.4      & 16.9,23.8      & [19.3] & 18.1       & [17.1]    & [17.4$\pm$1.0] & ...\\
BD+63 869  &   17.8             &17.7           & 17.8           & 17.9 & 17.7       & 17.8    & 17.8$\pm$ 0.1 & ...\\
AN Sex     &   ...              &21.3           & 21.3, 10.7     & [21.9] & 21.3,10.7  & [22.0]    & [21.1 $\pm$ 0.4] & 0.5, 0.5\\
Lalande 21185&      97.5             & ...           & 20.2           & ...& ...        & ...   & no period & ...\\
BD+44 2051A&   ...              & 137.9         & ...            & ... & ...        & ...   & no period & ...\\
BD+11 2576 &   ...              &30.1           & [30.3,62.5]    & ... & [30.5]     & [30.3]      & [30.3$\pm$] 0.2 & 2.3\\
HD 119850  &   ...              &47.9           & 48.3           & ...  & ...        & ...   & no period & ...\\
HD 147379  &   ...              &20.9           & 21.3           & 21.6 & 21.3       & 21.4      & 21.4$\pm$ 0.4 & ...\\
HD 199305  &   ...              &53.1           & ...            & ...  & ...        & ...   & no period & ...\\
HD 216899  &   38.7             &89.2,38.7,44.8 & 45.7           & [40.1] & ...        & [40.6]      & [39.2$\pm$0.4] & 0.3,1.7\\

\noalign{\smallskip}
\hline

\end{tabular}
\normalsize

\end{table*}

\begin{table}

  \caption{\label{significanceGP} Schwarz criterion $S$ for the GP modelling. }
\footnotesize
\begin{tabular}[h!]{lcccc}
\hline
\hline
\noalign{\smallskip}
 Star  & $S_{I(H\alpha)}$& $S_{I(\ion{Ca}{ii})}$ & S$_{\mathrm{pEW}(H\alpha) \mathrm{A1}}$  & S$_{\mathrm{pEW}(\mathrm{Ca}) \mathrm{A1}}$\\

\noalign{\smallskip}
\hline
\noalign{\smallskip}
GX And    & -8.0 & -6.4 & -1.1 & -2.9\\
BD+61 195 &  9.0 &  5.1 &  4.8 & 11.6\\
BD+21 652 & -0.9 &  7.9 & -0.5 & 3.2 \\
BD+52 857 & -5.5 &  4.8 & -3.8 & 1.4 \\
HD 36395  &  9.7 & 14.9 & 10.0 & 8.2\\
V2689 Ori &  5.6 &  4.5 &  6.5 & 3.1 \\
HD 79211  & -4.6 & -3.1 & -1.8 & 1.4 \\
BD+63 869 & 8.9  & 6.9  &  5.4 & 3.9 \\
AN Sex    & -4.1 & -3.6 & -2.6 & -7.3\\
Lalande 21185&-3.2&-12.7 & -4.8 & -6.8 \\
BD+44 2051A& -6.9 & -3.8 & -3.1 & -3.1\\ 
BD+11 2576& -4.0 & -3.0 & -1.9 &  0.7\\
HD 119850 & -5.5 & -5.4 & -5.2 & -2.5\\
HD 147379&   2.7 & 17.1 & -0.2 & 14.8\\
HD 199305 & -2.3 & -2.9 & -3.6 & -4.9\\
HD 216899 & -2.0 & -1.5 & -2.2 & -1.2\\

\noalign{\smallskip}
\hline

\end{tabular}\\
  \end{table}

In our analysis using GLS periodograms we also investigated the window function to
identify frequencies with high power introduced
by the temporal sampling of the observations. We did not find any significant peak in the power
of the window function. The highest peak therein is around two days for many of our stars, which
would lead to aliases around two days again for the found rotation periods.
Therefore, the sampling seems to be no problem here.

While the period search based on the H$\alpha$ and Ca~IRT lines provides encouraging results,
the measurement of the rotation periods using the
\ion{Na}{i}~D line indices turned out to be less successful. We elaborate on the 
possible reasons for this finding in Sect.~\ref{sec:nad}.
While the string length method and the phase dispersion minimisation perform worst for our sample,
the GP modelling is the clear winner among the studied algorithms in terms of detection performance;
see Sect. \ref{Sec:comparisonmethod}.

\section{Discussion}\label{sec:discussion}

\subsection{Connection of activity properties to period finding}
Since measurements of the rotation are only possible if the star exhibits a
non-uniform pattern of H$\alpha$ or \ion{Ca}{ii} plages that imprints itself
in periodic variations in the EW of the chromospheric lines, it is not
possible to measure rotation if the amplitude of these variations is not high enough.
For the case of low- or medium-active stars  the activity variations are not
dominated by micro-flaring and other intrinsic variability, but should be dominated by
plage evolution and rotation.
Therefore the activity properties of the individual stars should
provide some insight into whether or not it is probable that a period will be found.

Inspecting the median($I_{\rm Ca,1}$) we notice that for the seven stars with
values of 0.21 or smaller there was always
a period found (though for some stars only tentative or in only one index).
For the four stars with values of 0.26 or higher we do not find a period. For the
five stars with intermediate values, the situation is unclear.
For the median($I_{\rm H\alpha}$), the situation is also unclear since it can vary quite a lot
for the same values in median($I_{\rm Ca,1}$), which may be caused by different filling factors
of plages and filaments. 
This is in line with the finding that for the three stars with the worst
correlation between H$\alpha$ and \ion{Ca}{ii} IRT indices we cannot establish a period.

Moreover, we could find a (tentative) period in all cases where the MAD(Ca) is 0.004 or higher.
For the nine stars with a lower MAD(Ca) one may or may not find periods.
Also, for the MAD(H$\alpha$) the situation is less clear but there is the tendency to find a period for high
MAD(H$\alpha$), while for lower values finding a period becomes harder and harder.
Interestingly the low correlation between H$\alpha$ and \ion{Ca}{ii} IRT indices is only  coupled to the lowest MAD in \ion{Ca}{ii} as well as in H$\alpha$ only for GX~And. If there is no variation,
no correlation can be found either.

A period
could be detected in all stars with a correlation between H$\alpha$ and \ion{Na}{i} D.
A correlation between these two lines, one formed in the upper chromosphere, the other
in the lower chromosphere, may indicate plage configurations which are especially stable or of high
contrast or both.
For the low-activity stars in our sample it appears that a period is more likely to be found if the
star is among the more active ones as measured by median($I_{\rm Ca,1}$), MAD(Ca), and MAD(H$\alpha$), since
in these rather inactive stars the variability is probably dominated by rotational effects. In contrast,
for the most
active mid-type M dwarfs the opposite is expected, and the chances of finding a period
are lower because flaring variability  will dominate over rotational variability for these stars.

\subsection{Comparing the width of the H$\alpha$ central wavelength band}

Since the width of the H$\alpha$ line can vary a lot, different widths of the
central waveband are used in the literature to account for this problem. We therefore
 investigate whether or not these different widths have an influence on the period finding.

\begin{figure}
\begin{center}
\includegraphics[width=0.5\textwidth, clip]{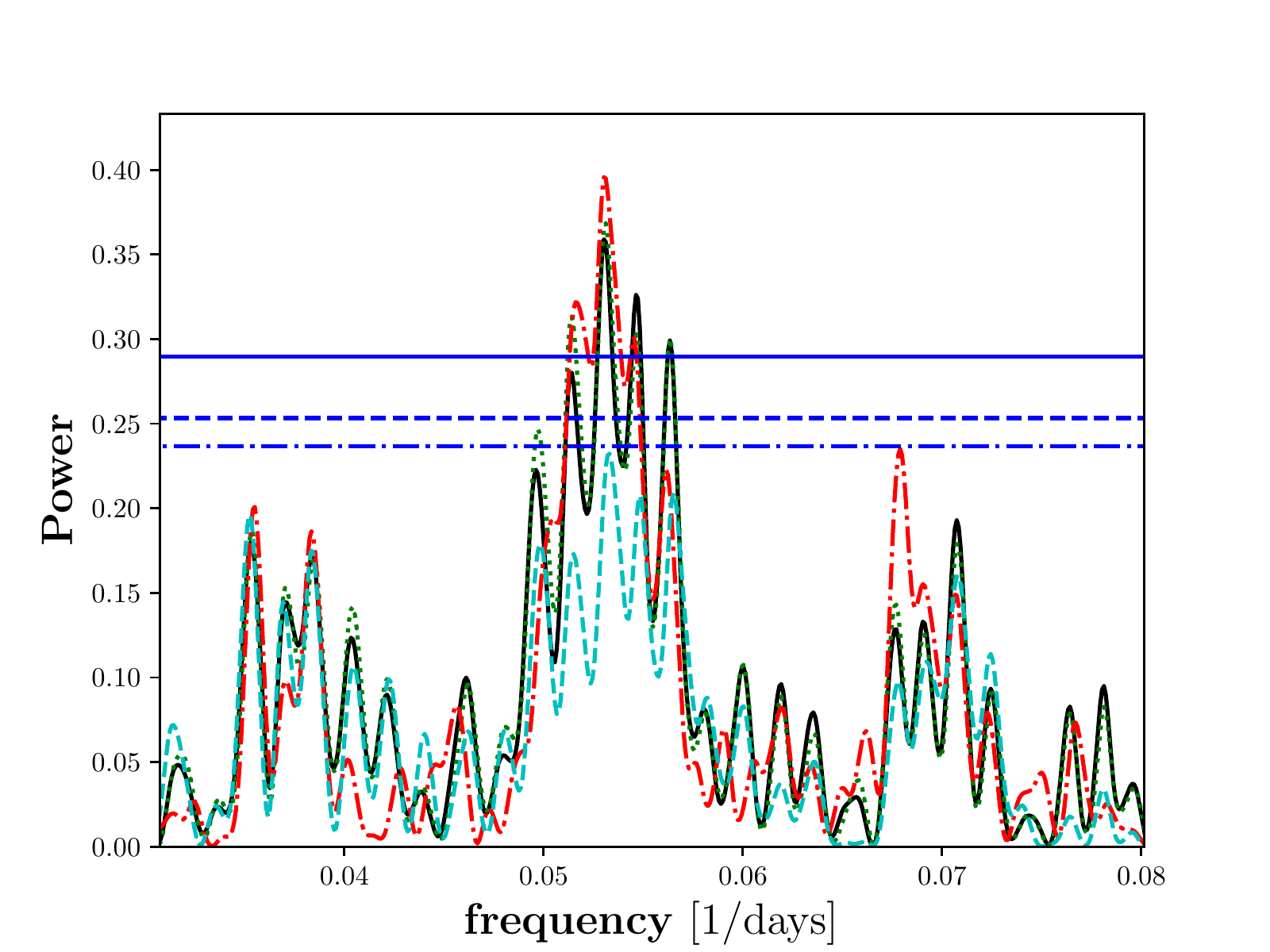}\\
\caption{\label{halphadiffcentralwavebands} Periodogram comparing
  the different central wavelength ranges of H$\alpha$ for BD+61~195. The pEW
  A1
  is designated by the black line, the core pEW A2 is in green/dotted, the
  pEW with broad central wavelength range A3 is in cyan/dashed, and the pEW with variable
  width A4 is in red/dot-dashed. The horizontal blue lines show the FAP levels of 0.1 (dash-dotted), 0.05(dashed), and
0.01 (solid).}
\end{center}
\end{figure}

\begin{figure}
\begin{center}
\includegraphics[width=0.5\textwidth, clip]{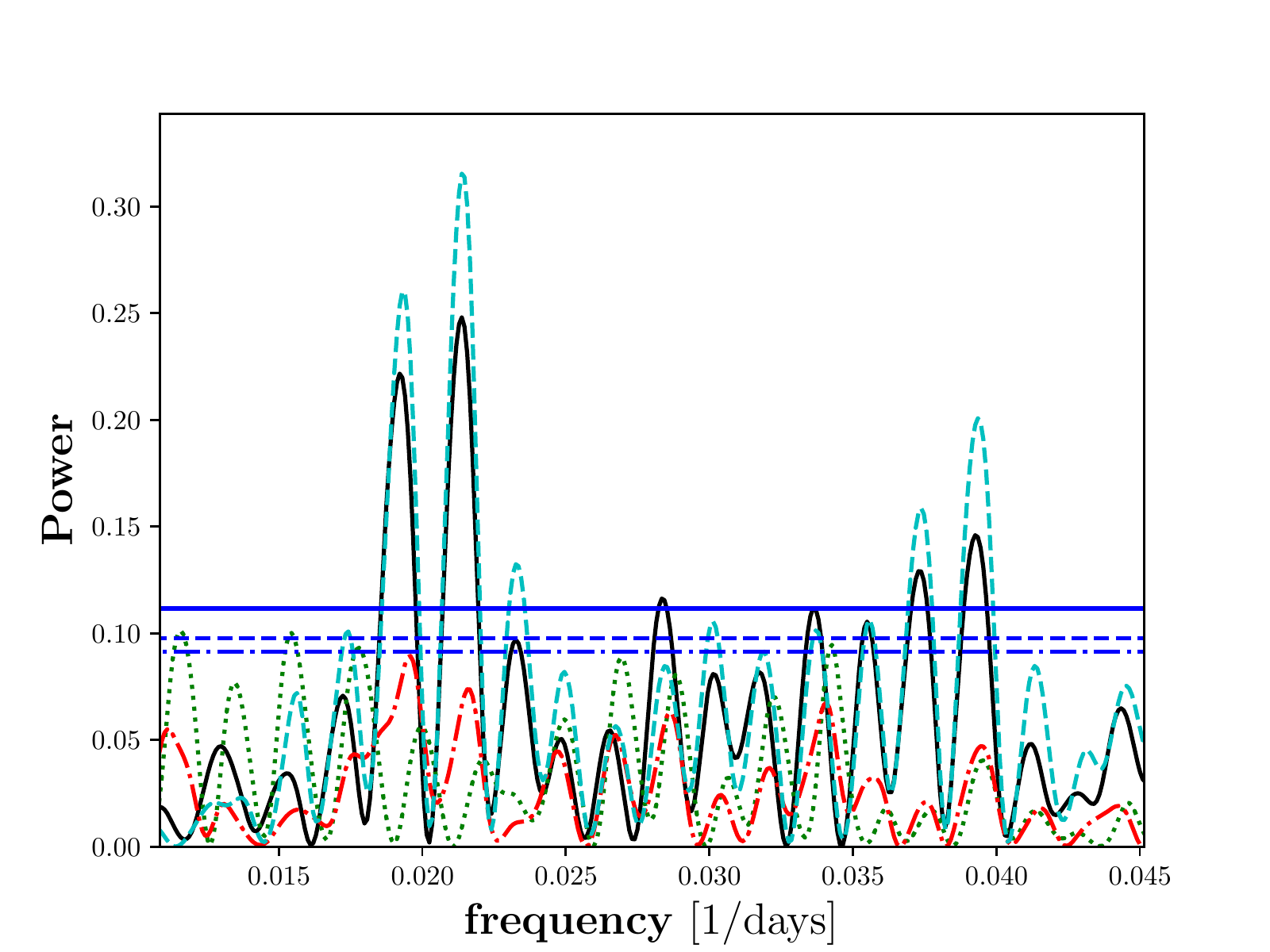}\\
\caption{\label{halphacorepew} Periodogram comparing
  the central wavelength ranges of H$\alpha$ for HD~119850 and with and without detrending.
  The pEW with a width of 1.6 \AA\,
  is designated by the green/dotted line for no detrending and by the red/dash-dotted line in case of detrending.
  The pEW of the line core is in black/solid for no detrending and in cyan/dashed for detrending.
  The horizontal blue lines show the FAP levels of 0.1 (dash-dotted), 0.05 (dashed), and
0.01 (solid).}
\end{center}
\end{figure}

In Fig. \ref{halphadiffcentralwavebands} we show a comparison using a GLS periodogram
and no detrending of the data of BD+61~195. While the alternative pEW A1,
the pEW with variable width A4, and the core pEW A2, all show  high significance
at the known period of $\sim$19 days, the pEW with the broadest central wavelength band (A3) does not.
This lower performance
of A3 is quite typical for the whole sample. Additionally it may
lead to spurious peaks at periods higher than 100 days. A1 and A4
normally perform about equally well. The A2 core pEW
shows quite different sensitivity to rotation and sometimes performs better,
but also sometime much worse than the other two. This may be expected since it is
more sensitive to smaller variations in H$\alpha$ which normally occur first
in the core. This leads to a better performance of A2 in cases of 
variations caused by rotational modulation.
We demonstrate this using the example of HD~119850 in Fig. \ref{halphacorepew}.
This star has a photometrically determined period of 52.3$\pm$ 1.3 days \citep{SM15}, but we do not find any
period. The star is among the most inactive of our sample.
We compare  A2 and  A1 with and without
detrending. While using A1, the period is not found regardless of whether detrending is used, whereas in the A2 data a signal corresponding to the photometric period is clearly seen.
Unfortunately, this does not work for the other very inactive stars, GX~And, Lalande 21185, and BD+44~2051A.

\subsection{The \ion{Na}{i} D lines}\label{sec:nad}

\begin{figure}
\begin{center}
\includegraphics[width=0.5\textwidth, clip]{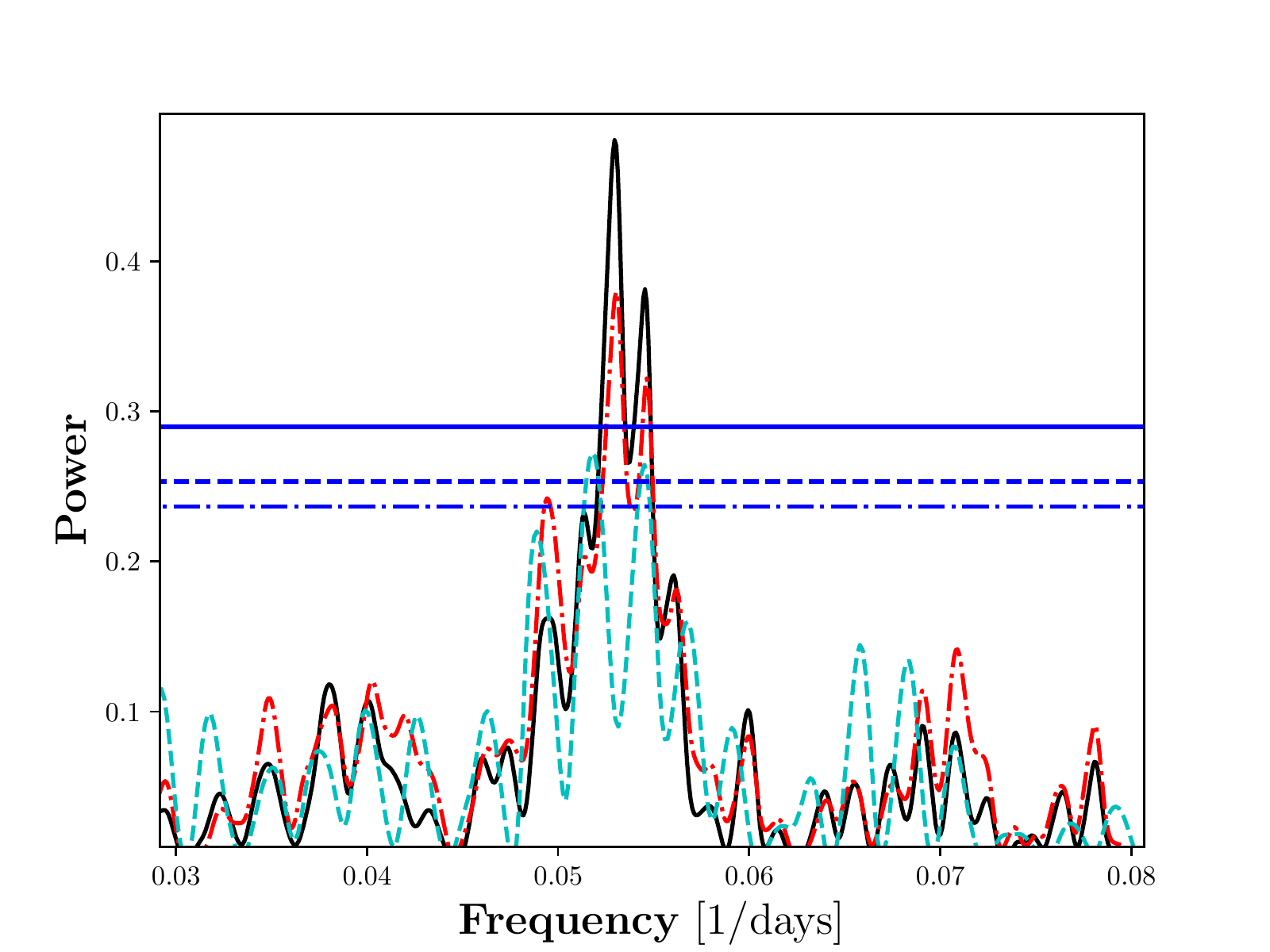}\\
\caption{\label{sodium} GLS periodogram for BD+61~195 comparing H$\alpha$
  in black/solid, \ion{Ca}{ii} IRT in red/dash-dotted, and \ion{Na}{i} D in cyan/dashed.
  The horizontal blue lines show the FAP levels of 0.1 (dash-dotted) , 0.05 (dashed), and
0.01 (solid).}
\end{center}
\end{figure}

Besides H$\alpha$ and \ion{Ca}{ii,} we considered the \ion{Na}{i} D lines for
our period search. For mid M dwarfs the \ion{Na}{i} D lines are 
known to develop emission cores and are therefore clearly also sensitive to
activity. A big obstacle in their usage for period analysis is the \ion{Na}{i} D
airglow lines. This terrestrial emission is variable in time and can
be relatively strong. 
%Moreover, there may be a contribution by street lamps.
Therefore, we did not try to correct this emission,
but we excluded all spectra from the analysis where the airglow may influence
the central wavelength band of the \ion{Na}{i} D line. Unfortunately this leads
to rather small numbers of available spectra (see Table \ref{activityproperties})
for all stars where we found rotation periods besides BD+21~652. For this star the
significance in the GLS periodogram for \ion{Na}{i} D unfortunately stays below a FAP of 0.1 at the 
period found. One should nevertheless keep in mind that BD+21652 is a rather inactive star; in fact
the most inactive star for which we could establish a period.
Generally, the low correlation of
H$\alpha$ and \ion{Na}{i} D  and the low MAD(Na) values may indicate that the \ion{Na}{i} D
line is not as sensitive to activity as H$\alpha$ and therefore could only be used for
more active stars. For a more active star on the other
hand, namely BD+61~195, where we have only 37 usable spectra
for the \ion{Na}{i} D periodogram computation, the highest peak is nearly
significant as can be seen in Fig. \ref{sodium}. This may promise
that for stars with higher activity level (as indicated e.g. by the MAD(Ca))
an analysis using \ion{Na}{i} D can be accomplished once enough spectra
not contaminated by airglow are available.

\subsection{Influence of detrending, flaring activity, and multiple observations per night}\label{multiobs}\label{Sec:flaring}
The handling of detrending, flaring activity, and multiple observations per night
  is theoretically expected to potentially hinder the period search, since all three
  have a negative impact on the systematic `noise' of the time series.

Regarding the detrending, it can be seen in Fig. \ref{halphacorepew} that detrending often
leads to a higher significance of the period. In cases of weaker trends in the time series
we do not, however, find an influence.
Detrending may also lead
to small shifts in the period, with the largest shift we found from 33.0 without
to 34.2 days with detrending (in HD~36395).
We conclude that detrending the data
leads normally to an equally good or better performance of the GLS periodogram.

We want to make clear nonetheless that these trends removed by detrending are interesting
in themselves as they may represent parts of activity cycles. Unfortunately the time baseline of the CARMENES
data up to now is not long enough for a reasonable search of such long-period signals.
If one looks at the index time series directly, striking examples of nearly linear trends in both $I_{H\alpha}$ and
$I_{Ca IRT,1}$ may be found in BD+52~857, BD+11~2576, and HD~147379, while BD+21~652, HD~36395, and V2689~Ori
show more parabolic trends.

%\begin{figure}
%\begin{center}
%\includegraphics[width=0.5\textwidth, clip]{detrending.pdf}\\
%\caption{\label{detrending} Periodogram comparing
%  the central wavelength ranges of H$\alpha$ for HD~36395 with and without detrending.
%  The pEW A1,
%  is shown by the black line for no detrending and by the red line in case of detrending.
%  The A2 is in green for no detrending and in blue for detrending.
%  The horizontal dashed lines show the FAP levels of 0.1, 0.05, and
%0.01.}
%\end{center}
%\end{figure}

%\subsection{Influence of flaring activity}\label{Sec:flaring}

\begin{figure}
\begin{center}
\includegraphics[width=0.5\textwidth, clip]{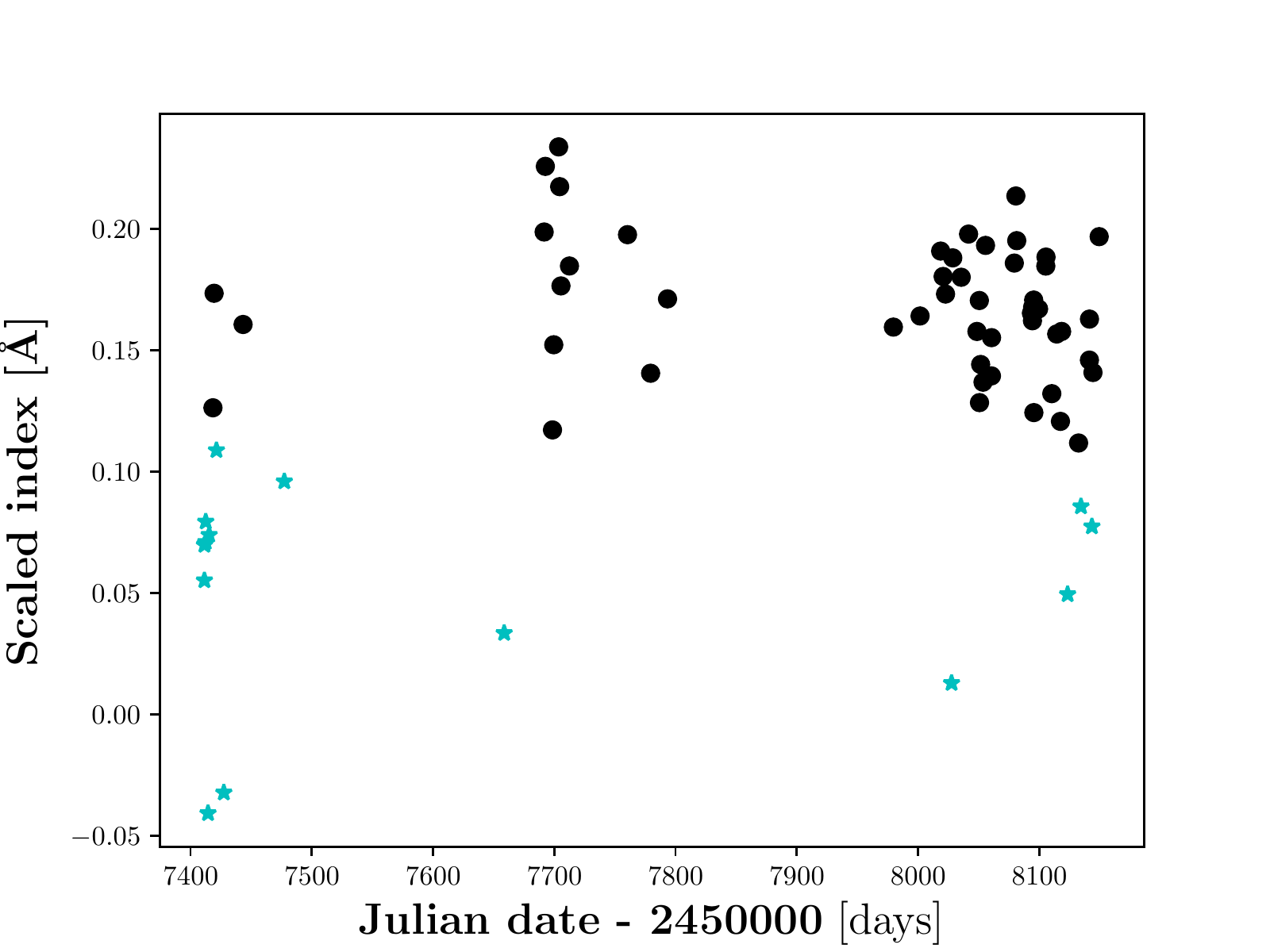}\\%\vspace{-0.4mm}\\
\includegraphics[width=0.5\textwidth, clip]{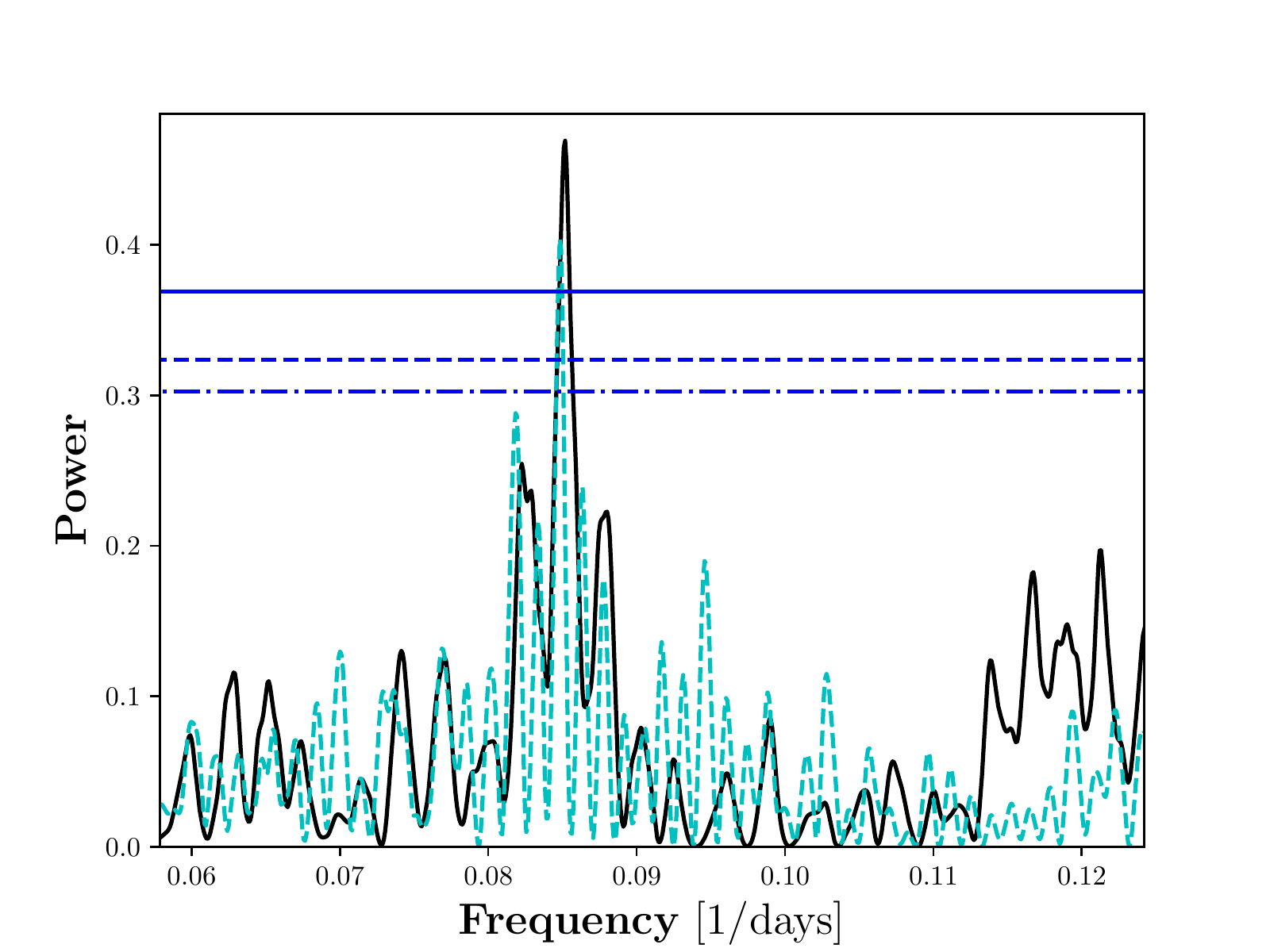}%\vspace{-0.4mm}\\
\caption{\label{fig:flaring} Influence of flaring spectra for V2689~Ori. Top: Index time series of $I(H\alpha)$ with (cyan) and without (black) flare-flagged spectra.
Bottom: Comparison of the GLS with inclusion of flare-contaminated spectra
(cyan/dashed) and without (black/solid). The horizontal blue lines show the FAP levels of 0.1 (dash-dotted) , 0.05 (dashed), and
0.01 (solid).}
\end{center}
\end{figure}

In the case of flaring one can test the sensitivity of the search algorithm with respect to outliers.
For the star V2689~Ori we had to exclude a significant number (22\,\%) of spectra flagged as taken when the star was flaring.
To infer how important the exclusion of such outliers is,
we recomputed the period of the star using the different methods on the
full index time series including the spectra flagged due to flaring activity. Surprisingly we could recover the period
without problems. The string length and phase dispersion method both had the problem of heavy aliasing as
when using the clipped index time series. The double period is found again significantly while even higher
multiples and the
period itself give lower significance. As an example we show the recomputed GLS in Fig. \ref{fig:flaring}.
Interestingly,
the power computed including the spectra affected by flares shows a peak nearly as high as the power computed without
them.
For the \ion{Ca}{ii} IRT this effect is even more remarkable since the highest peak becomes statistically significant
with FAP < 0.01 when including the flare spectra.
This leads us to the hypothesis that for V~2689 Ori our flare criterion is not working well and instead a high amplitude in rotational variation triggers our flare criterion so that we
exclude spectra not flaring but being the peaks in the rotational index time series.
If one were to examine the index time series in Fig. \ref{fig:flaring}
by eye, we would mark only three to four spectra as flaring -- a number so
low that we would expect the period search algorithms to work correctly despite these outliers. 

Since V2689~Ori is our most active star and the only one where we
flagged such a high number of spectra as flaring, we regard it a special case in our sample. However, this also
leads to the conclusion that the flare criterion has to be defined with care.

%\subsection{Multiple observations in single nights}\label{multiobs}
Having examined the influence of detrending and flaring, we 
now discuss multiple observations on single nights.
Such observations may be problematic, if the star is intrinsically variable on short timescales and especially
if the amplitude of these changes in the chromospheric lines are of the same order as the rotational changes.
All stars have at least one night when more than one spectrum was taken as given in Table \ref{stars},
but for most of the stars this is a minor aspect, and so intrinsic variability can again be seen as a small number of outliers.
On the other hand, we have one extreme case in our sample: GX~And with more than half
of the spectra taken on only a few nights and up to 22 spectra taken on one night. Since for this star 
intra-night variations are easily noticed, the questions arises of whether this flickering has an amplitude hampering the
period ana\-lysis. We therefore recomputed our period ana\-lysis after removing all spectra but one from each night with
multiple observations, maintaining only the spectrum with the median H$\alpha$
index.  Figure 9 shows the comparison between the ana\-lyses including multiple
observations per night, and those using only single observations, for the GLS periodogram for GX~And and for the more
typically behaving star
BD~21652 (which is nevertheless among the four stars with many multiple observations on single nights).

\begin{figure}
\begin{center}
\includegraphics[width=0.5\textwidth, clip]{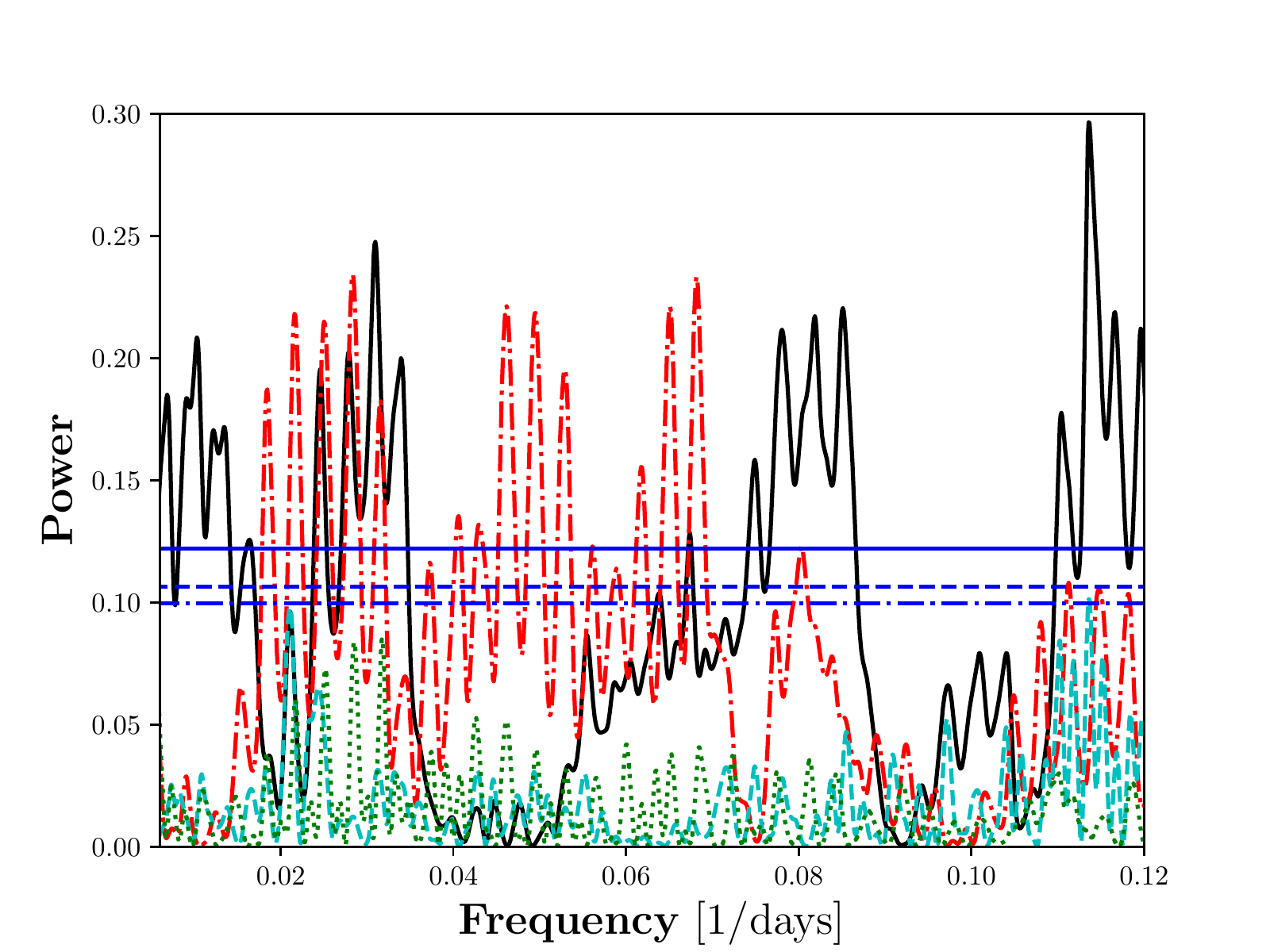}\\%\vspace{-0.4mm}\\
\includegraphics[width=0.5\textwidth, clip]{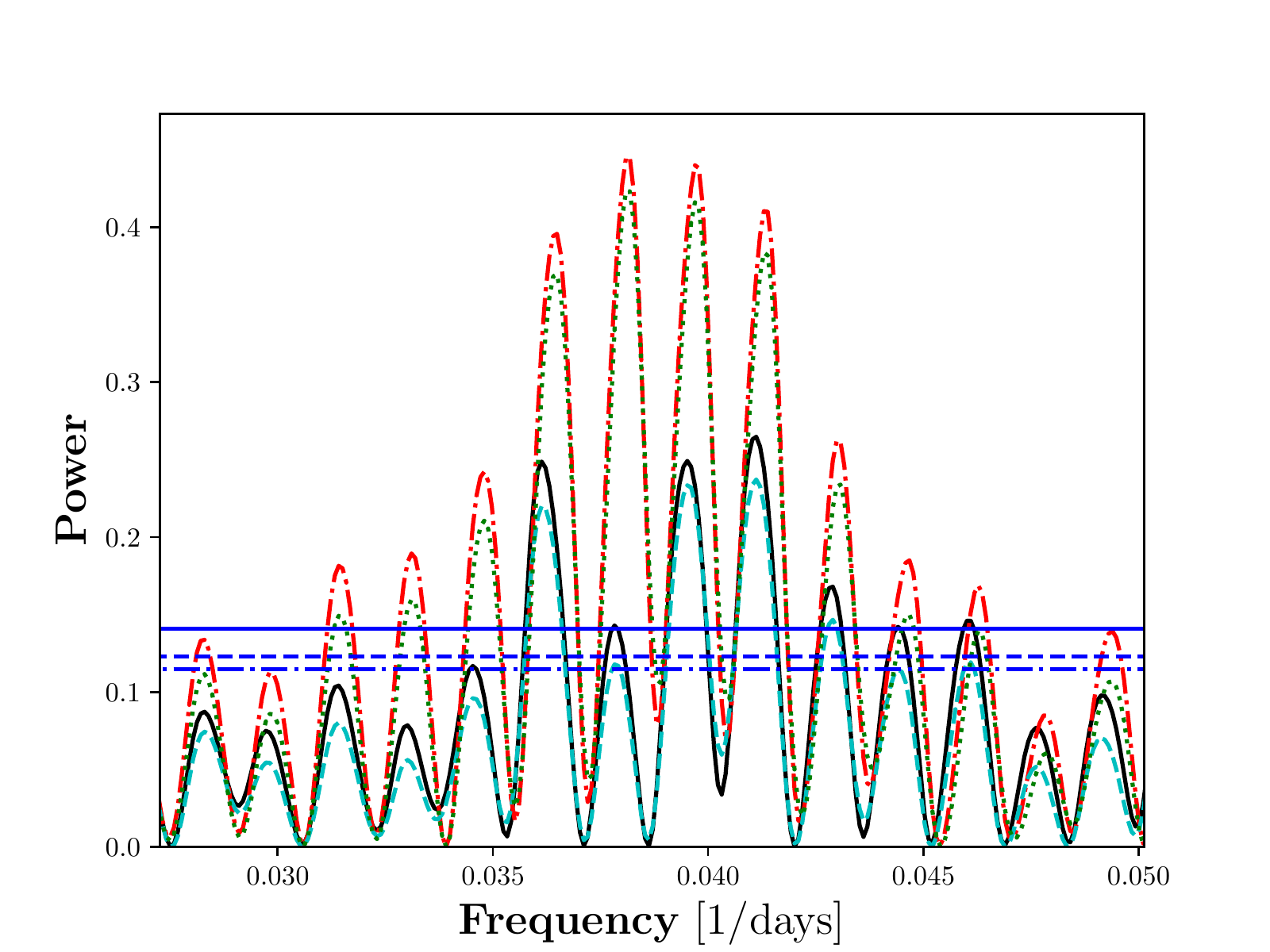}%\vspace{-0.4mm}\\
\caption{\label{fig:multinight} Influence of multiple spectra taken on one night on the GLS periodogram. Black/solid and red/dash-dotted denote the
  power for $I_{\rm{H}\alpha}$ and $I_{\rm Ca,1}$, respectively,  including multiple spectra. Cyan/dashed and
  green/dotted denote the power for $I_{\rm{H}\alpha}$ and $I_{\rm Ca,1}$, respectively,  excluding multiple spectra.
  The horizontal blue lines show the FAP levels of 0.1 (dash-dotted) , 0.05 (dashed), and
  0.01 (solid).
  Top: The extreme case of GX~And. 
Bottom: The typical case of BD~21652. }
\end{center}
\end{figure}

The GP is slightly more affected in analyses using multiple spectra from one night. For the majority of
stars, the period stays the same or is only shifted by less than 5\,\% when the multi-observations are
removed. Nevertheless, instead of the period of V2689~Ori, the threefold multiple is found and
for the two stars BD~52857 and GX~And significant periods of 21.8 and 46.2 days are measured if the multiple
observations on single nights are excluded. 

\subsection{Comparison to other algorithms}
There is ample literature on period searching and many algorithms have been introduced accordingly. Below, we discuss two variations of the methods tested here, which have shown
promising results in previous studies.

The first is the conditional entropy (CE) algorithm by \citet{Graham1}, which has been shown to perform well in a variety of astrophysical period-finding problems \citep{Graham}.
The CE algorithm belongs to the family of phase-folding algorithms and is thus related to the string length and the phase dispersion algorithms. The specific idea here is to arrange the data in a unity square of phase and
  normalised variation and quantify the degree of order in the resulting distribution via
  CE, which is to be minimized. We tested this algorithm
  on our $I_{\rm H\alpha}$ data using the same significance criterion as for
  the string length method.
  We were able to recover
  the periods of BD+61~195, HD~36395, and BD+63~869. We also found periods consistent with our previous results for BD+21~652, V2689~Ori, BD+11~2576, and HD~216899; however, these findings remained formally insignificant.
  Moreover, a significant period of 145~days was detected for Lalande~21185, which did not show up earlier in our analysis. The CE method clearly shows a sensitivity profile, that is similar but not identical to those of the string length and phase dispersion minimisation algorithms. On our data set, we found the performance of the CE method to be comparable to
  that of the string length and the phase dispersion algorithms. We note, however, that superior performance may still be
  achieved in other problems such as those studied by \citet{Graham}.

Secondly, we test the performance of a different quasi-periodic kernel in the GP modelling. 
In particular, we apply the kernel also adopted by \citet{Angus2017} in their periodicity study of \texttt{Kepler} light curves\footnote{See also \citet{Rasmussen2005} for a discussion of this kernel.}. Here we used the \texttt{python} module \texttt{george}, because
% As regarding the GP we wanted to test additionally what influence the
% choice of the quasi-periodic kernel has. Since the quasi-periodic kernel used by \citet{Angus2017} worked well on
%   simulated and on real , we decided to use it
%   alongside the above described \texttt{celerite} kernel as a cross-check.
this kernel cannot be combined with the fast matrix-inversion method applied by \texttt{celerite}.
% and we used it therefore in the context of \texttt{george} as \citet{Angus2017}
% did.
  Based on the H$\alpha$ data and the alternative kernel, we could reproduce
  all periods found with \texttt{celerite} with the exception of those for V2689~Ori and AN~Sex.
  Three periods identified tentatively using \texttt{celerite}, namely those for BD+21~652, HD~79211,
  and HD~147379, showed up as significant
  using the alternative kernel. Moreover, a significant period of 50.5 days was found for
  HD~119850, where we were not previously capable of assigning a period, although a spectroscopically
  determined period of 52.3 days is known \citep{SM15}. For the \ion{Ca}{ii}
  IRT data the situation is similar: the period for HD~119850 was reproduced again,
  the periods for V2689~Ori and AN~Sex were not found, for HD~216899 half the period
  showed up significantly, and for HD~36395 a period neither consistent with our previous results nor
 with  the literature was found, which we consider likely spurious.
  Again, we conclude, that the results do not allow us to claim superiority for one or the other kernel. In fact, the two kernels perform similarly well but show an individual sensitivity
  profile on our data set.

At least as far as our data are concerned, the above results indicate that the concept behind the algorithm
  affects their performance more strongly than the details of their implementation. However, as shown for instance by
  \citet{Graham1}, more pronounced differences may well show up when other data sets are studied. Therefore, we caution
  that this conclusion is not generally applicable.

\subsection{Comparison to another CARMENES study}
\citet{Patrick} studied activity indicators for  332 M stars of the CARMENES sample. The main focus was on the question of which line indices could be used
best as activity indicators and which ones are also suited for period searches.
Therefore, using a GLS periodogram and different
line/molecular indices in the optical and infrared a period search was
conducted for 154 stars
with known periods. More than 10\,\% of the known periods could be
recovered using the molecular
indices of TiO 7050 \AA\, and TiO 8430 \AA, H$\alpha$, and the second line of the
\ion{Ca}{ii} IRT. The bluest and the reddest lines of the \ion{Ca}{ii} IRT
were used with less success, as were the \ion{Na}{i} D lines and a number of other
indices for chromospheric active lines. Nevertheless, a direct comparison to our study
shows that of the four periods we found in H$\alpha,$ all were
detected in the GLS periodogram and all were also found by \citet{Patrick}. For the blue line of the \ion{Ca}{ii}
IRT they recover periods for all stars where we also found periods besides BD+61~195.
Therefore our results agree quite well with \citet{Patrick} for the stellar sample used here, though
they used a quite different definition of the activity indices with
a\, broad central wavelength range of  5 \AA\  in combination with the subtraction
of the spectrum of an inactive star.

\subsection{Comparison of the methods}\label{Sec:comparisonmethod}
To determine which of the methods used here works best, we counted (i) the false positive detections, which are
either detections where no (tentative) period was assigned in the end or the values of the detected periods
were not within 10\,\%  of the assigned period (also counting double and half periods)
and (ii) period non-detections where in the end a (tentative) period was assigned. The numbers
of false positives and of non-detections can be found in Table \ref{methodcomparison}. 
These numbers imply that the string length method and phase dispersion minimisation are the least powerful
methods used in this study, whereas the GP is clearly the best. Moreover, we sum all non-detections
with all methods in the H$\alpha$ line (which are 5) and compare them to \ion{Ca}{ii} IRT (with a sum of 6
non-detections). These numbers are almost equal. We do the same with the false positives and find that the sum
using the H$\alpha$ line is 17, while using the \ion{Ca}{ii} IRT line index one finds a sum of 13.
Therefore, the \ion{Ca}{ii} IRT index seems to perform slightly better in the period search, but these numbers
are dominated by the less powerful methods. For the GP, whether one activity indicator is better
than another is totally unclear
 because they both perform very well, that is, without incorrectly predicted periods and with all periods being found.

\begin{table}
  \caption{\label{methodcomparison} Numbers of false positives and non-detections for the different methods. }
\footnotesize
\begin{tabular}[h!]{lllll}
\hline
\hline
\noalign{\smallskip}
Method  & \multicolumn{2}{c}{False positive} & \multicolumn{2}{c}{Period not found}\\
        & H$\alpha$ & \ion{Ca}{ii} IRT & H$\alpha$ & \ion{Ca}{ii} IRT\\
\noalign{\smallskip}
\hline
\noalign{\smallskip}
SL & 3 & 2 & 4 & 5 \\
PD & 6 & 4 & 0 & 0 \\
GLS & 4 & 5 & 0 & 0\\
GP & 0 & 0 & 0&0\\
GLS A1 & 4 & 2 & 1 & 1 \\
GP A1 & 0 & 0 & 0 & 0 \\
\noalign{\smallskip}
\hline

\end{tabular}\\
\end{table}

\section{Conclusions}\label{sec:conclusion}

Out of the more than 10\,000 high-resolution spectra that  CARMENES have obtained so far, 
we compiled a list of 1861 spectra from 16 stars with low to moderate 
activity without strong flaring and with at least
50 observations per star.
We define five spectral chromospheric indices: for H$\alpha$ at 6564.6 \AA, 
\ion{Ca}{ii} at 8500.35  and 8544.44 \AA, and \ion{Na}{i} at 5891.58 and 5897.56 \AA, 
and measure them in each spectrum, allowing us to construct 
80 spectroscopic time series, five for each of
the 16 targets, to which different period search methods were applied: the generalized
Lomb-Scargle periodogram, the phase dispersion minimisation, the string length
method, and the Gaussian process regression with quasi-periodic kernel .

We have determined rotation periods for four stars in our sample using H$\alpha$
and for five stars using the \ion{Ca}{ii} IRT; in addition, we found tentative periods for another four and five stars using those respective lines. We did not find any fast rotator in our sample, V2689~Ori being the fastest 
rotator with $P_{\rm rot}$ = 11.9\,d. Nor did we find any especially slow rotators with
$P_{\rm rot} >$ 100\,d.

Two photometrically known periods, namely of  HD~119850 and Lalande~21185, could not be recovered.
This can be explained by their low amplitude
variations as measured by the MAD.

Comparing the different period search methods we conclude that
the string length method and the phase dispersion minimisation are the weakest methods applied here, leading
to the highest numbers of false positives and non-detections compared to the other methods.
While the generalized Lomb-Scargle periodogram performs slightly better than these two methods, GP
 modelling clearly leads
to the fewest false positives and non-detections. We therefore recommend this method for period
searches in chromospheric emission line indices for M dwarfs. Furthermore, GP modelling confers the additional advantage that
the long-term variations do not have to be subtracted. For all other methods, we suggest 
to double check results applying different methods or indices.

Likewise, the indices of the \ion{Na}{i} D lines are not well suited
for period searches for two reasons: first, these lines are affected by strong and highly variable
airglow, which either needs a large effort to correct or leads to the exclusion of many spectra. Second,
the \ion{Na}{i} D lines seem to be less sensitive to variability, and therefore the
amplitudes of variability are lower, which means the periods are more difficult to recover.
The use of the H$\alpha$ or \ion{Ca}{ii} IRT indices leads to similar
results in most cases. However, the  $I_{\ion{Ca}{ii},1}$  index leads to slightly more (tentative) period detections and also
to fewer false positives with the weaker methods.
This indicates that the \ion{Ca}{ii} IRT lines can be recommended for rotation measurements,
as was also found by \citet{Mittag2017} for late F to mid K dwarfs.

\begin{acknowledgements}
  B.~F. acknowledges funding by the DFG under Cz \mbox{222/1-1}.
  CARMENES is an instrument for the Centro Astron\'omico Hispano-Alem\'an de
  Calar Alto (CAHA, Almer\'{\i}a, Spain). 
  CARMENES is funded by the German Max-Planck-Gesellschaft (MPG), 
  the Spanish Consejo Superior de Investigaciones Cient\'{\i}ficas (CSIC),
  the European Union through FEDER/ERF FICTS-2011-02 funds, 
  and the members of the CARMENES Consortium 
  (Max-Planck-Institut f\"ur Astronomie,
  Instituto de Astrof\'{\i}sica de Andaluc\'{\i}a,
  Landessternwarte K\"onigstuhl,
  Institut de Ci\`encies de l'Espai,
  Institut f\"ur Astrophysik G\"ottingen,
  Universidad Complutense de Madrid,
  Th\"uringer Landessternwarte Tautenburg,
  Instituto de Astrof\'{\i}sica de Canarias,
  Hamburger Sternwarte,
  Centro de Astrobiolog\'{\i}a and
  Centro Astron\'omico Hispano-Alem\'an), 
  with additional contributions by the Spanish Ministry of Economy, 
  the German Science Foundation through the Major Research Instrumentation 
    Programme and DFG Research Unit FOR2544 ``Blue Planets around Red Stars'', 
  the Klaus Tschira Stiftung, 
  the states of Baden-W\"urttemberg and Niedersachsen, 
  and by the Junta de Andaluc\'{\i}a.

\end{acknowledgements}

\bibliographystyle{aa}
\bibliography{papers}

\begin{appendix}

\section{Graphical display of period search for each star}\label{appendix}

\begin{figure*}[h!]
\includegraphics[width=0.45\textwidth, clip]{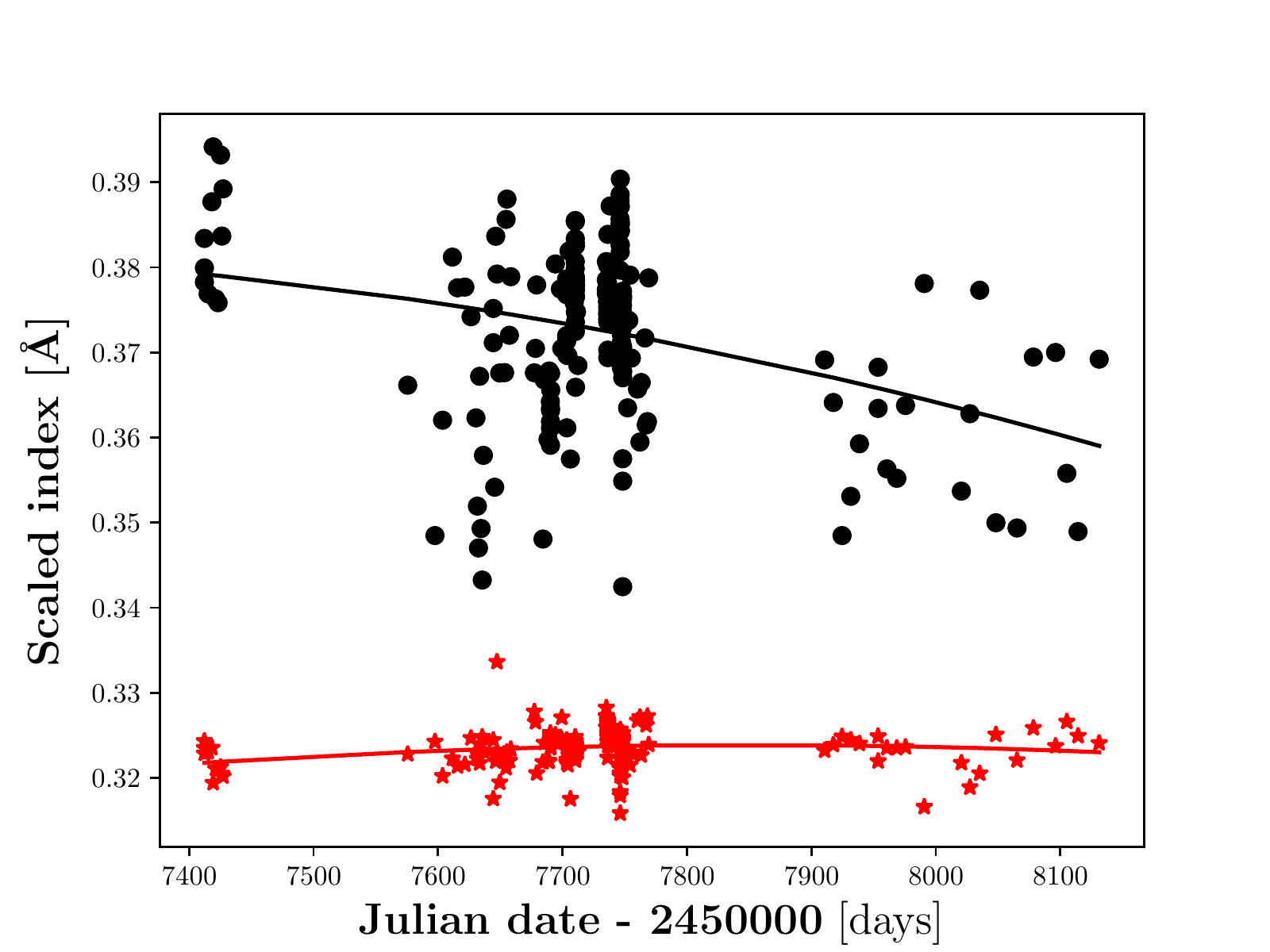}\\%\vspace{-0.4mm}\\

\includegraphics[width=0.45\textwidth, clip]{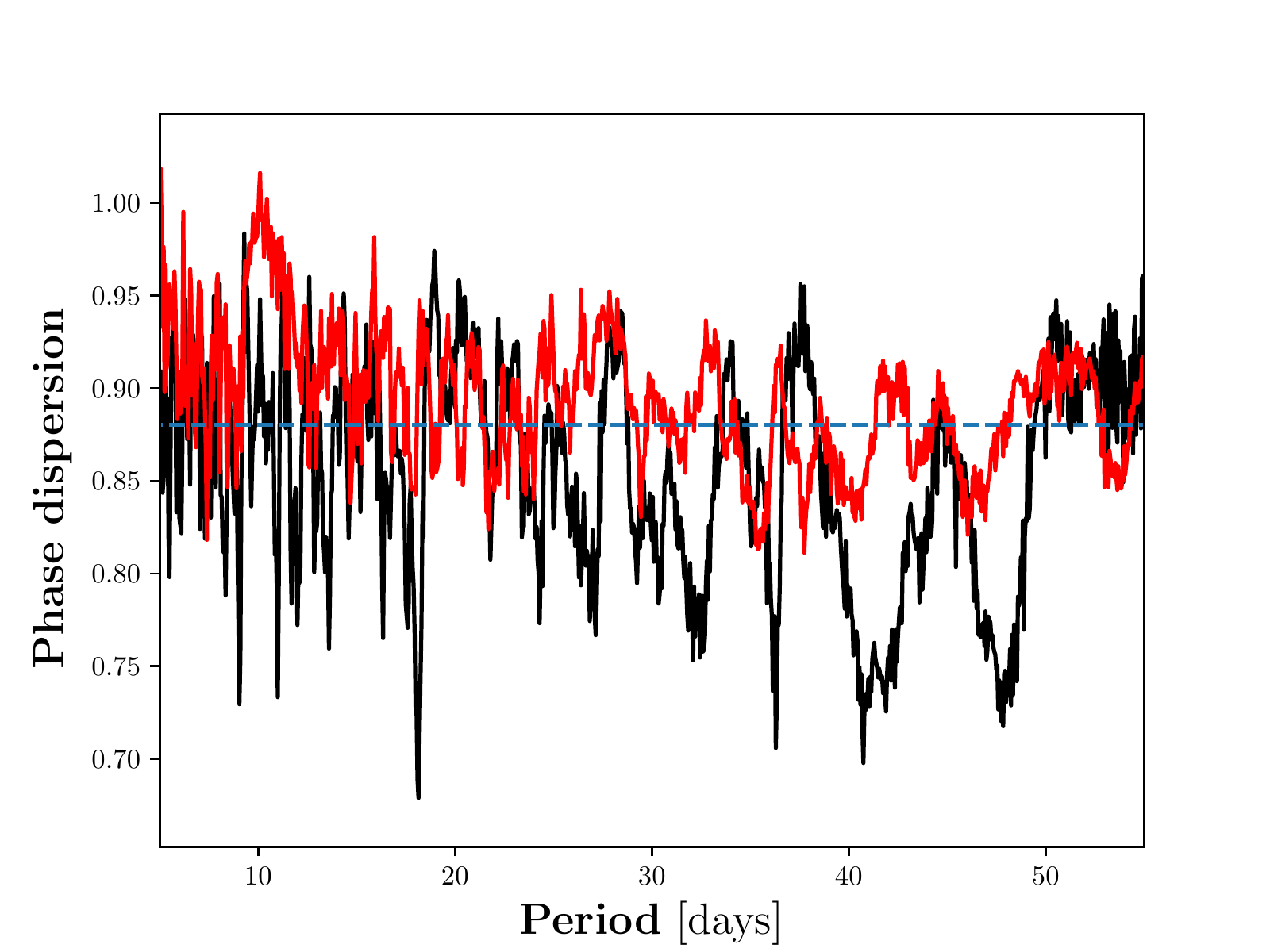}%\vspace{-0.4mm}\\
\includegraphics[width=0.45\textwidth, clip]{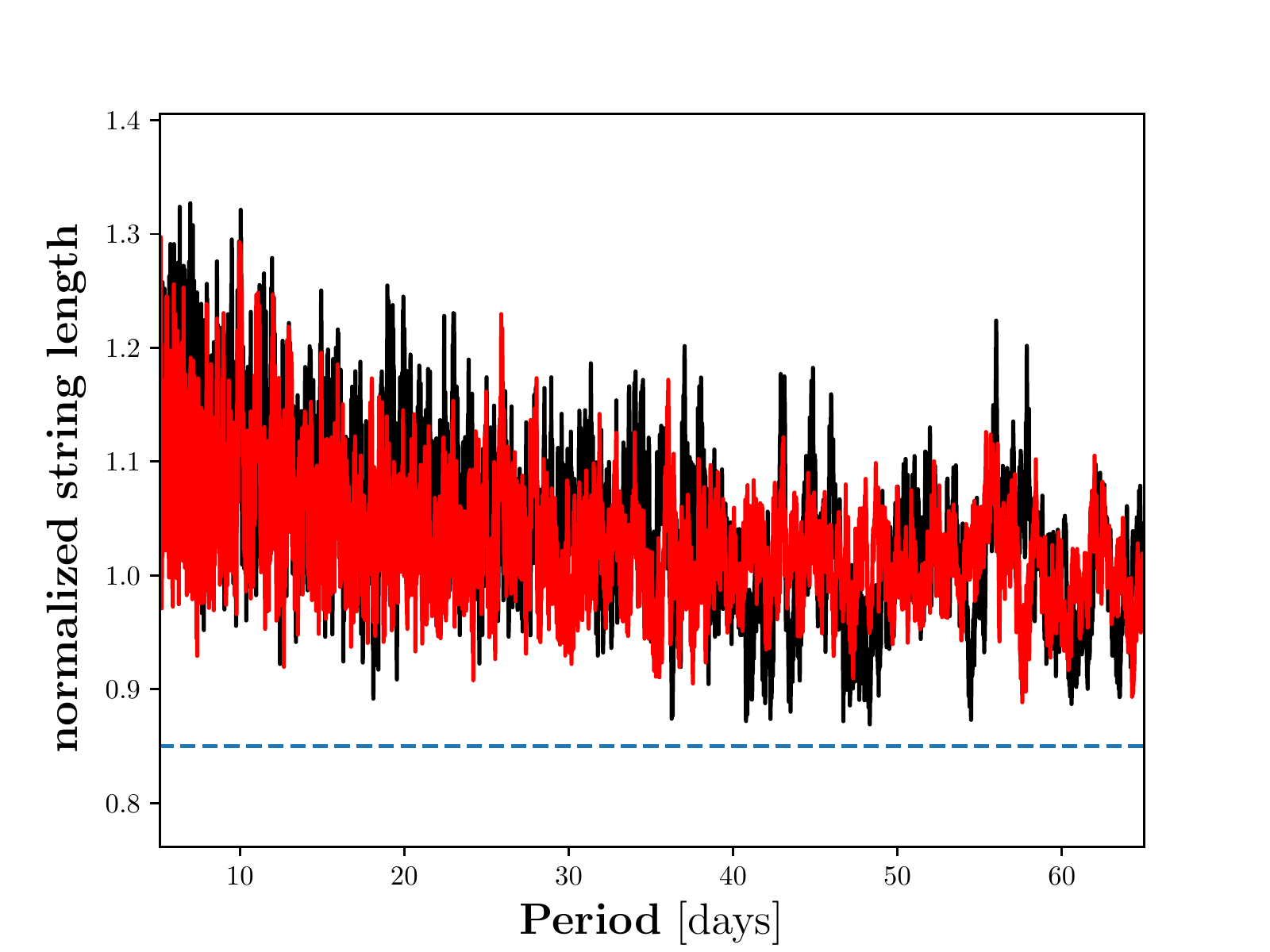}\\
\includegraphics[width=0.45\textwidth, clip]{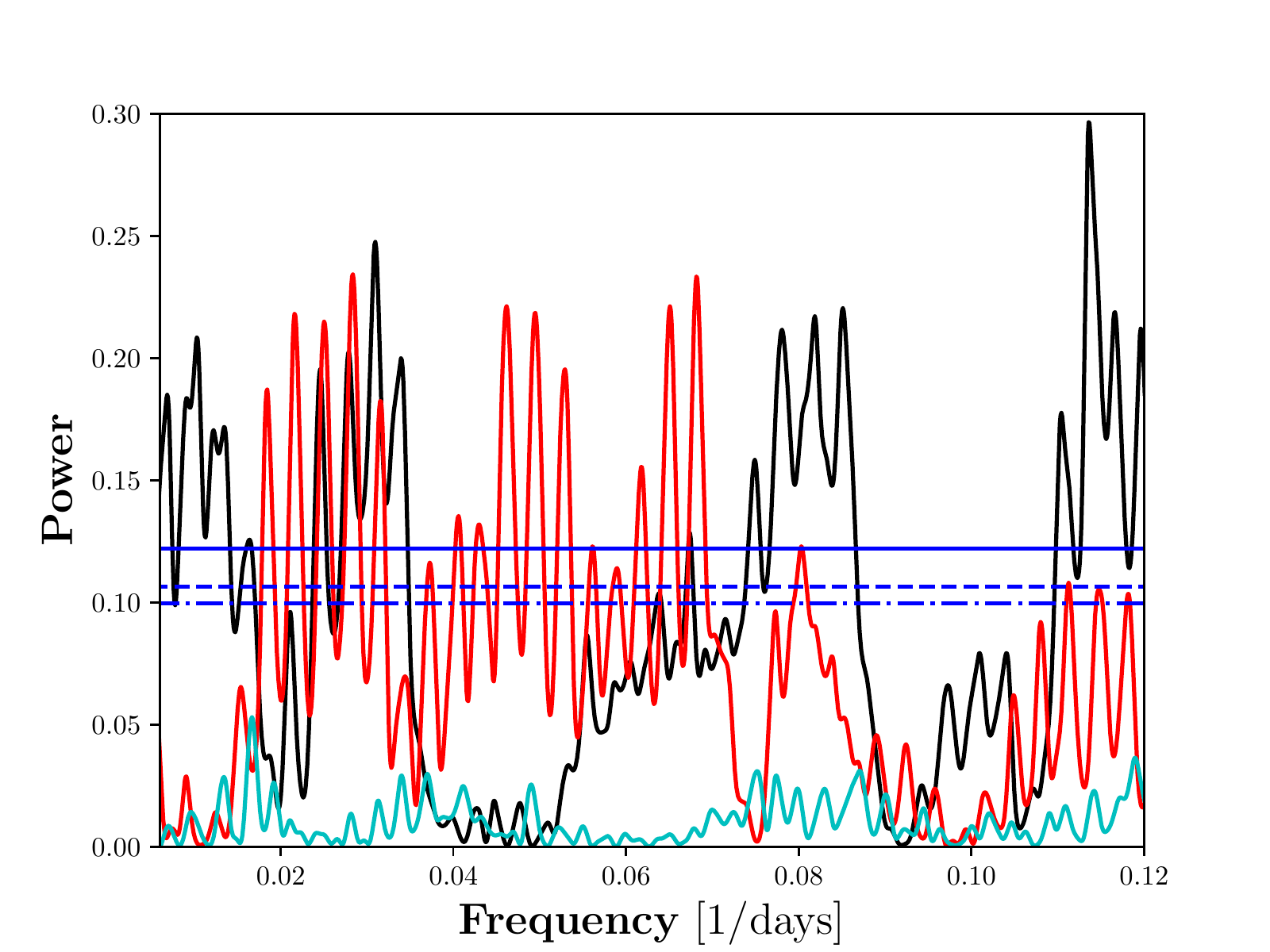}%\vspace{-0.4mm}\\
\includegraphics[width=0.45\textwidth, clip]{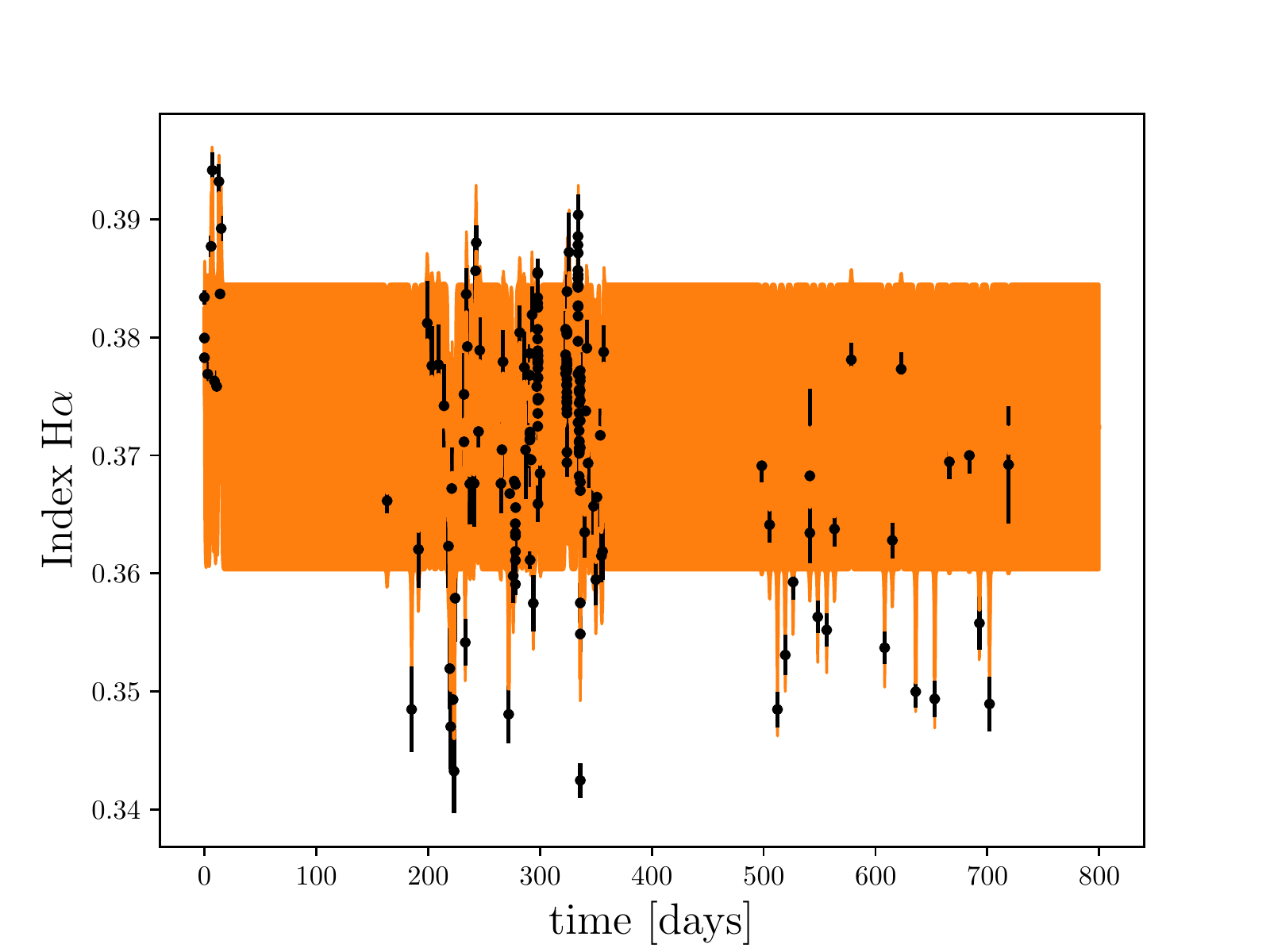}%\vspace{-0.4mm}\\
\caption{\label{gxand} Results for GX And. Top left: Index time series in black dots for $I_{\rm{H}\alpha}$ and in red asterisks for scaled $I_{\rm Ca,1}$.
The corresponding solid lines represent the best second-degree polynomial fit.
Top right: Phase folded index time series of $I_{\rm{H}\alpha}$ (only present in case we found a
period using H$\alpha$).
Second row left:
Phase dispersion minimisation. The black line denotes $\Theta$ computed for $I_{\rm{H}\alpha}$, while
the red line denotes $\Theta$  for $I_{\rm Ca,1}$. The dashed horizontal
line marks the 1\,\% FAP level. 
Second row  right: String length method. Again, the black line denotes the string length $L$ for $I_{\rm{H}\alpha}$
and the red line is also the string length but computed for $I_{\rm Ca,1}$.
Third row left: 
GLS periodogram with FAP levels of 10\,\%, 5\,\%, and 1\,\% in blue (dash-dotted, dashed, and solid),
respectively. In the periodogram the black line denotes the normalised power calculated with $I_{\rm{H}\alpha}$,
while the red line again is the power of $I_{\rm Ca,1}$,
and the cyan line the window function. Third row right: Index time series of the H$\alpha$ indices with the GP best-fit
model in orange and its standard deviation shaded in light orange.}
\end{figure*}

\begin{figure*}
\includegraphics[width=0.45\textwidth, clip]{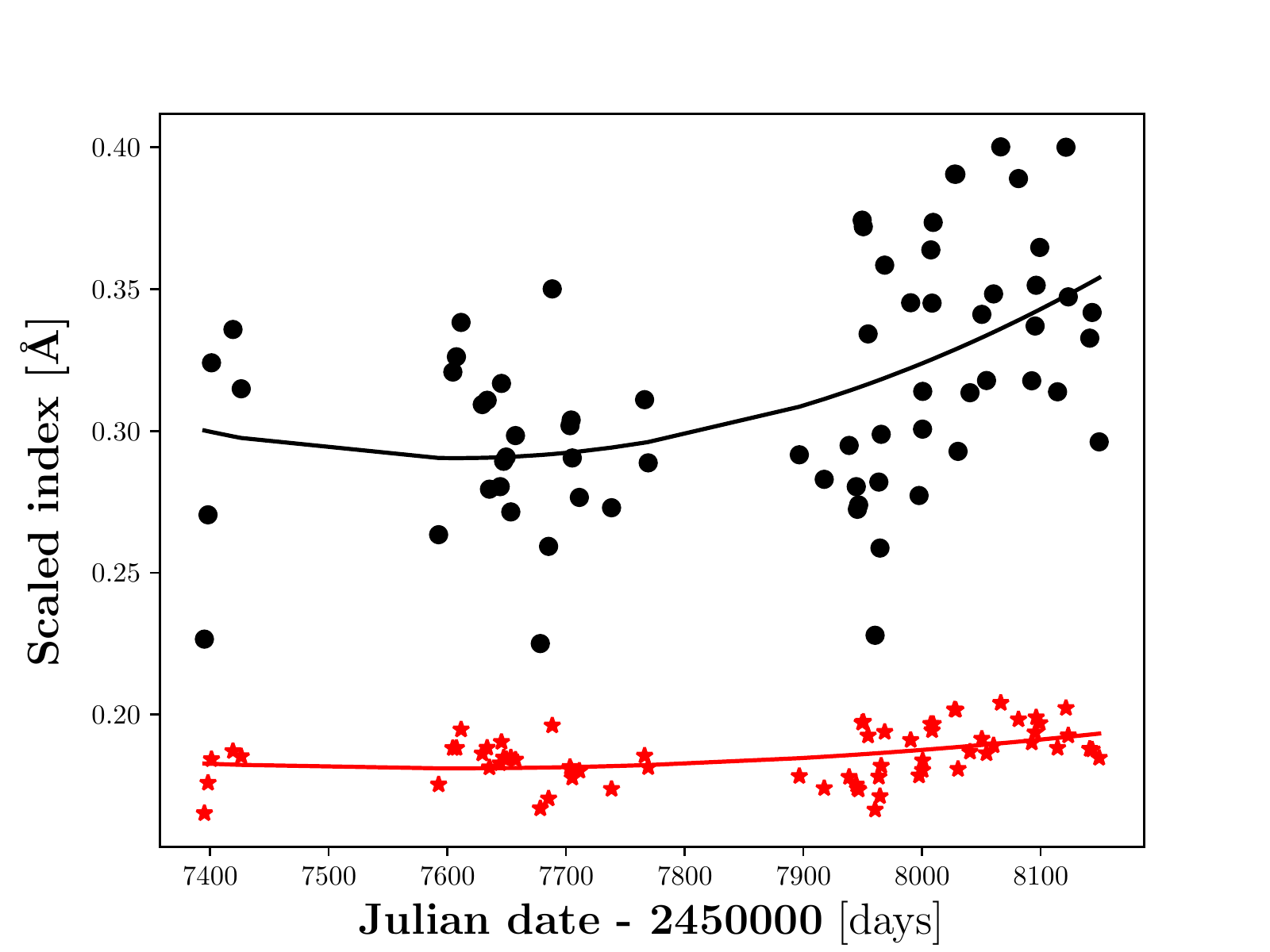}%\vspace{-0.4mm}\\
\includegraphics[width=0.45\textwidth, clip]{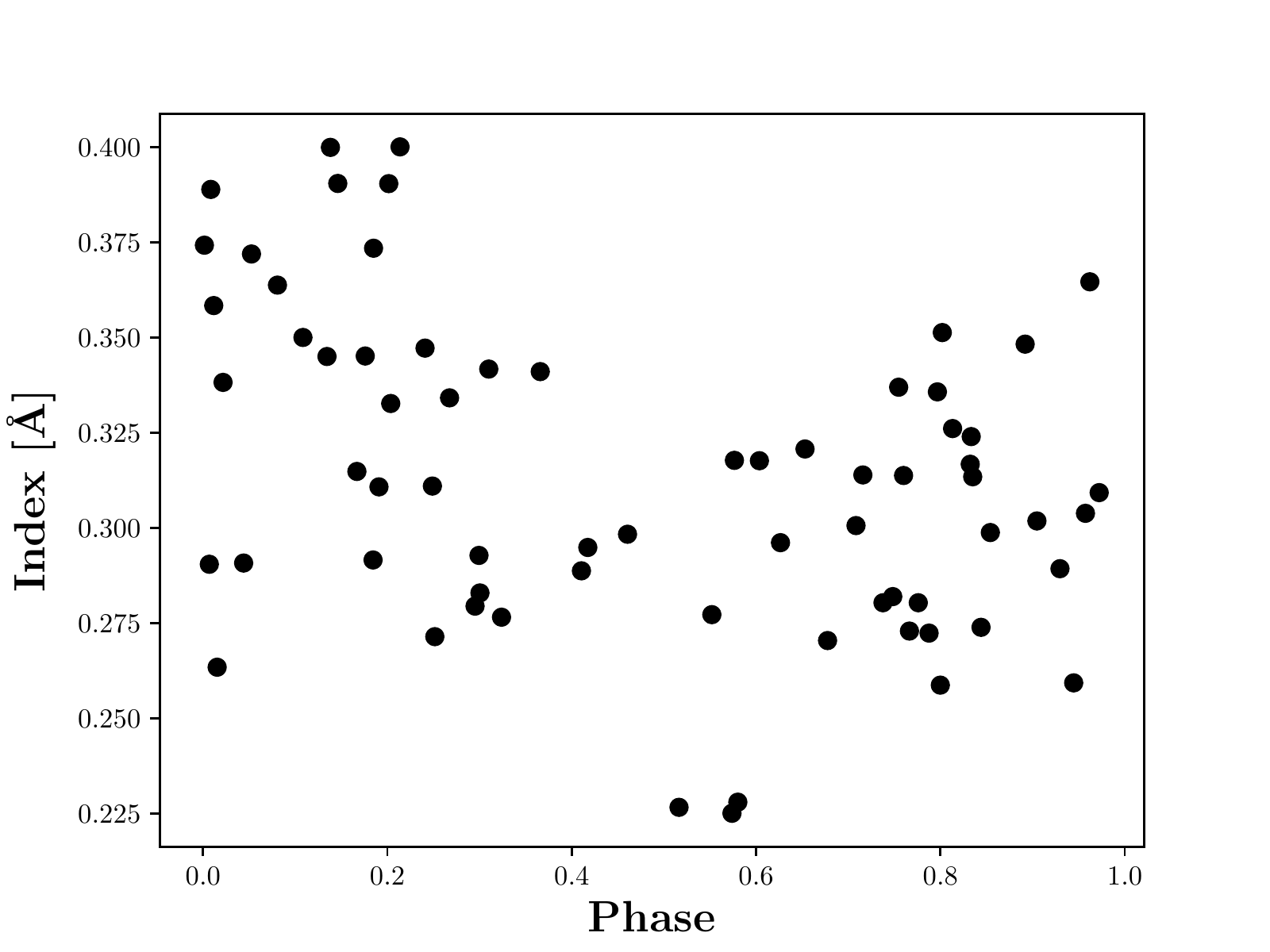}\\
\includegraphics[width=0.45\textwidth, clip]{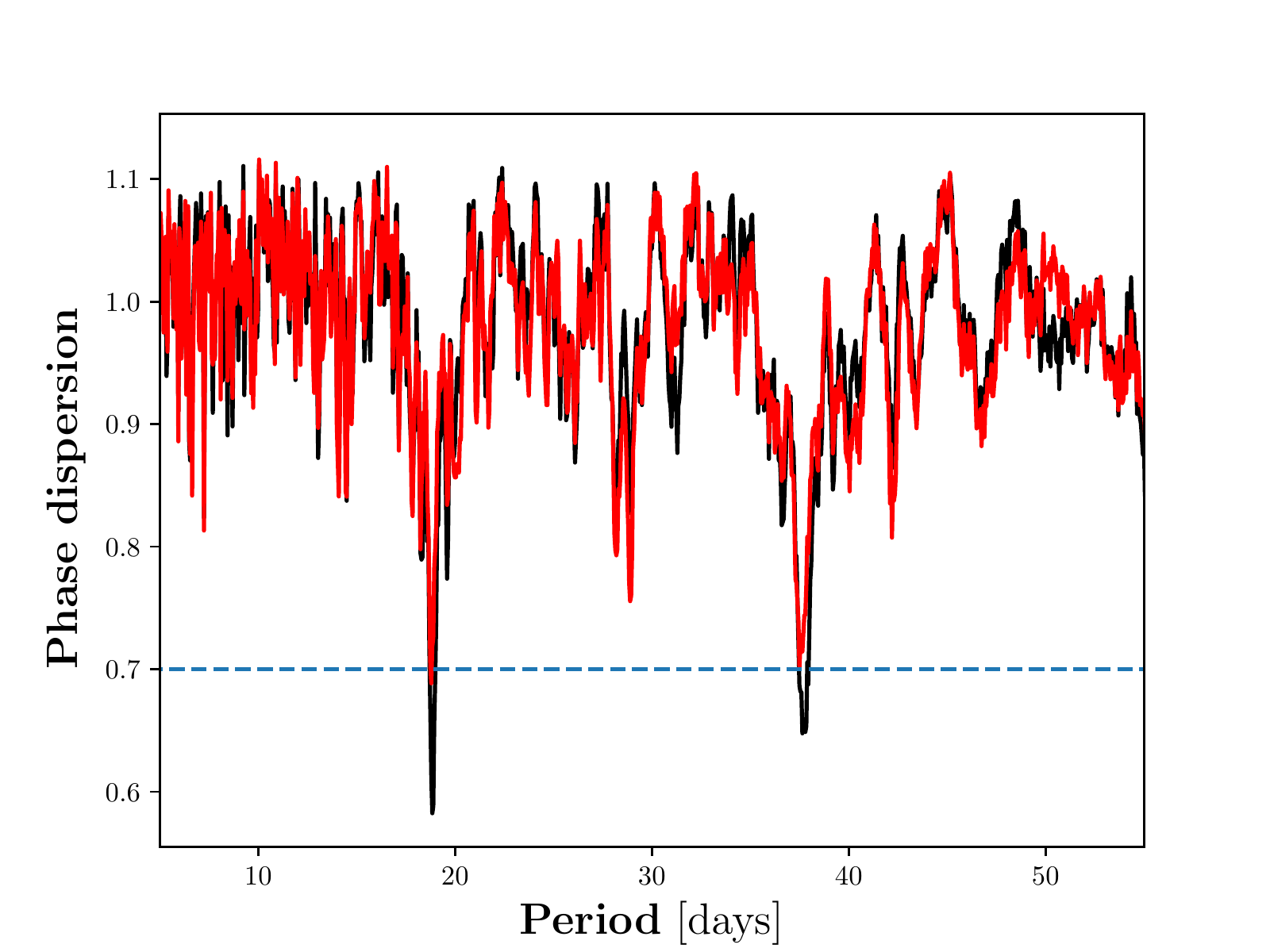}%\vspace{-0.4mm}\\
\includegraphics[width=0.45\textwidth, clip]{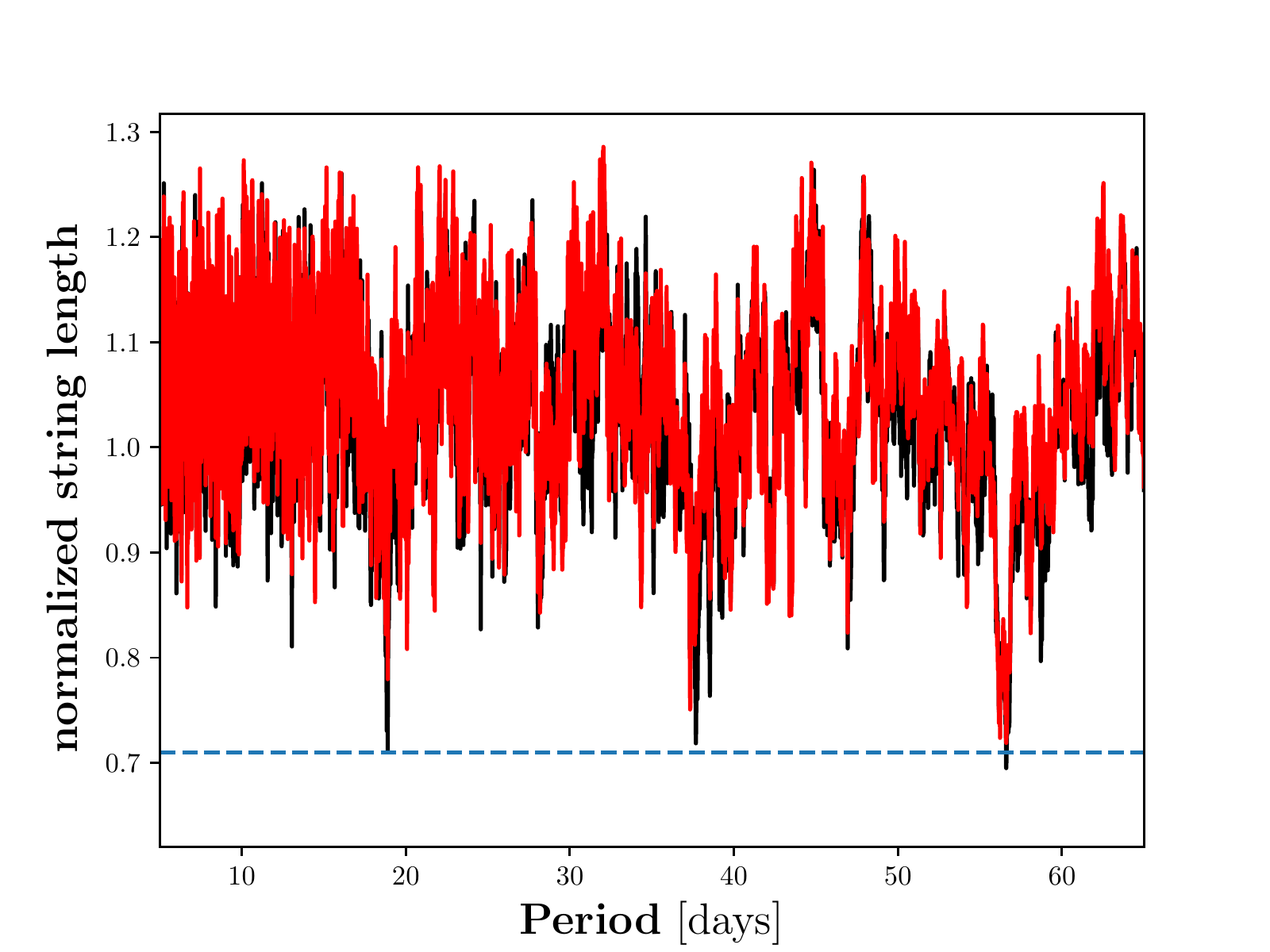}\\
\includegraphics[width=0.45\textwidth, clip]{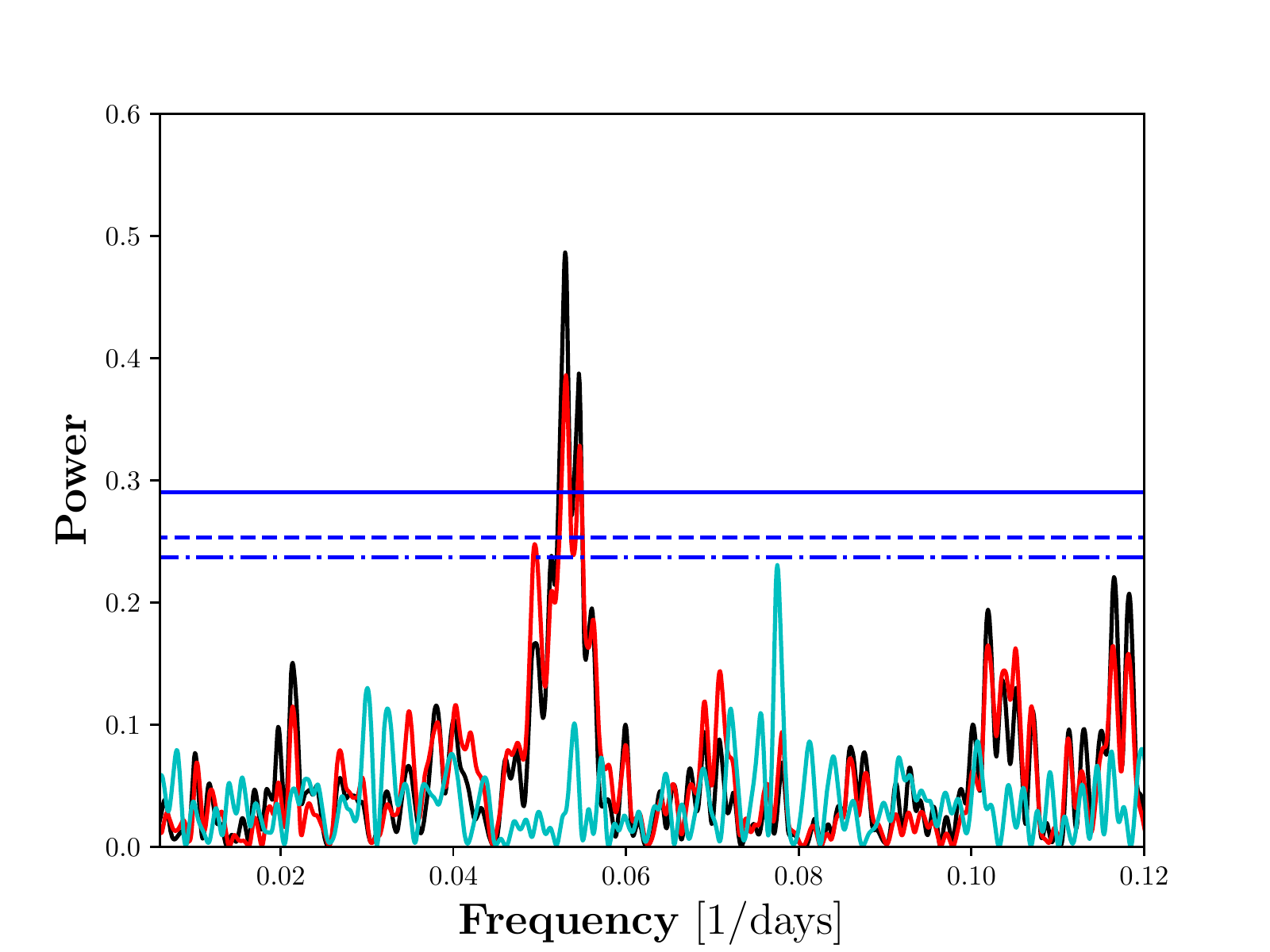}%\vspace{-0.4mm}\\
\includegraphics[width=0.45\textwidth, clip]{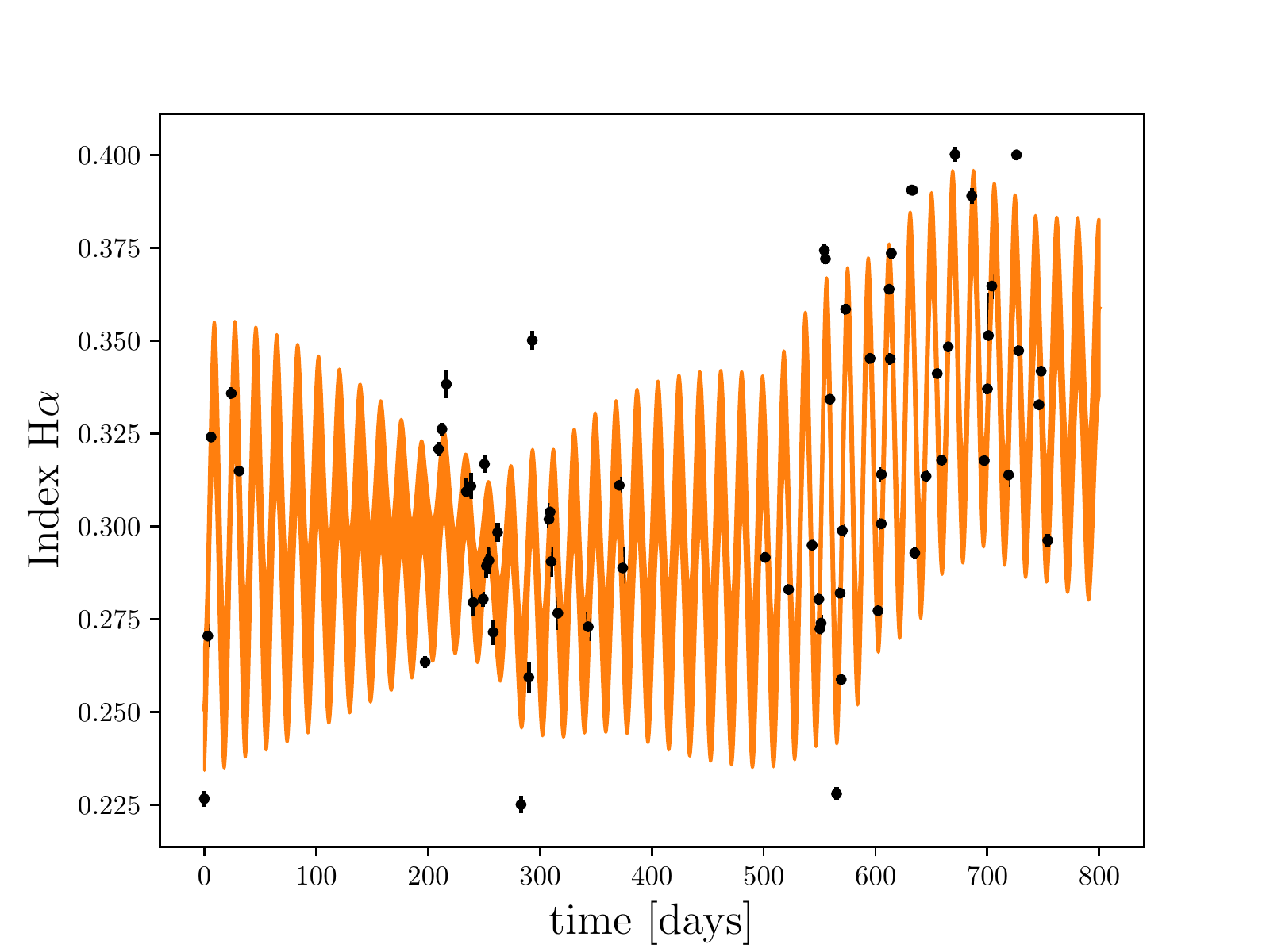}%\vspace{-0.4mm}\\
\caption{\label{bd61195} Results for BD+61 195 as explained in Fig. \ref{gxand}.}
\end{figure*}

\begin{figure*}
\includegraphics[width=0.45\textwidth, clip]{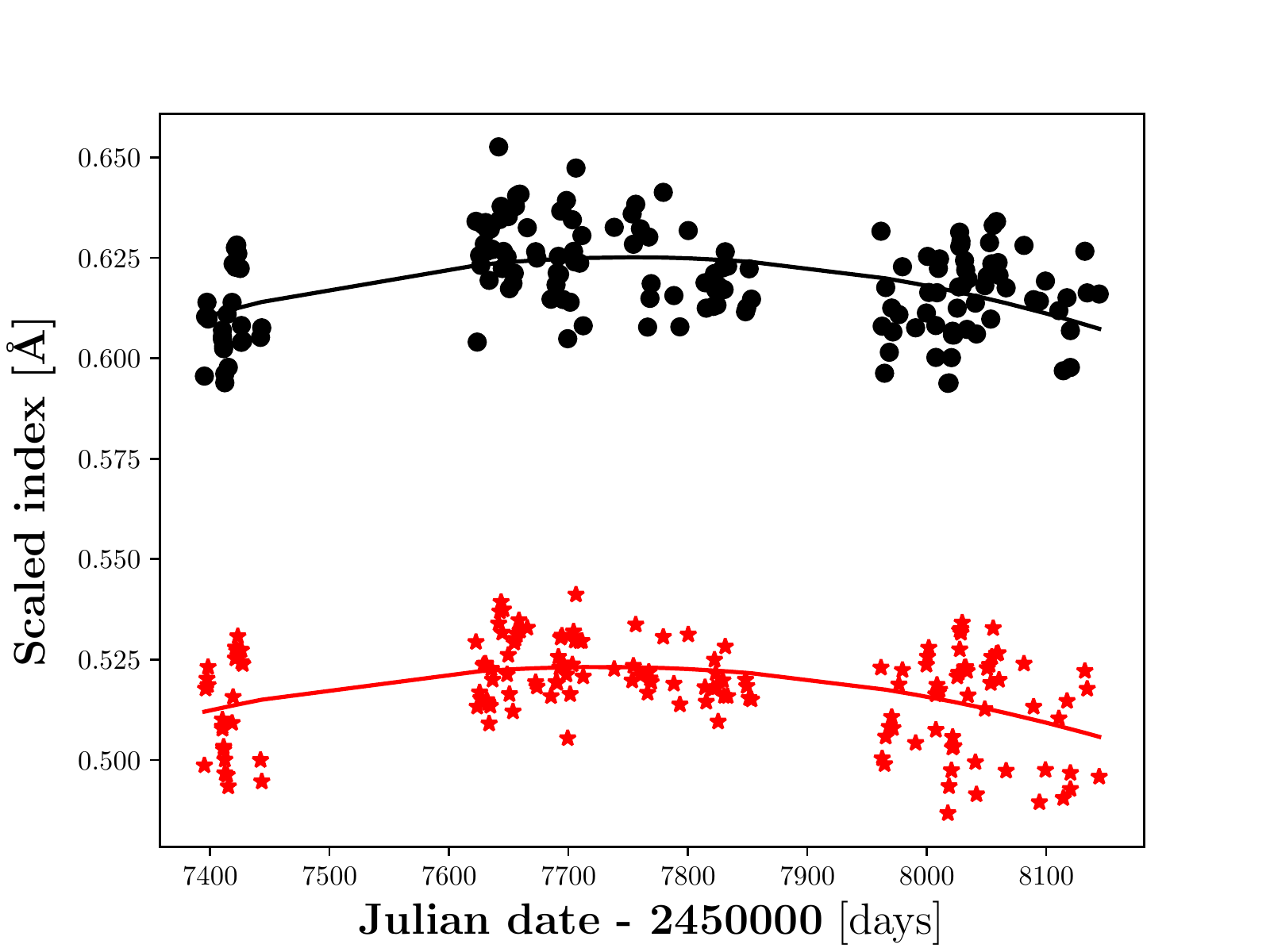}%\vspace{-0.4mm}\\
\includegraphics[width=0.45\textwidth, clip]{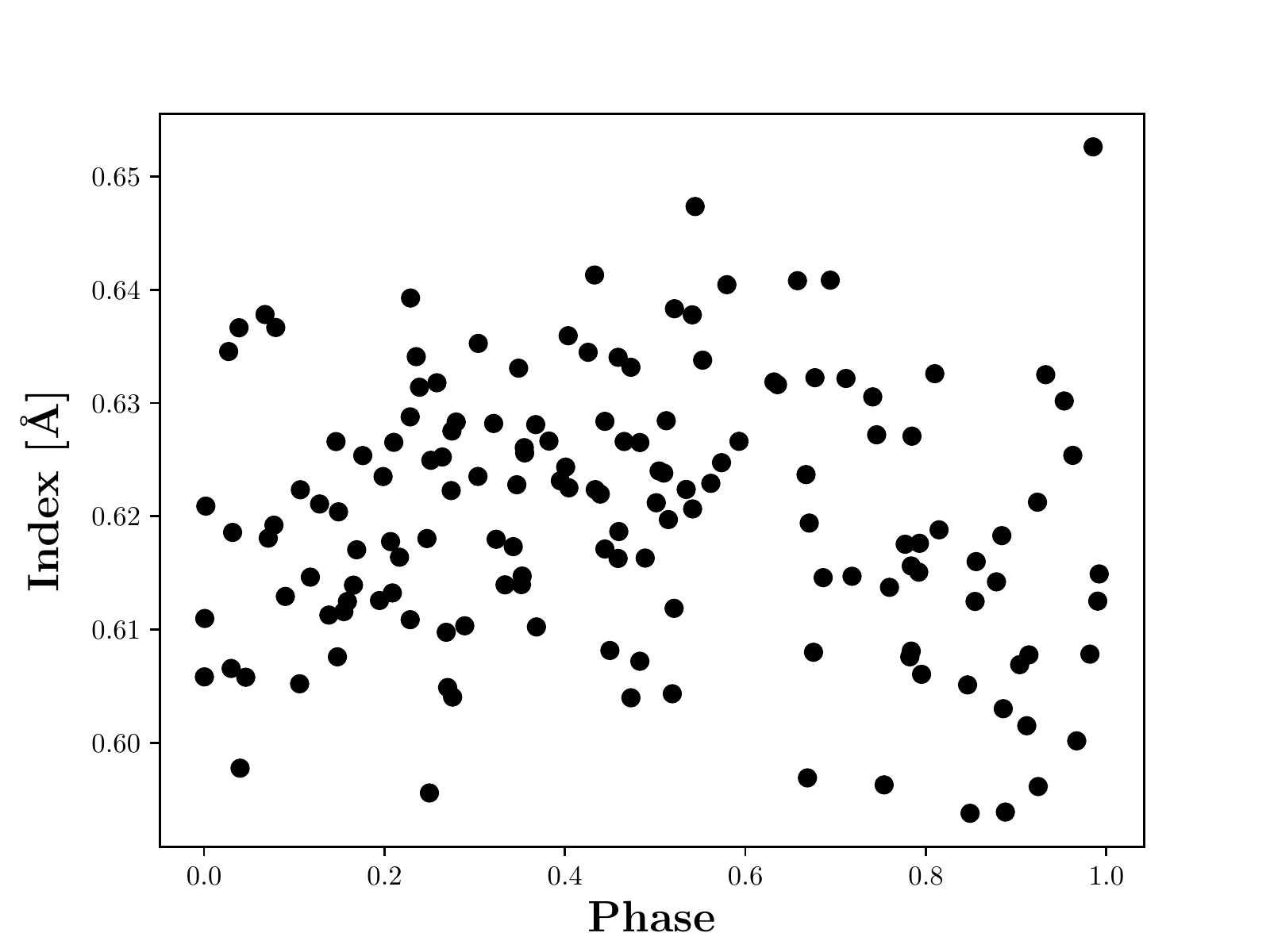}\\
\includegraphics[width=0.45\textwidth, clip]{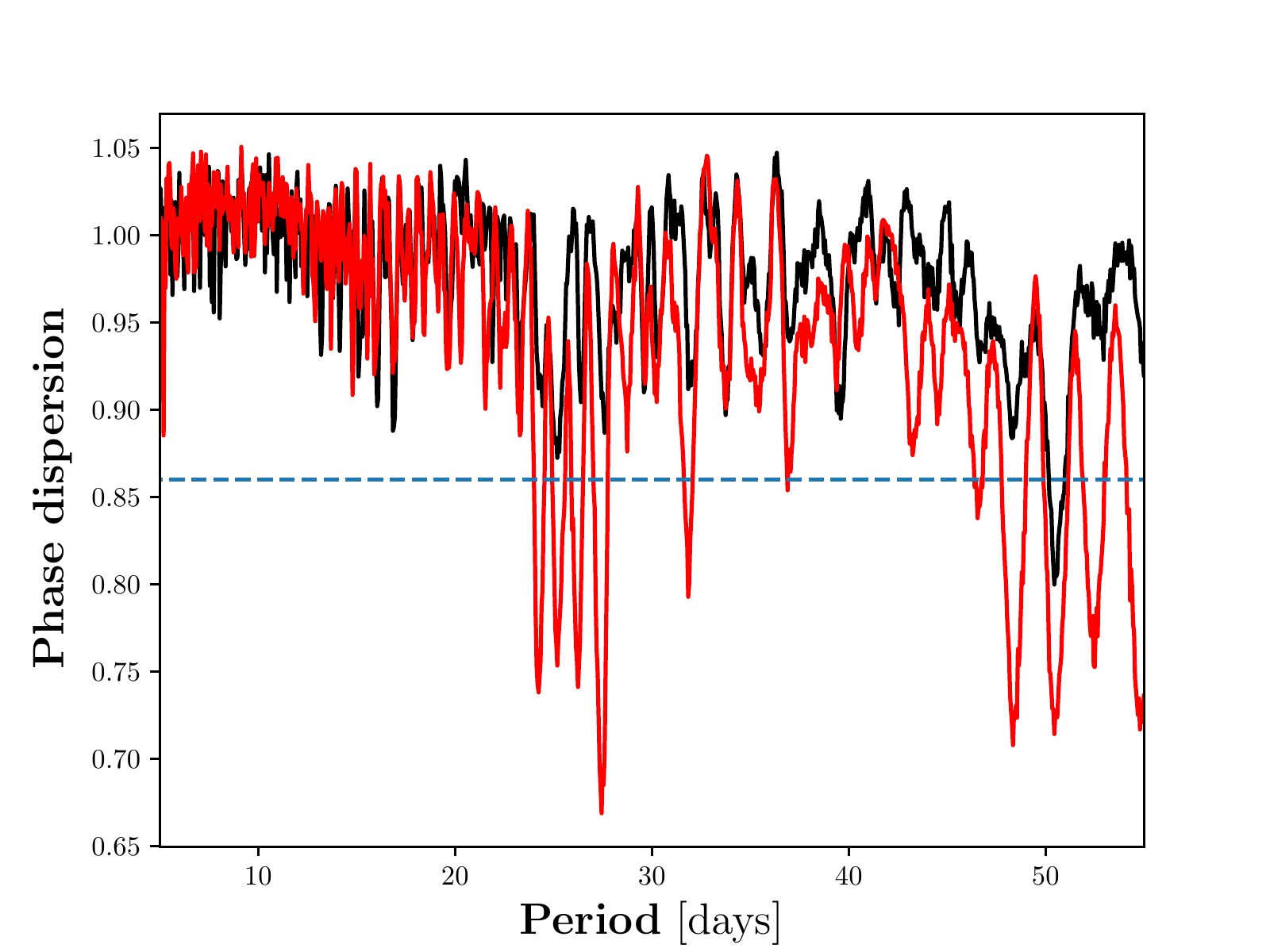}%\vspace{-0.4mm}\\
\includegraphics[width=0.45\textwidth, clip]{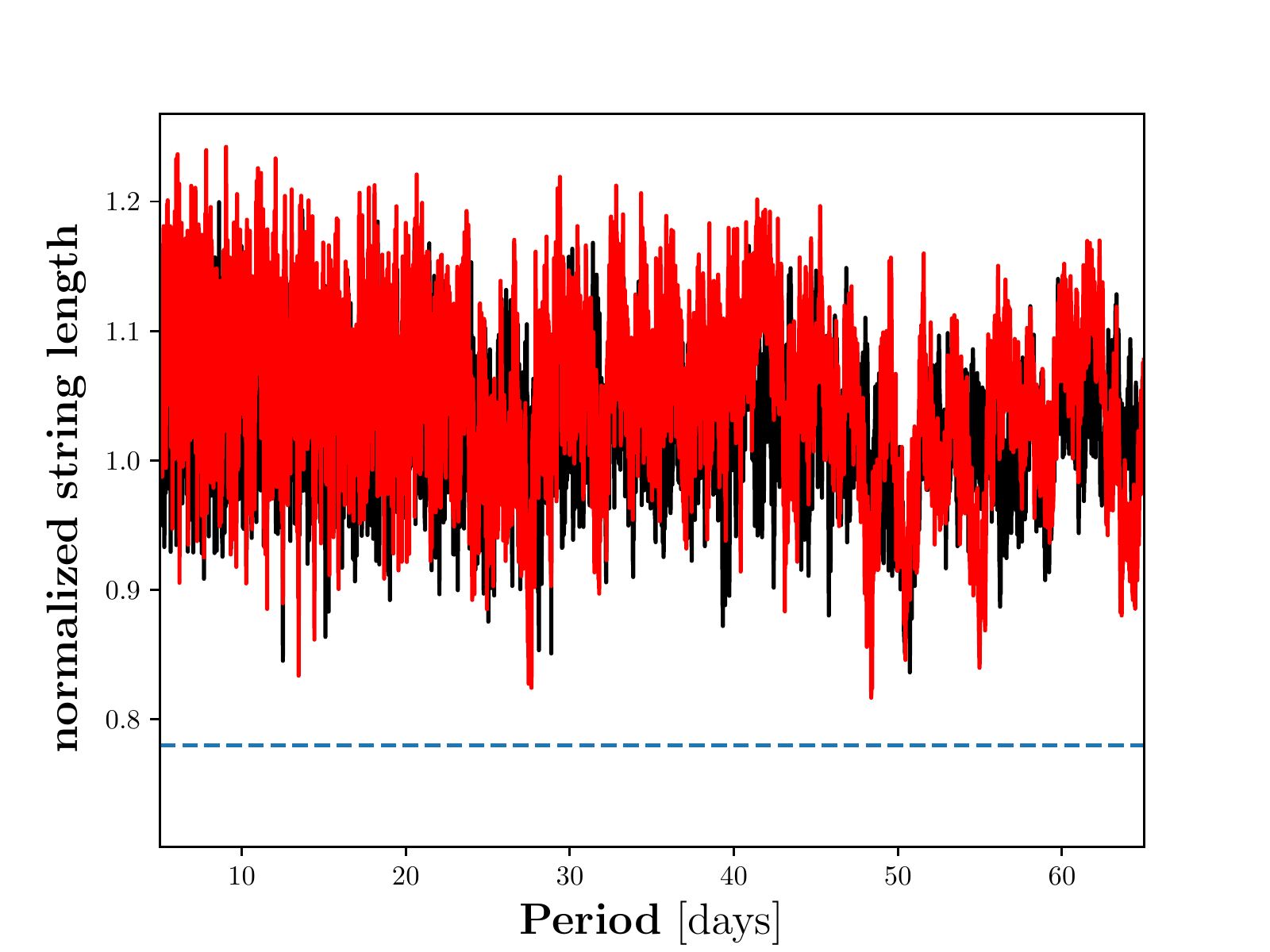}\\
\includegraphics[width=0.45\textwidth, clip]{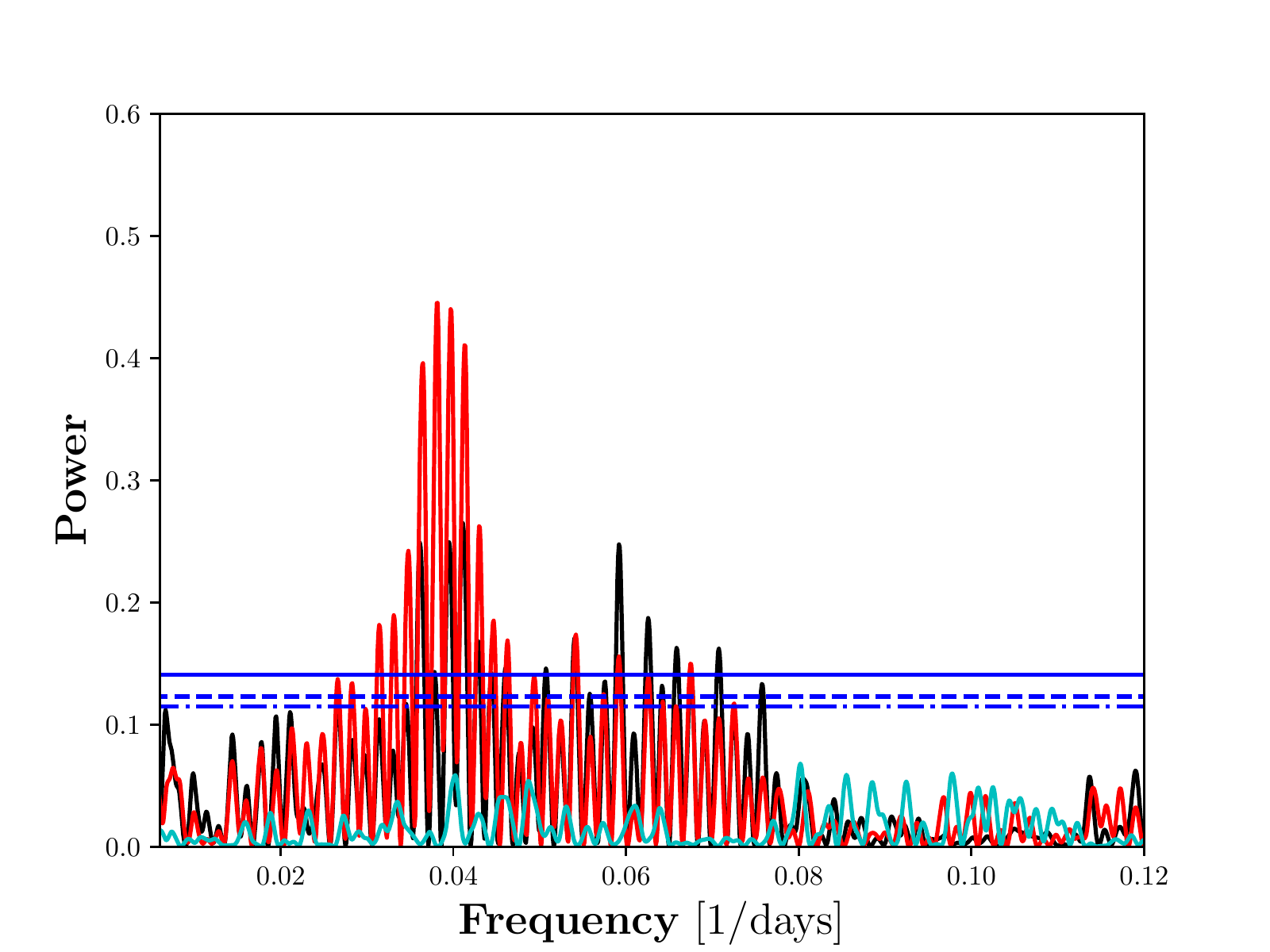}%\vspace{-0.4mm}\\
\includegraphics[width=0.45\textwidth, clip]{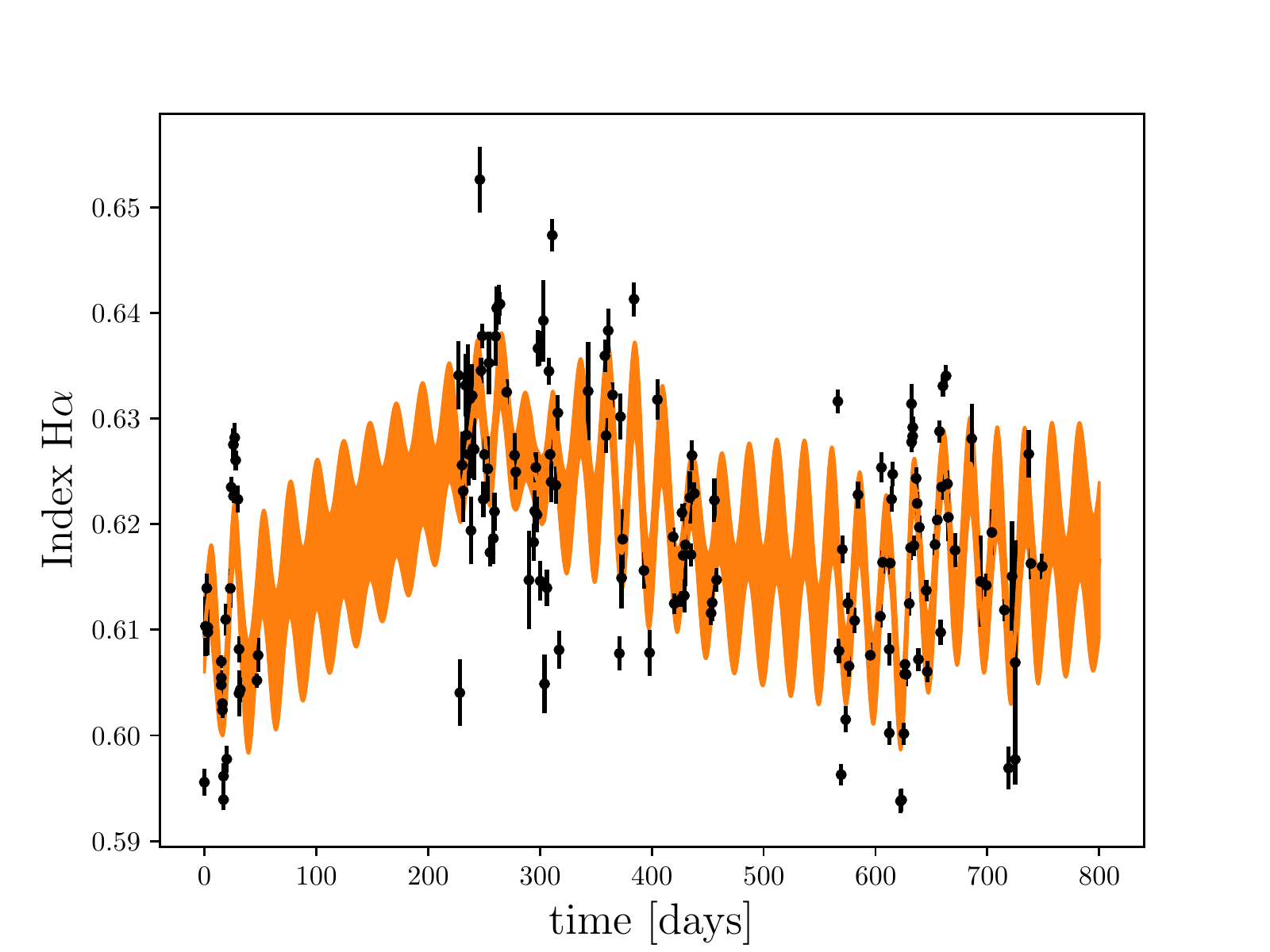}%\vspace{-0.4mm}\\
\caption{\label{bd21652} Results for BD+21 652 as explained in Fig. \ref{gxand}.}
\end{figure*}

\begin{figure*}
\includegraphics[width=0.45\textwidth, clip]{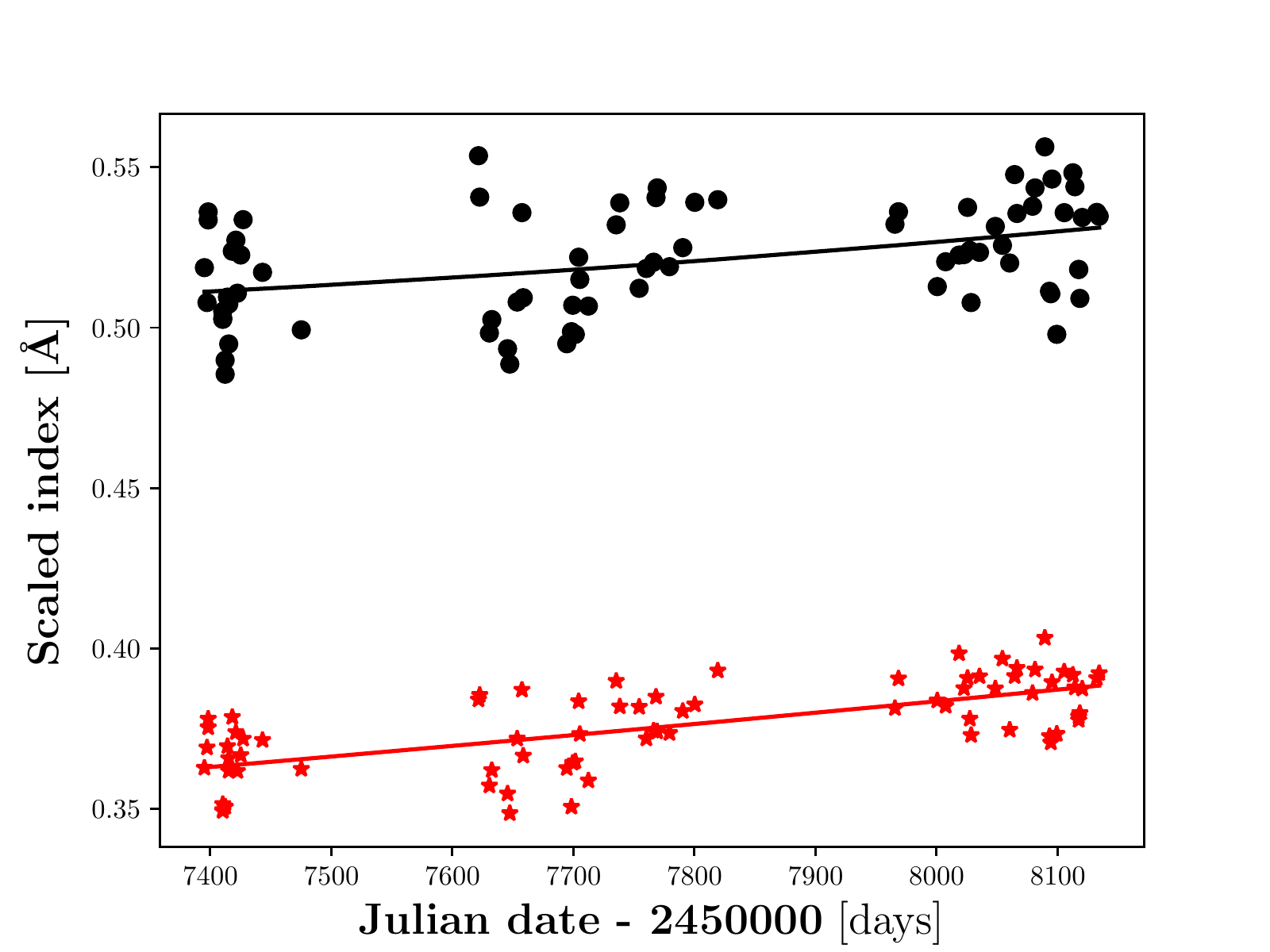}\\%\vspace{-0.4mm}\\
\includegraphics[width=0.45\textwidth, clip]{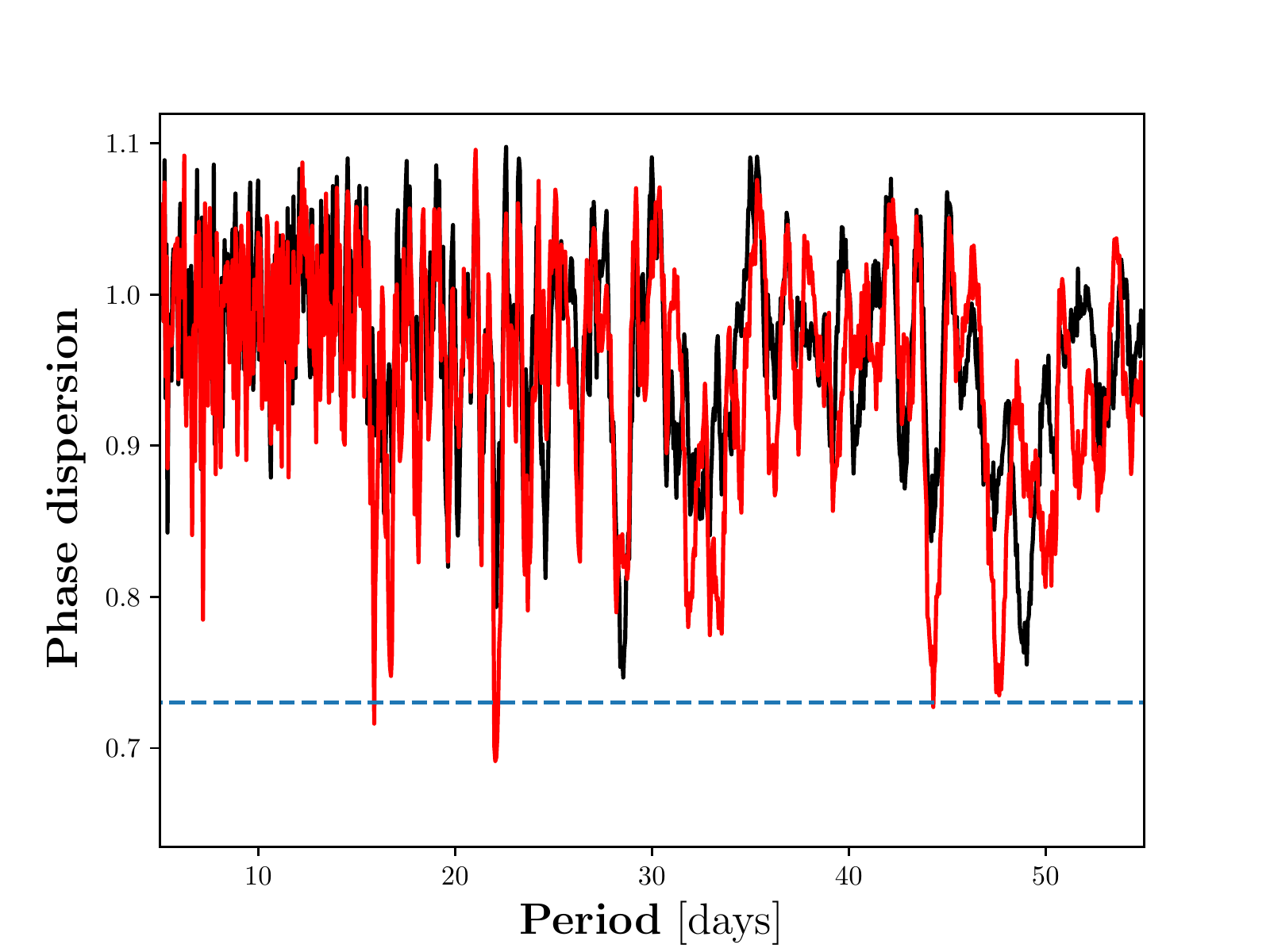}%\vspace{-0.4mm}\\
\includegraphics[width=0.45\textwidth, clip]{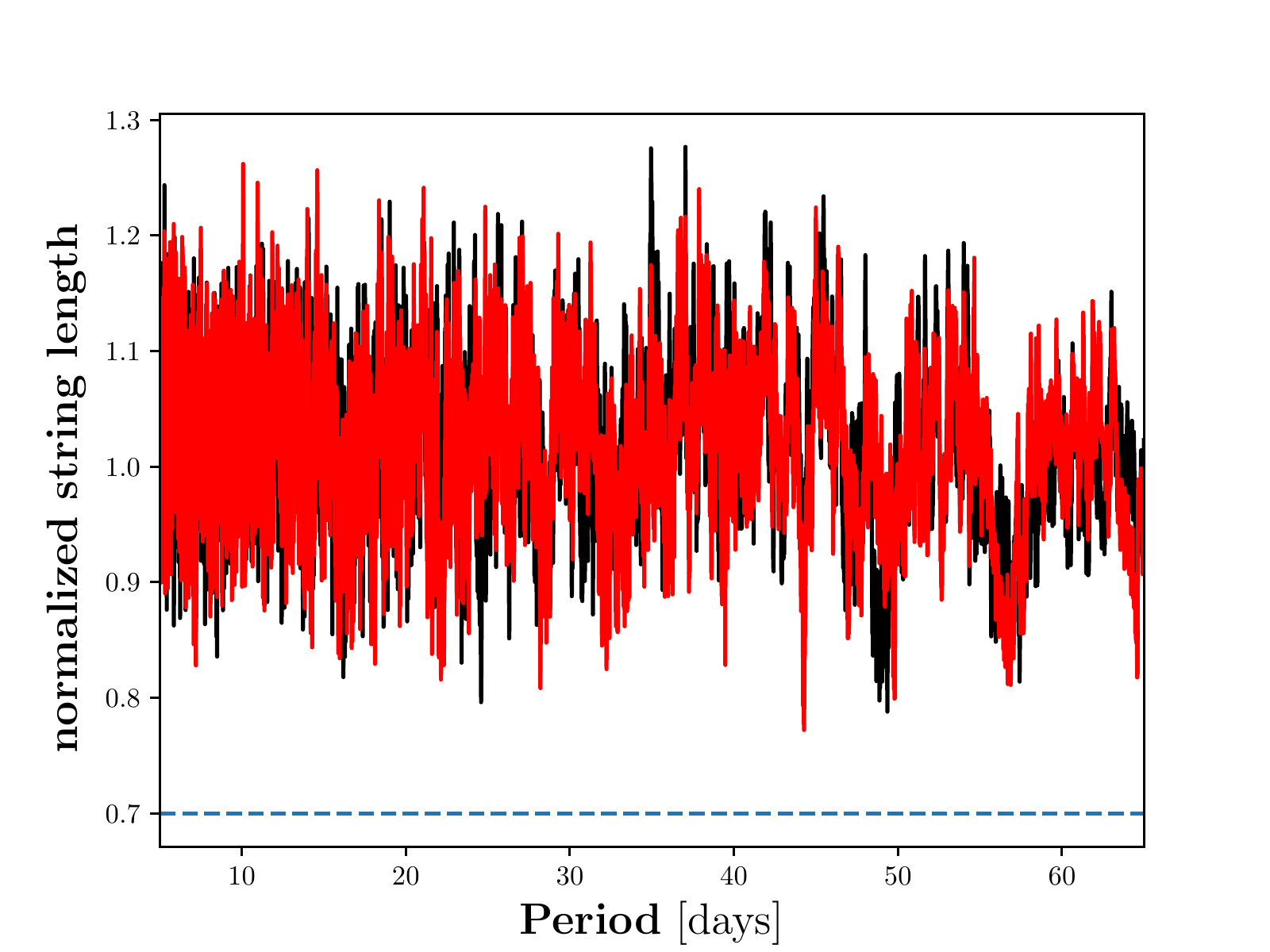}\\
\includegraphics[width=0.45\textwidth, clip]{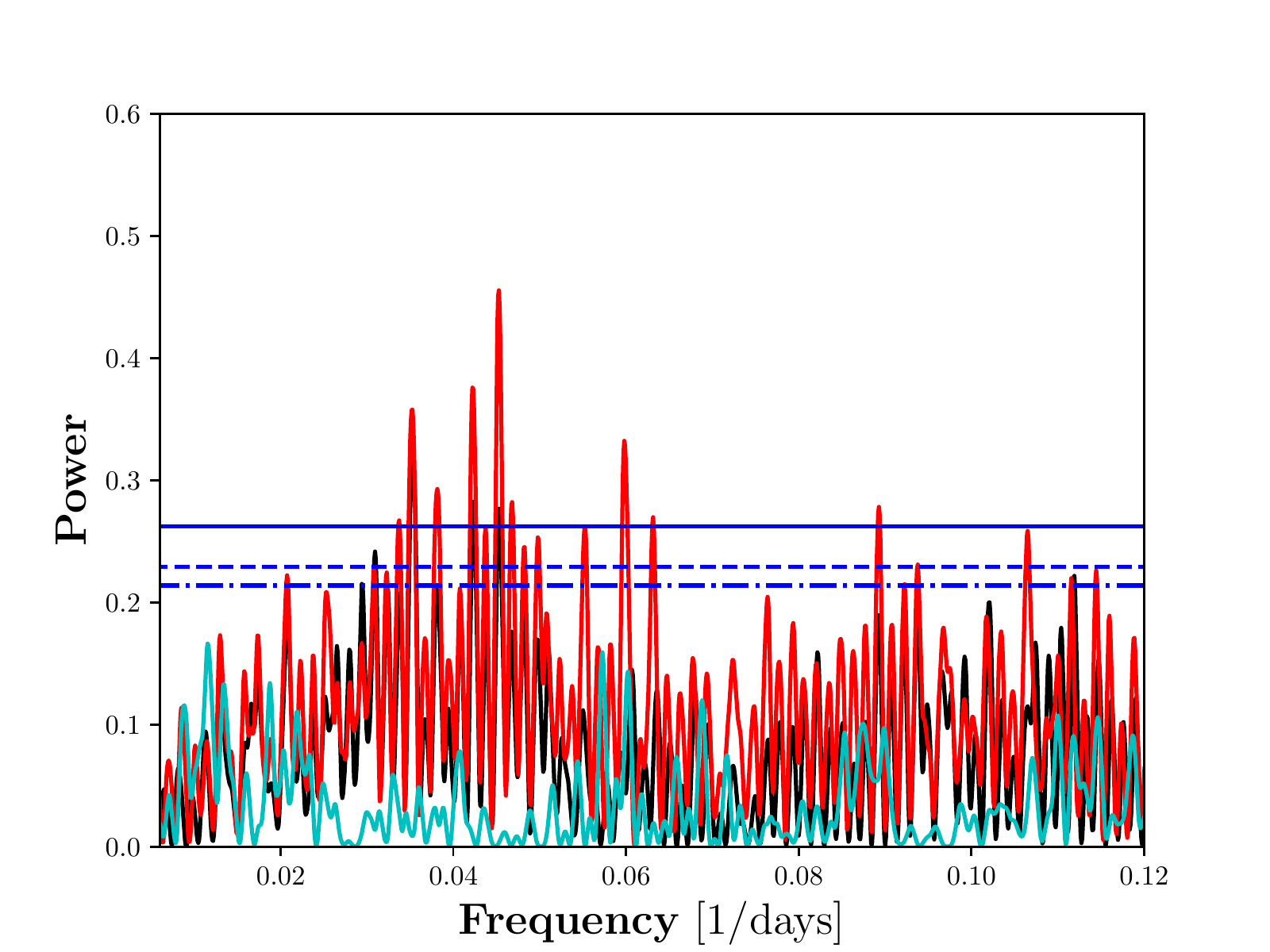}%\vspace{-0.4mm}\\
\includegraphics[width=0.45\textwidth, clip]{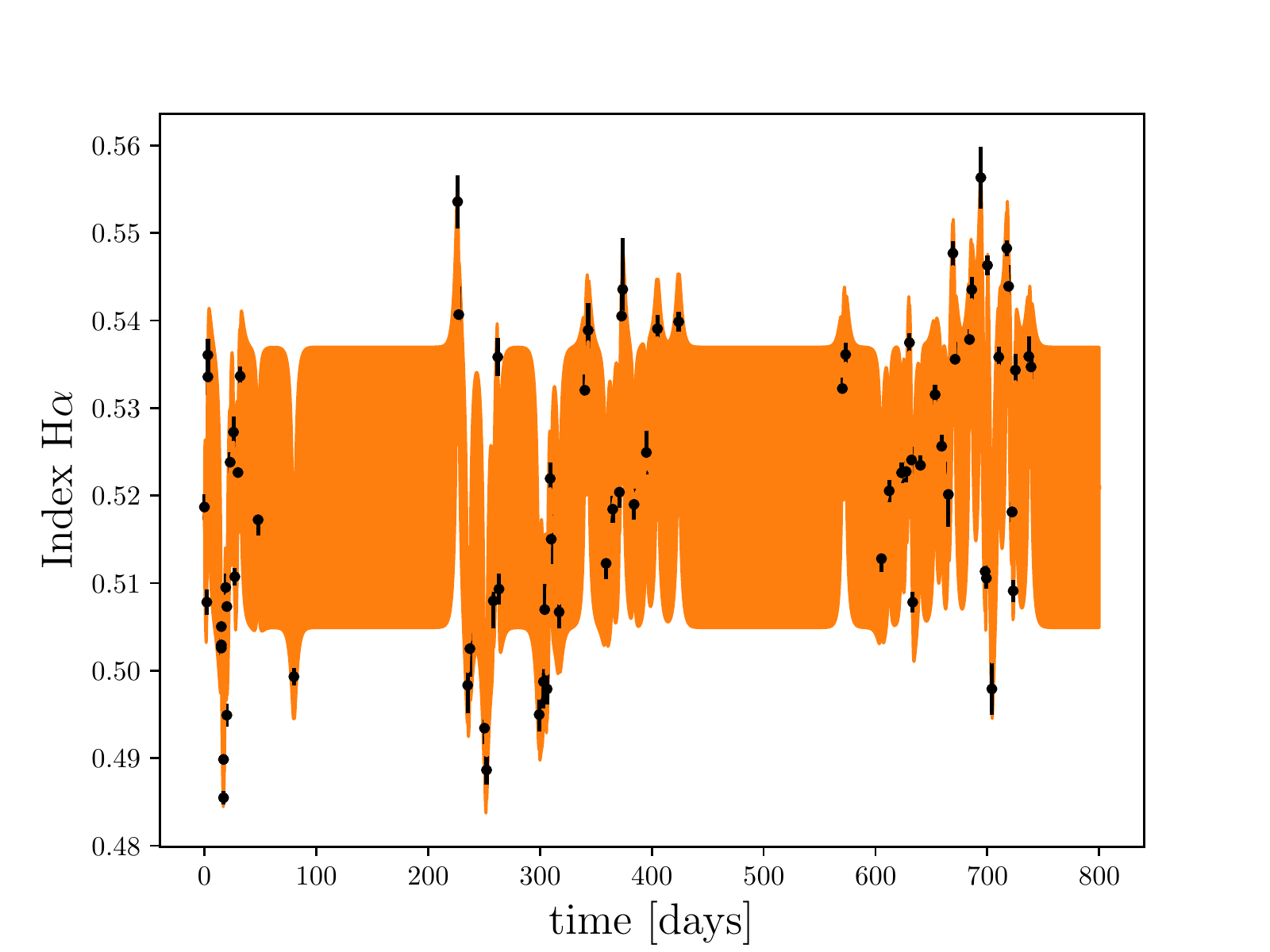}%\vspace{-0.4mm}\\
\caption{\label{bd52857} Results for BD+52 857 as explained in Fig. \ref{gxand}.}
\end{figure*}

\begin{figure*}
\includegraphics[width=0.45\textwidth, clip]{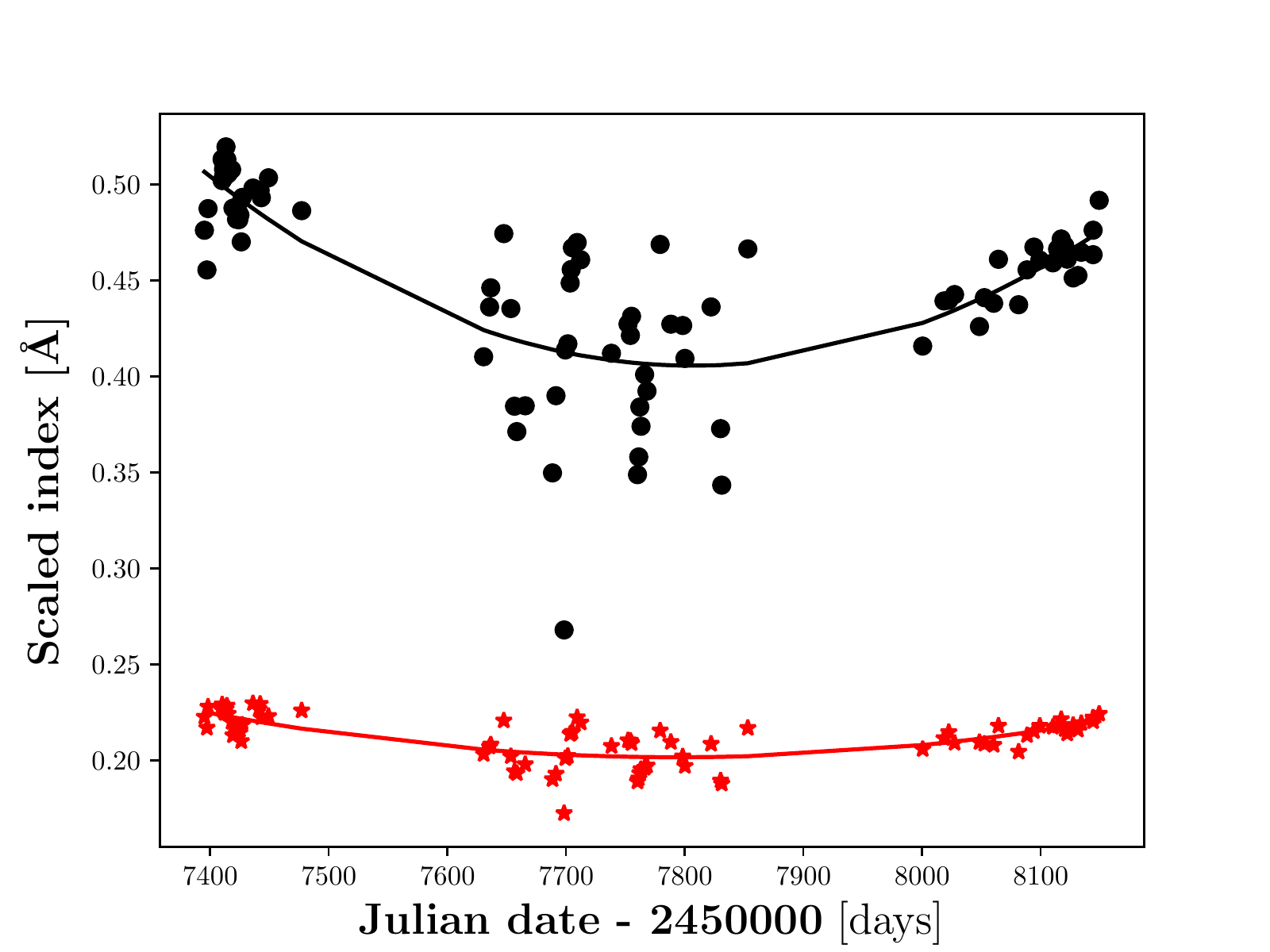}%\vspace{-0.4mm}\\
\includegraphics[width=0.45\textwidth, clip]{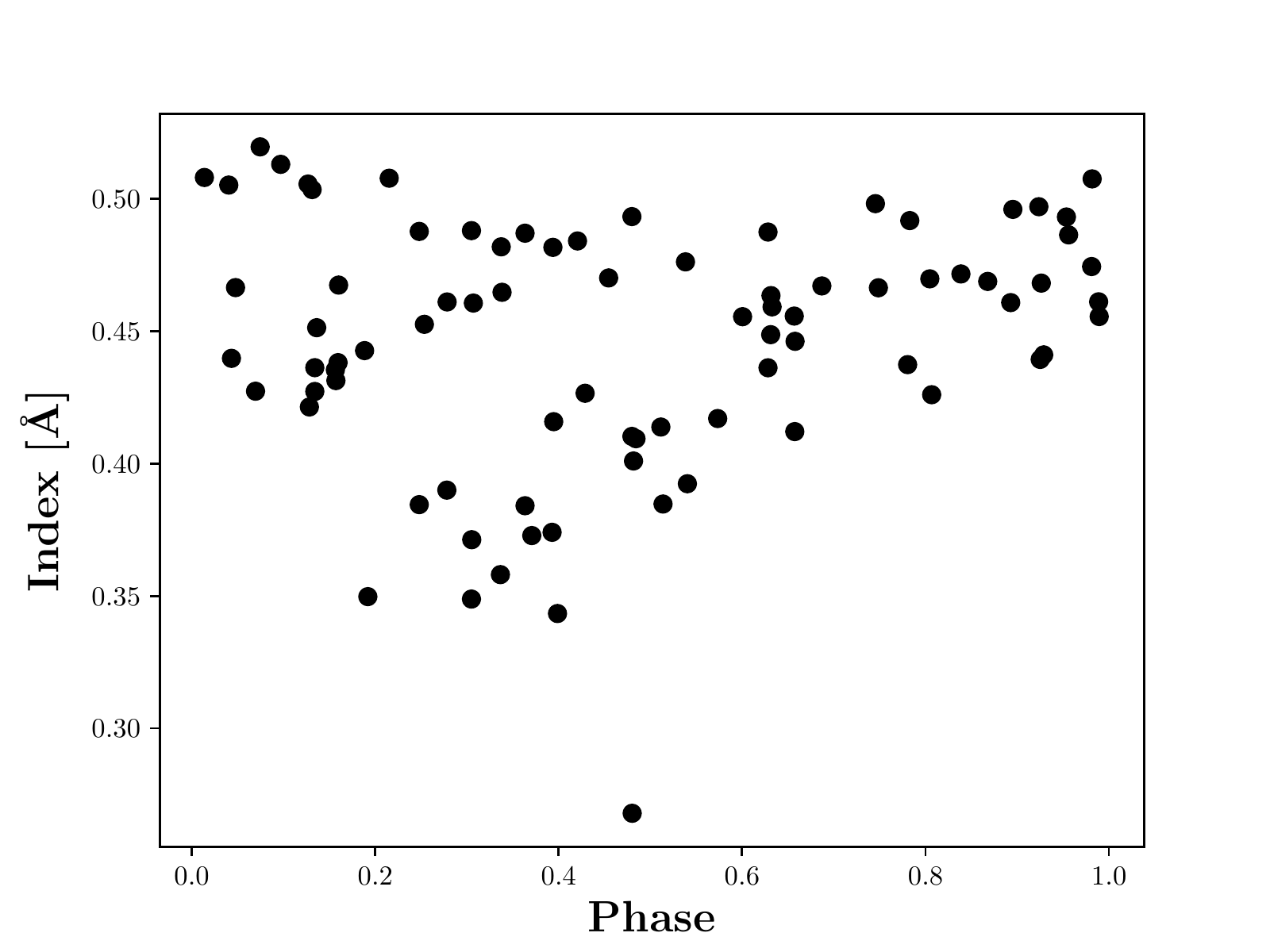}\\
\includegraphics[width=0.45\textwidth, clip]{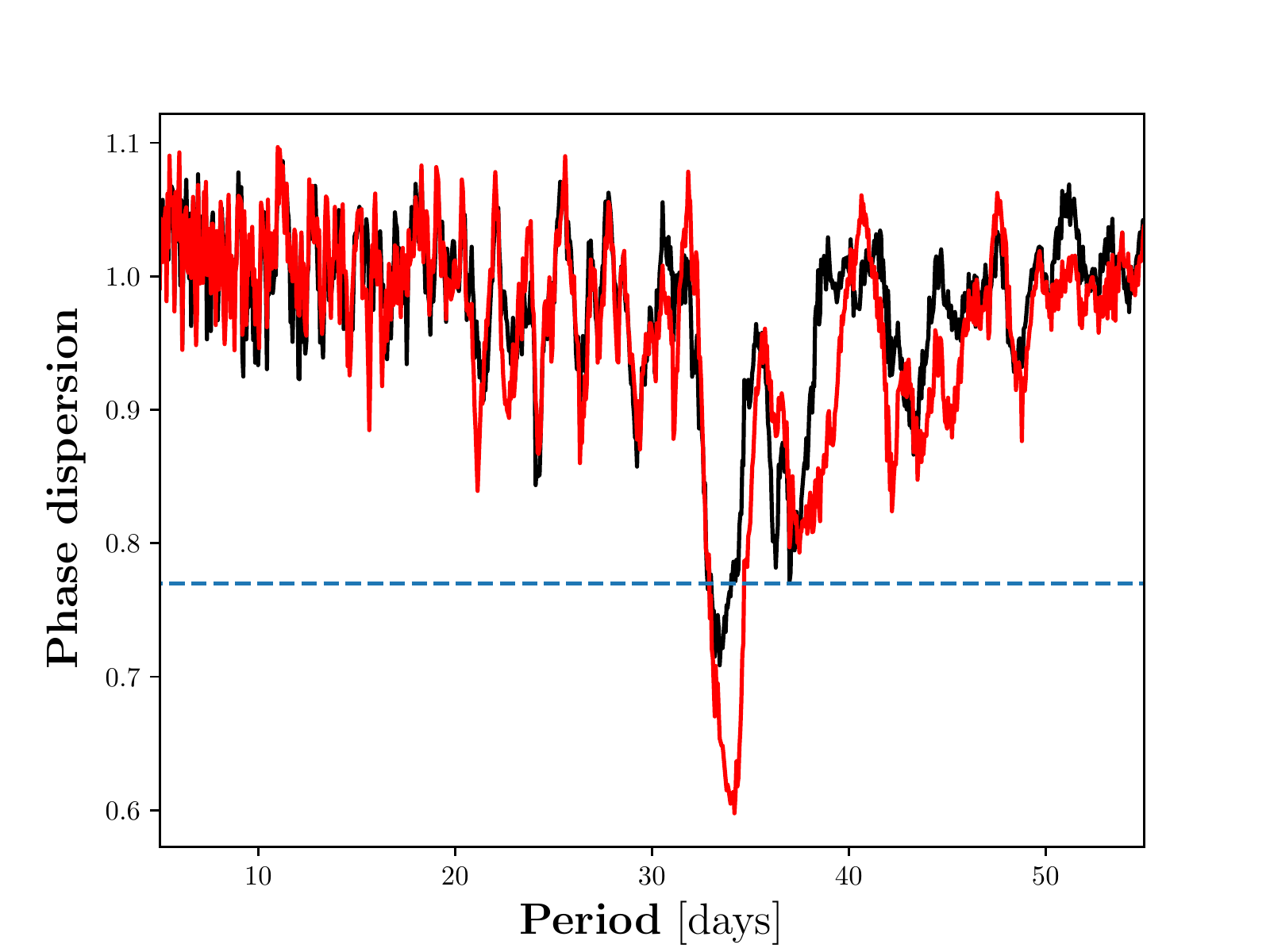}%\vspace{-0.4mm}\\
\includegraphics[width=0.45\textwidth, clip]{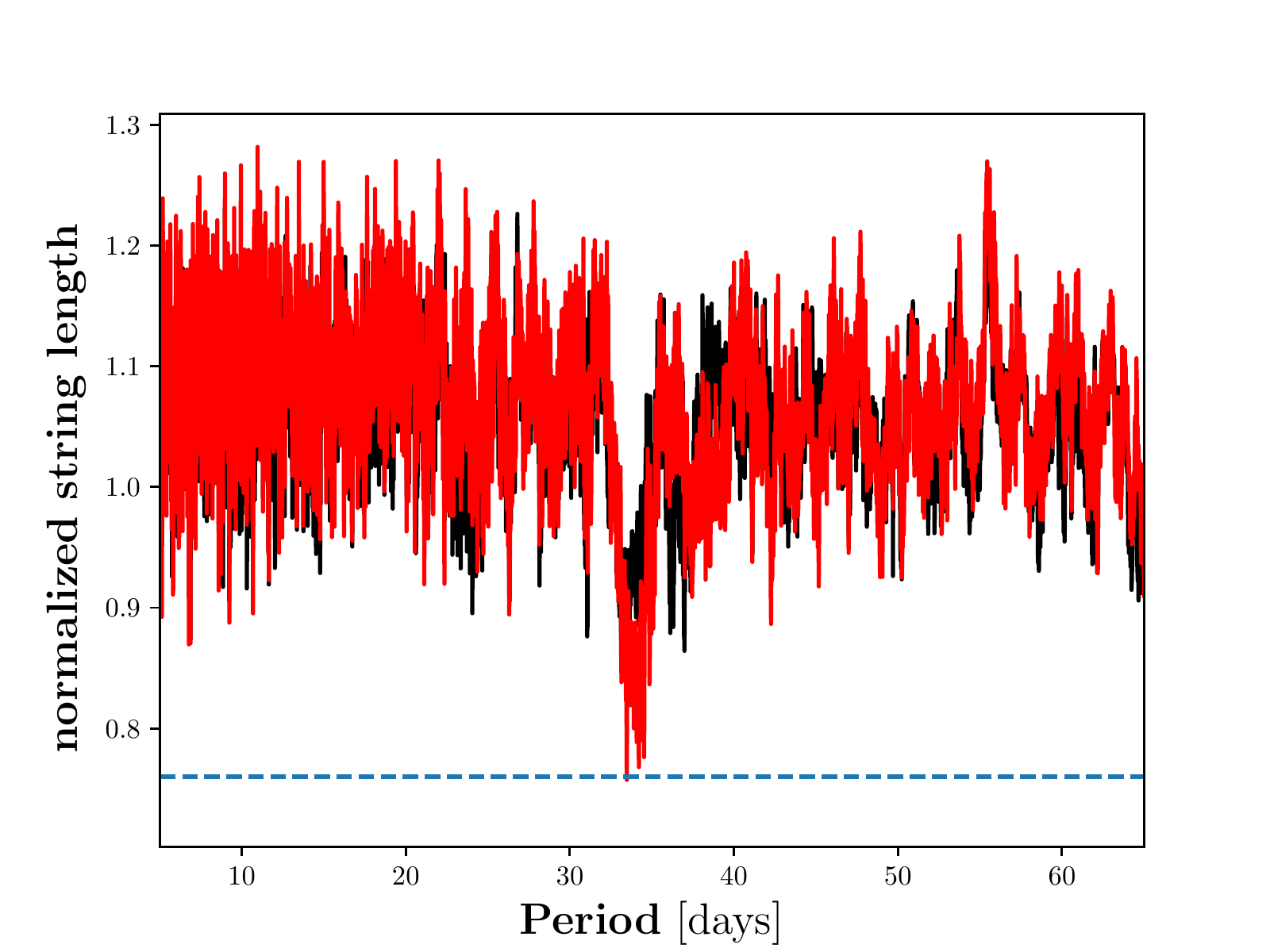}\\
\includegraphics[width=0.45\textwidth, clip]{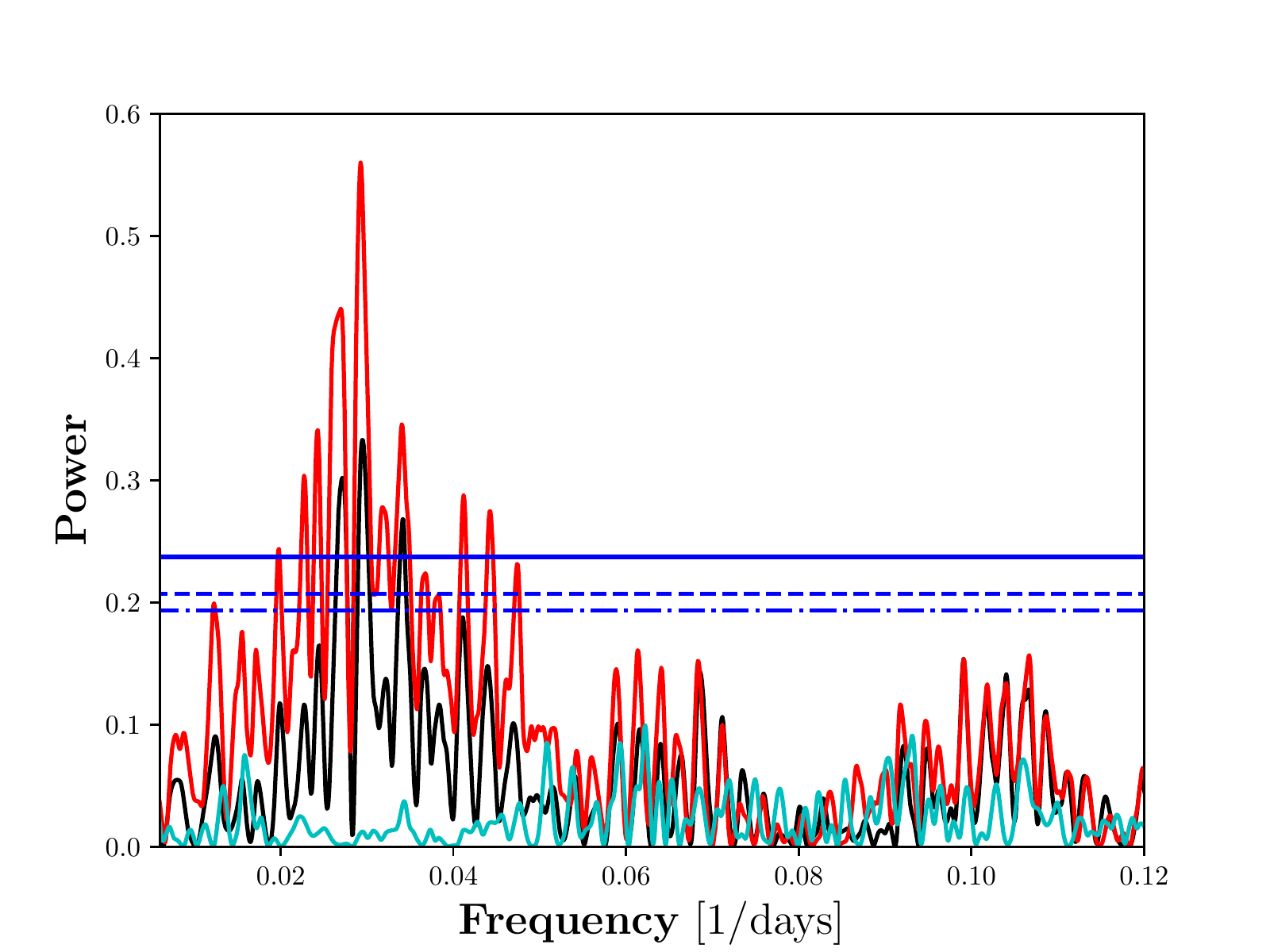}%\vspace{-0.4mm}\\
\includegraphics[width=0.45\textwidth, clip]{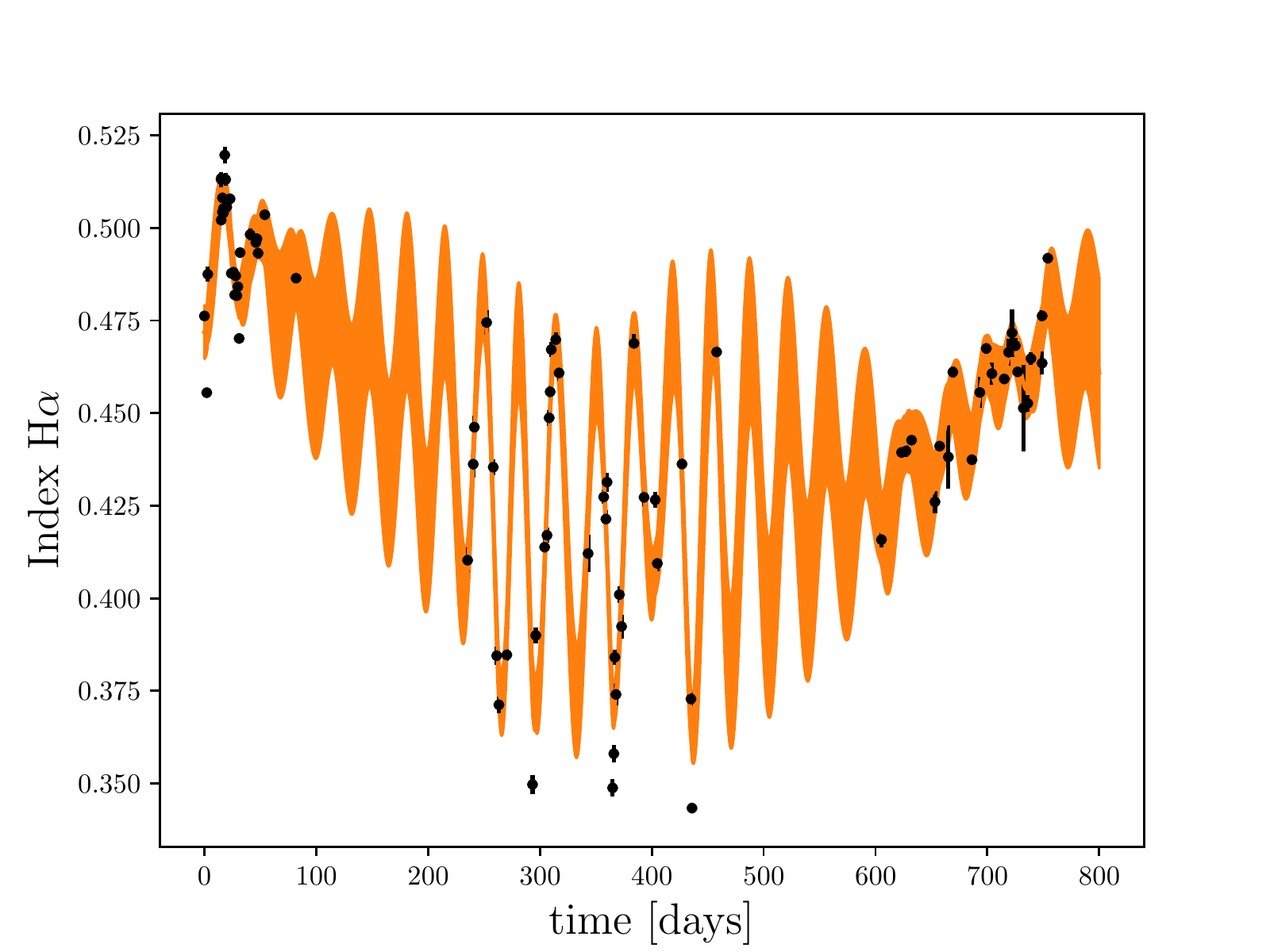}%\vspace{-0.4mm}\\
\caption{\label{hd36395} Results for HD 36395 as explained in Fig. \ref{gxand}.}
\end{figure*}

\begin{figure*}
\includegraphics[width=0.45\textwidth, clip]{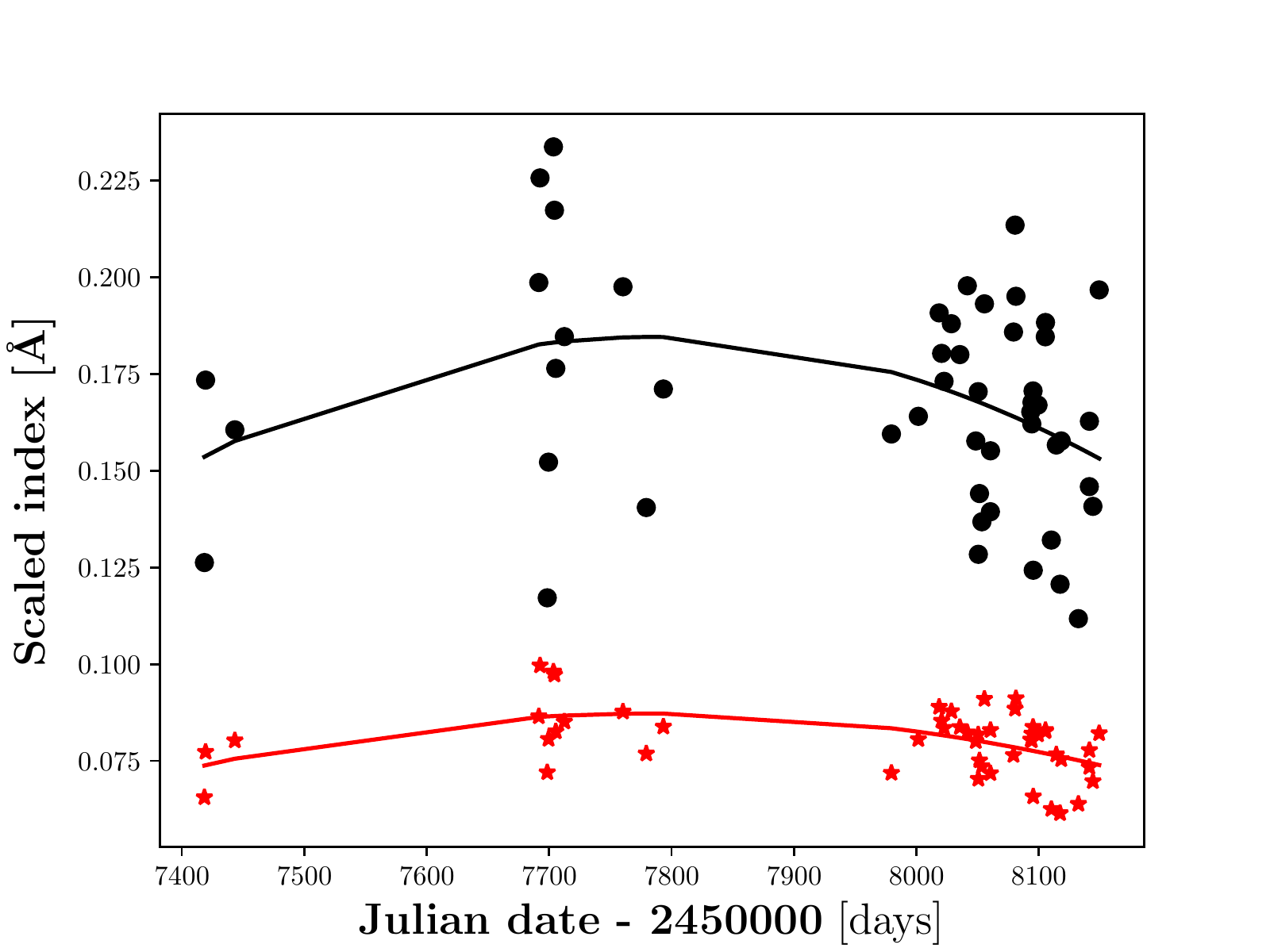}%\vspace{-0.4mm}\\
\includegraphics[width=0.45\textwidth, clip]{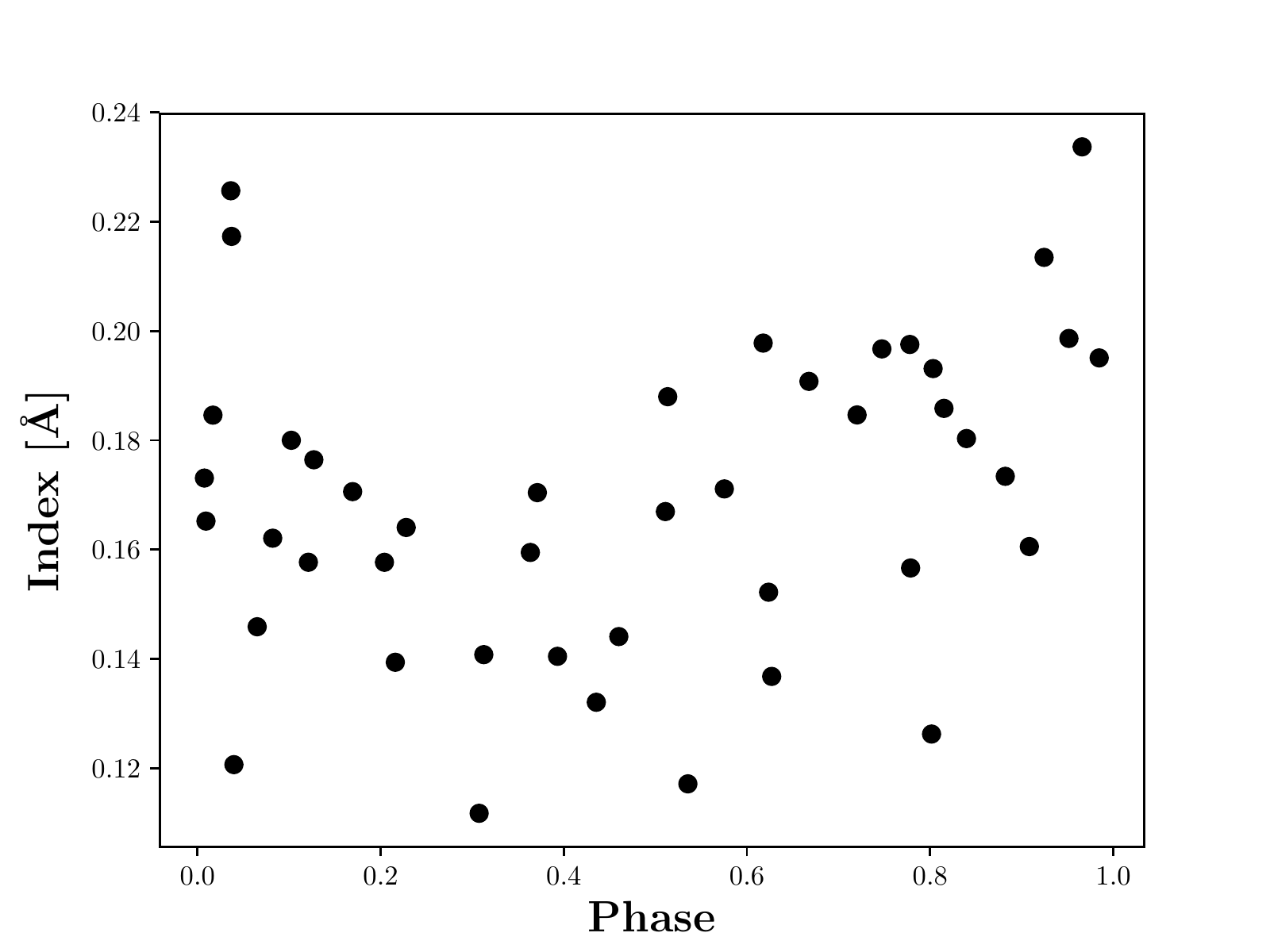}\\
\includegraphics[width=0.45\textwidth, clip]{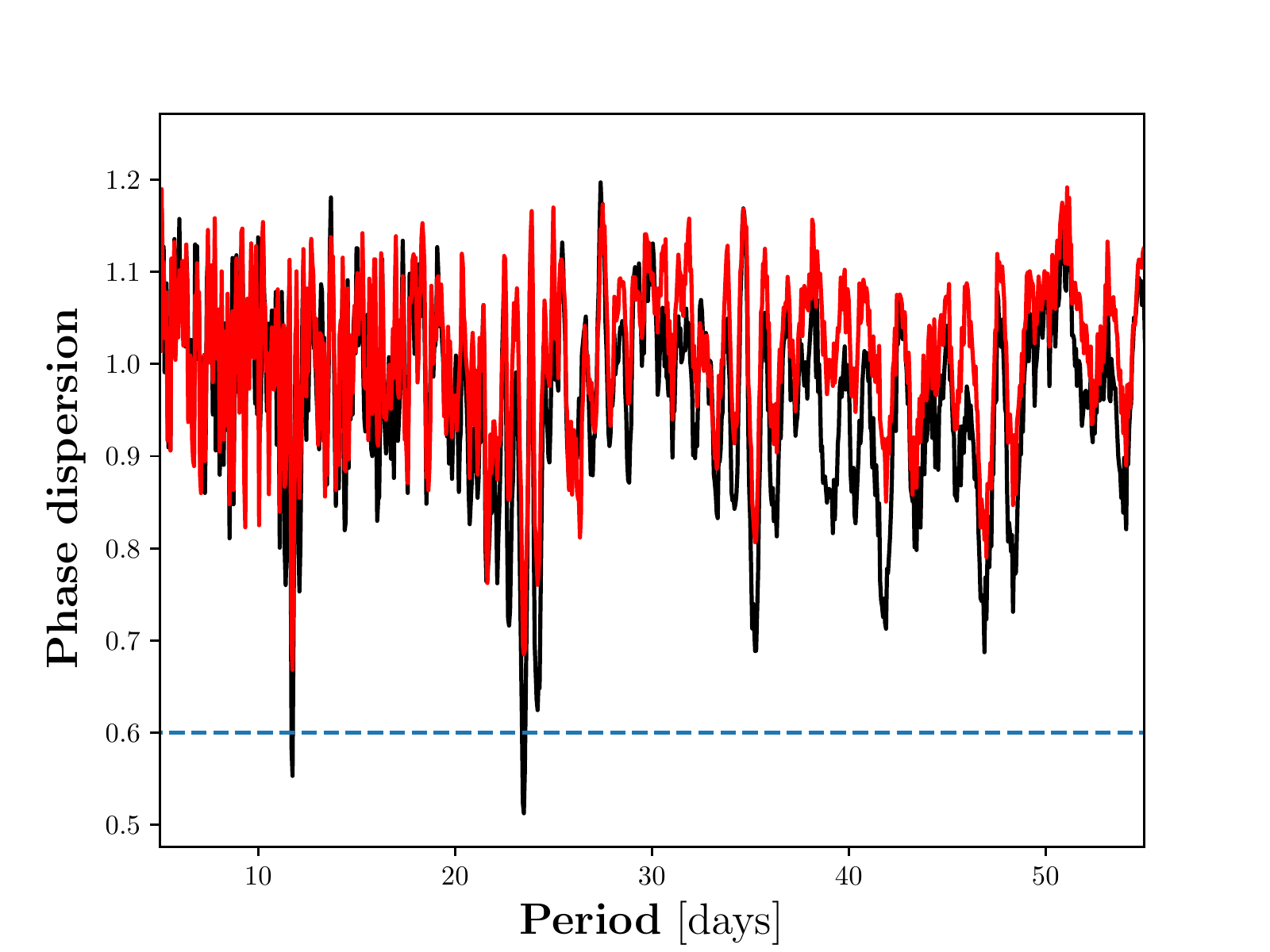}%\vspace{-0.4mm}\\
\includegraphics[width=0.45\textwidth, clip]{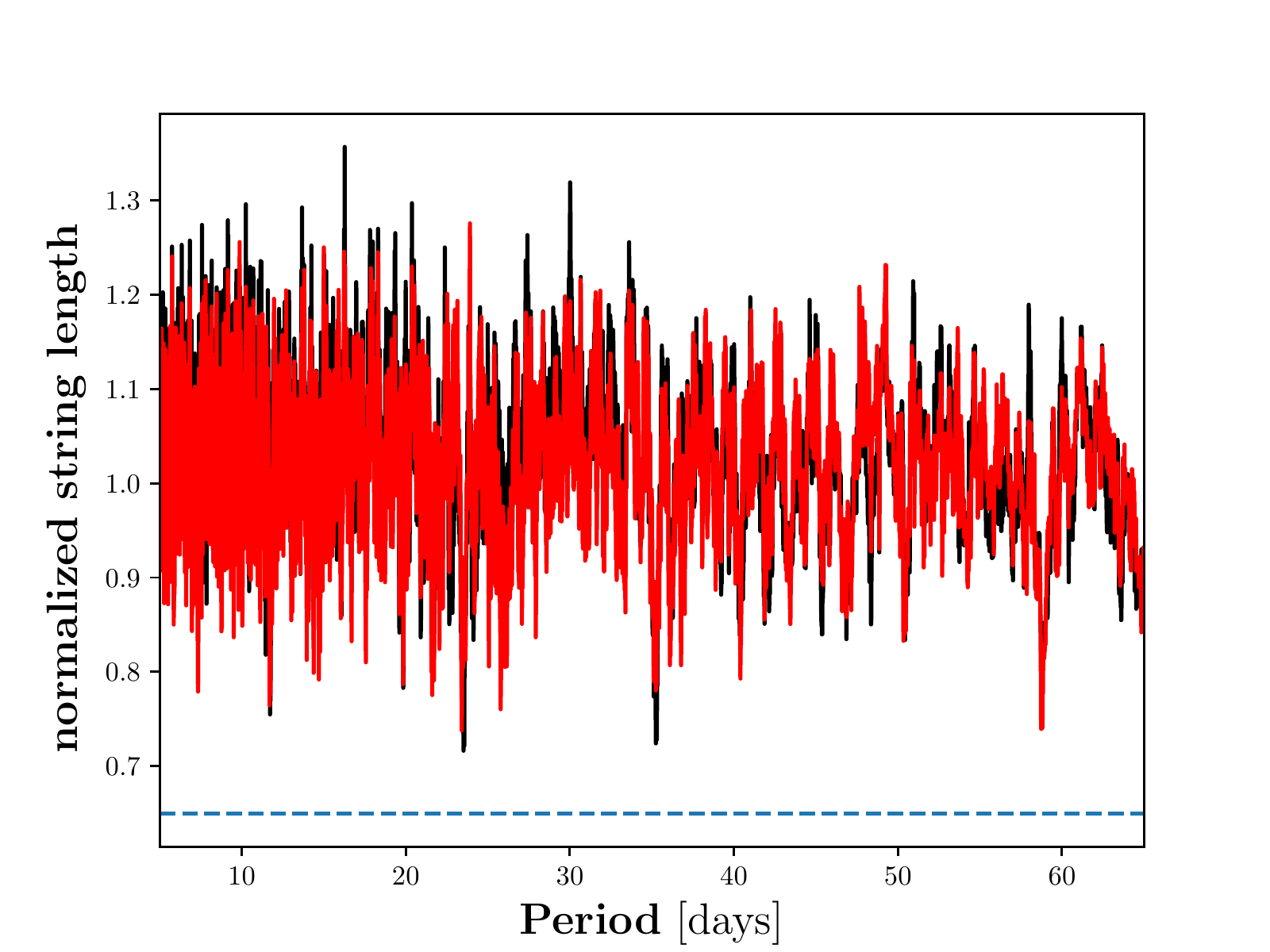}\\
\includegraphics[width=0.45\textwidth, clip]{gls_v2689ori.pdf}%\vspace{-0.4mm}\\
\includegraphics[width=0.45\textwidth, clip]{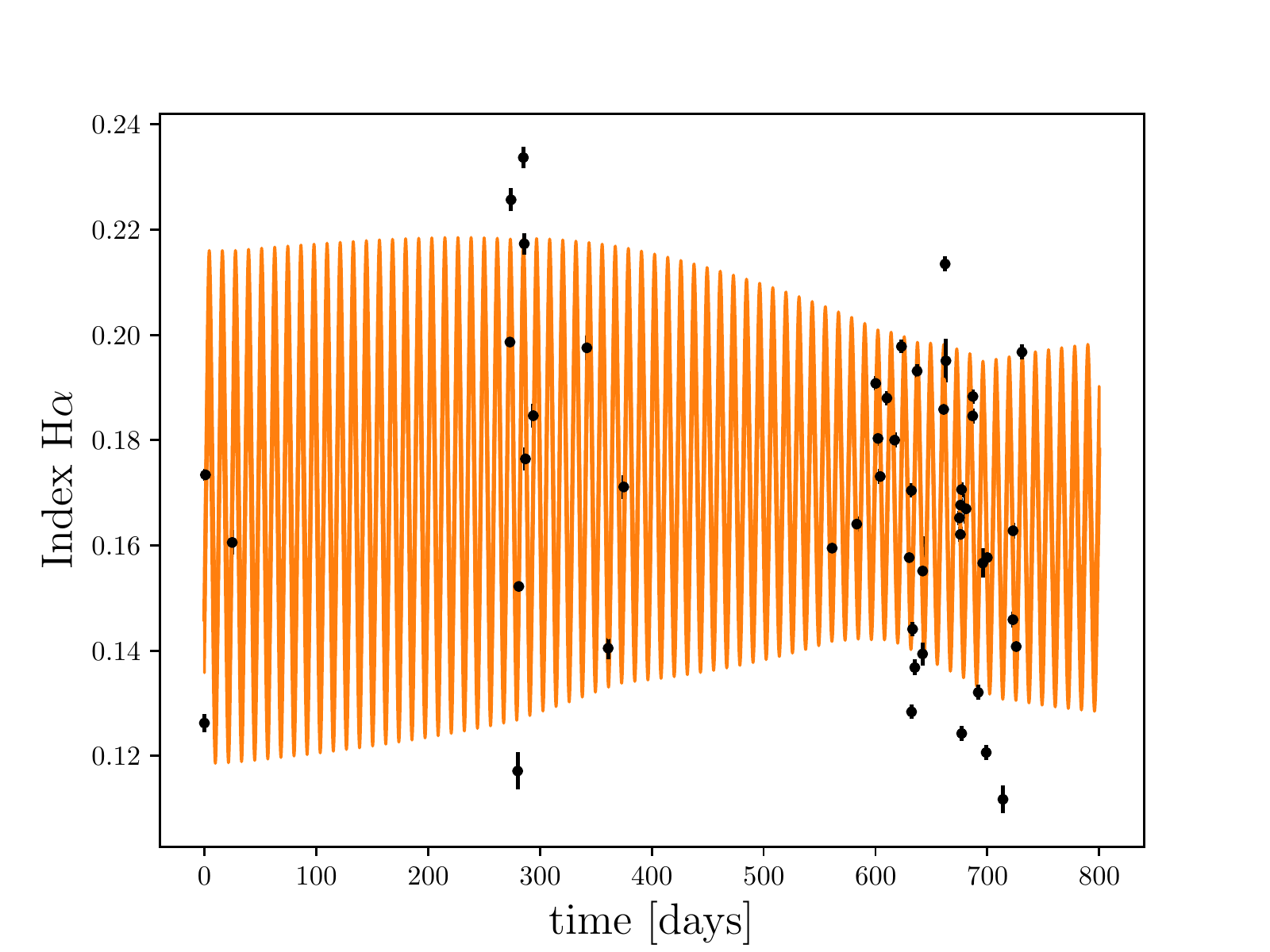}%\vspace{-0.4mm}\\
\caption{\label{v2689ori} Results for V2689 Ori as explained in Fig. \ref{gxand}.}
\end{figure*}

\begin{figure*}
\includegraphics[width=0.45\textwidth, clip]{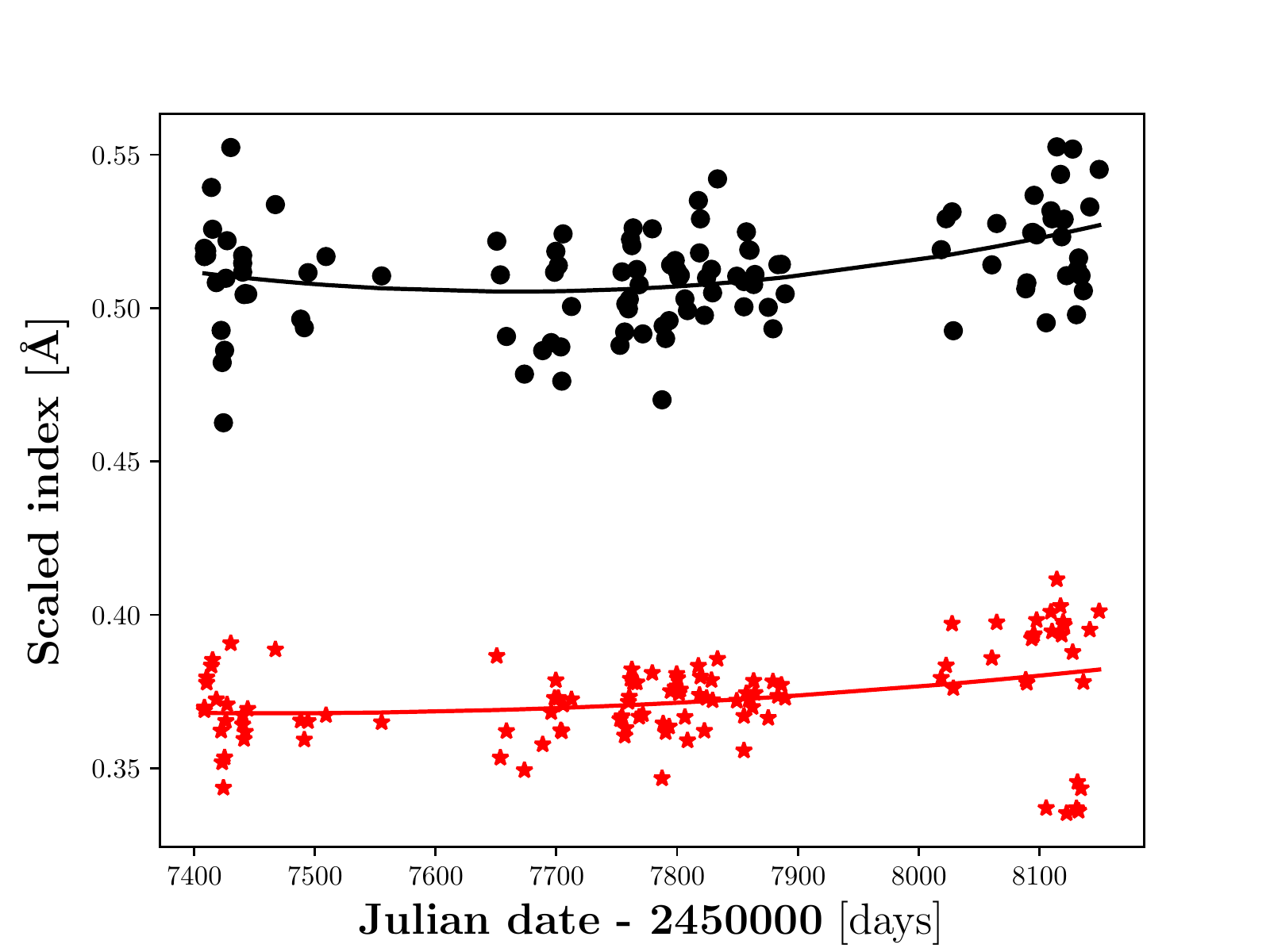}%\vspace{-0.4mm}\\
\includegraphics[width=0.45\textwidth, clip]{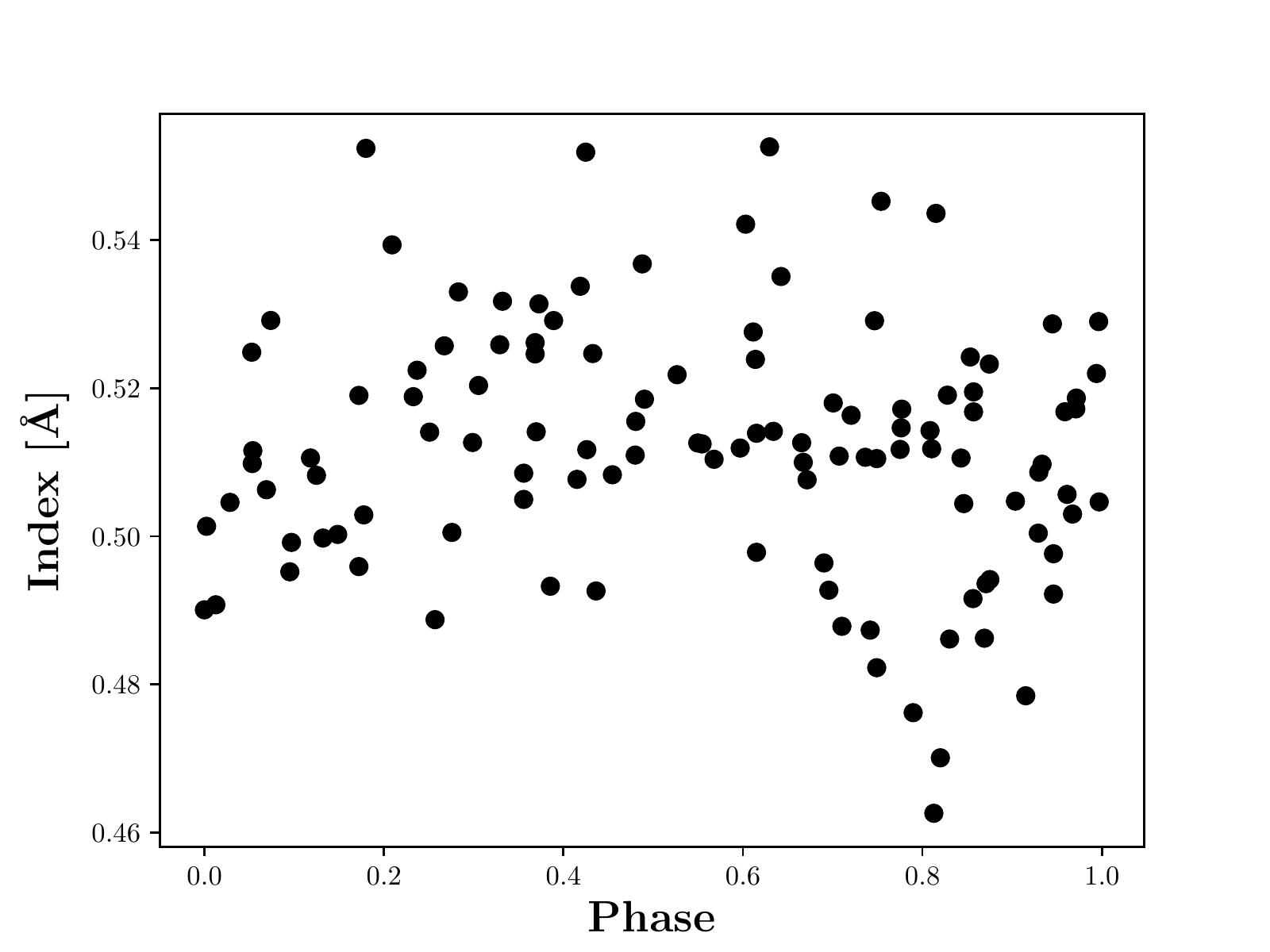}\\
\includegraphics[width=0.45\textwidth, clip]{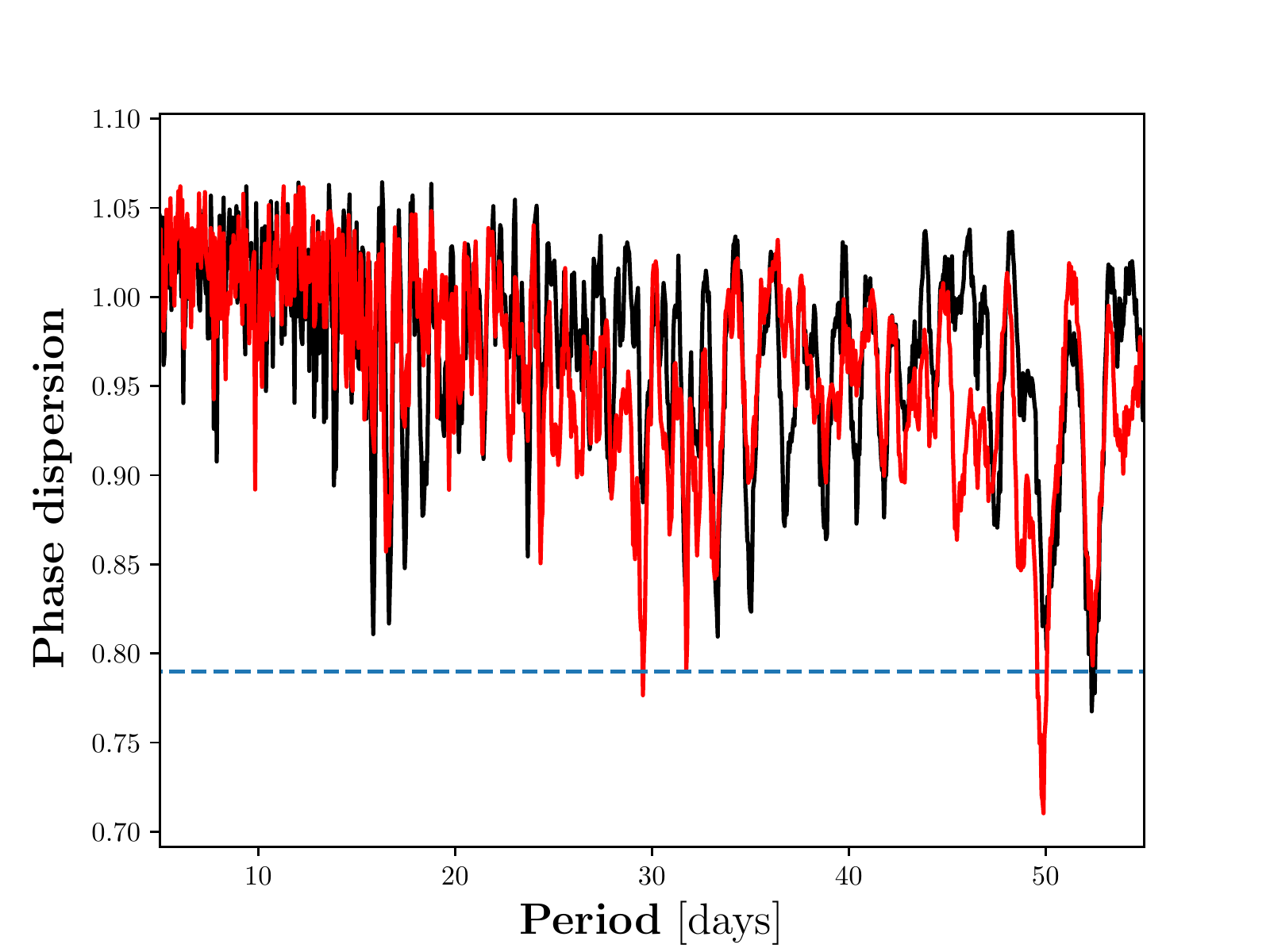}%\vspace{-0.4mm}\\
\includegraphics[width=0.45\textwidth, clip]{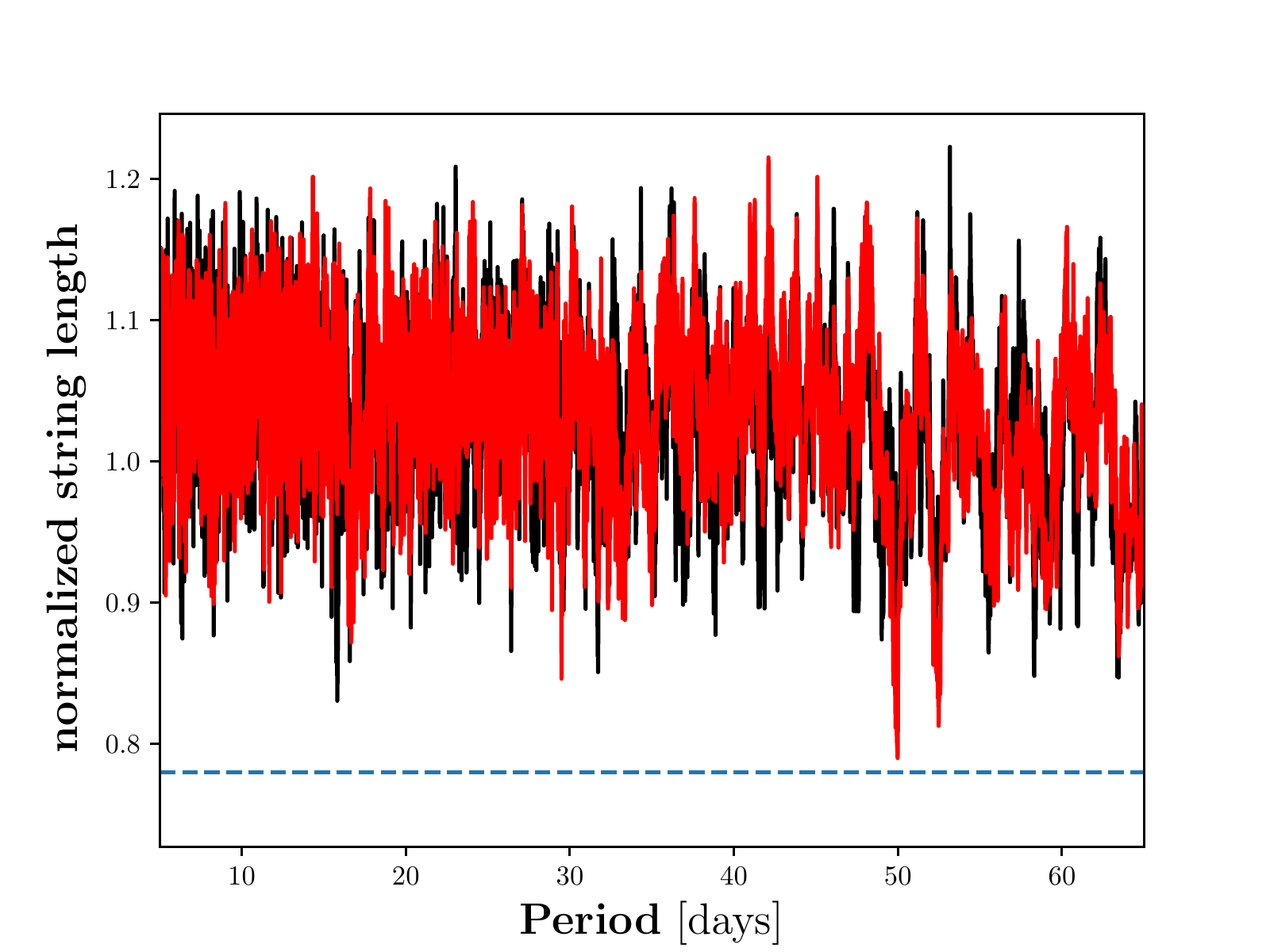}\\
\includegraphics[width=0.45\textwidth, clip]{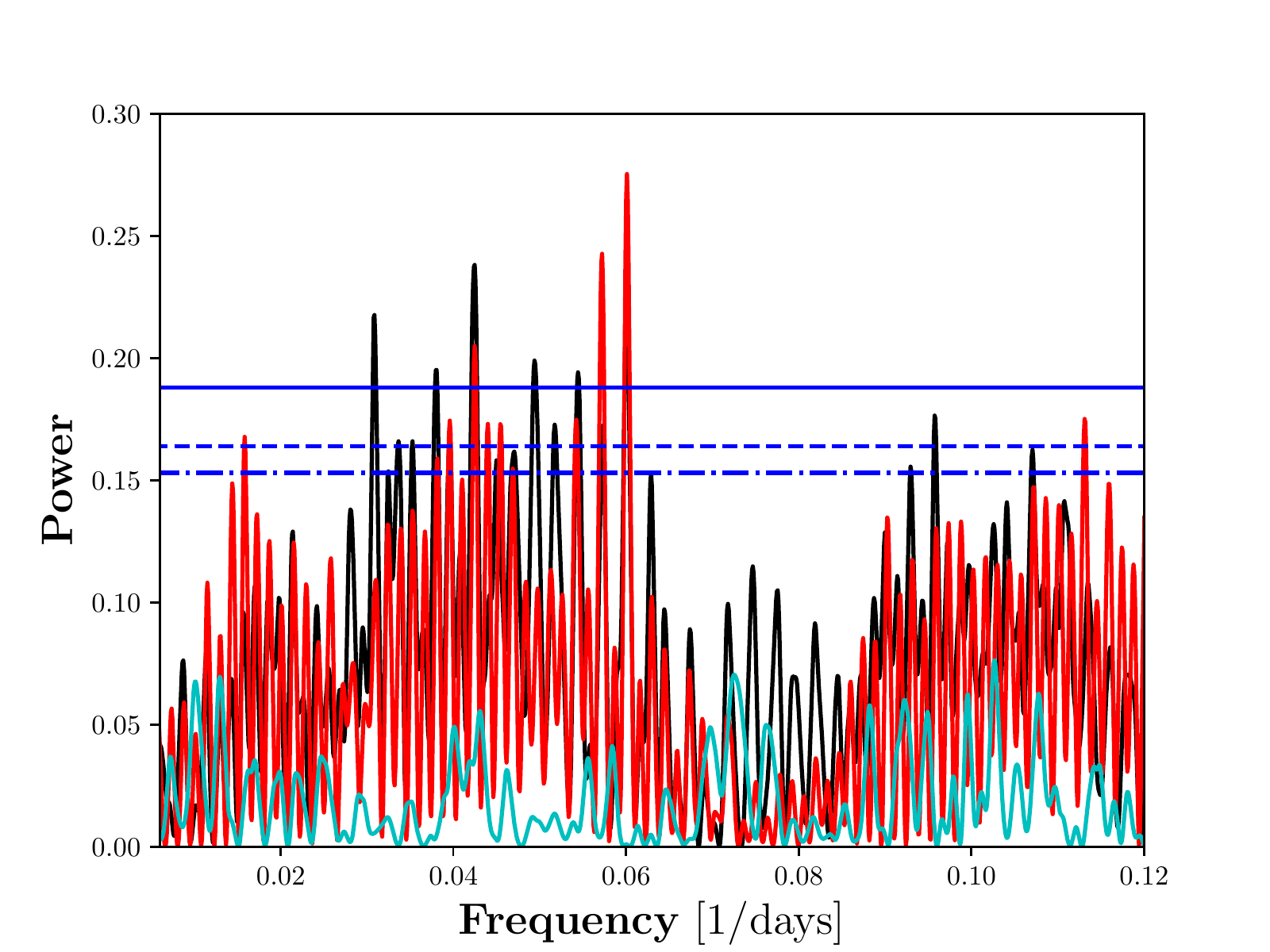}%\vspace{-0.4mm}\\
\includegraphics[width=0.45\textwidth, clip]{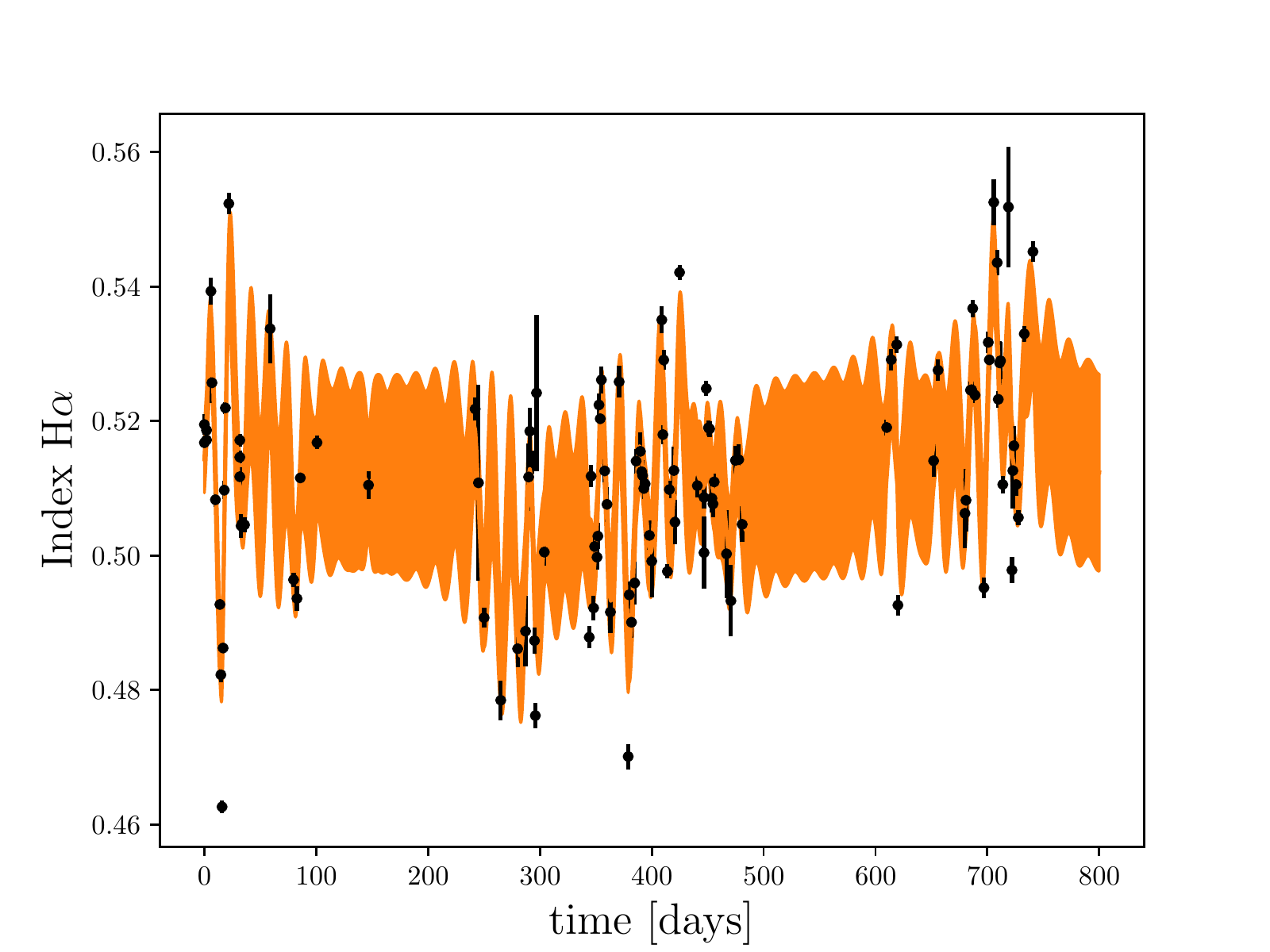}%\vspace{-0.4mm}\\
\caption{\label{hd79211} Results for HD 79211 as explained in Fig. \ref{gxand}.}
\end{figure*}

\begin{figure*}
\includegraphics[width=0.45\textwidth, clip]{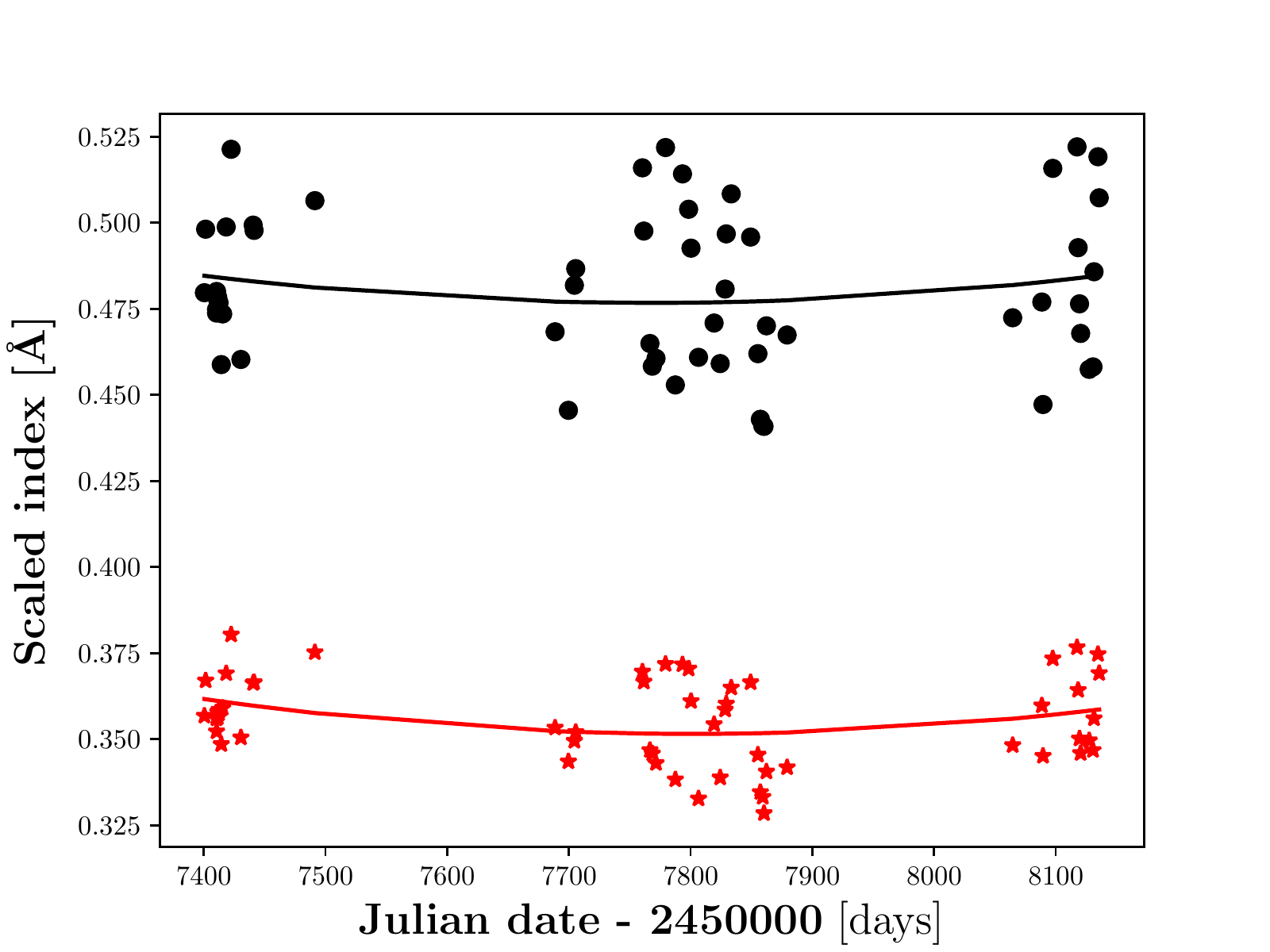}%\vspace{-0.4mm}\\
\includegraphics[width=0.45\textwidth, clip]{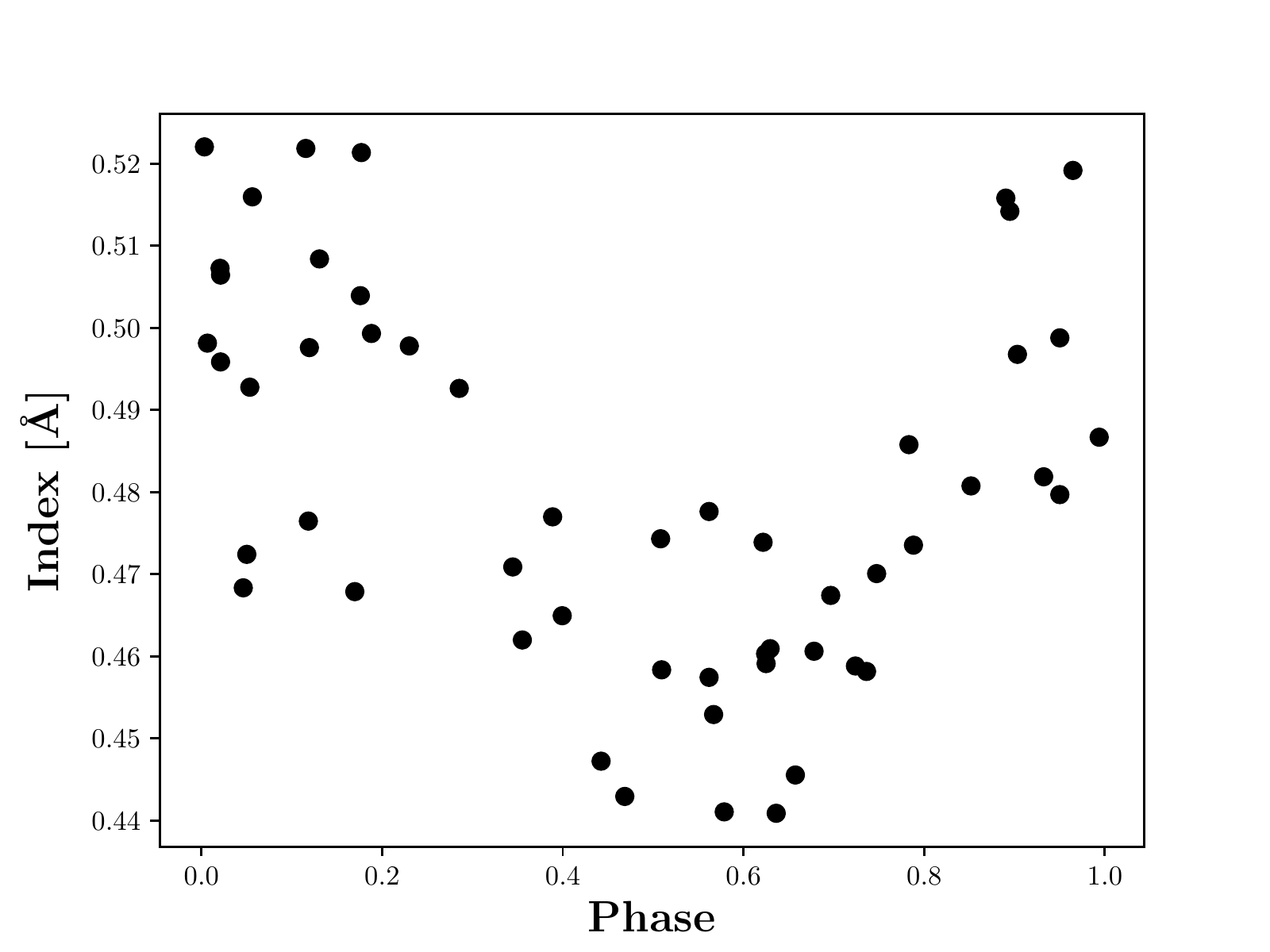}\\
\includegraphics[width=0.45\textwidth, clip]{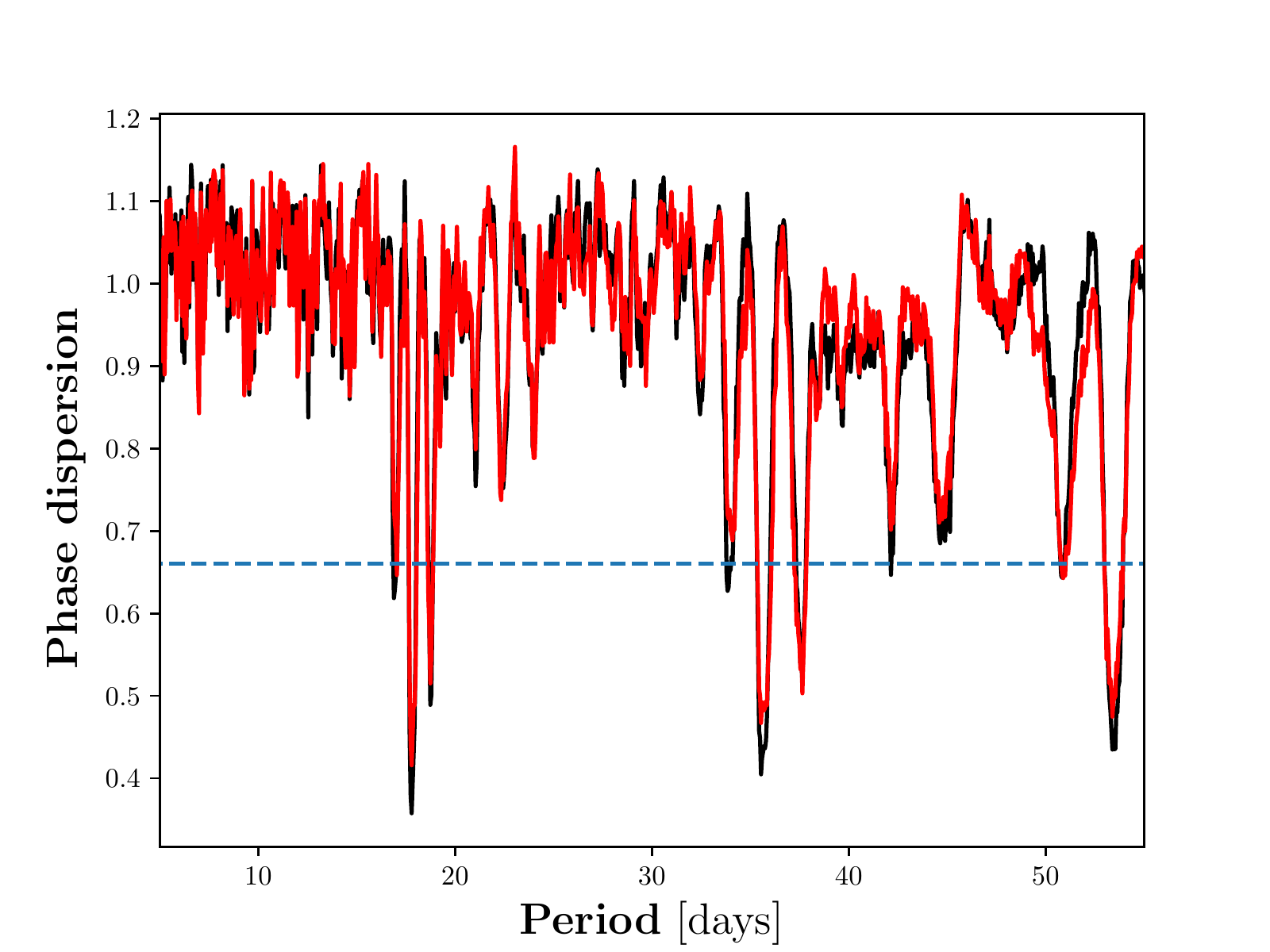}%\vspace{-0.4mm}\\
\includegraphics[width=0.45\textwidth, clip]{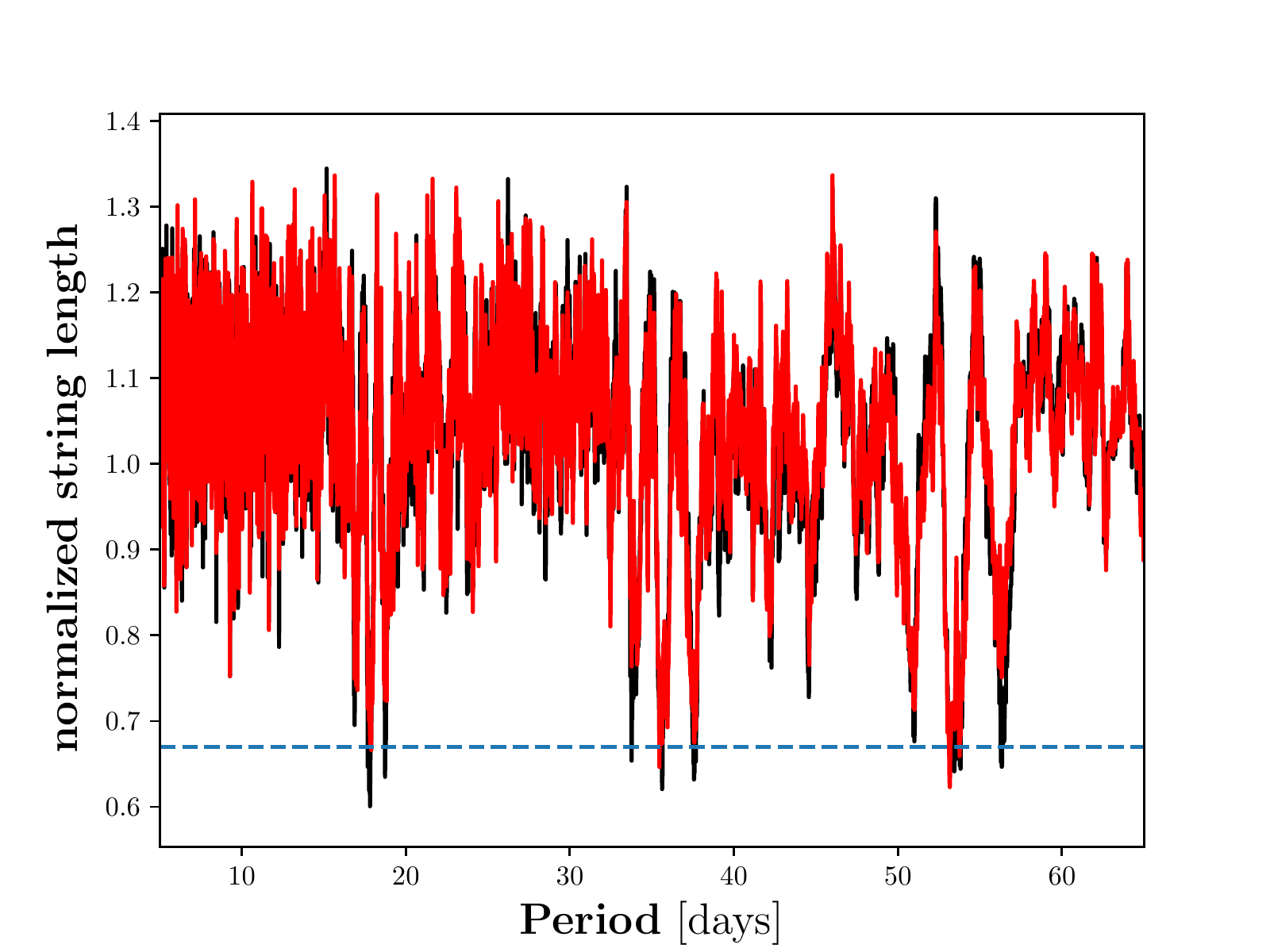}\\
\includegraphics[width=0.45\textwidth, clip]{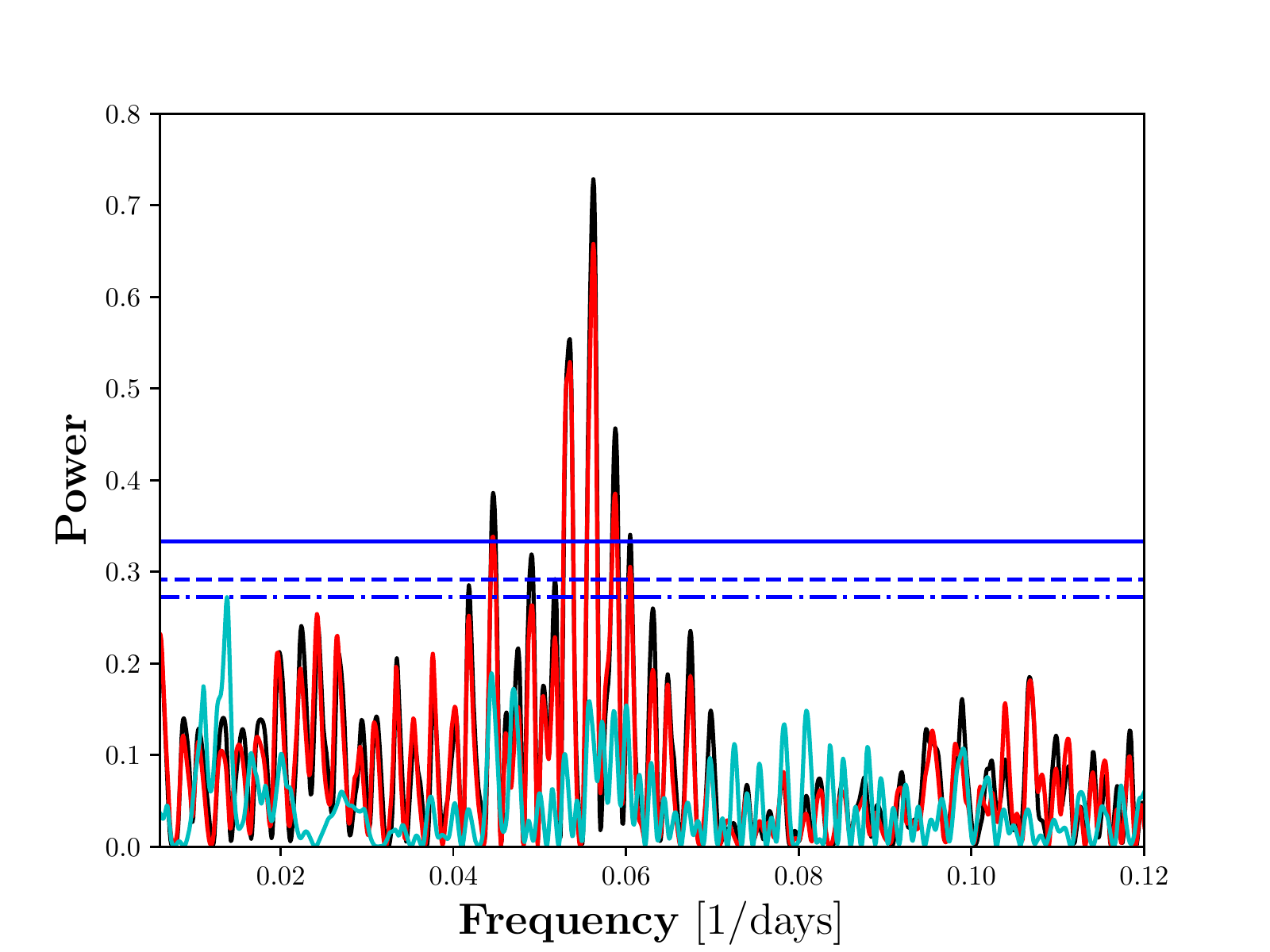}%\vspace{-0.4mm}\\
\includegraphics[width=0.45\textwidth, clip]{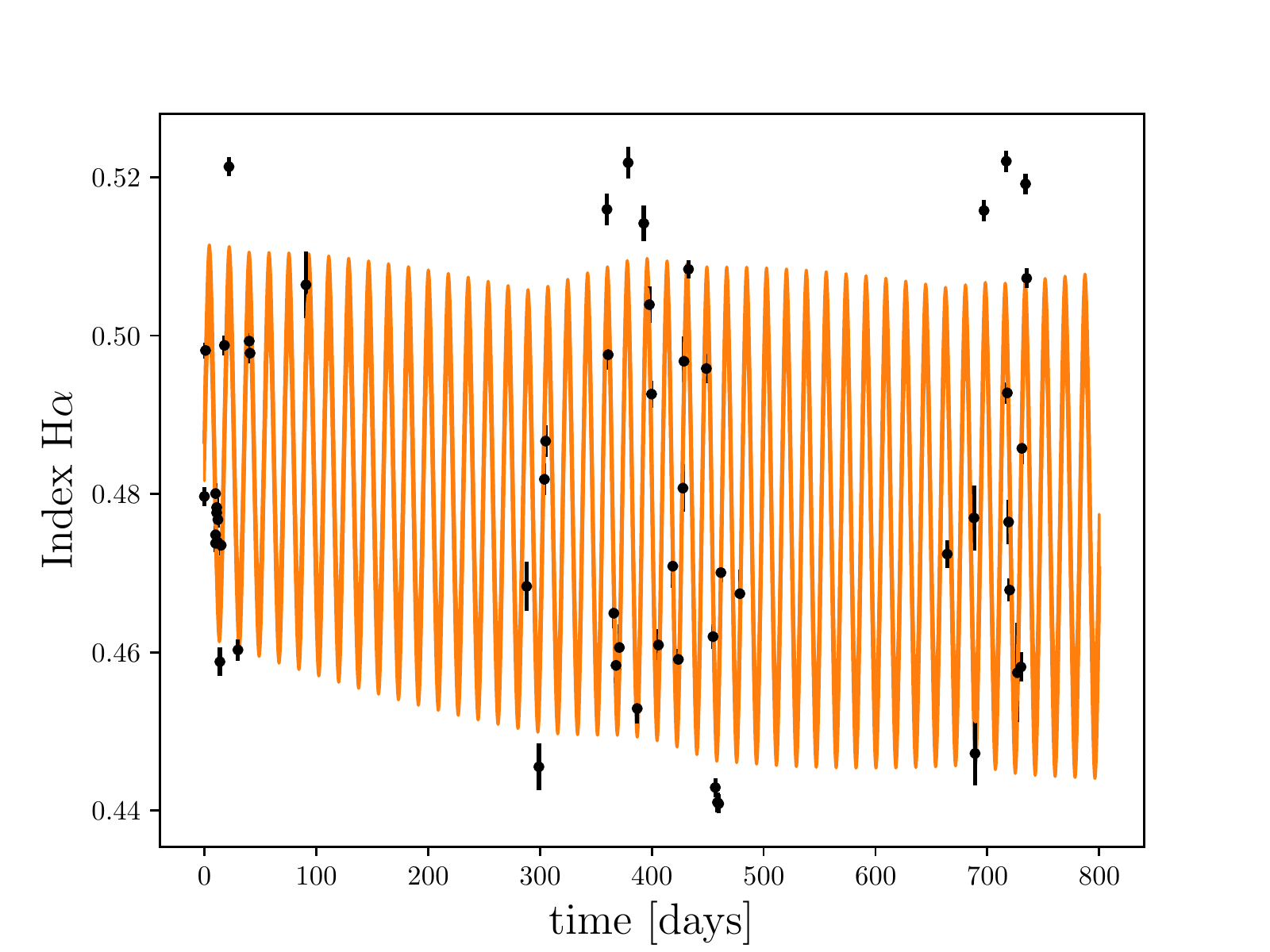}%\vspace{-0.4mm}\\
\caption{\label{bd63869} Results for BD+63 869 as explained in Fig. \ref{gxand}.}
\end{figure*}

\begin{figure*}
\includegraphics[width=0.45\textwidth, clip]{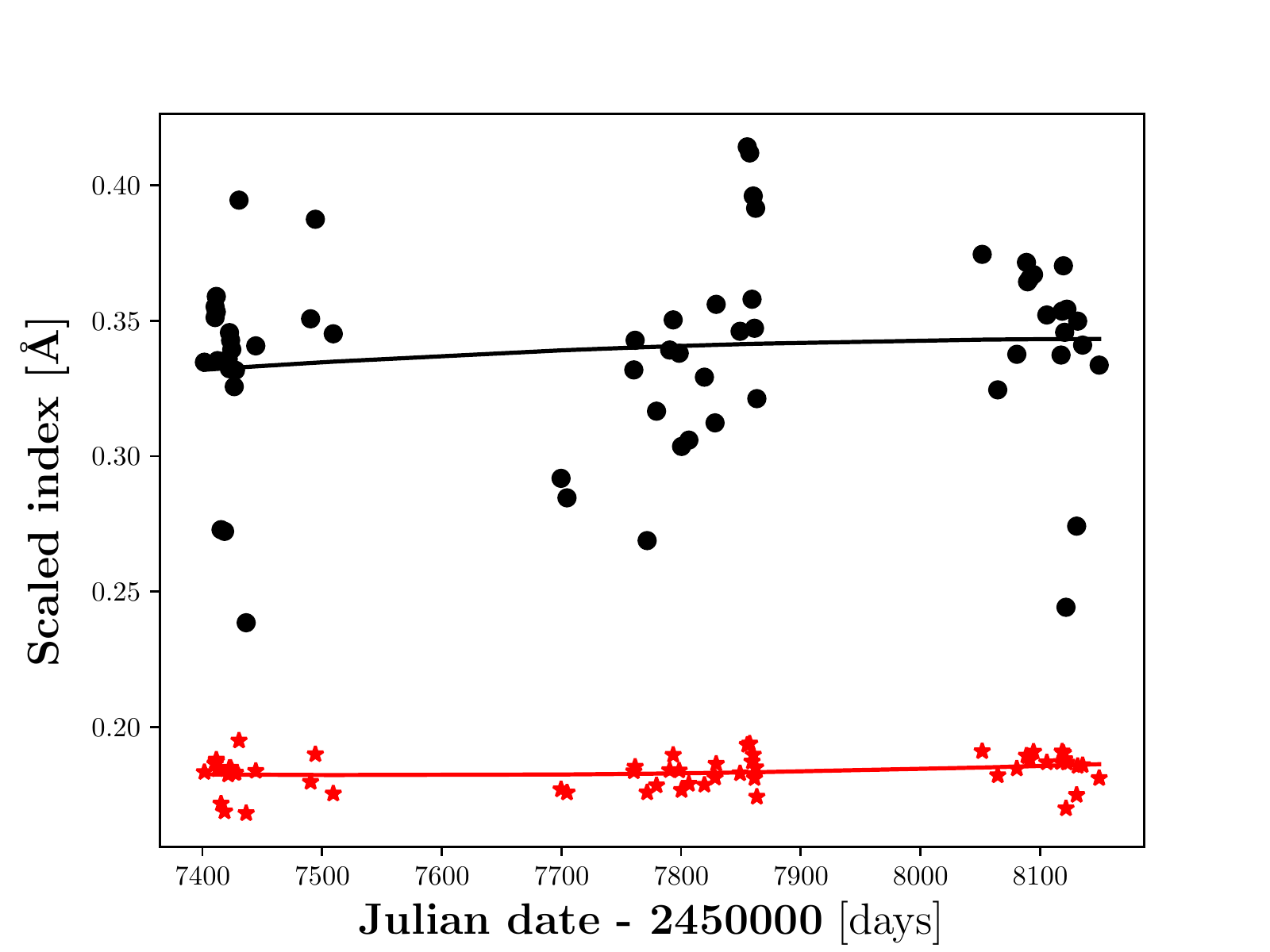}%\vspace{-0.4mm}\\
\includegraphics[width=0.45\textwidth, clip]{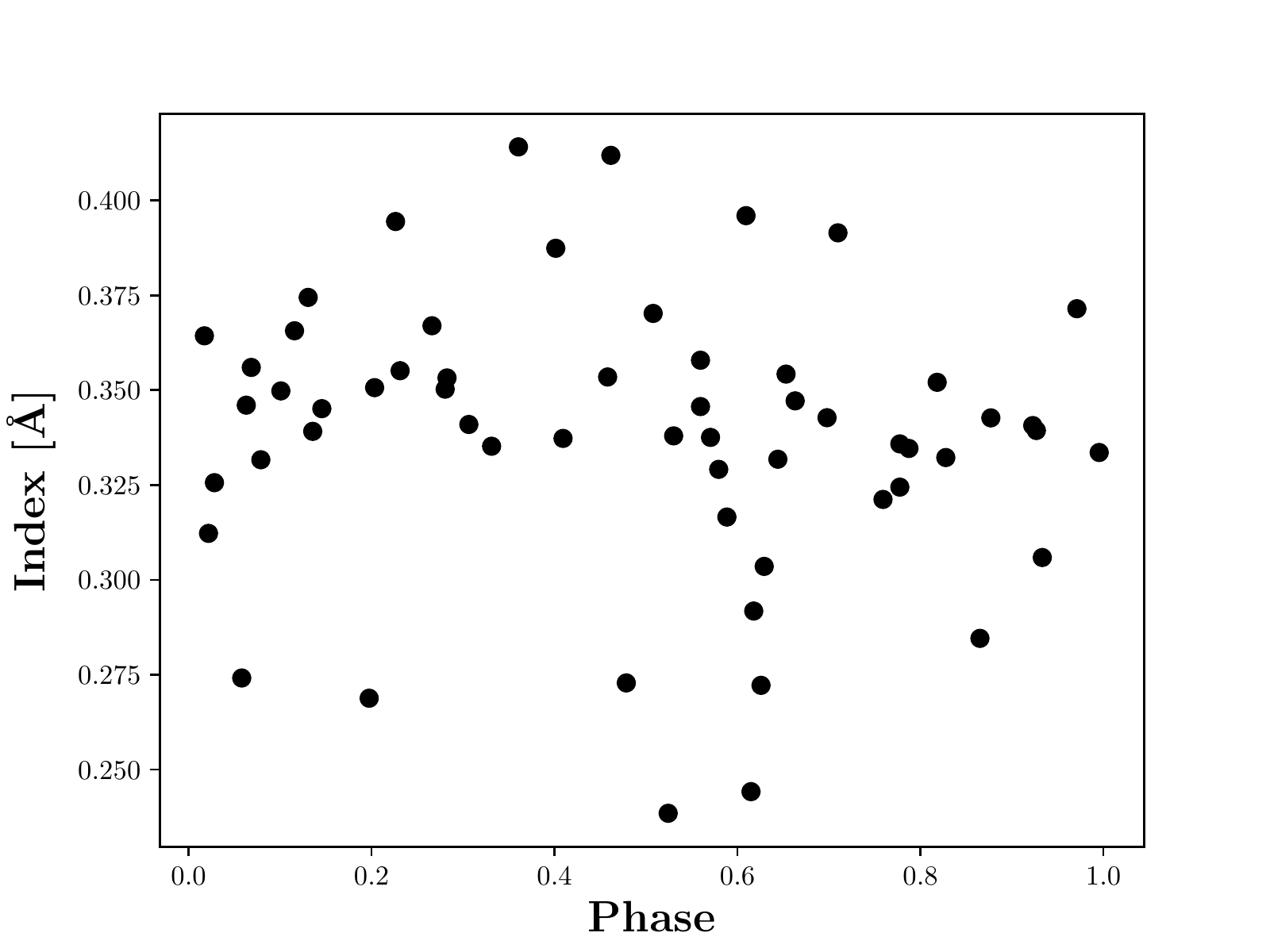}\\
\includegraphics[width=0.45\textwidth, clip]{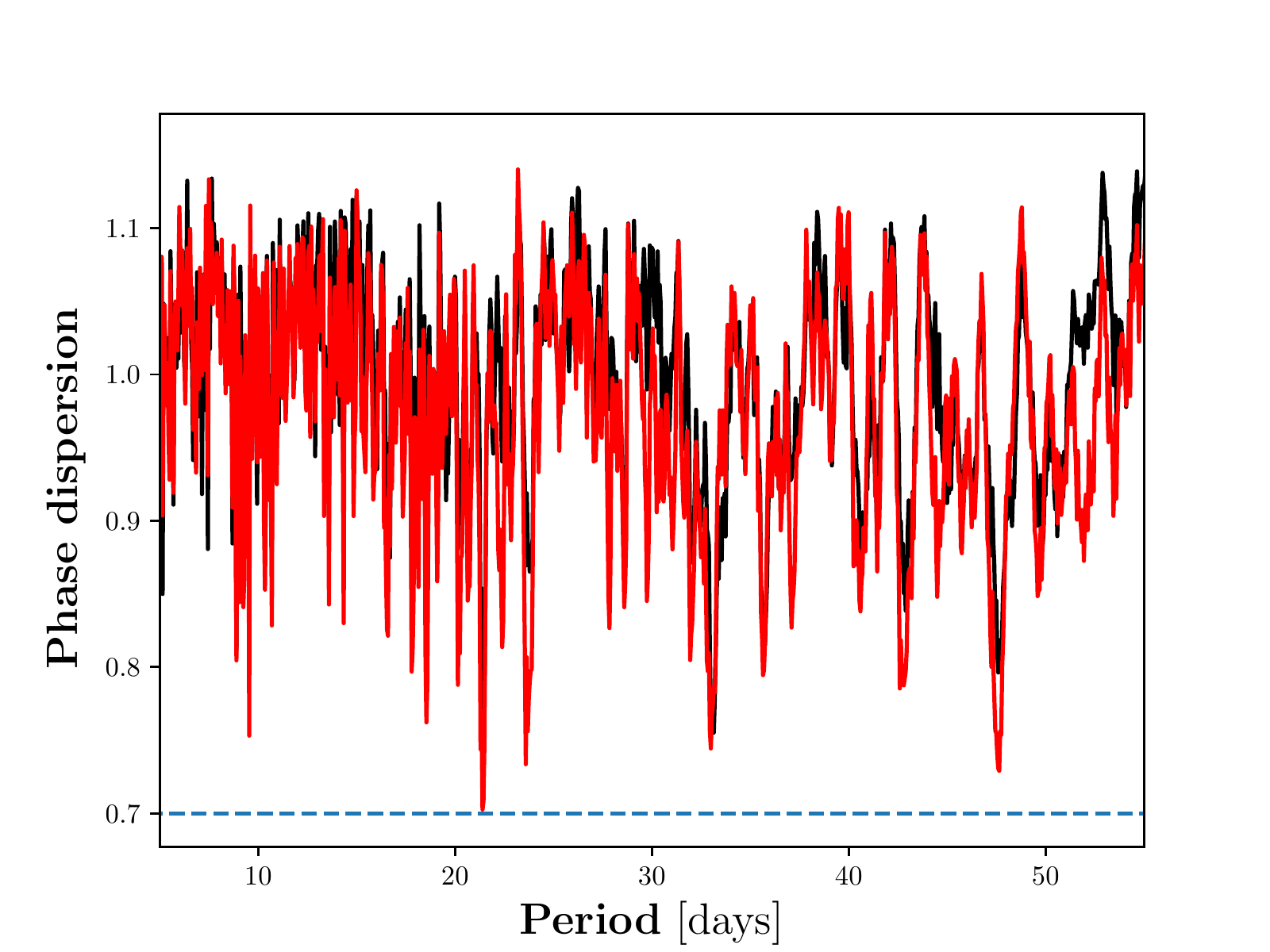}%\vspace{-0.4mm}\\
\includegraphics[width=0.45\textwidth, clip]{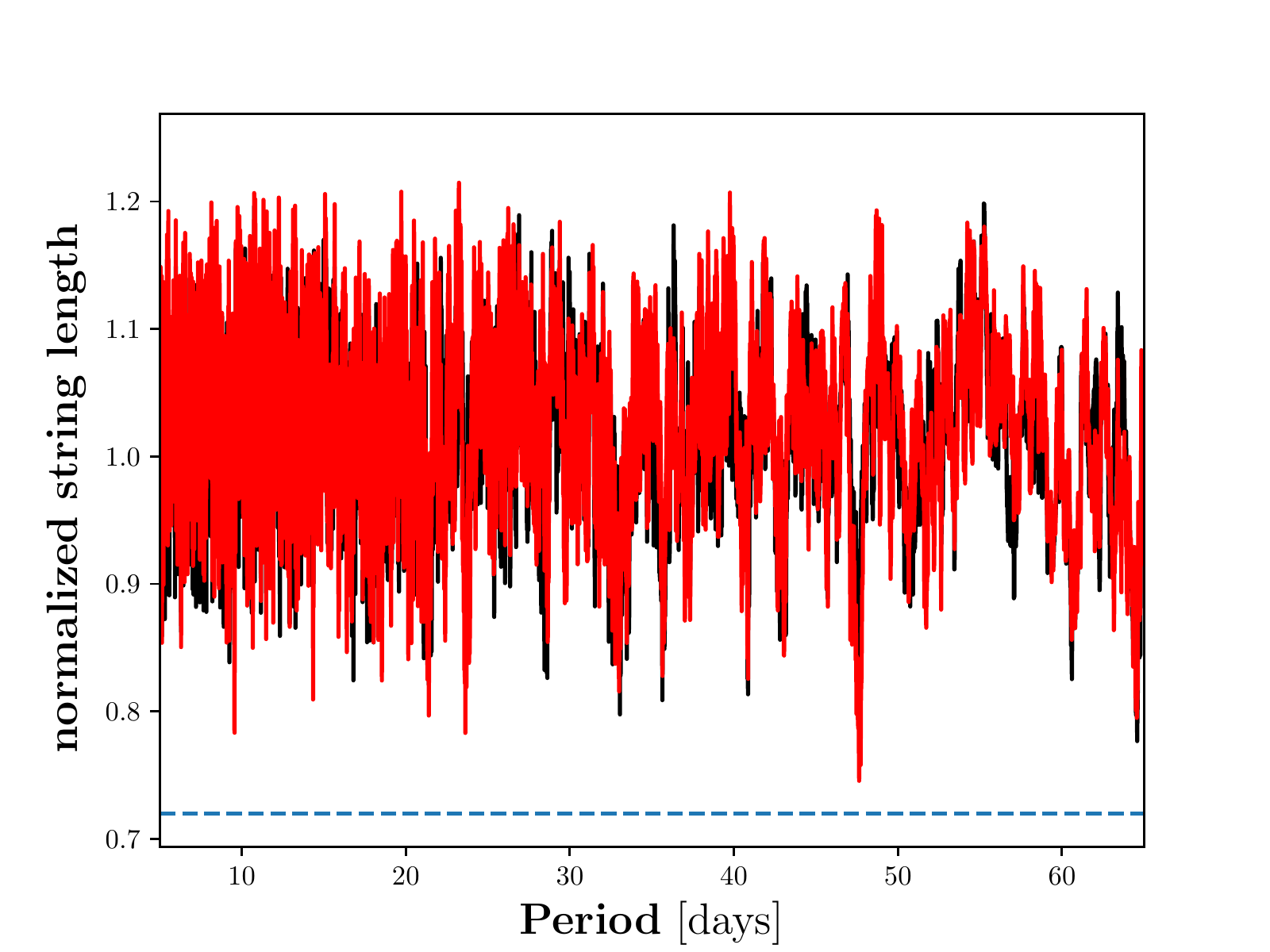}\\
\includegraphics[width=0.45\textwidth, clip]{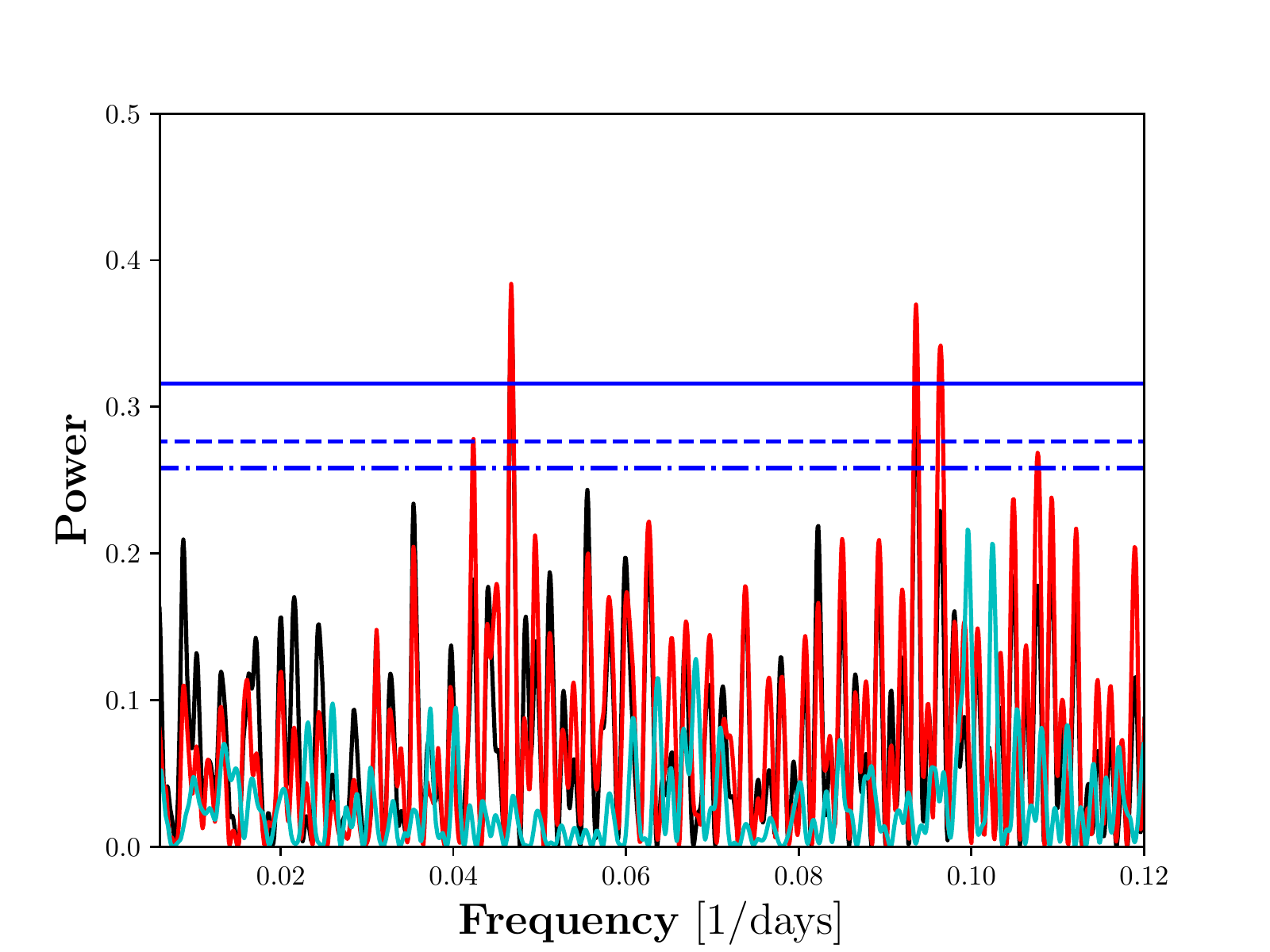}%\vspace{-0.4mm}\\
\includegraphics[width=0.45\textwidth, clip]{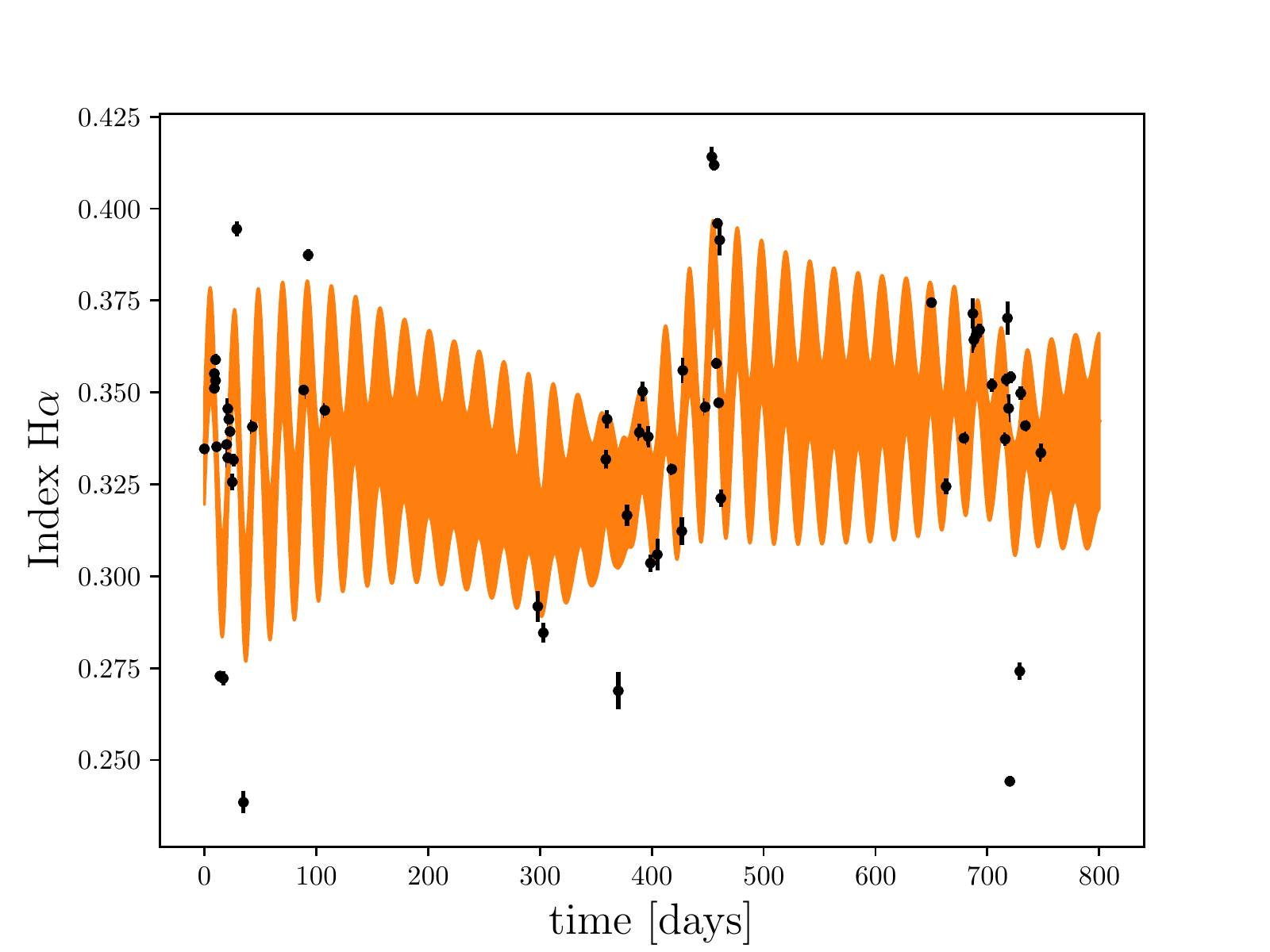}%\vspace{-0.4mm}\\
\caption{\label{ansex} Results for AN Sex as explained in Fig. \ref{gxand}.}
\end{figure*}

\begin{figure*}

\includegraphics[width=0.45\textwidth, clip]{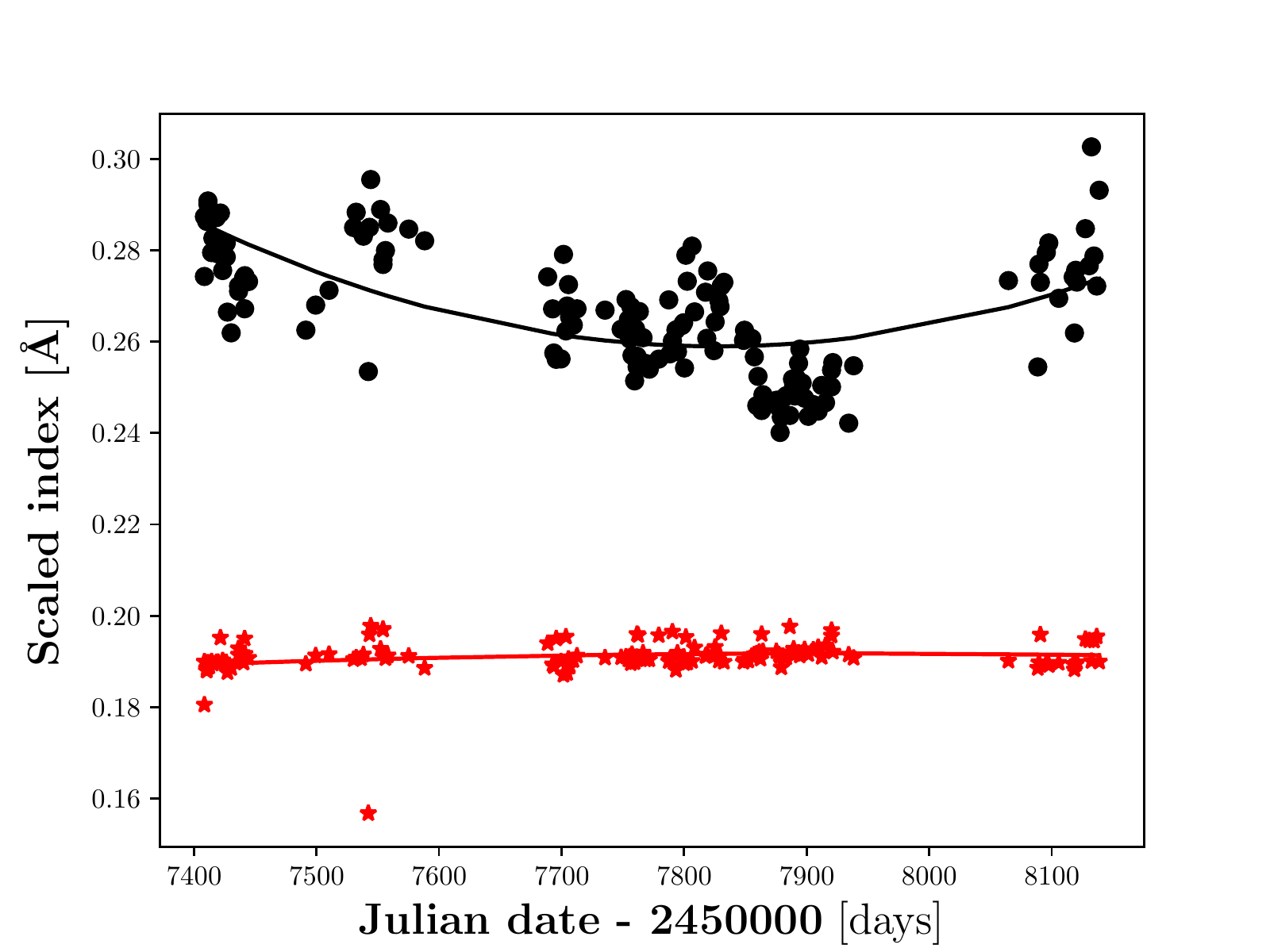}\\%\vspace{-0.4mm}\\
\includegraphics[width=0.45\textwidth, clip]{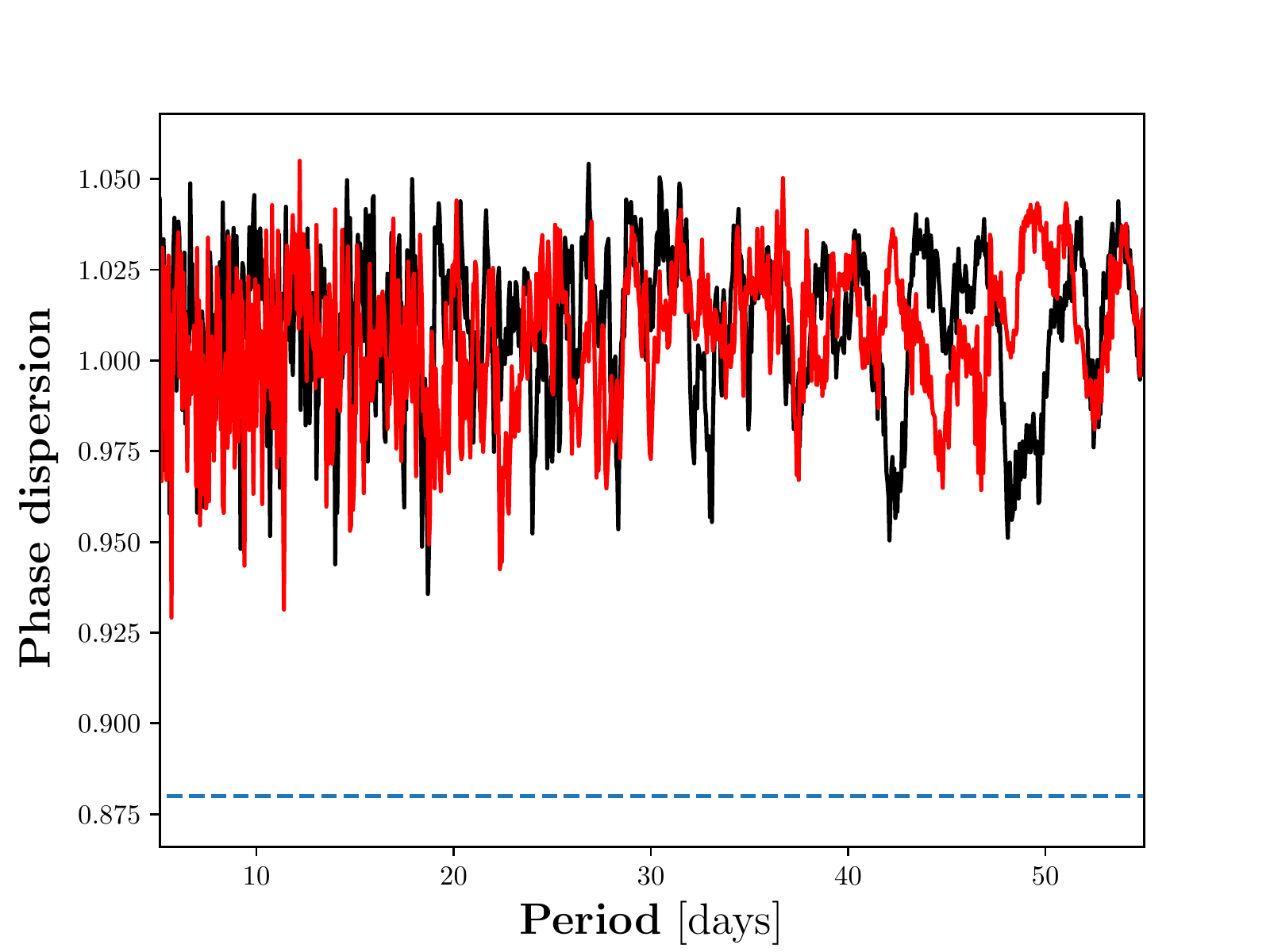}%\vspace{-0.4mm}\\
\includegraphics[width=0.45\textwidth, clip]{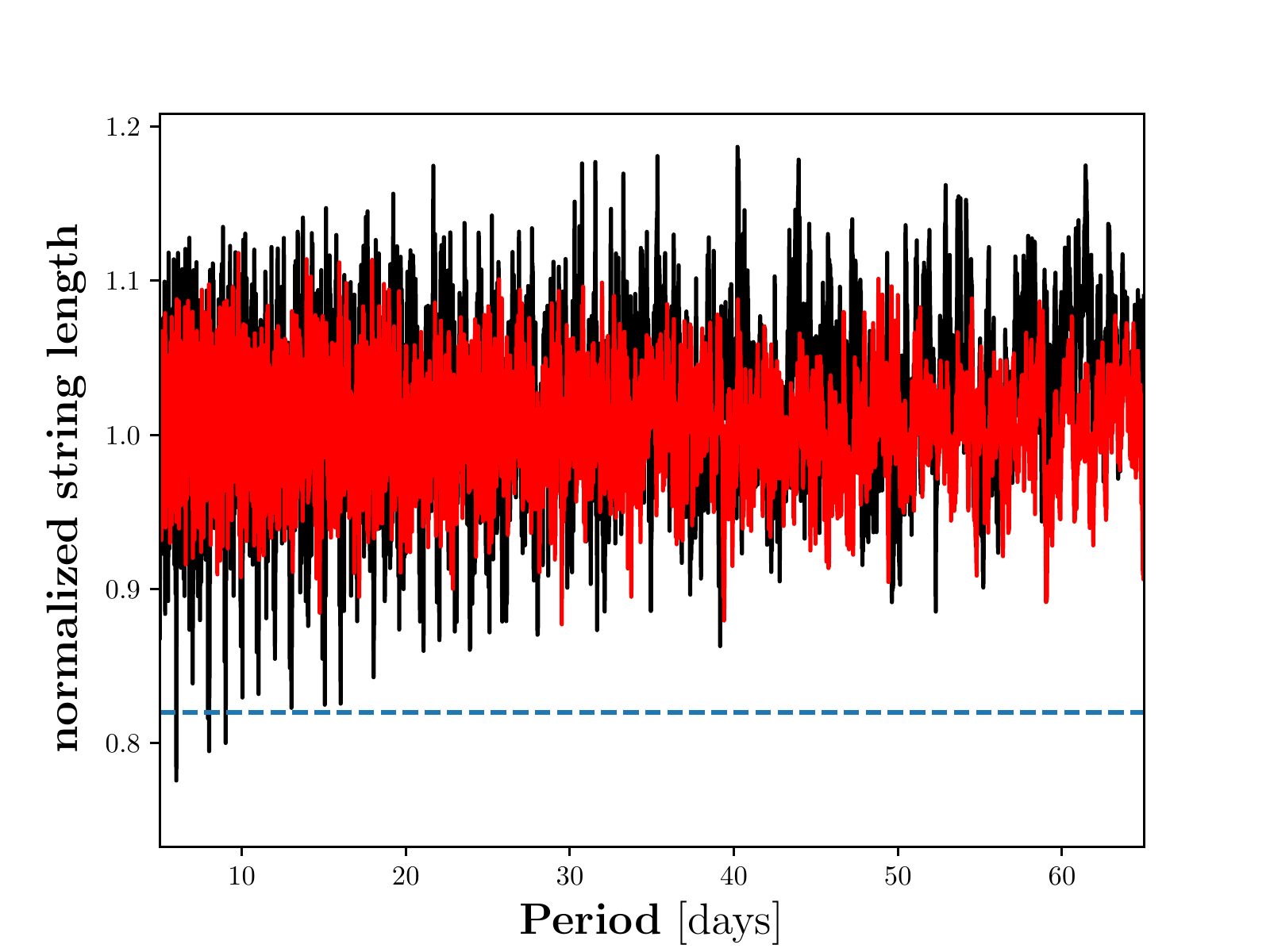}\\
\includegraphics[width=0.45\textwidth, clip]{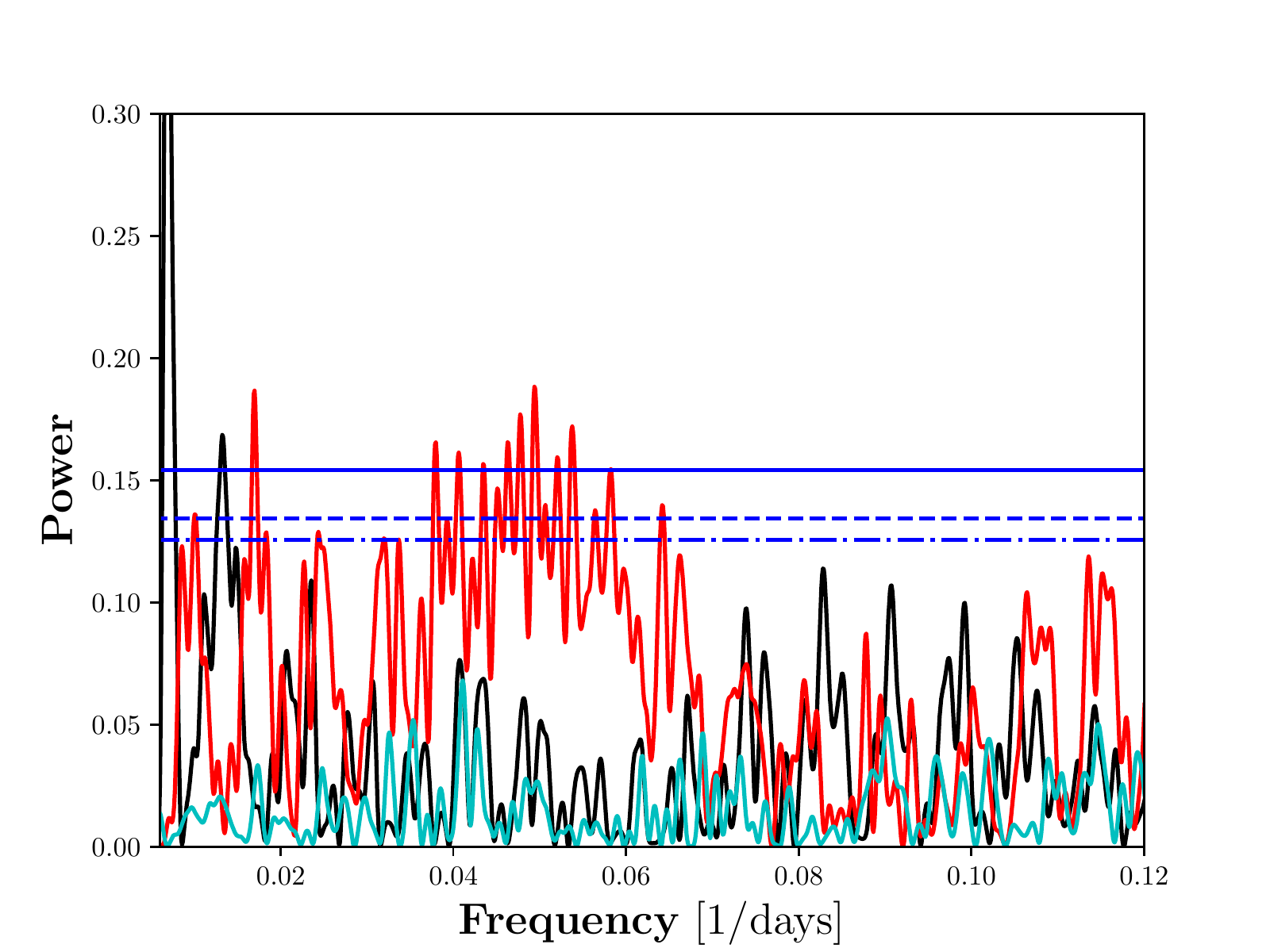}%\vspace{-0.4mm}\\
\includegraphics[width=0.45\textwidth, clip]{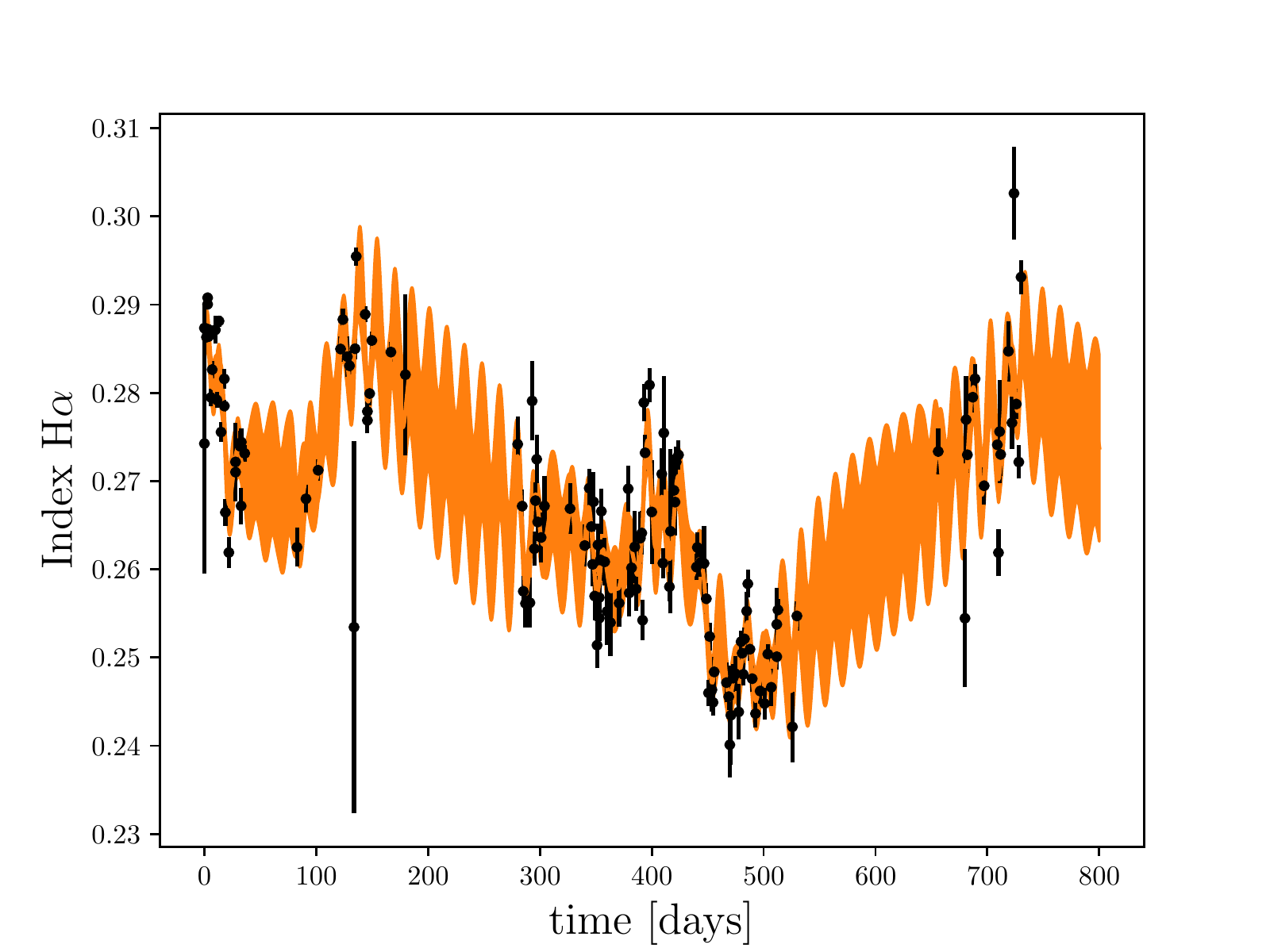}%\vspace{-0.4mm}\\
\caption{\label{lalande} Results for Lalande 21185 as explained in Fig. \ref{gxand}.}
\end{figure*}

\begin{figure*}
\includegraphics[width=0.45\textwidth, clip]{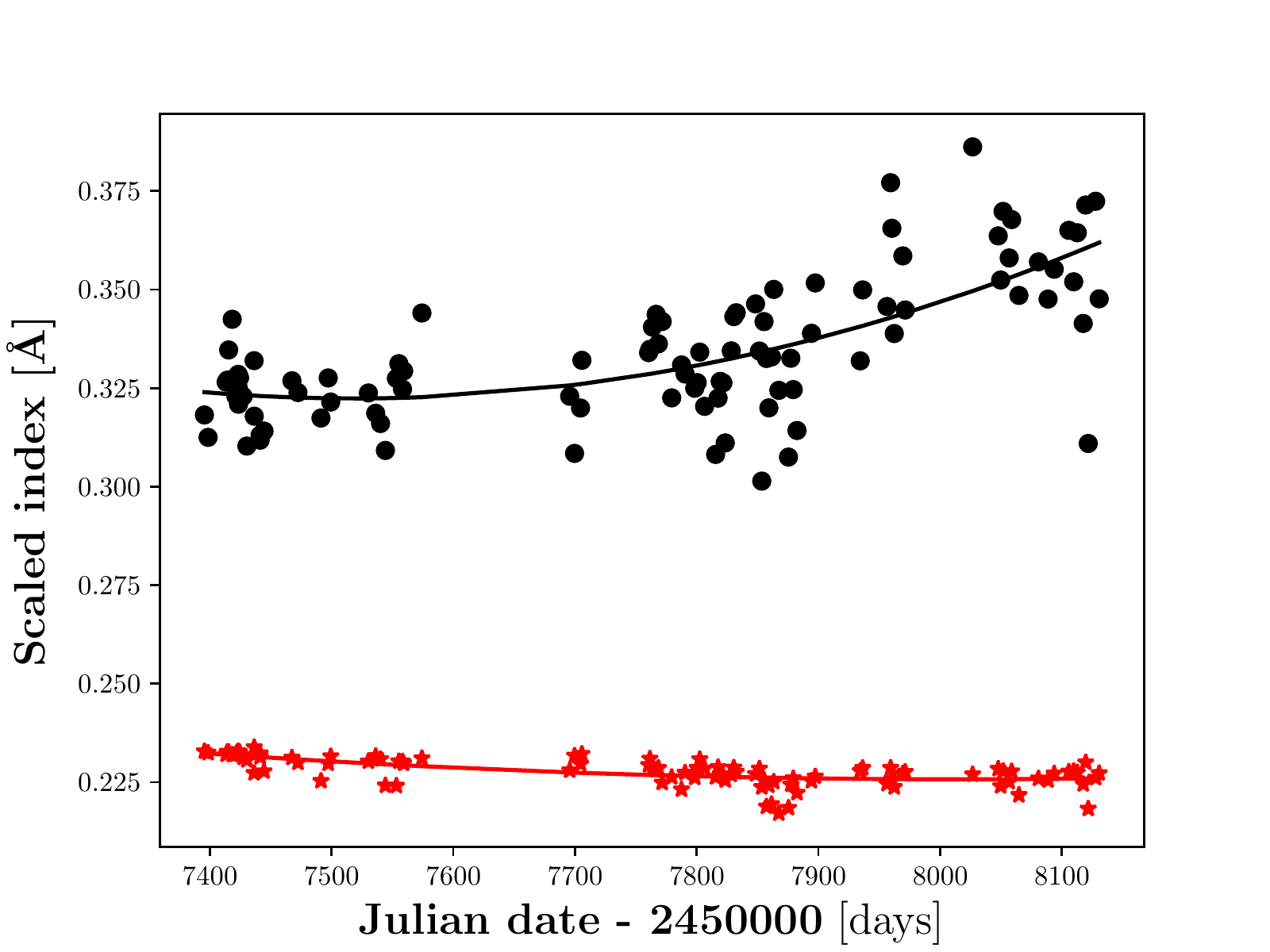}\\%\vspace{-0.4mm}\\
\includegraphics[width=0.45\textwidth, clip]{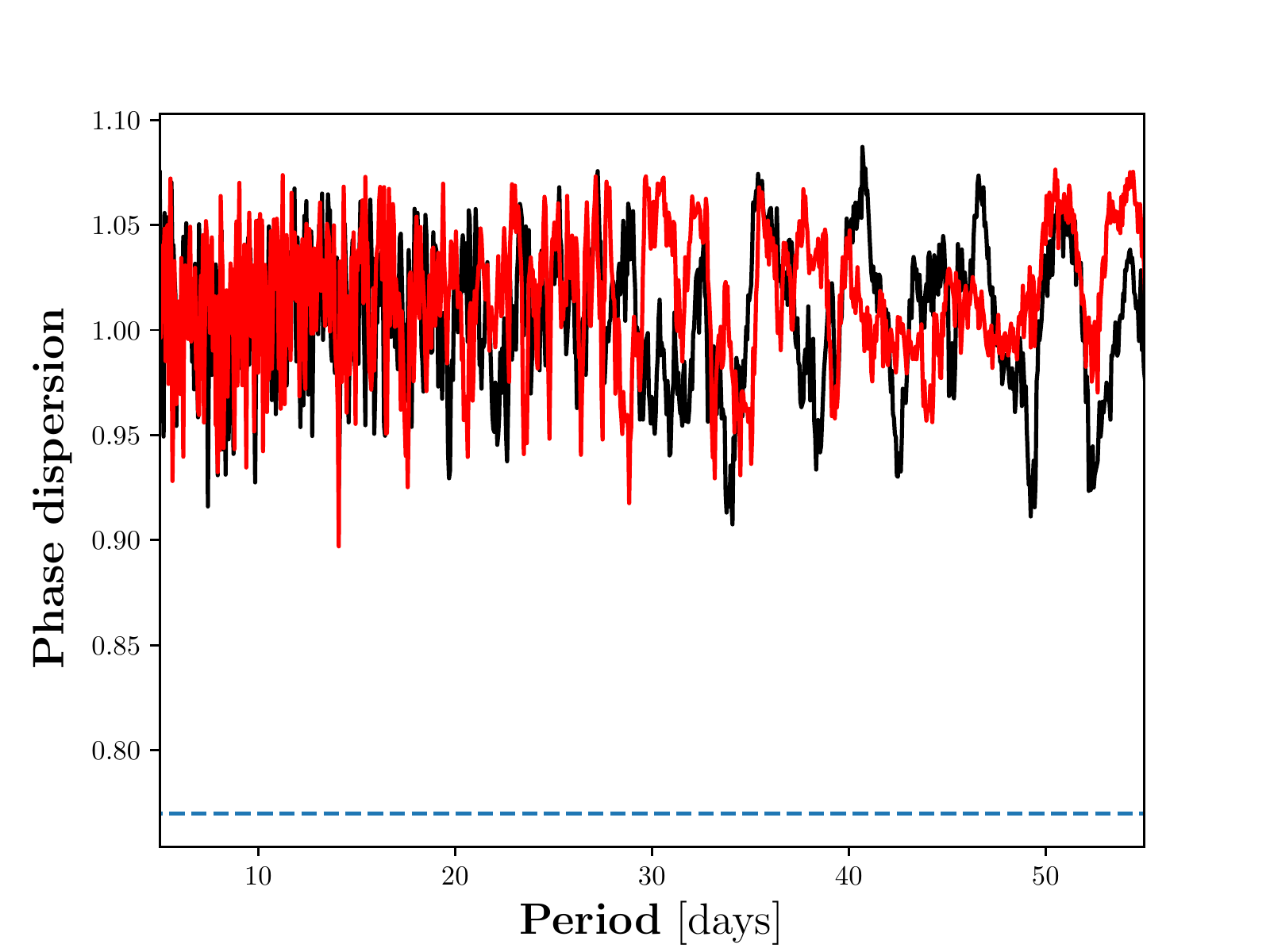}%\vspace{-0.4mm}\\
\includegraphics[width=0.45\textwidth, clip]{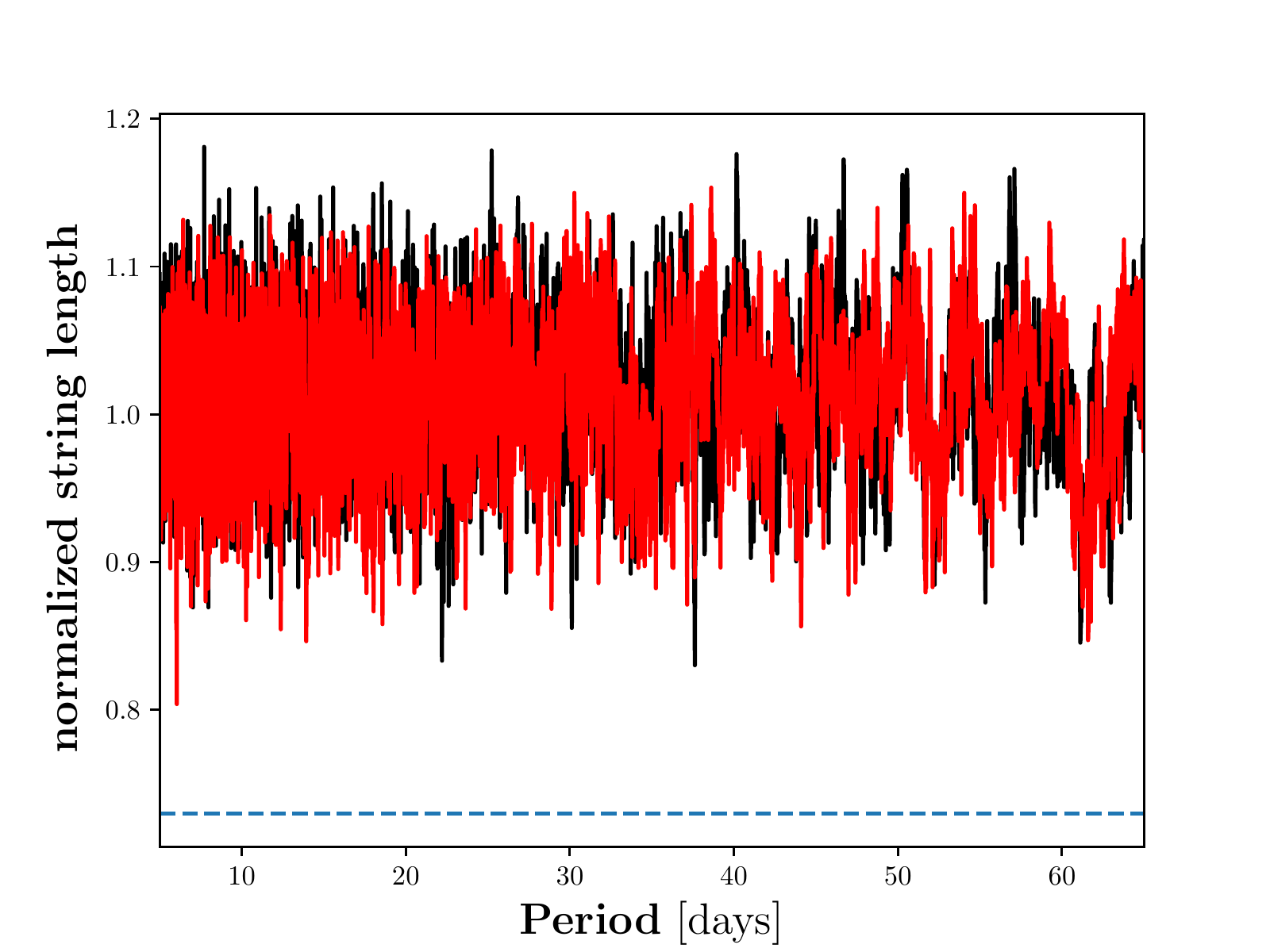}\\
\includegraphics[width=0.45\textwidth, clip]{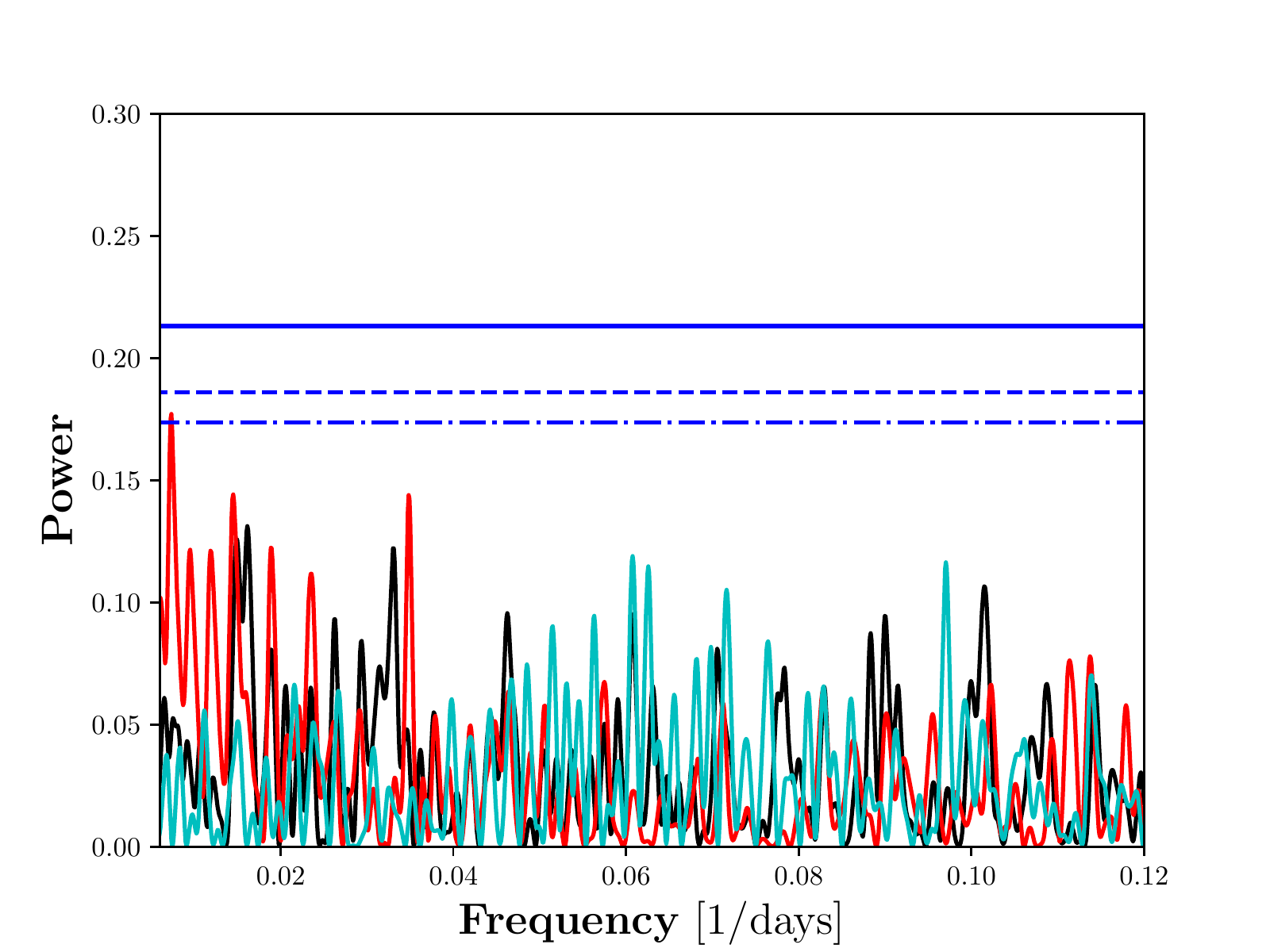}%\vspace{-0.4mm}\\
\includegraphics[width=0.45\textwidth, clip]{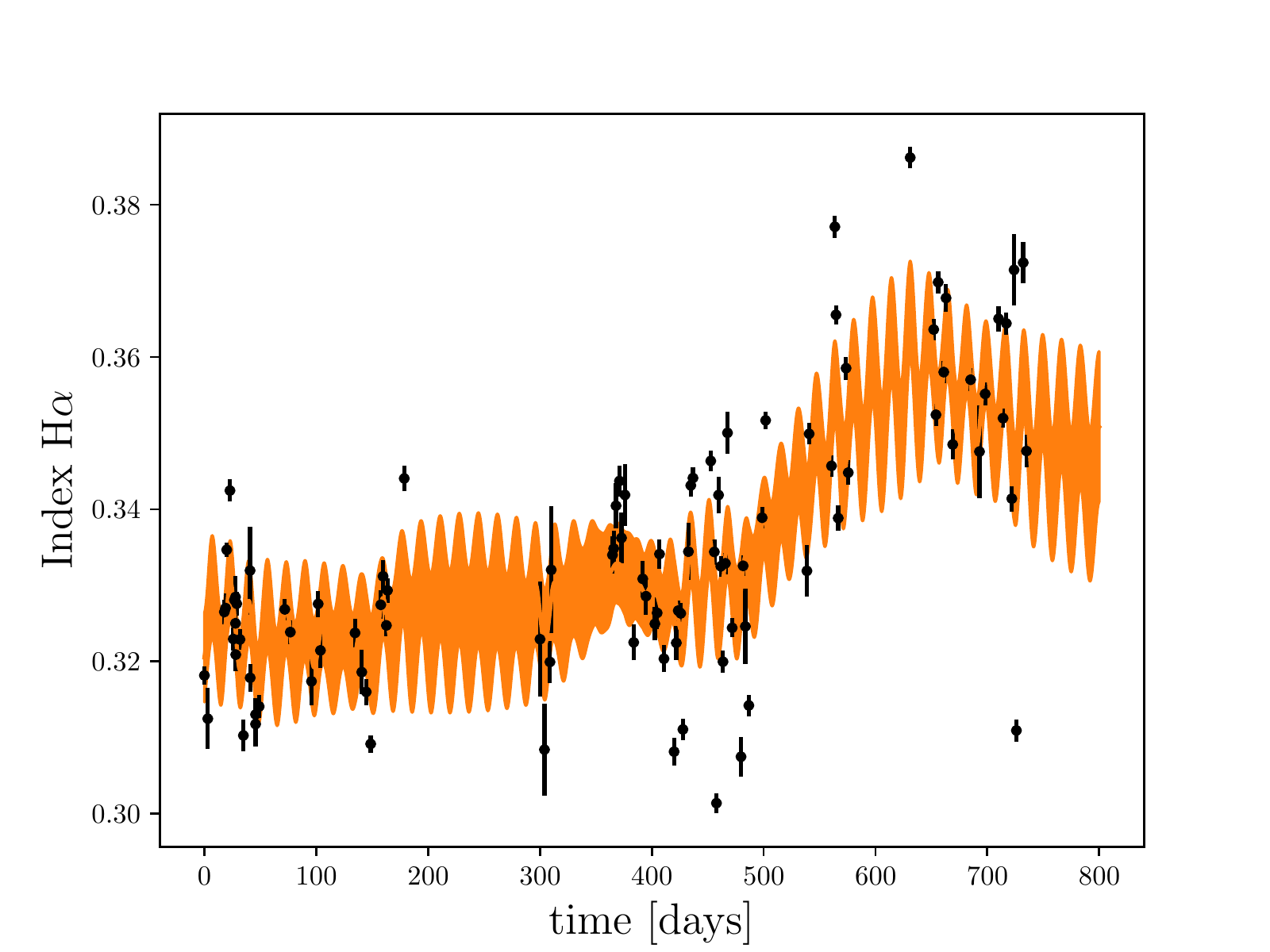}%\vspace{-0.4mm}\\
\caption{\label{bd442051} Results for BD+44 2051A as explained in Fig. \ref{gxand}.}
\end{figure*}

\begin{figure*}
\includegraphics[width=0.45\textwidth, clip]{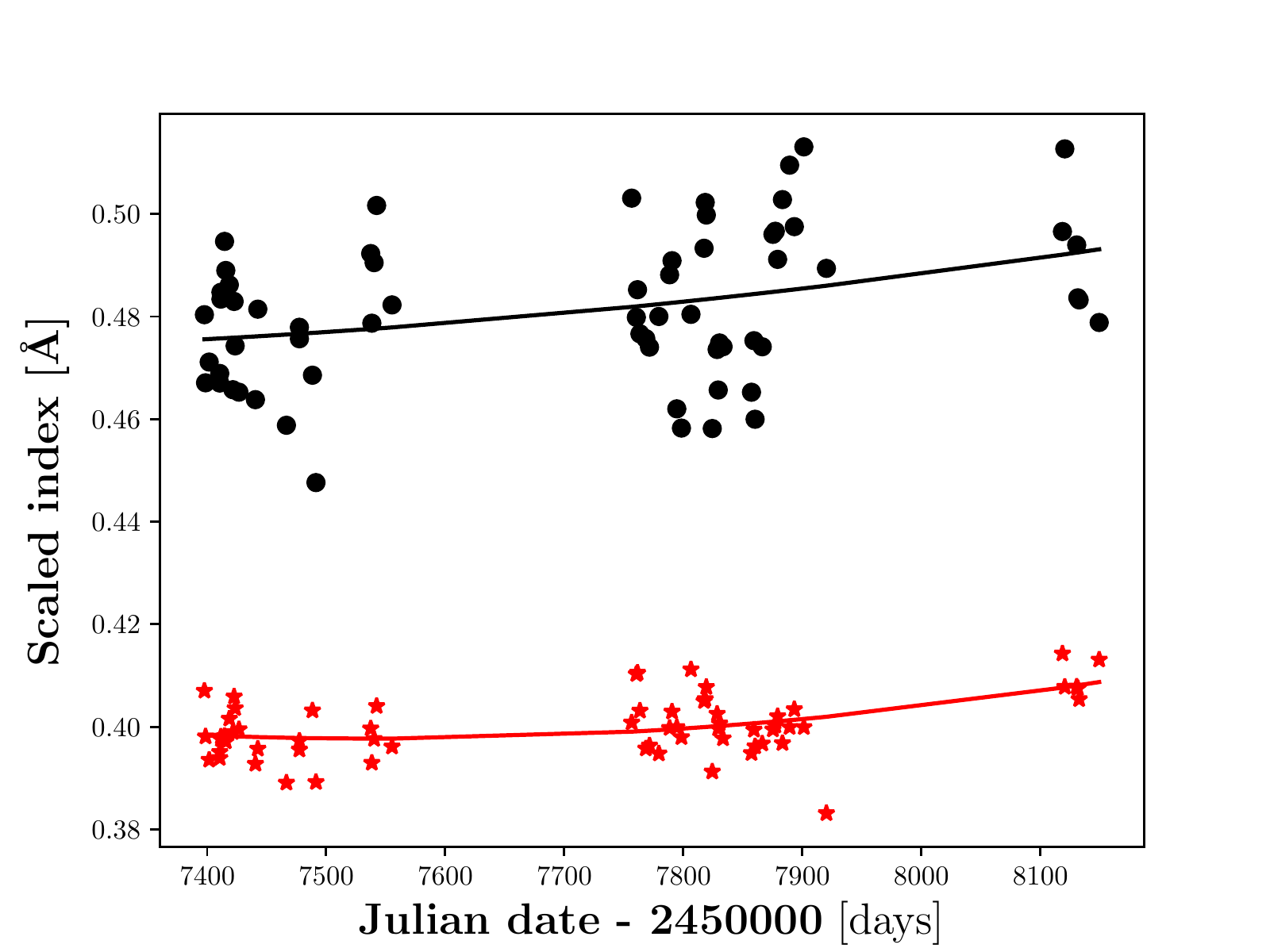}%\vspace{-0.4mm}\\
\includegraphics[width=0.45\textwidth, clip]{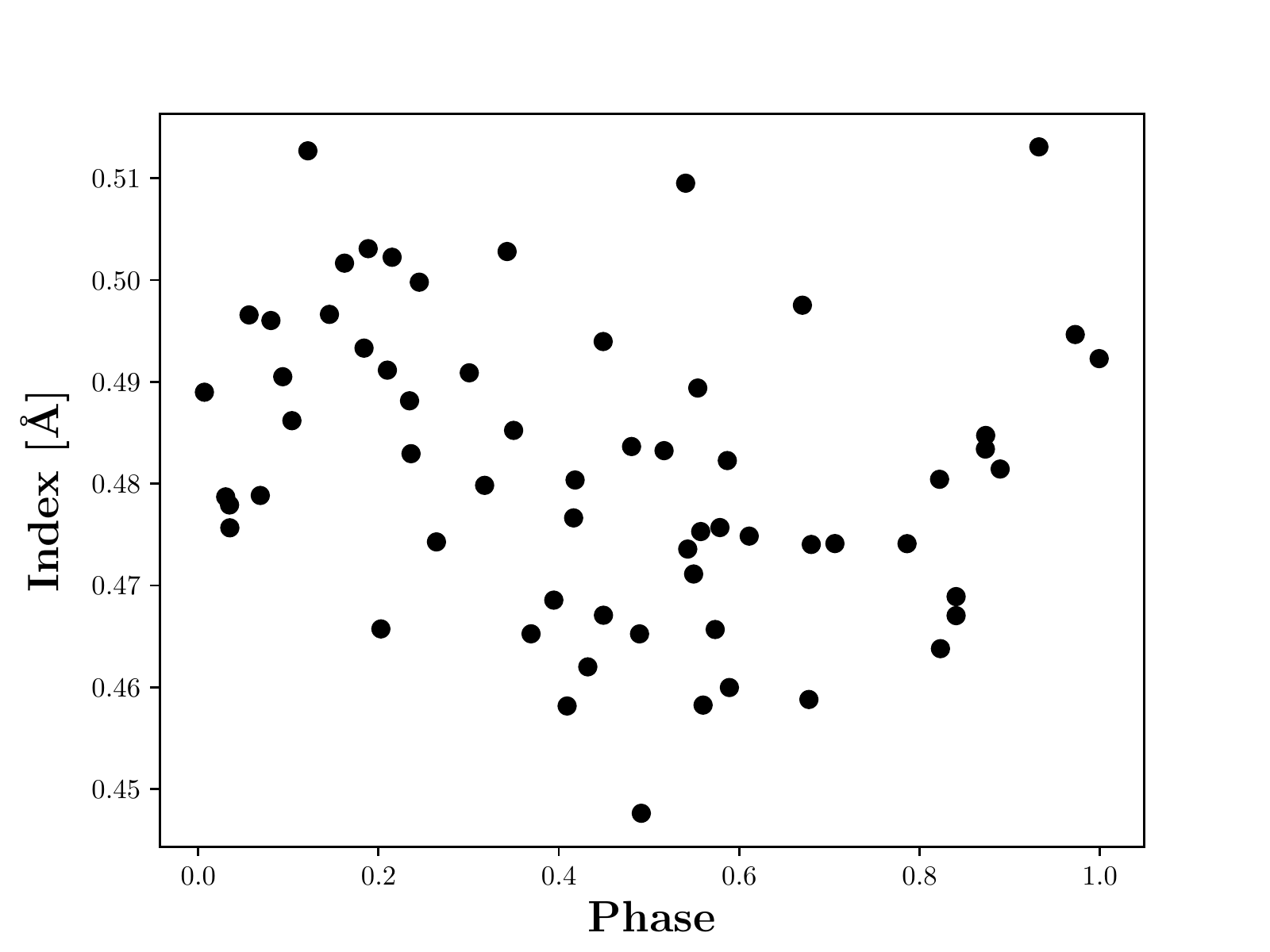}\\
\includegraphics[width=0.45\textwidth, clip]{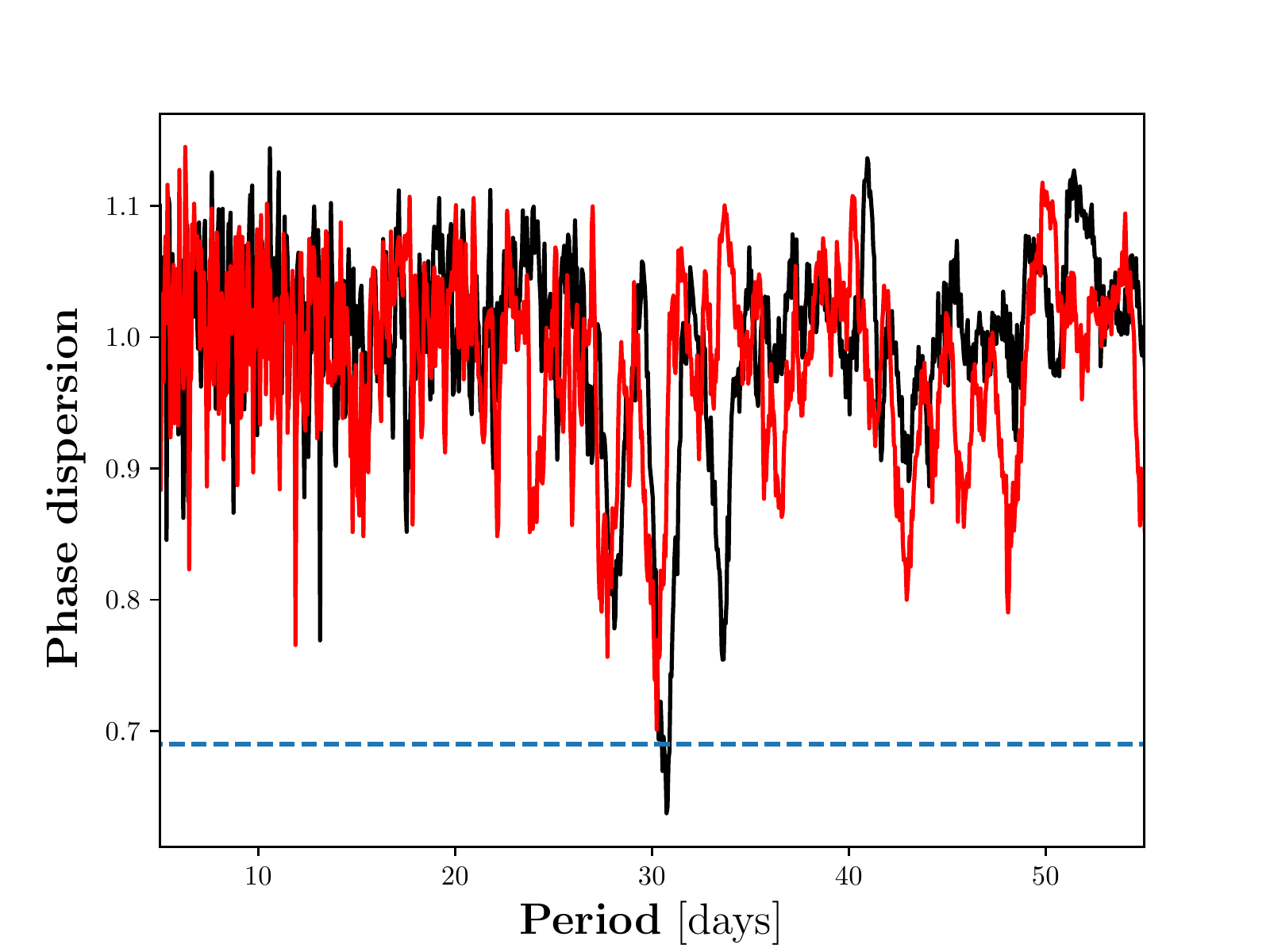}%\vspace{-0.4mm}\\
\includegraphics[width=0.45\textwidth, clip]{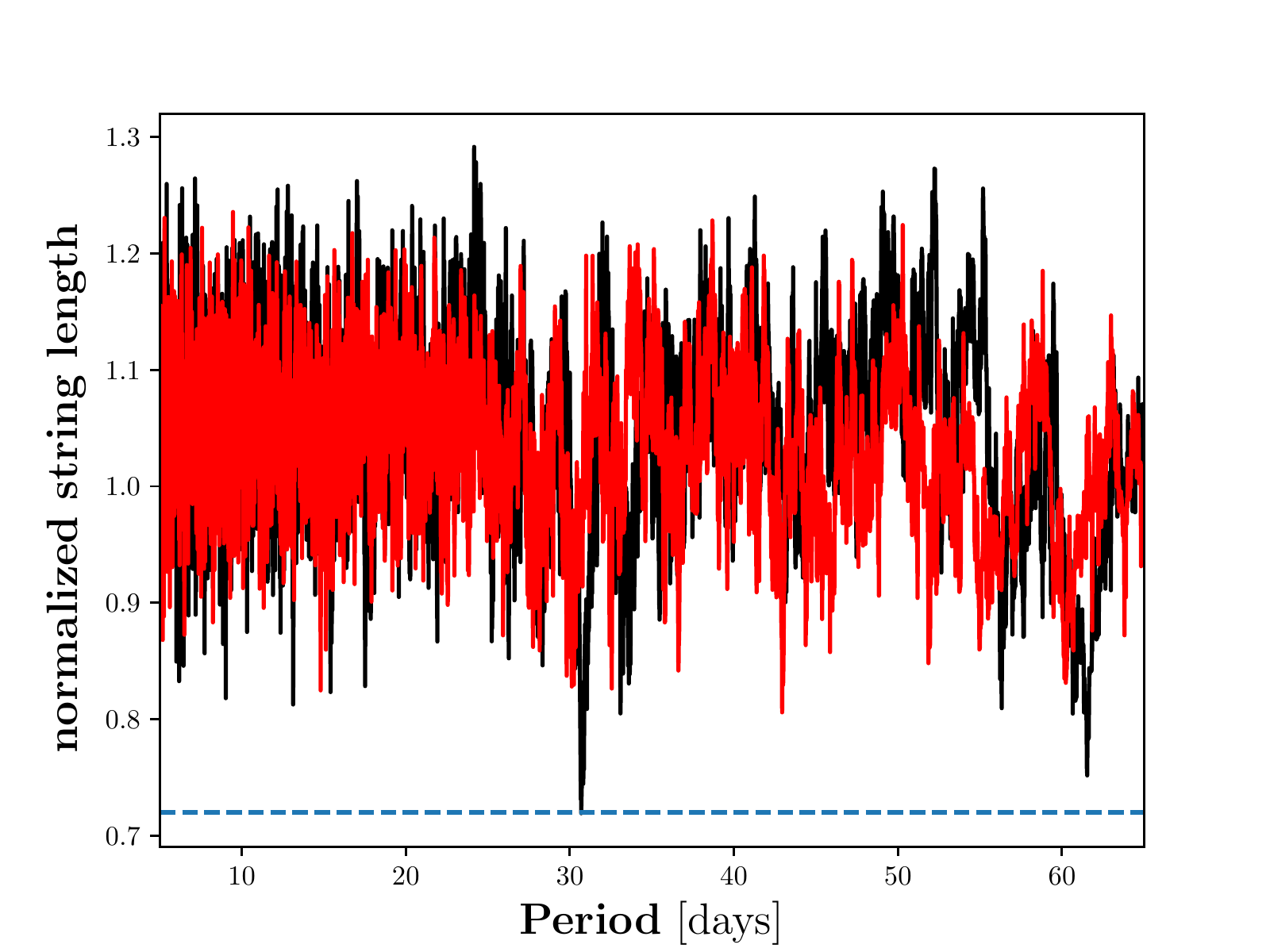}\\
\includegraphics[width=0.45\textwidth, clip]{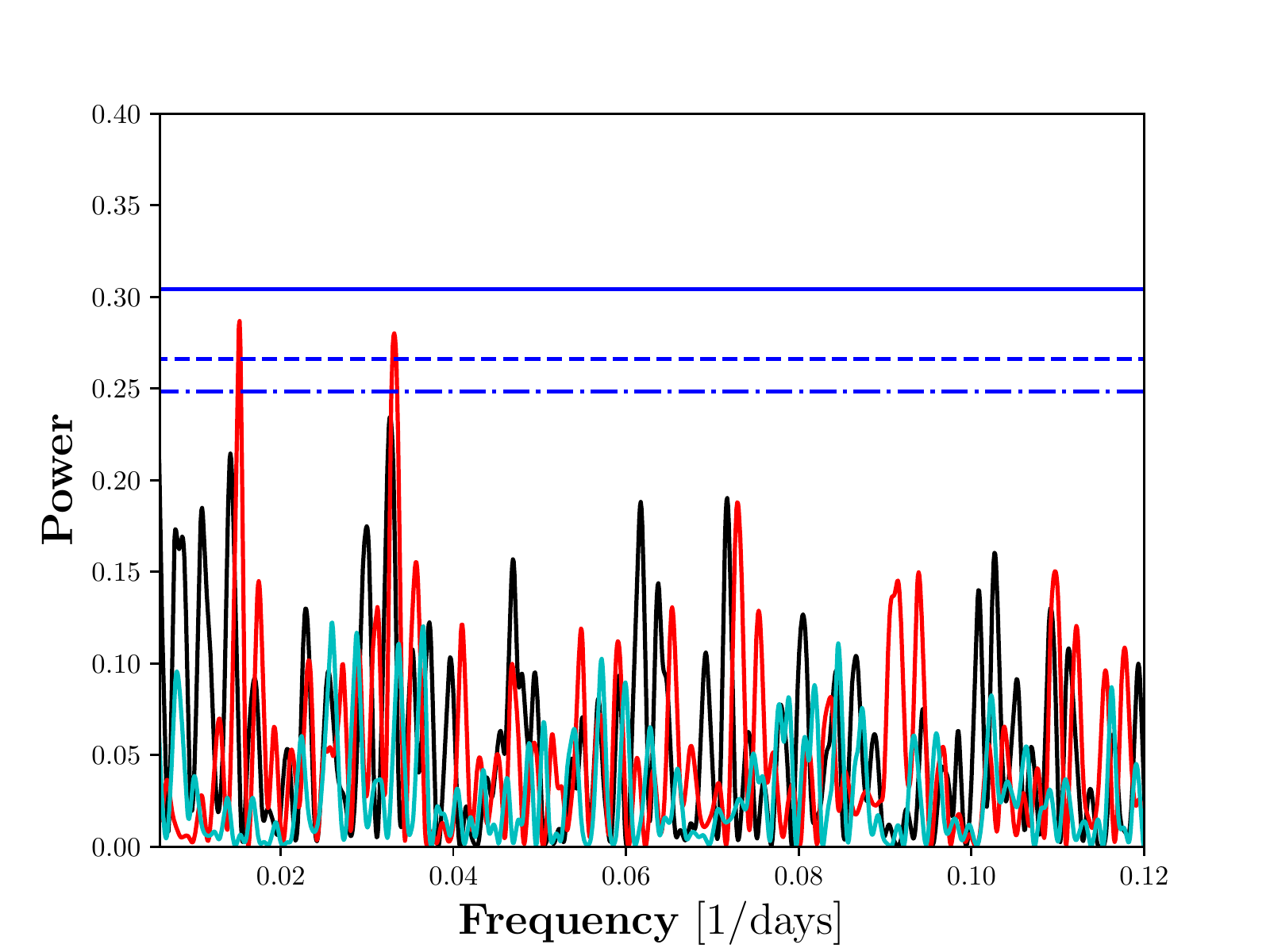}%\vspace{-0.4mm}\\
\includegraphics[width=0.45\textwidth, clip]{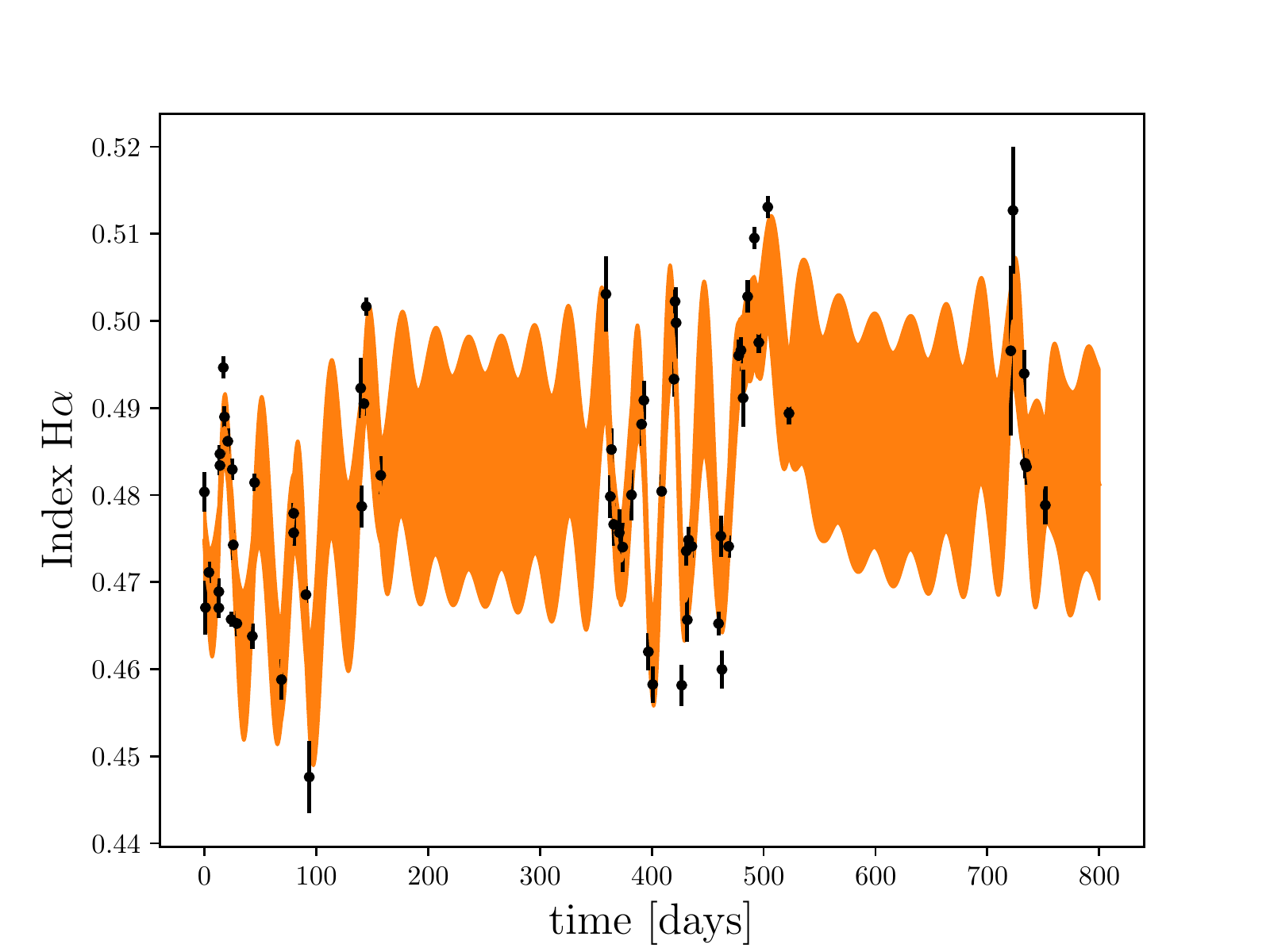}%\vspace{-0.4mm}\\
\caption{\label{bd112576} Results for BD+11 2576 as explained in Fig. \ref{gxand}.}
\end{figure*}

\begin{figure*}
\includegraphics[width=0.45\textwidth, clip]{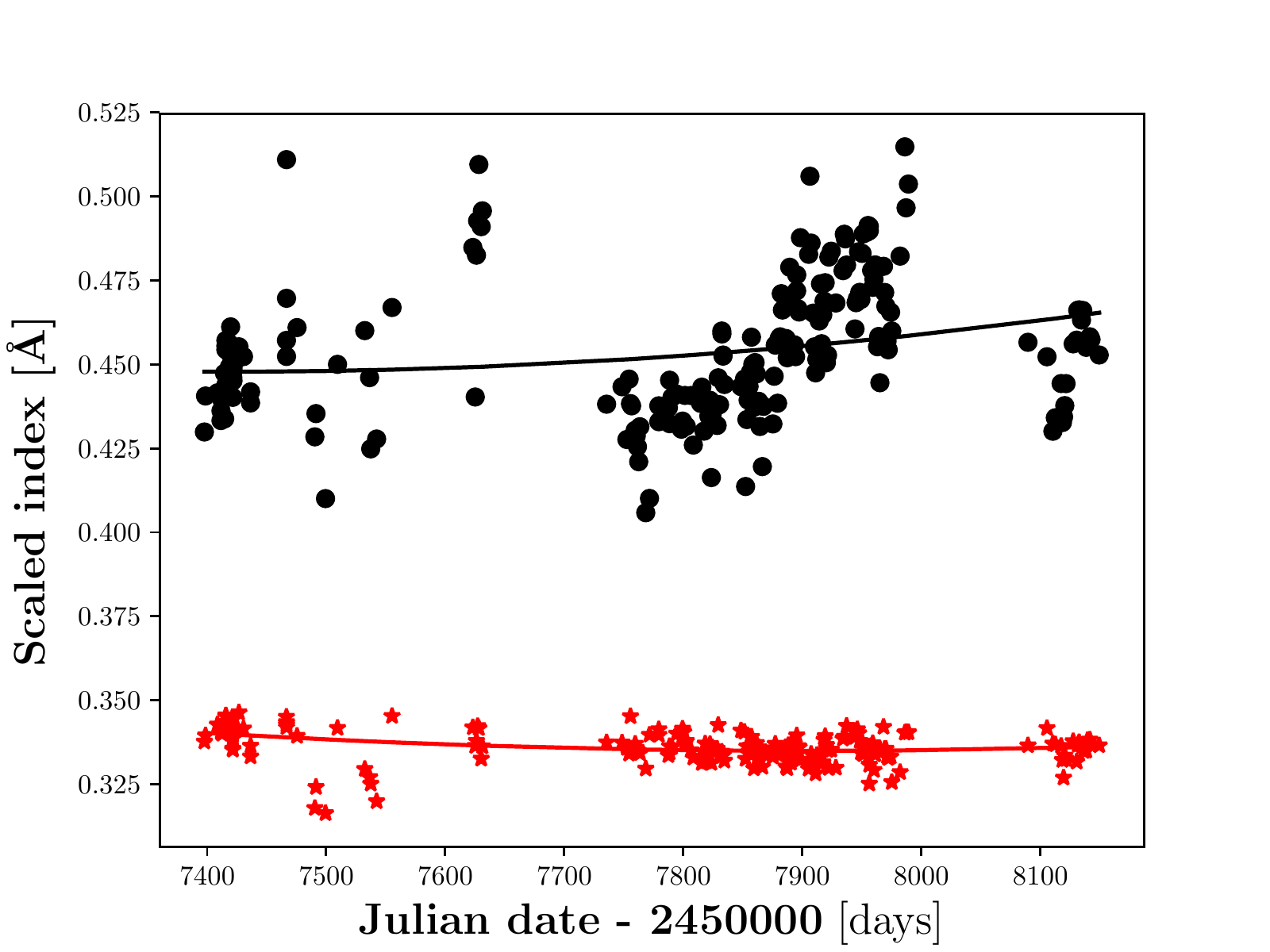}\\%\vspace{-0.4mm}\\
\includegraphics[width=0.45\textwidth, clip]{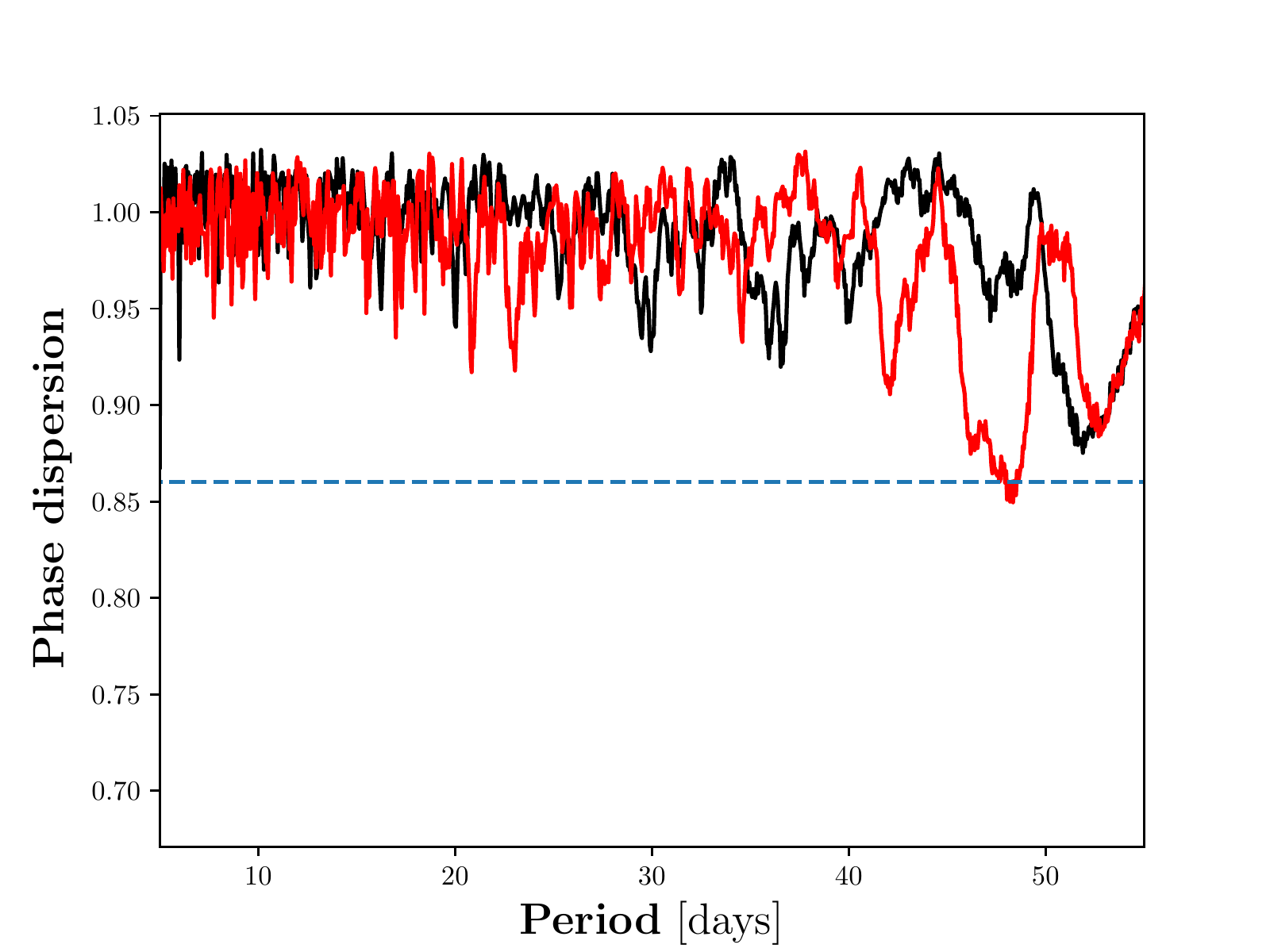}%\vspace{-0.4mm}\\
\includegraphics[width=0.45\textwidth, clip]{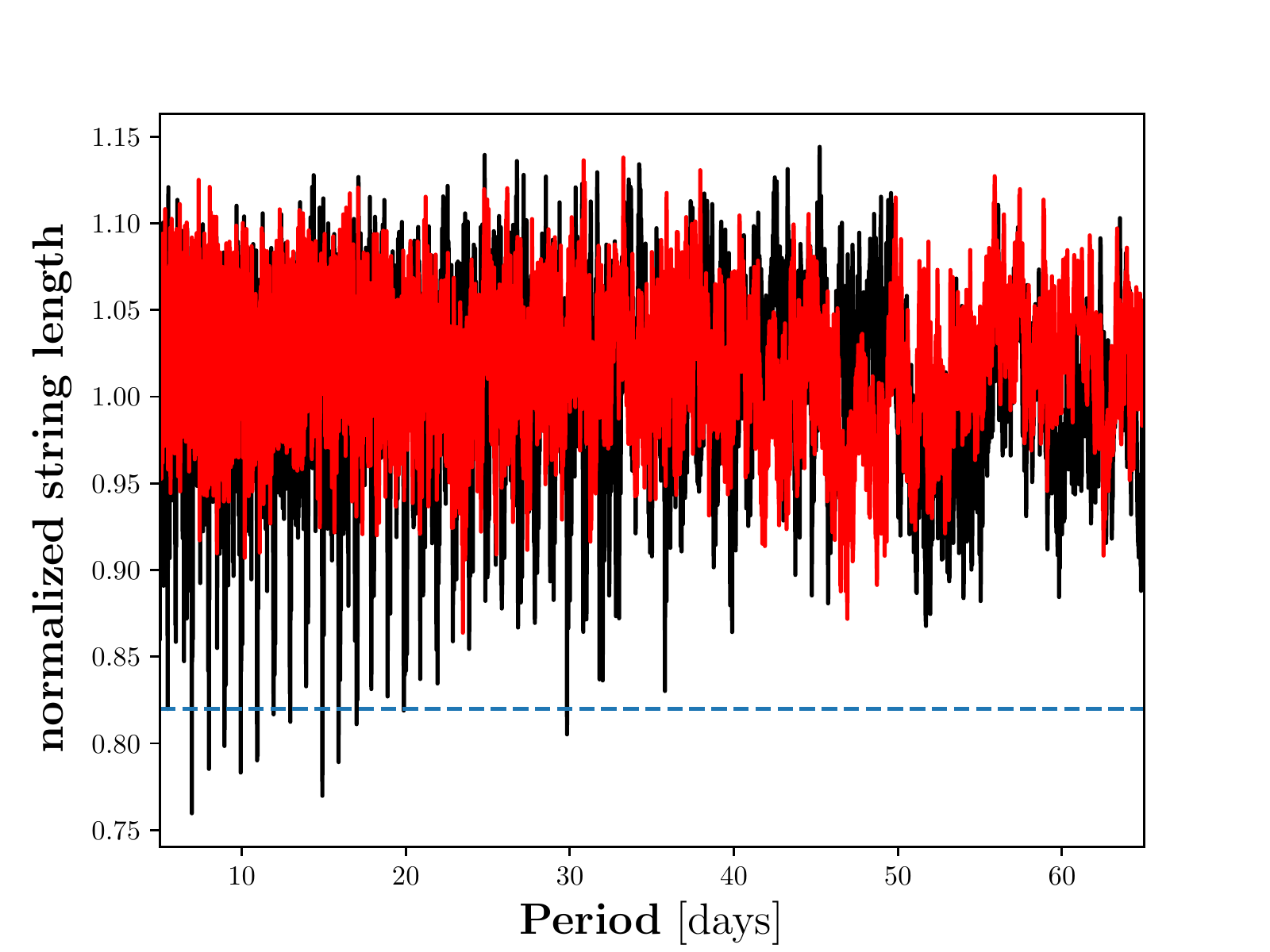}\\
\includegraphics[width=0.45\textwidth, clip]{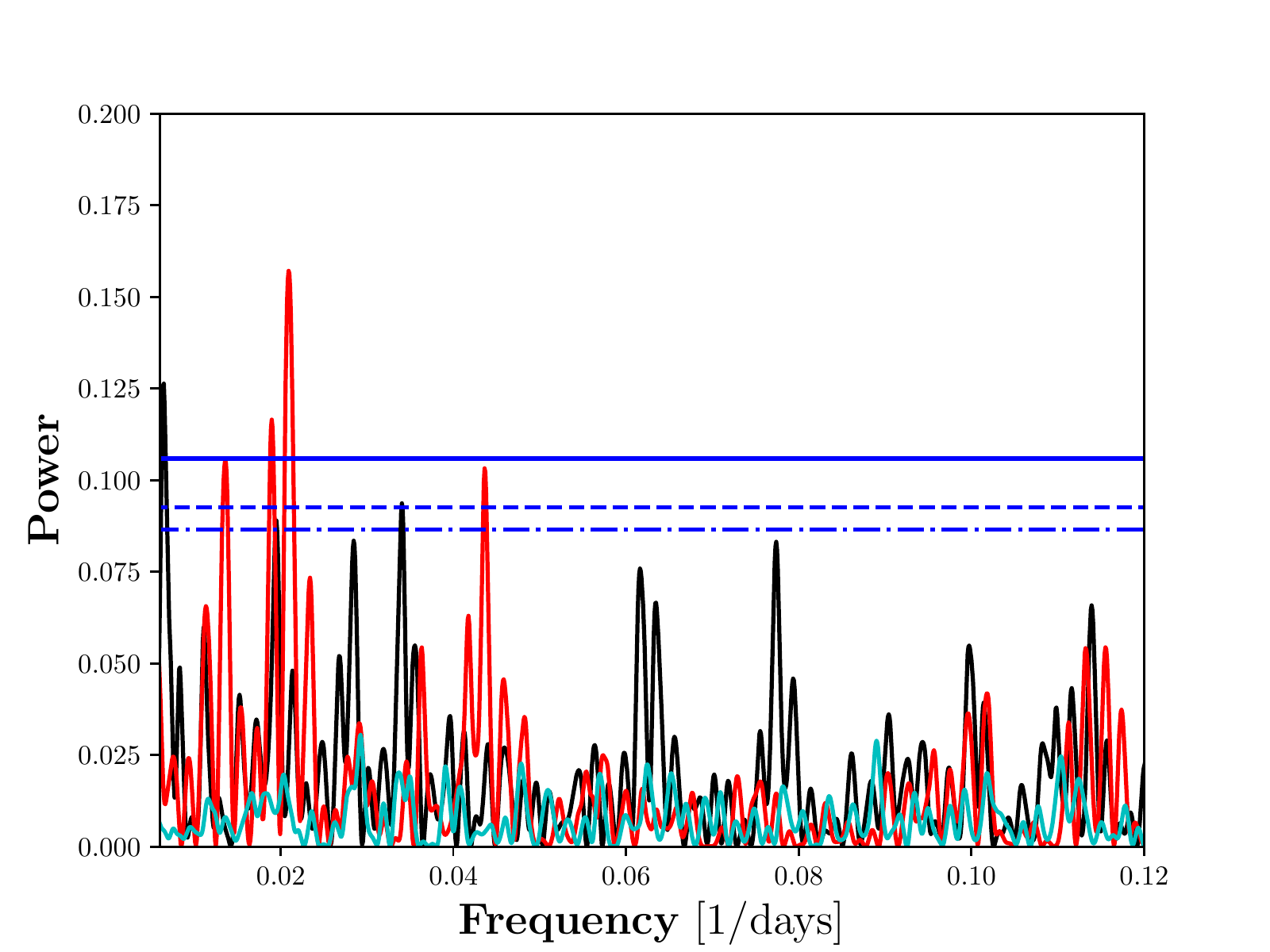}%\vspace{-0.4mm}\\
\includegraphics[width=0.45\textwidth, clip]{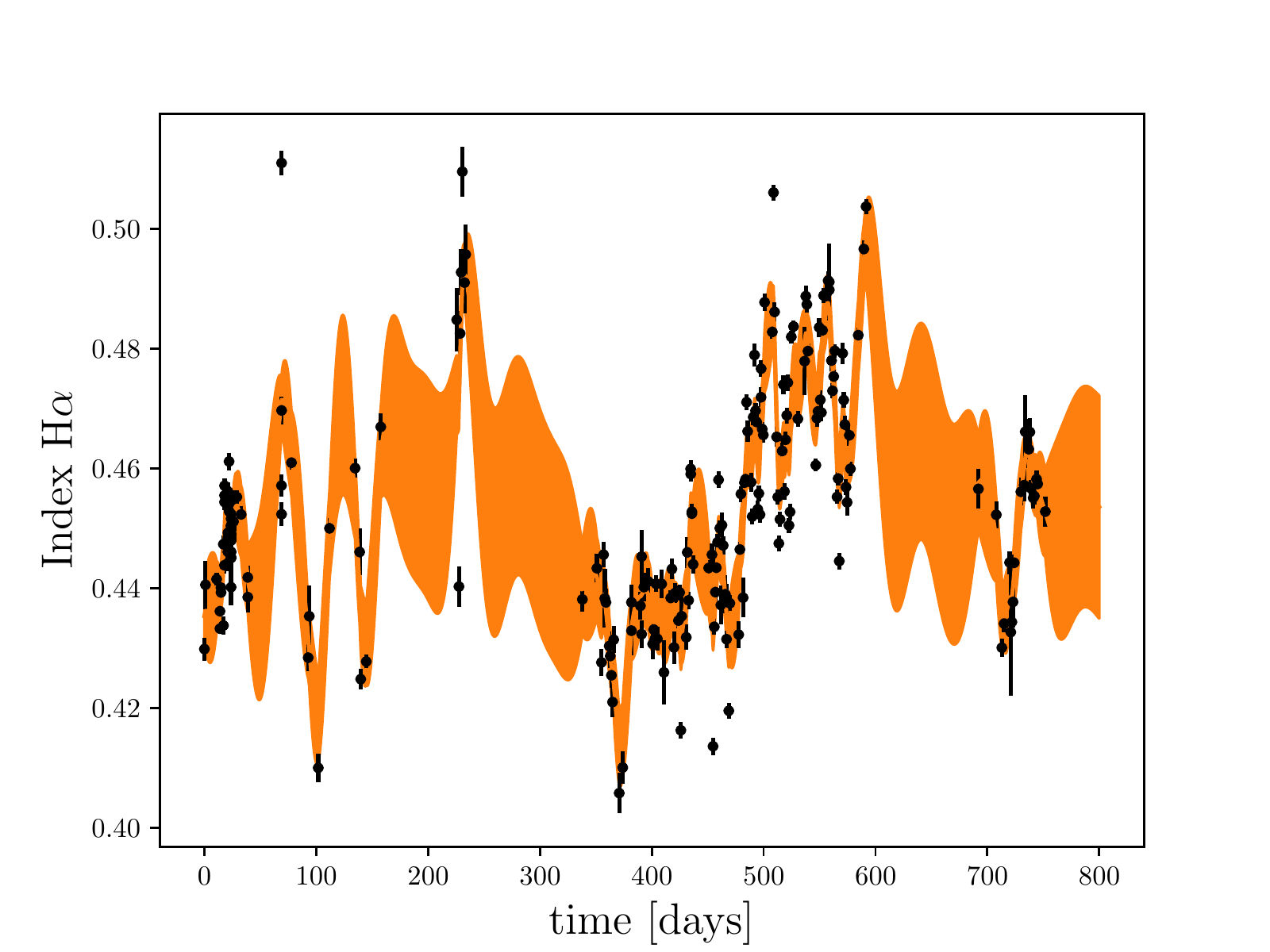}%\vspace{-0.4mm}\\
\caption{\label{hd119850} Results for HD 119850 as explained in Fig. \ref{gxand}.}
\end{figure*}

\begin{figure*}
\includegraphics[width=0.45\textwidth, clip]{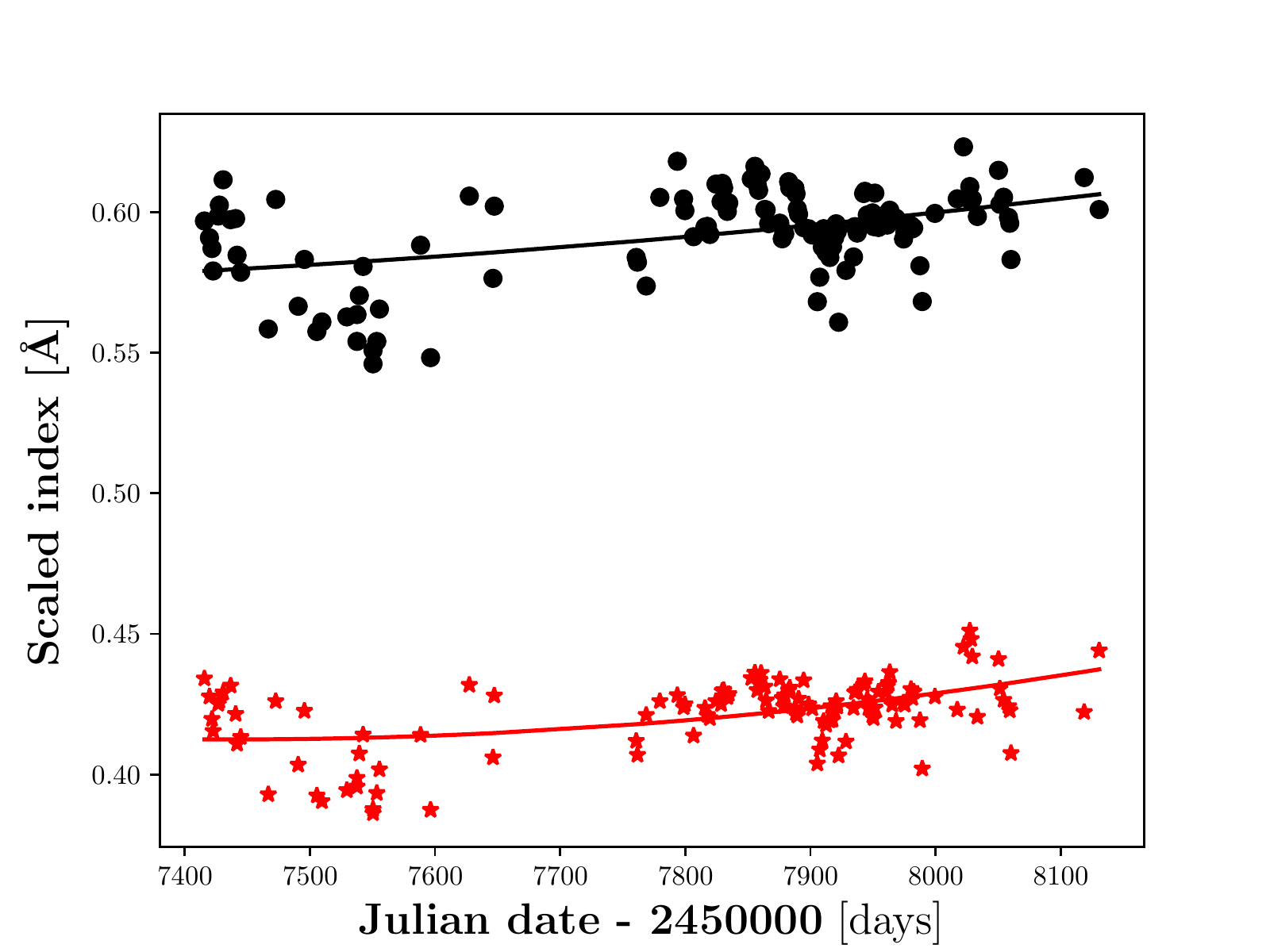}\\%\vspace{-0.4mm}\\
\includegraphics[width=0.45\textwidth, clip]{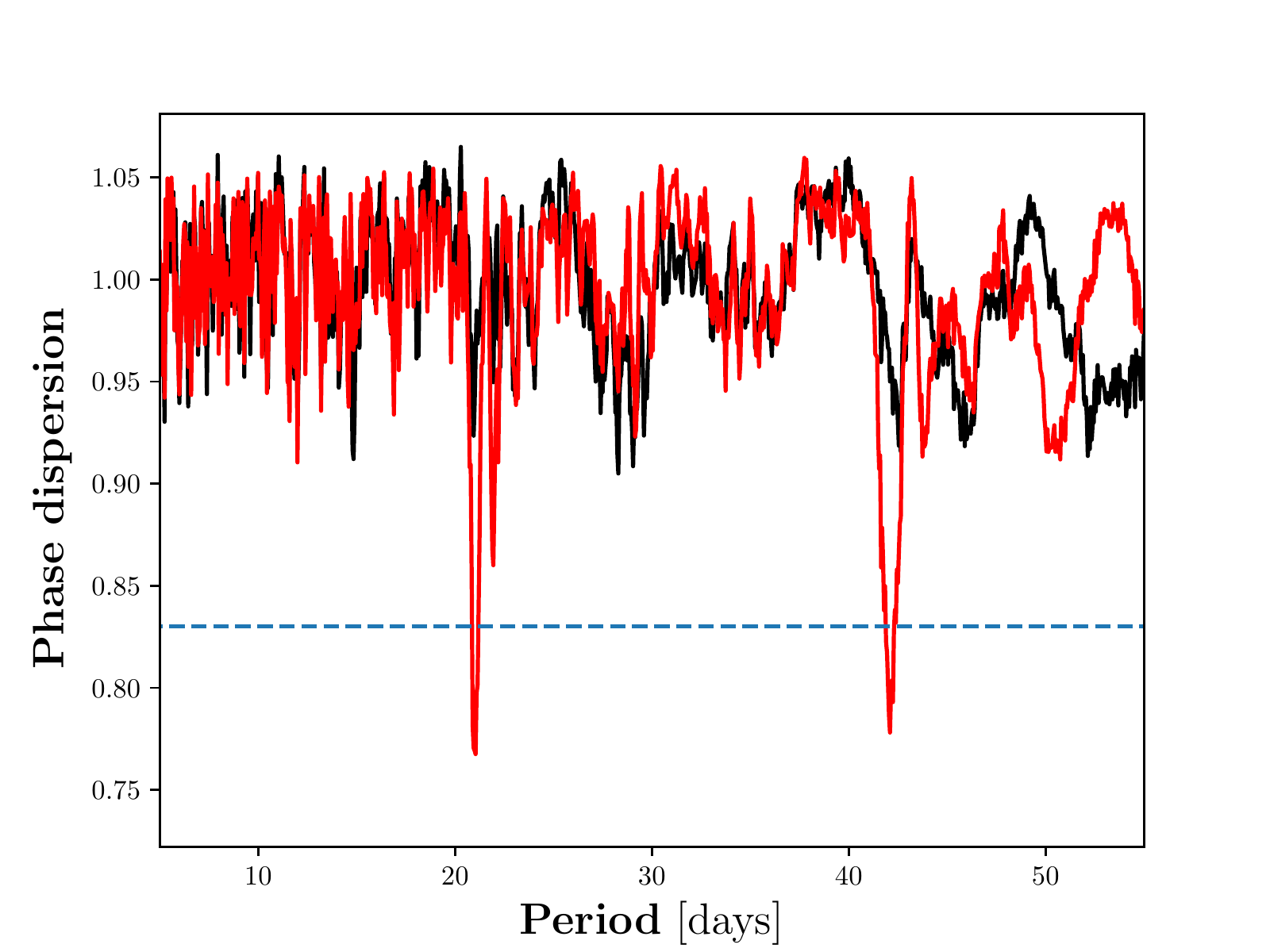}%\vspace{-0.4mm}\\
\includegraphics[width=0.45\textwidth, clip]{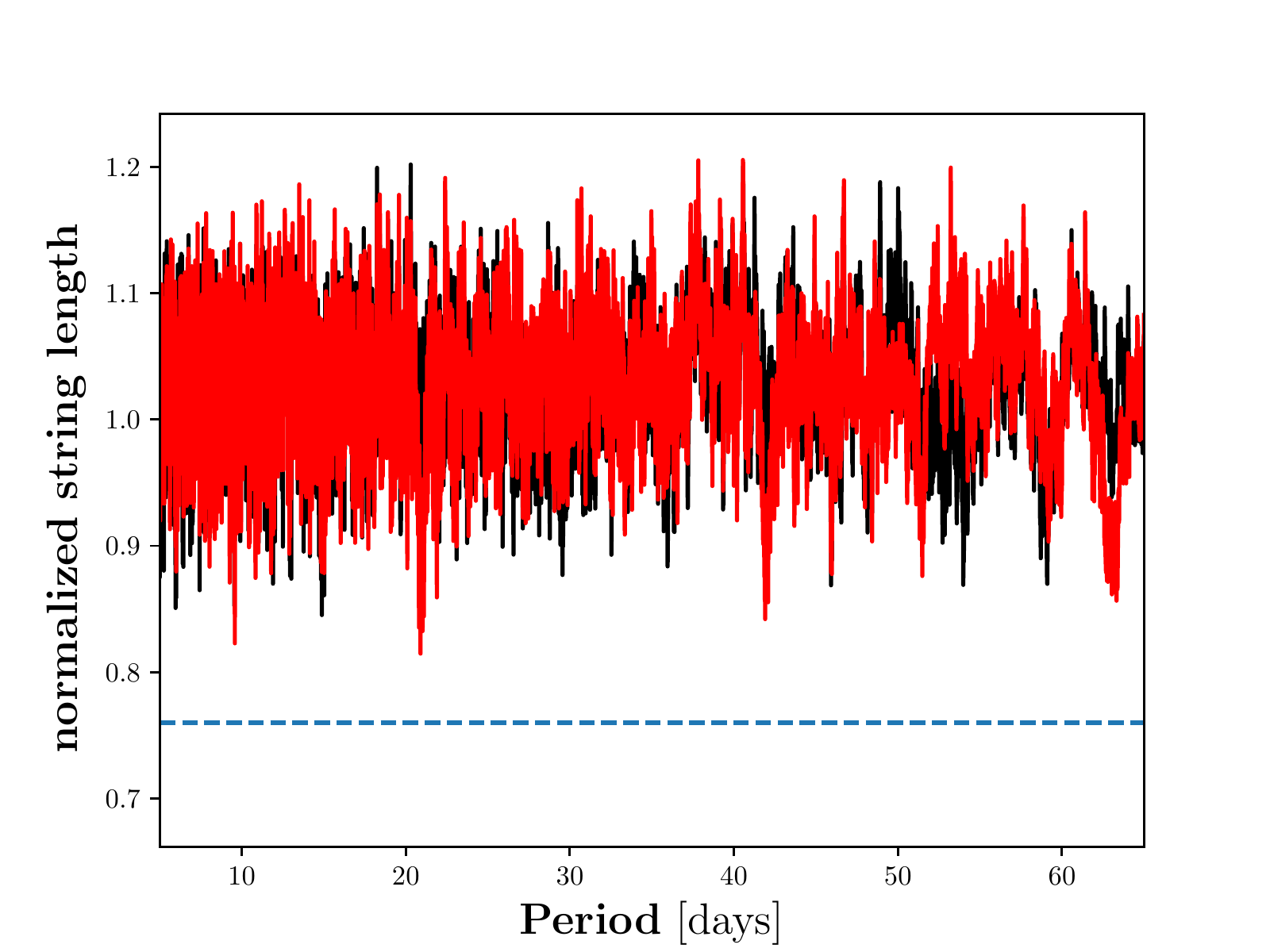}\\
\includegraphics[width=0.45\textwidth, clip]{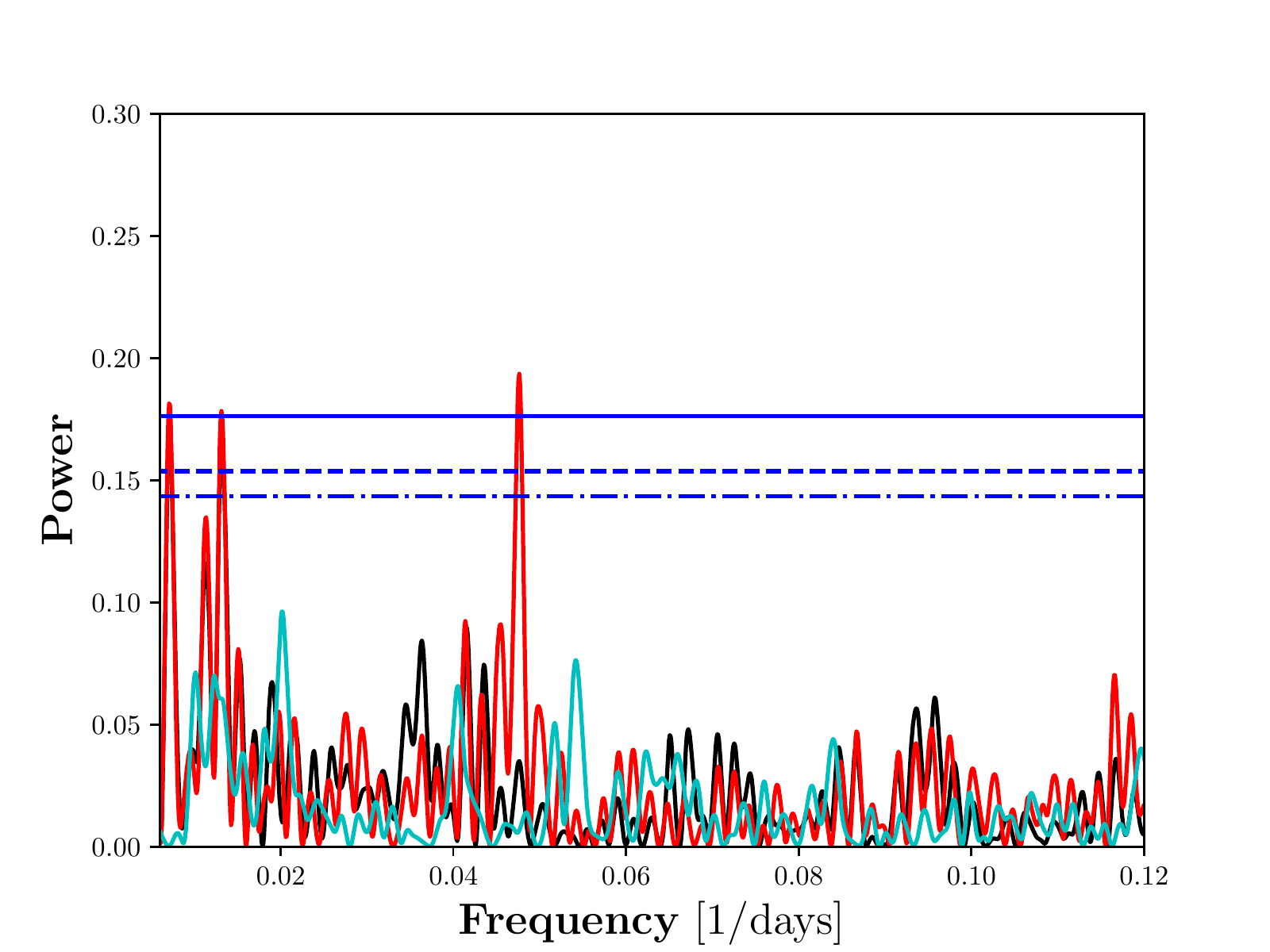}%\vspace{-0.4mm}\\
\includegraphics[width=0.45\textwidth, clip]{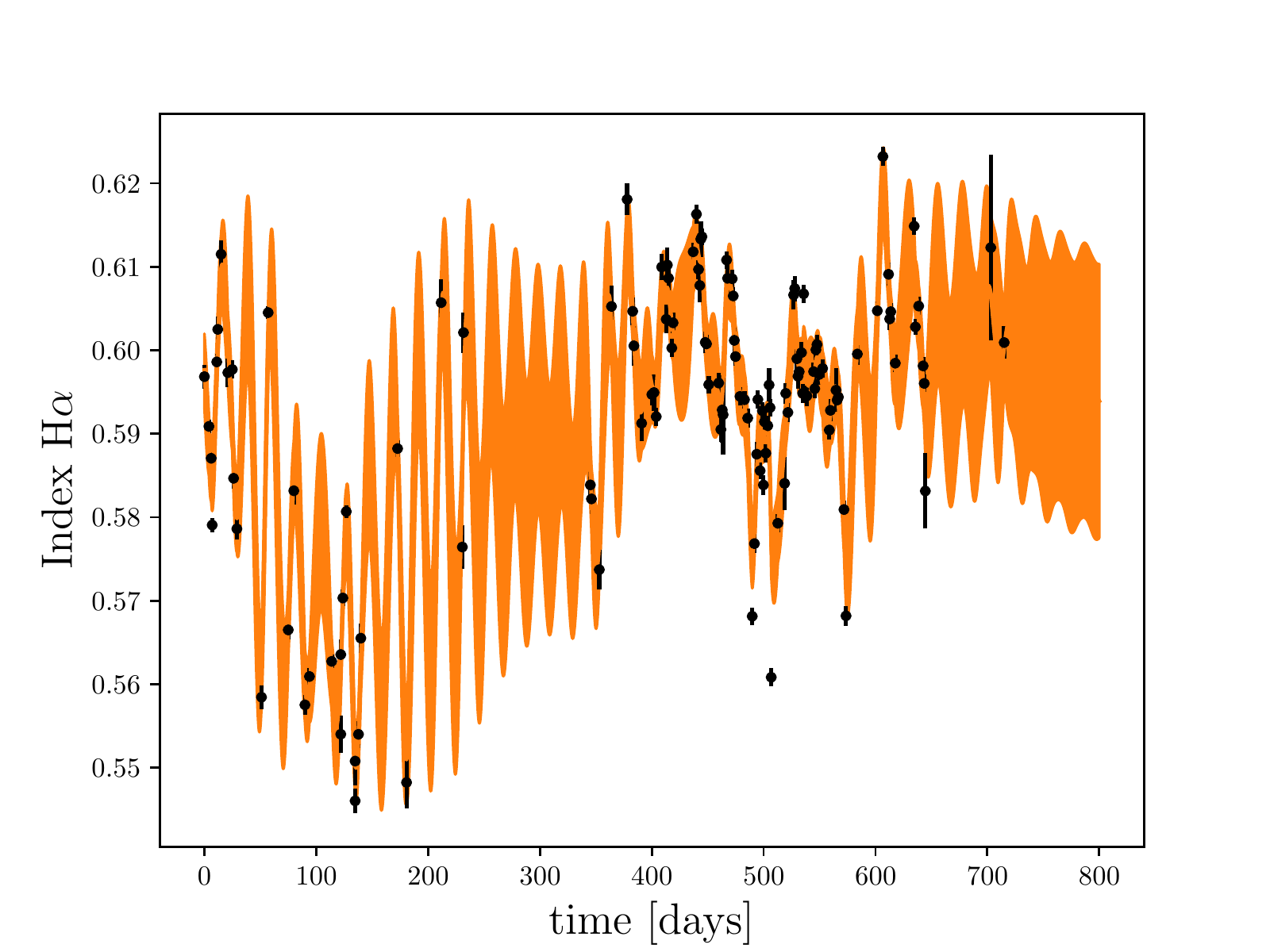}%\vspace{-0.4mm}\\
\caption{\label{hd117379} Results for HD 147379 as explained in Fig. \ref{gxand}.}
\end{figure*}

\begin{figure*}
\includegraphics[width=0.45\textwidth, clip]{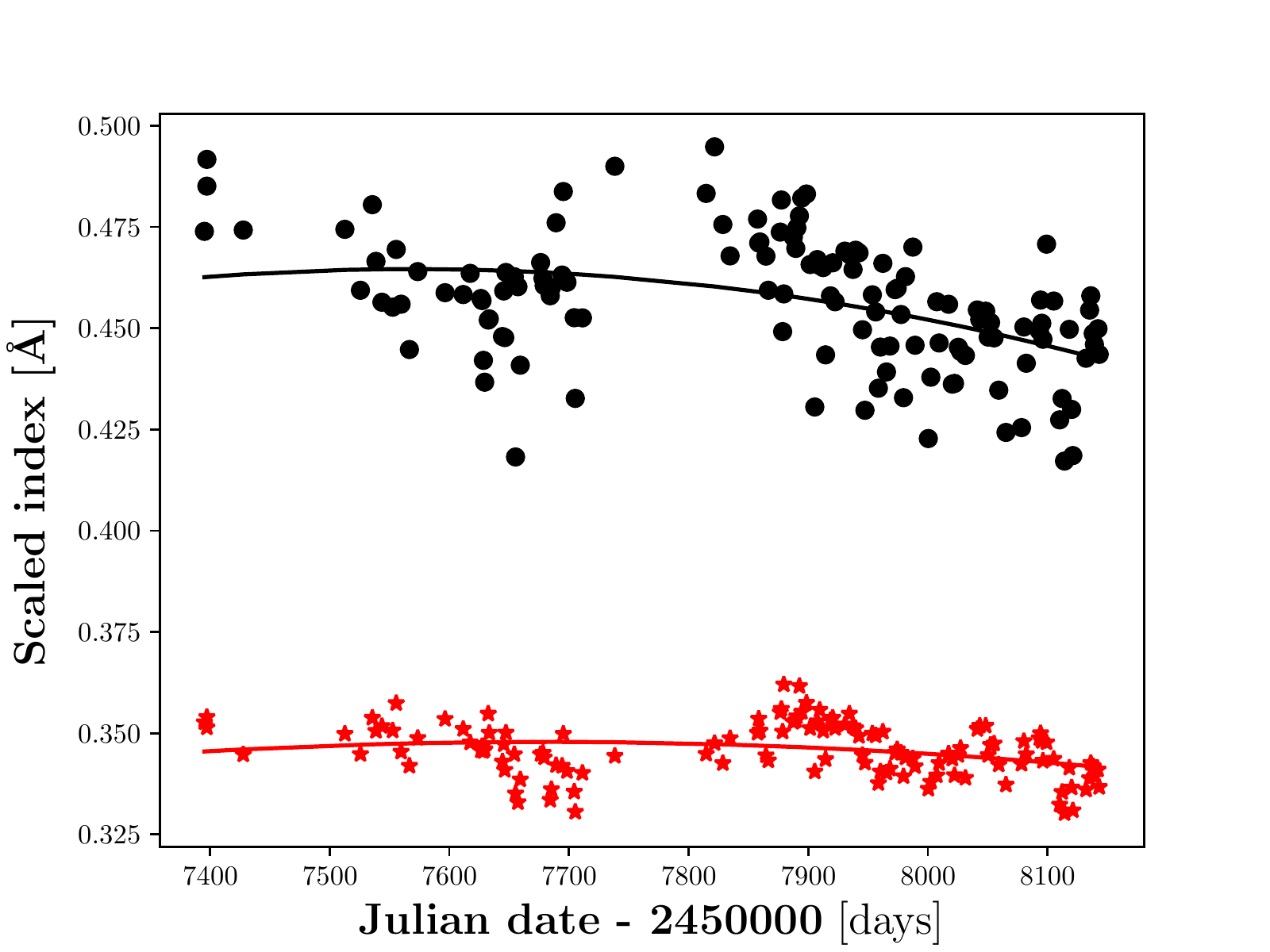}\\%\vspace{-0.4mm}\\
\includegraphics[width=0.45\textwidth, clip]{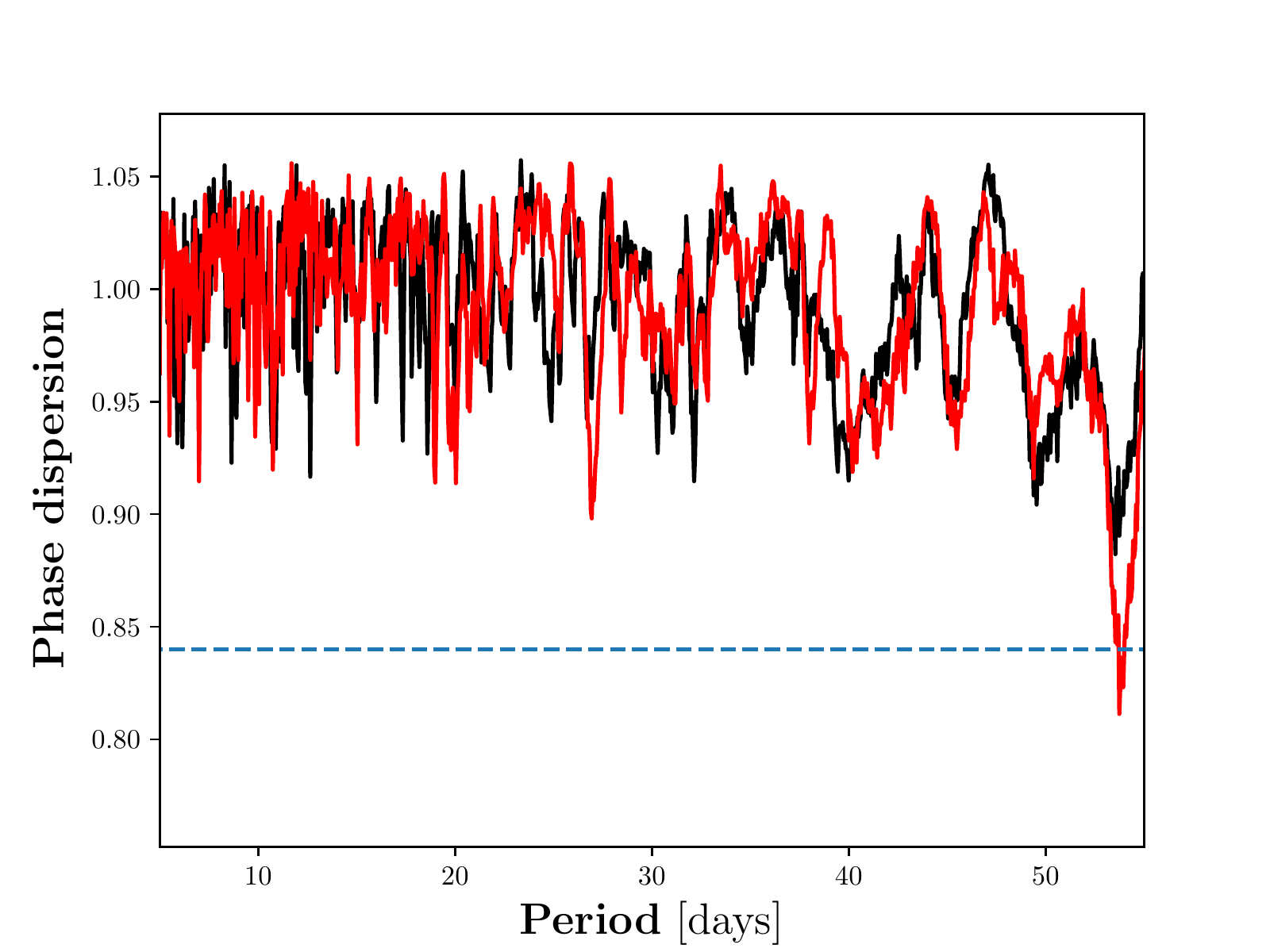}%\vspace{-0.4mm}\\
\includegraphics[width=0.45\textwidth, clip]{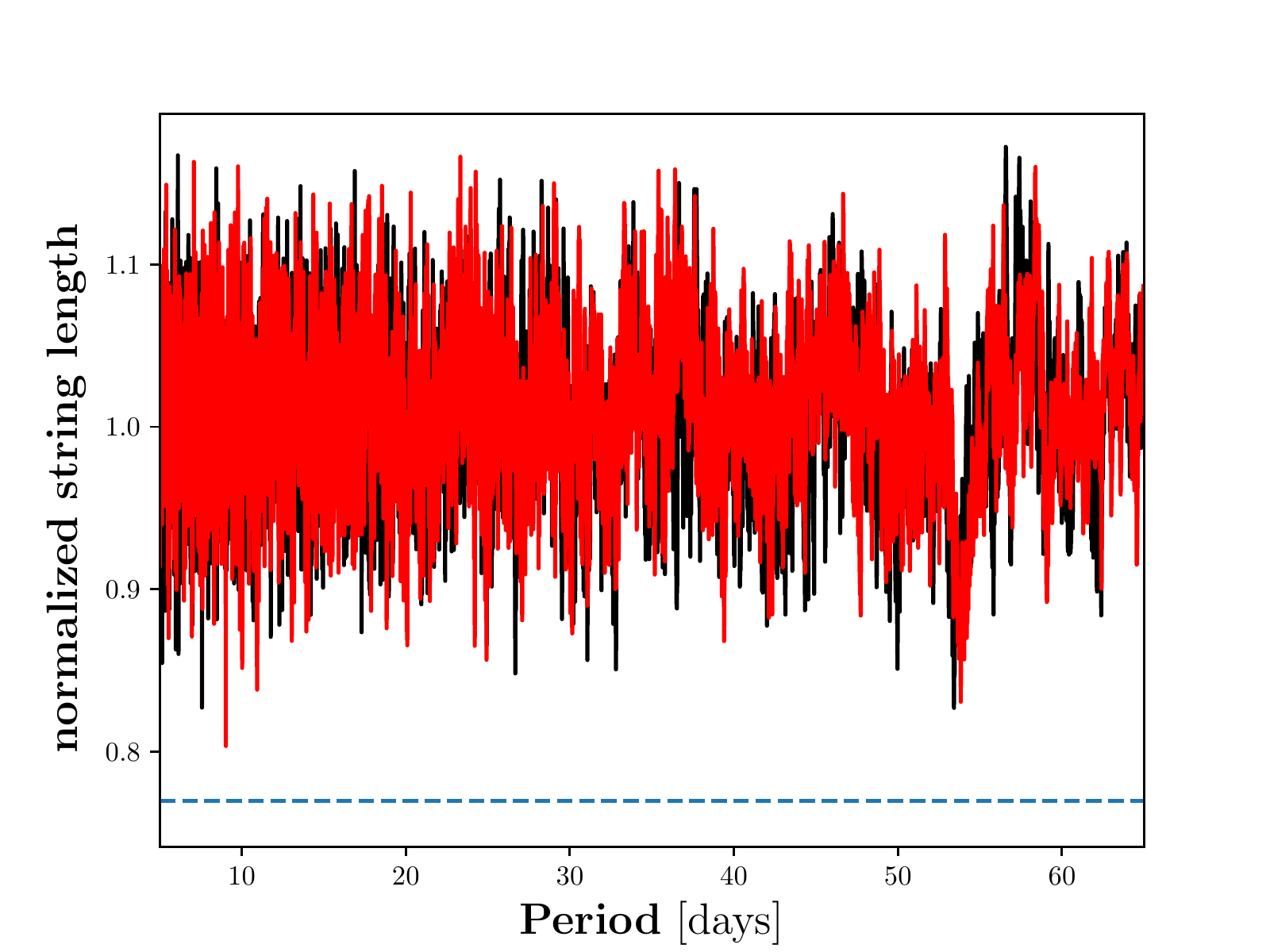}\\
\includegraphics[width=0.45\textwidth, clip]{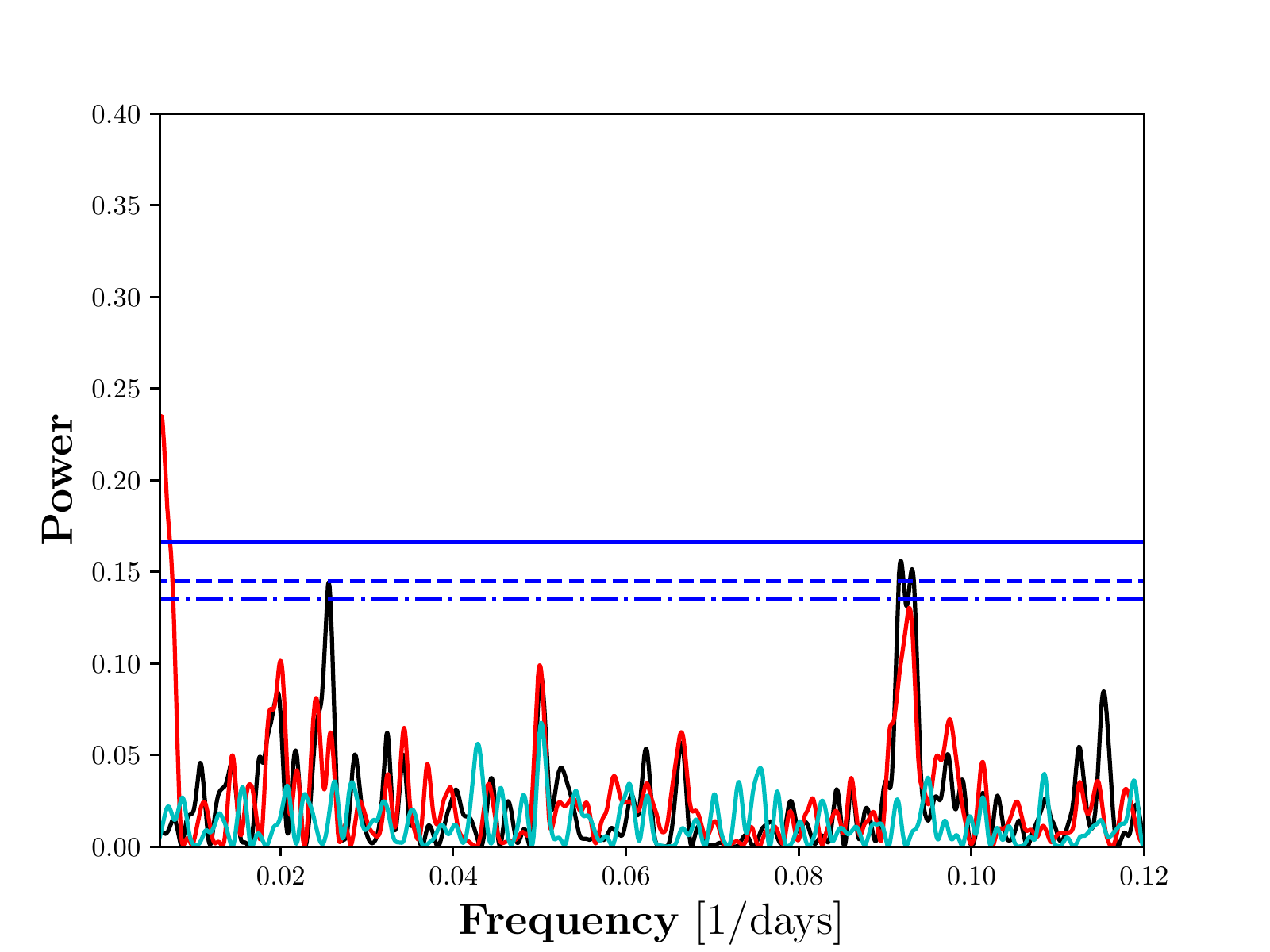}%\vspace{-0.4mm}\\
\includegraphics[width=0.45\textwidth, clip]{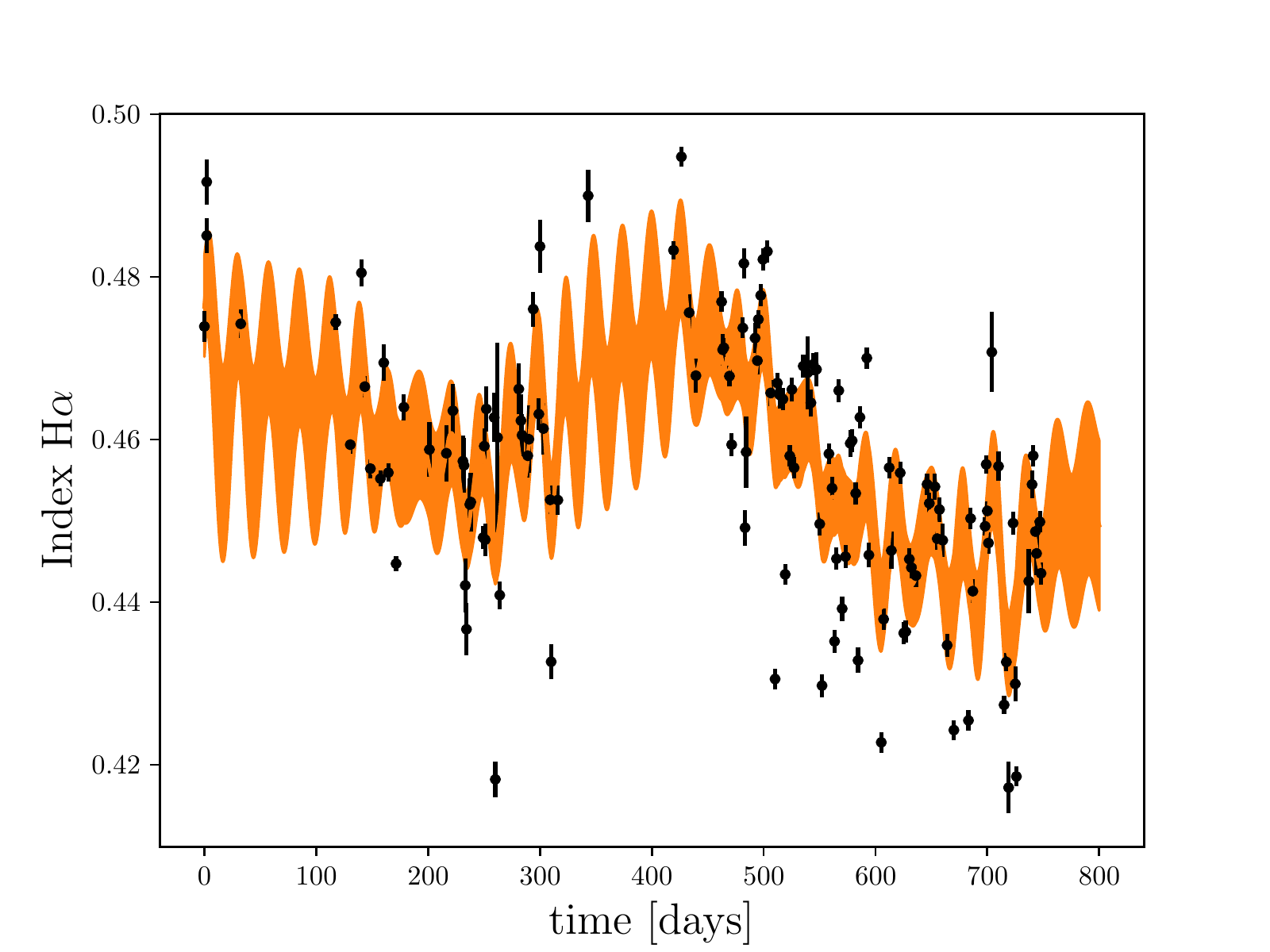}%\vspace{-0.4mm}\\
\caption{\label{hd199305} Results for HD 199305 as explained in Fig. \ref{gxand}.}
\end{figure*}

\begin{figure*}
\includegraphics[width=0.45\textwidth, clip]{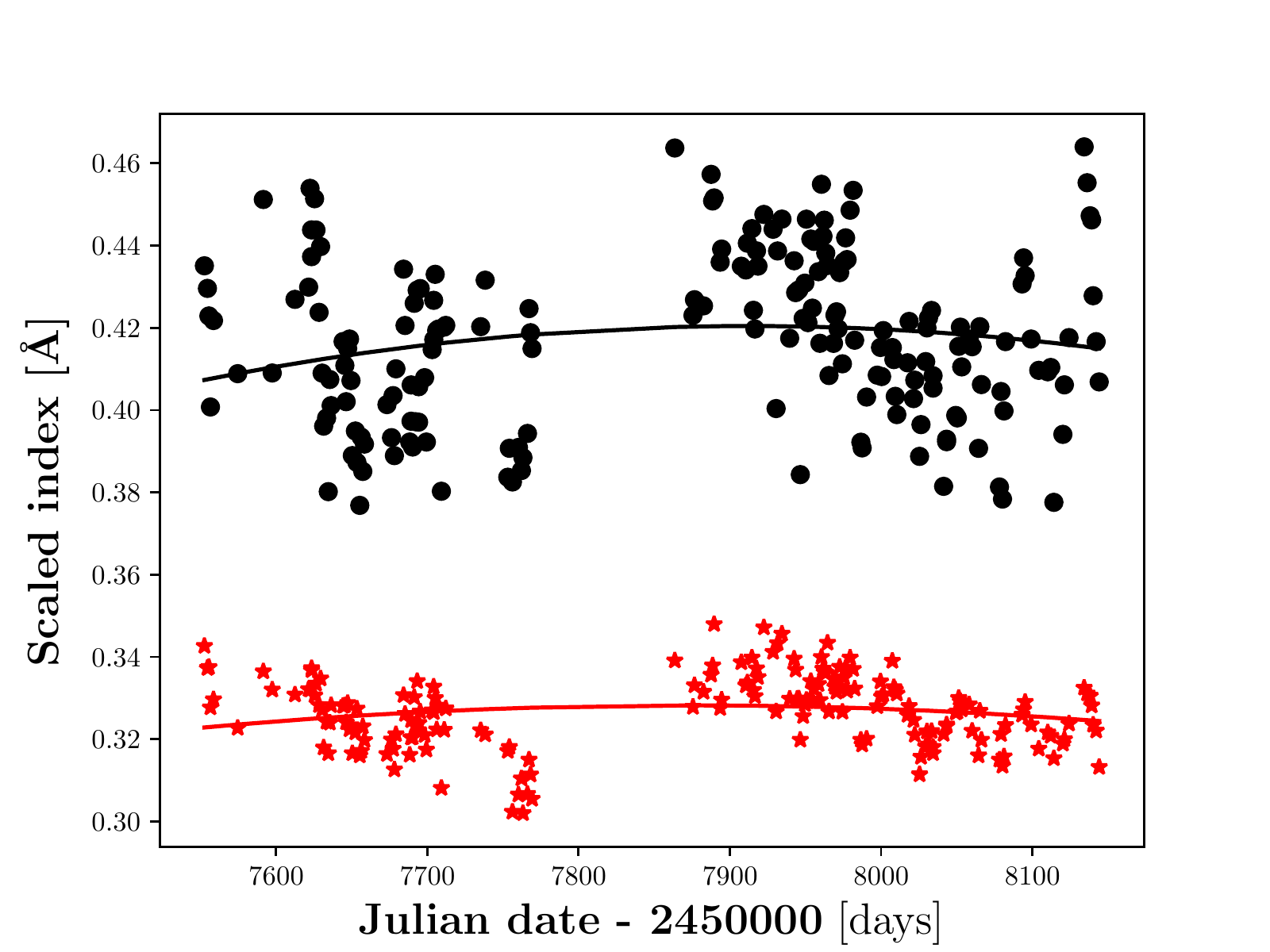}\\%\vspace{-0.4mm}\\
\includegraphics[width=0.45\textwidth, clip]{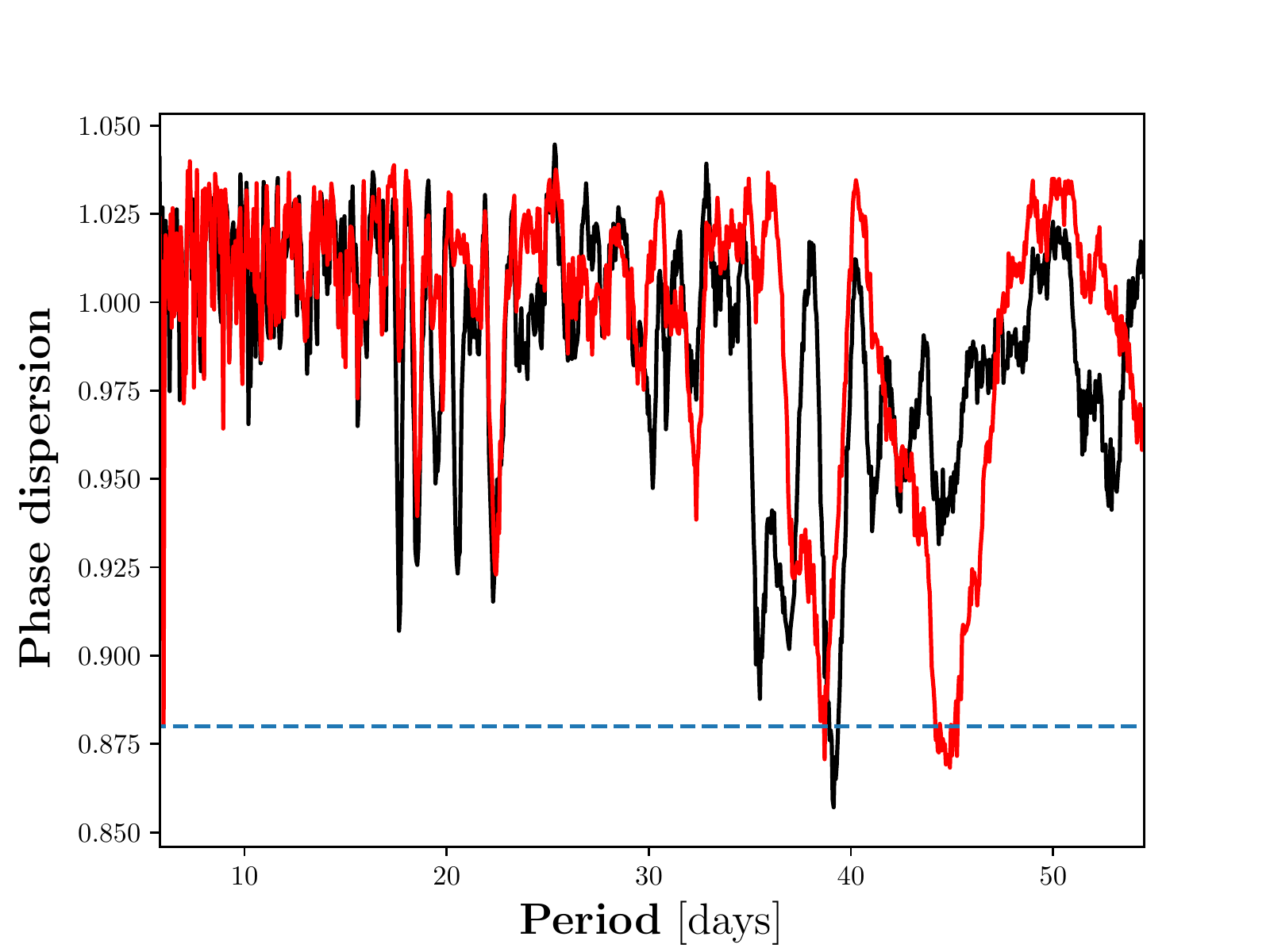}%\vspace{-0.4mm}\\
\includegraphics[width=0.45\textwidth, clip]{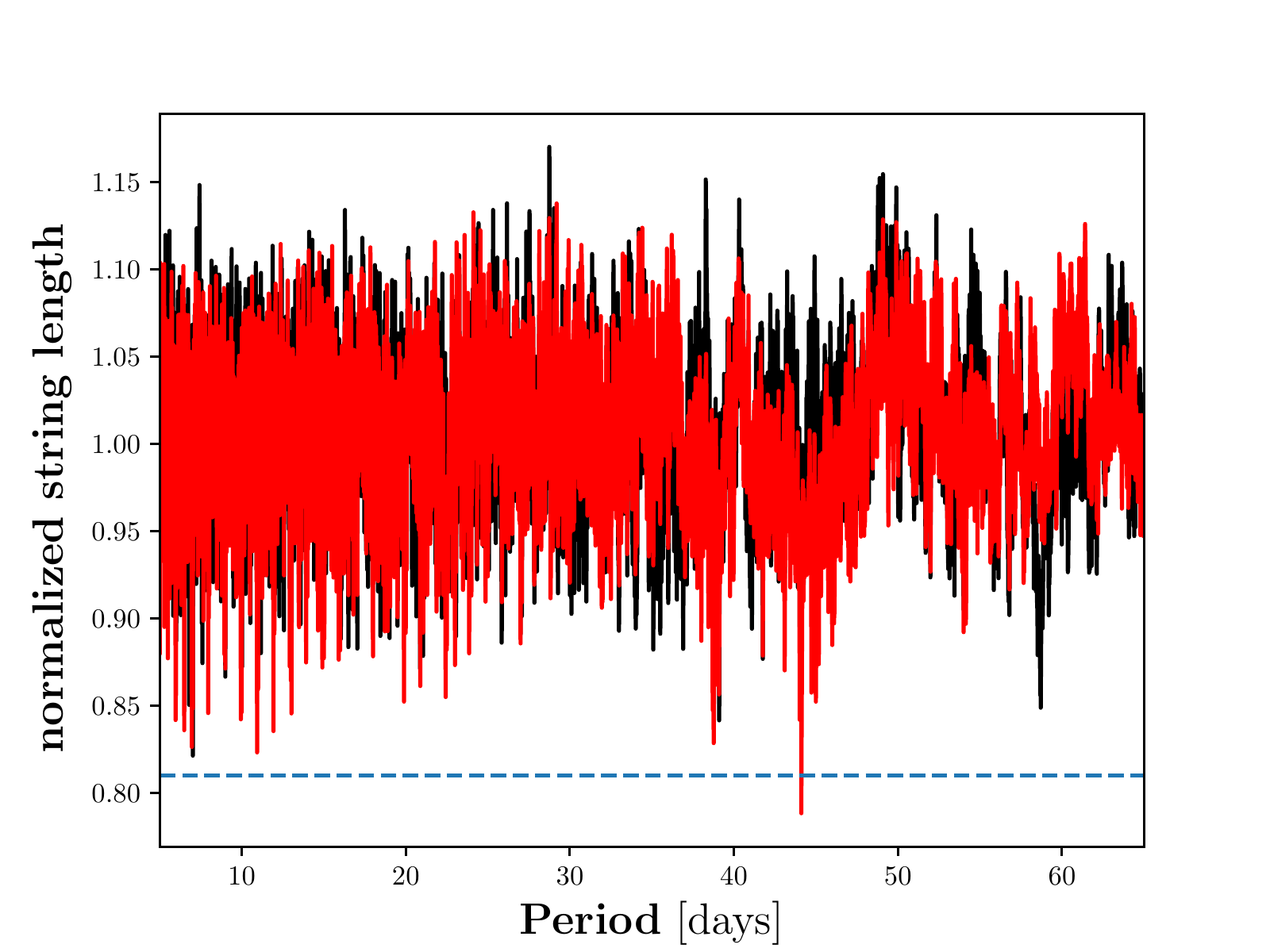}\\
\includegraphics[width=0.45\textwidth, clip]{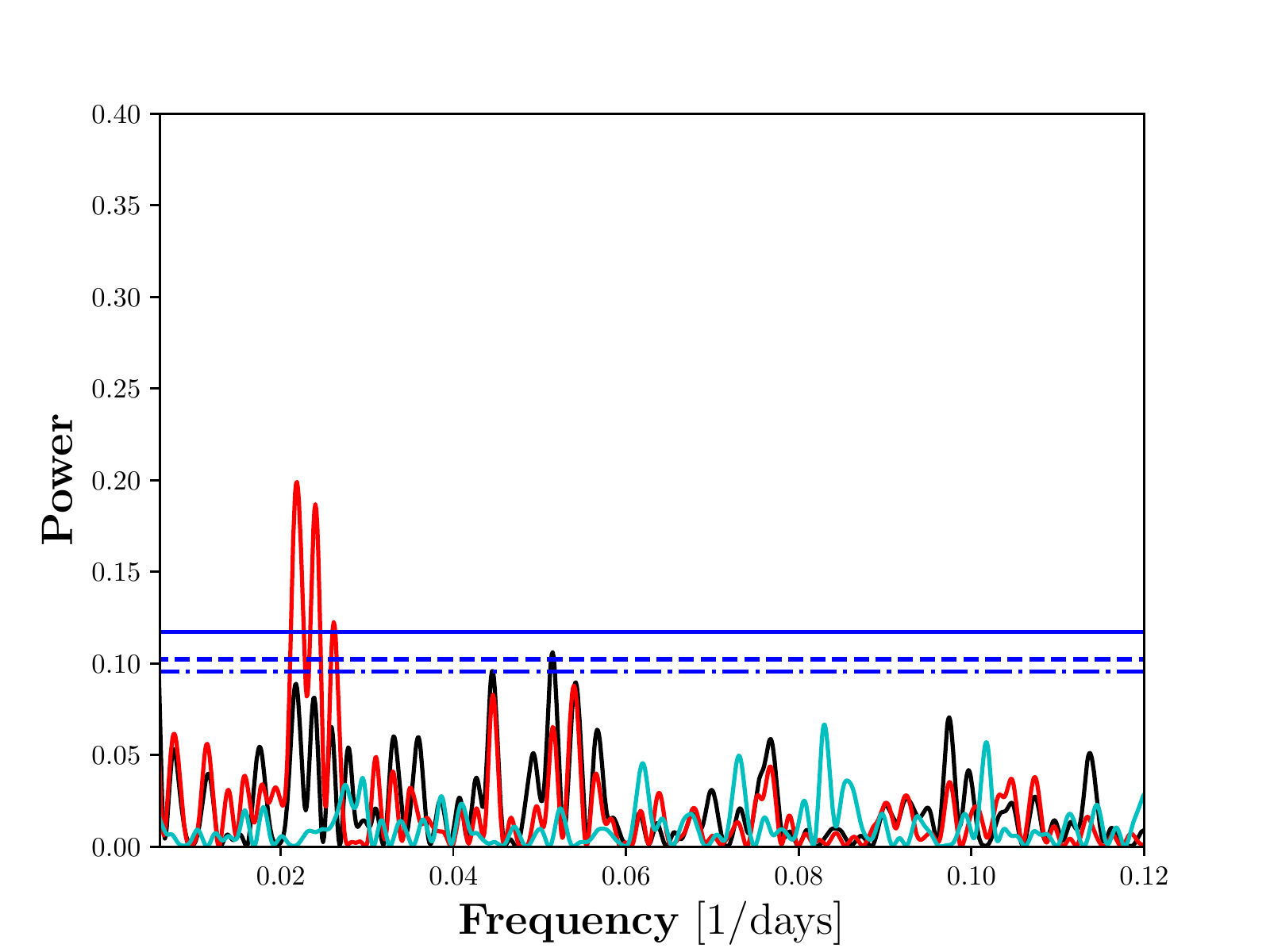}%\vspace{-0.4mm}\\
\includegraphics[width=0.45\textwidth, clip]{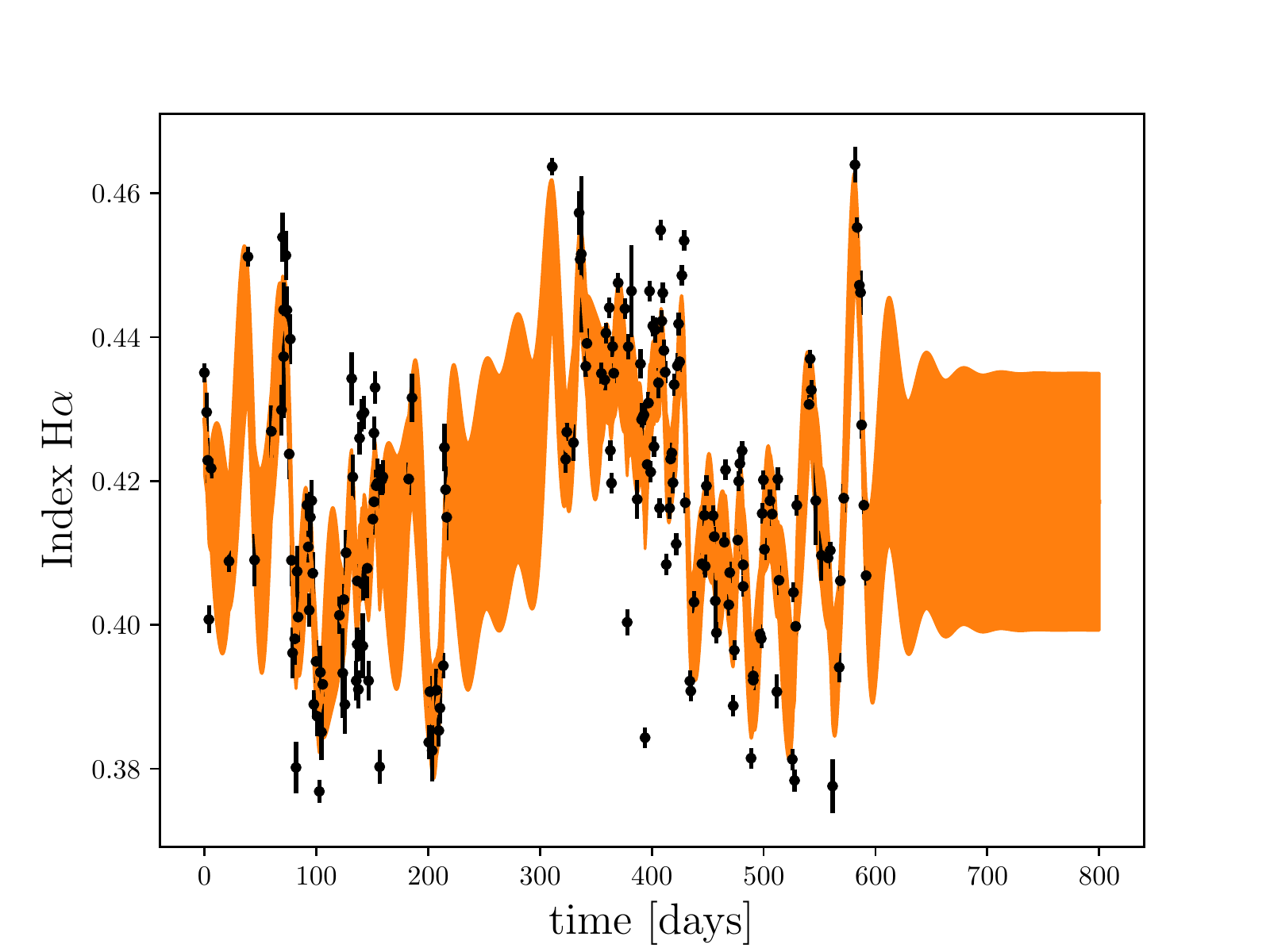}%\vspace{-0.4mm}\\
\caption{\label{hd216899} Results for HD 216899 as explained in Fig. \ref{gxand}.}
\end{figure*}

\end{appendix}
\end{document}